\documentclass[longauth]{aa}  
\usepackage{color}
\usepackage{longtable,arydshln}
\usepackage{supertabular,multicol}
\usepackage{lscape}
\usepackage[varg]{txfonts}
\usepackage{graphicx,amssymb, url}
\usepackage{natbib,twoopt}
\makeatother

\begin{document}

\font\fiverm=cmr5       \font\fivebf=cmbx5      \font\sevenrm=cmr7     
\def\lof{LOFAR}

\title {Wide-band, low-frequency pulse profiles of \\ 
100 radio pulsars with LOFAR}

\author{
M.~Pilia\inst{1,2}\and 
J.~W.~T.~Hessels\inst{1,3}\and 
B.~W.~Stappers\inst{4}\and 
V.~I.~Kondratiev\inst{1,5}\and 
M.~Kramer\inst{6,4}\and 
J.~van~Leeuwen\inst{1,3}\and 
P.~Weltevrede\inst{4}\and
A.~G.~Lyne\inst{4}\and 
K.~Zagkouris\inst{7}\and 
T.~E.~Hassall\inst{8}\and 
A.~V.~Bilous\inst{9}\and 
R.~P.~Breton\inst{8}\and 
H.~Falcke\inst{9,1}\and 
J.-M.~Grie\ss{}meier\inst{10,11}\and 
E.~Keane\inst{12,13}\and 
A.~Karastergiou\inst{7}\and 
M.~Kuniyoshi\inst{14}\and 
A.~Noutsos\inst{6}\and 
S.~Os{\l}owski\inst{15,6}\and 
M.~Serylak\inst{16}\and 
C.~Sobey\inst{1}\and 
S.~ter~Veen\inst{9}\and  
A.~Alexov\inst{17}\and 
J.~Anderson\inst{18}\and 
A.~Asgekar\inst{1,19}\and 
I.~M.~Avruch\inst{20,21}\and 
M.~E.~Bell\inst{22}\and 
M.~J.~Bentum\inst{1,23}\and 
G.~Bernardi\inst{24}\and 
L.~B\^{i}rzan\inst{25}\and 
A.~Bonafede\inst{26}\and 
F.~Breitling\inst{27}\and 
J.~W.~Broderick\inst{7,8}\and 
M.~Br\"uggen\inst{26}\and 
B.~Ciardi\inst{28}\and 
S.~Corbel\inst{29,11}\and 
E.~de Geus\inst{1,30}\and 
A.~de Jong\inst{1}\and 
A.~Deller\inst{1}\and 
S.~Duscha\inst{1}\and 
J.~Eisl\"offel\inst{31}\and 
R.~A.~Fallows\inst{1}\and 
R.~Fender\inst{7}\and 
C.~Ferrari\inst{32}\and 
W.~Frieswijk\inst{1}\and 
M.~A.~Garrett\inst{1,25}\and 
A.~W.~Gunst\inst{1}\and 
J.~P.~Hamaker\inst{1}\and 
G.~Heald\inst{1}\and 
A.~Horneffer\inst{6}\and 
P.~Jonker\inst{20}\and 
E.~Juette\inst{33}\and 
G.~Kuper\inst{1}\and 
P.~Maat\inst{1}\and 
G.~Mann\inst{27}\and 
S.~Markoff\inst{3}\and 
R. McFadden\inst{1}\and 
D.~McKay-Bukowski\inst{34,35}\and 
J.~C.~A.~Miller-Jones\inst{36}\and 
A.~Nelles\inst{9}\and 
H.~Paas\inst{37}\and 
M.~Pandey-Pommier\inst{38}\and 
M.~Pietka\inst{7}\and 
R.~Pizzo\inst{1}\and 
A.~G.~Polatidis\inst{1}\and 
W.~Reich\inst{6}\and 
H.~R\"ottgering\inst{25}\and 
A.~ Rowlinson\inst{22}\and 
D.~Schwarz\inst{15}\and 
O.~Smirnov\inst{39,40}\and 
M.~Steinmetz\inst{27}\and 
A.~Stewart\inst{7}\and 
J.~D.~Swinbank\inst{41}\and 
M.~Tagger\inst{10}\and 
Y.~Tang\inst{1}\and 
C.~Tasse\inst{42}\and 
S.~Thoudam\inst{9}\and 
M.~C.~Toribio\inst{1}\and 
A.~J.~van~der~Horst\inst{3}\and 
R.~Vermeulen\inst{1}\and 
C. Vocks\inst{27}\and 
R.~J.~van Weeren\inst{24}\and 
R.~A.~M.~J.~Wijers\inst{3}\and 
R.~Wijnands\inst{3}\and 
S.~J.~Wijnholds\inst{1}\and 
O.~Wucknitz\inst{6}\and 
P.~Zarka\inst{42}
}

\institute{
ASTRON, The Netherlands Institute for Radio Astronomy, Postbus 2, 7990 AA Dwingeloo, The Netherlands \and
INAF - Osservatorio Astronomico di Cagliari, via della Scienza 5, I-09047 Selargius (CA), Italy   \and
Anton Pannekoek Institute, University of Amsterdam, Postbus 94249, 1090 GE Amsterdam, The Netherlands \and
Jodrell Bank Center for Astrophysics, School of Physics and Astronomy, The University of Manchester, Manchester M13 9PL,UK \and
Astro Space Center of the Lebedev Physical Institute, Profsoyuznaya str. 84/32, Moscow 117997, Russia \and
Max-Planck-Institut f\"{u}r Radioastronomie, Auf dem H\"ugel 69, 53121 Bonn, Germany \and
Astrophysics, University of Oxford, Denys Wilkinson Building, Keble Road, Oxford OX1 3RH \and
School of Physics and Astronomy, University of Southampton, Southampton, SO17 1BJ, UK \and
Department of Astrophysics/IMAPP, Radboud University Nijmegen, P.O. Box 9010, 6500 GL Nijmegen, The Netherlands \and
LPC2E - Universite d'Orleans/CNRS \and
Station de Radioastronomie de Nancay, Observatoire de Paris - CNRS/INSU, USR 704 - Univ. Orleans, OSUC , route de Souesmes, 18330 Nancay, France \and
Centre for Astrophysics and Supercomputing, Swinburne University of Technology, Mail H29, PO Box 218, VIC 3122, Australia \and
ARC Centre of Excellence for All-sky astrophysics (CAASTRO), Sydney Institute of Astronomy, University of Sydney Australia \and
NAOJ Chile Observatory, National Astronomical Observatory of Japan, 2-21-1 Osawa, Mitaka, Tokyo 181-8588, Japan      \and
Fakult\"{a}t f\"{u}r Physik, Universit\"{a}t Bielefeld, Postfach 100131, D-33501, Bielefeld, Germany \and
Department of Physics \& Astronomy, University of the Western Cape, Private Bag X17, Bellville 7535, South Africa \and
Space Telescope Science Institute, 3700 San Martin Drive, Baltimore, MD 21218, USA \and
Helmholtz-Zentrum Potsdam, DeutschesGeoForschungsZentrum GFZ, Department 1: Geodesy and Remote Sensing, Telegrafenberg, A17, 14473 Potsdam, Germany \and
Shell Technology Center, Bangalore, India \and
SRON Netherlands Insitute for Space Research, PO Box 800, 9700 AV Groningen, The Netherlands \and
Kapteyn Astronomical Institute, PO Box 800, 9700 AV Groningen, The Netherlands \and
CSIRO Australia Telescope National Facility, PO Box 76, Epping NSW 1710, Australia \and
University of Twente, The Netherlands \and
Harvard-Smithsonian Center for Astrophysics, 60 Garden Street, Cambridge, MA 02138, USA \and
Leiden Observatory, Leiden University, PO Box 9513, 2300 RA Leiden, The Netherlands \and
University of Hamburg, Gojenbergsweg 112, 21029 Hamburg, Germany \and
Leibniz-Institut f\"{u}r Astrophysik Potsdam (AIP), An der Sternwarte 16, 14482 Potsdam, Germany \and
Max Planck Institute for Astrophysics, Karl Schwarzschild Str. 1, 85741 Garching, Germany \and
Laboratoire AIM (CEA/IRFU - CNRS/INSU - Universit\'e Paris Diderot), CEA DSM/IRFU/SAp, F-91191 Gif-sur-Yvette, France \and
SmarterVision BV, Oostersingel 5, 9401 JX Assen \and
Th\"{u}ringer Landessternwarte, Sternwarte 5, D-07778 Tautenburg, Germany \and
Laboratoire Lagrange, UMR7293, Universit\`{e} de Nice Sophia-Antipolis, CNRS, Observatoire de la C\'{o}te d'Azur, 06300 Nice, France \and
Astronomisches Institut der Ruhr-Universit\"{a}t Bochum, Universitaetsstrasse 150, 44780 Bochum, Germany \and
Sodankyl\"{a} Geophysical Observatory, University of Oulu, T\"{a}htel\"{a}ntie 62, 99600 Sodankyl\"{a}, Finland \and
STFC Rutherford Appleton Laboratory,  Harwell Science and Innovation Campus,  Didcot  OX11 0QX, UK \and
International Centre for Radio Astronomy Research - Curtin University, GPO Box U1987, Perth, WA 6845, Australia \and
Center for Information Technology (CIT), University of Groningen, The Netherlands \and
Centre de Recherche Astrophysique de Lyon, Observatoire de Lyon, 9 av Charles Andr\'{e}, 69561 Saint Genis Laval Cedex, France \and
Department of Physics and Elelctronics, Rhodes University, PO Box 94, Grahamstown 6140, South Africa \and
SKA South Africa, 3rd Floor, The Park, Park Road, Pinelands, 7405, South Africa \and
Department of Astrophysical Sciences, Princeton University, Princeton, NJ 08544, USA \and
LESIA, Observatoire de Paris, CNRS, UPMC, Universit\`{e} Paris-Diderot, 5 place Jules Janssen, 92195 Meudon, France 
}

\date{Received ...; accepted ...}

\abstract 
{LOFAR offers the unique capability of observing pulsars across the
  $10-240$\,MHz frequency range with a fractional bandwidth of roughly 50\%.
  This spectral range is well suited for studying the frequency
  evolution of pulse profile morphology caused by both intrinsic and
  extrinsic effects such as changing emission altitude in the pulsar magnetosphere or scatter
  broadening by the interstellar medium, respectively.}
{The magnitude of most of these effects increases rapidly towards low
  frequencies.  LOFAR can thus address a number of open questions
  about the nature of radio pulsar emission and its propagation
  through the interstellar medium. }
{We present the average pulse profiles of 100 pulsars observed in the
  two LOFAR frequency bands: high band (120--167\,MHz, 100 profiles)
  and low band (15--62\,MHz, 26 profiles). We compare them with
  Westerbork Synthesis Radio Telescope (WSRT) and Lovell Telescope observations at higher frequencies (350 and
  1400\,MHz) to study the profile evolution.  The profiles
  were aligned in absolute phase by folding with a new set of timing
  solutions from the Lovell Telescope, which we present along with precise
  dispersion measures obtained with LOFAR.}
{We find that the profile evolution with decreasing radio frequency
  does not follow a specific trend; depending on the geometry of
  the pulsar, new components can enter into or be hidden from view.
  Nonetheless, in general our observations confirm the widening of
  pulsar profiles at low frequencies, as expected from
  radius-to-frequency mapping or birefringence theories.
We offer this
  catalogue of low-frequency pulsar profiles in a user friendly way via
  the EPN Database of Pulsar Profiles\thanks{http://www.epta.eu.org/epndb/}.}
{}

\titlerunning{LOFAR 100 Pulsar Profiles}
\authorrunning{Pilia et al.}
\maketitle

\section{Introduction}

\begin{figure}
\centering
\includegraphics[width=0.4\textwidth]{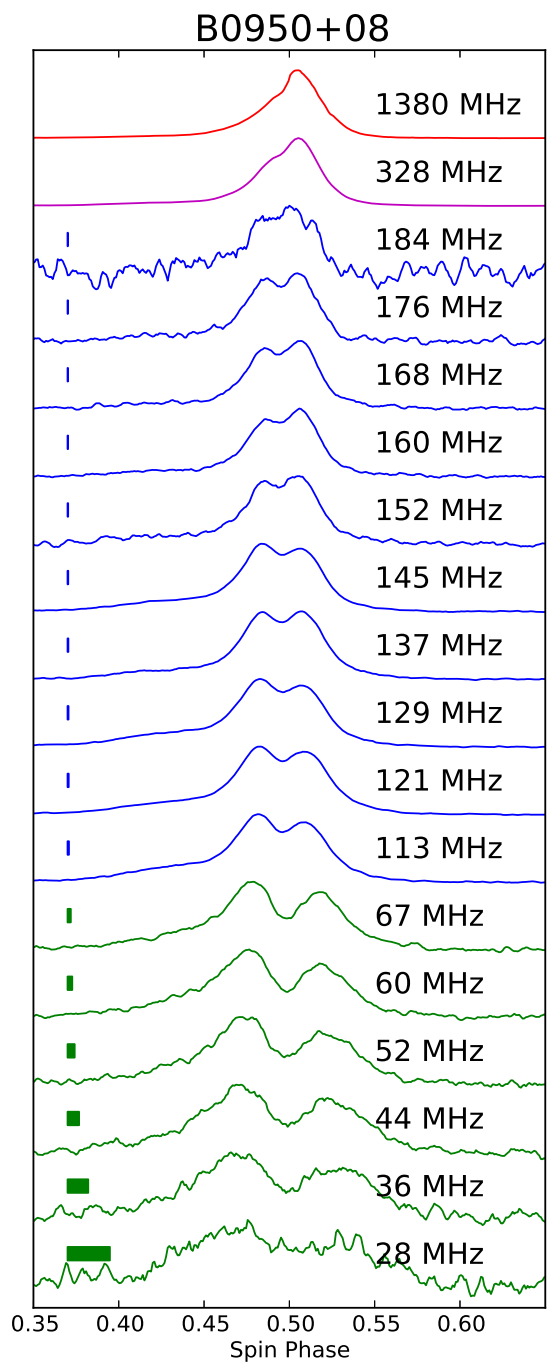}
\caption{\label{fig:prof_evol} Example of pulsar profile evolution for PSR~B0950+08, from 1400 MHz down to 30 MHz. It becomes more rapid at low frequencies. The bars on the left represent the
  intra-channel smearing due to uncorrected DM delay within a channel at each frequency. 
The profiles were aligned using a timing ephemeris (see text for details).}
\end{figure}

The cumulative (i.e. average) pulse profiles of radio pulsars are the
sum of hundreds to thousands of individual pulses, and are, loosely speaking, a
unique signature of each pulsar \citep{lorimer08}.  They are normally
stable in their morphology and are reproducible, although several types
of variation have been observed both for non-recycled pulsars (see
\citealt{helfand75}, \citealt{weisberg89}, \citealt{rathnasree95} and \citealt{lyne10}) and for millisecond pulsars \citep{liu12}.
For most pulsars, this cumulative pulse profile morphology often
varies (sometimes drastically, sometimes subtly) as a function of observing frequency because of a number
of intrinsic effects (e.g. emission location in the pulsar magnetosphere) and
extrinsic effects (i.e. due to propagation in the interstellar medium; ISM), see for instance \cite{cordes78}.  As Fig. \ref{fig:prof_evol} shows, pulse
profile evolution can become increasingly evident at the lowest observing
frequencies ($< 200$\,MHz).

Mapping profile evolution over a wide range of frequencies can aid in
modelling the pulsar radio emission mechanism itself (see
e.g. \citealt{rankin93} and related papers of the series) and
constraining properties of the ISM (e.g. \citealt{hassall12}).  Since
many of the processes that affect the pulse shape strongly
depend on the observing frequency,  observations at low frequencies provide valuable insights on them.  The LOw-Frequency ARray (LOFAR) is the first
telescope capable of observing nearly the entire radio spectrum in the
$10-240$\,MHz frequency range, the lowest 4.5 octaves of the `radio
window' \citep{vanhaarlem13} and \citep{stappers11}.

Low-frequency pulsar observations have previously been conducted by a number of
  telescopes, (e.g. 
Gauribidanur Radio Telescope (GEETEE): \citealt{asgekar99};  Large Phased Array Radio Telescope, Puschino:  \cite{kuzmin98}, \citealt{malov10}; Arecibo: \citealt{hanran10}; Ukrainian T-shaped Radio telescope, second modification (UTR-2): \citealt{zak13}),
and simultaneous efforts are being undertaken by other groups in parallel (e.g. Long Wavelength Array (LWA): \citealt{stovall15}, Murchison Widefield Array (MWA): \citealt{tremblay15}).
Nevertheless, LOFAR offers several advantages over the previous studies. 
Firstly, the large bandwidth
that can be recorded at any given time ($ 48$\,MHz in 16-bit mode and
$ 96$\,MHz in 8-bit mode) allows for continuous wide-band studies of
the pulse profile evolution, compared to studies using a number of
widely separated, narrow bands (e.g. the $5\times 32$ and $20 \times
32$\,kHz bands used at 102 MHz in \citealt{kuzmin98}).  Secondly,
LOFAR's ability to track sources is also an advantage, as many pulses
can be collected in a single observing session instead of having to
combine several short observations in the case of transit instruments
(e.g. \citealt{izvekova89}).  This eliminates systematic errors in
the profile that are due to imprecise alignment of the data from several
observing sessions.  Thirdly, LOFAR can achieve greater sensitivity by
coherently adding the signals received by individual stations,
giving a collecting area equivalent to
the sum of the collecting area of all stations (up to $57000$\,m$^2$
at HBAs and $75200$\,m$^2$ at LBAs, see \citealt{stappers11}).
Finally, LOFAR offers excellent frequency and time resolution. This is
necessary for dedispersing the data to resolve narrow features in the
profile.  LOFAR is also capable of coherently dedispersing the data,
although that mode was not employed here.

As mentioned above, there are two types of effects that LOFAR will
allow us to study with great precision. One of these are extrinsic
effects related to the ISM. Specifically, the ISM causes scattering
and dispersion.  Mean scatter-broadening (assuming a Kolmogorov
distribution of the turbulence in the ISM) scales with observing frequency as
$\nu^{-4.4}$. The scattering causes delays in the arrival time of
the emission at Earth, which can be modelled as having an exponentially
decreasing probability of being scattered back into our line-of-sight: this means that the intensity of the pulse is spread in an exponentially decreasing tail.
Dispersion scales as $\nu^{-2}$ and is mostly corrected for by
channelising and dedispersing the data (see e.g.
\citealt{lorimer04}).  Nonetheless, for filterbank (channelised) data
some residual dispersive smearing persists within each channel:
\begin{equation}
t_{DM} = {8.3 \cdot  {\rm{DM}}{{\Delta\nu} \over {\nu^{3}}}} ~\rm{\mu s},
\end{equation}
where DM is the dispersion measure in cm$^{-3}$ \,pc, $\Delta\nu$ is the
channel width in MHz, and $\nu$ the central observing frequency in
GHz. This does not significantly affect the profiles that we present
here (at least not at frequencies above $\sim 50$\,MHz) because the pulsars studied have low DMs and the data are
chanellised in narrow frequency channels (see Sect. \ref{sec:obs}).
Second-order effects in the ISM may also be present, but have yet to
be confirmed.  For instance, previous claims of
`super-dispersion', meaning a deviation from the $\nu^{-2}$ scaling law
(see e.g. \citealt{kuzmin08} and references therein), were not observed by
\cite{hassall12}, with an upper limit of $\lesssim 50$\,ns at a
reference frequency of 1400\,MHz.

The second type of effects under investigation are those intrinsic
to the pulsar. One of the most well-known intrinsic effects are pulse
broadening at low frequencies, which has been observed in many
pulsars (e.g. \citealt{hankins86} and \citealt{mrVII_02}), while others
show no evidence of this (e.g. Hassall et al. 2012).  One of the theories
explaining this effect is radius-to-frequency mapping (RFM,
\citealt{cordes78}): it postulates that the origin of the radio emission
in the pulsar's magnetosphere increases in altitude above the magnetic
poles towards lower frequencies.  RFM predicts that the pulse profile
 will increase in width towards lower observing frequency, since emission will
be directed tangentially to the diverging magnetic field lines of the magnetosphere that corotates with the pulsar.
An alternative interpretation \citep{mckinnon97} proposes
birefringence of the plasma above the polar caps as the cause for
broadening: the two propagation modes split at low frequencies,
causing the broadening, while they stay closer together at high frequencies, causing
depolarisation (this is investigated in the \lof\ work on pulsar polarisation, see \citealt{noutsos15}).

In this paper, we present the average pulse profiles of 100 pulsars
observed in two LOFAR frequency ranges: high band (119--167 MHz, 100
profiles) and low band (15--62 MHz, 26 out of the 100 profiles).  We
compare the pulse profile morphologies with those obtained around 350 and
1400\,MHz with the WSRT and Lovell telescopes, respectively, to study their evolution with respect to a magnetospheric origin and DM-induced variations.  We do not discuss here profile
evolution due to the effects of scattering in the ISM, which will be
the target of a future work (Zagkouris et al. in prep.).  In Sect. 2
we describe the LOFAR observational setup and parameters.  In Sect. 3 we
describe the analysis.  In Sect. 4 we discuss the results, and in Sect. 5 we
conclude with some discussion on future extensions of this work.

\section{Observations}
\label{sec:obs}
The observed sample of pulsars was loosely based on a selection of the brightest
objects in the LOFAR-visible sky (declination $ > -30 ^{\circ}$), using the ATNF
Pulsar
Catalog\footnote{http://www.atnf.csiro.au/people/pulsar/psrcat/}
\citep{manchester05} for guidance. Because pulsar flux and
spectral indices are typically measured at higher frequencies, we also based our
selection on the previously published data at low frequencies
\citep{malov10}.  Since the LOFAR dipoles have a sensitivity that
decreases as a function of the zenith angle, all sources were observed
as close to transit as possible, and only the very brightest sources
south of the celestial equator were observed.

We observed 100 pulsars using the high-band antennas (HBAs) in
the six central `Superterp' stations (CS002$-$CS007) of the \lof\ core\footnote{
The full LOFAR Core can now be used for
  observations and provides four times the number of stations
  available on the Superterp (and a proportional increase in
  sensitivity).}.
The observations were performed in tied-array mode, that is, a coherent sum of
the station signals using appropriate geometrical and instrumental
phase and time delays (see \citealt{stappers11} for a detailed
description of LOFAR's pulsar observing modes and
\citealt{vanhaarlem13} for a general description of LOFAR).  The
$119-167$-MHz frequency range was observed using 240 subbands of
195\,kHz each, synthesised at the station level, where the individual
HBA tiles were combined to form station beams.  Using the LOFAR Blue
Gene/P correlator, each subband was further channelised into
16 channels, formed into a tied-array beam. The linear polarisations
were summed in quadrature (pseudo-Stokes I), and the signal intensity was written out as
245.76\,$\mu$s samples.  The integration time of each observation
was at least 1020\,s.  This was chosen to provide an adequate
number of individual pulses, so as to average out the absolute scale
of the variance associated with pulse-phase `jitter' to the
cumulative profile. The jitter, also termed stochastic wide-band
impulse modulated self-noise (SWIMS, as in \citealt{oslowski11}), is
the variation in individual pulse intensity and position with respect
to the average pulse profile (see also \citealt{cordes93} and
\citealt{liu12} and references therein).  
This variation does not significantly affect the measurements that have been carried out for the scope of this paper (i.e. pulse widths, peak heights), but we have checked that the resulting
profile was stable on the considered time scales by dividing each observation into shorter sections and
comparing the shapes of the resulting profiles with the overall profile. In the cases where stability was not achieved, we used longer
integration times.  
Regardless, in almost all cases the profile
evolution with observing frequency is a significantly stronger effect at low
frequencies.

Twenty-six of the brightest pulsars were also observed using the
Superterp low-band antennas (LBAs) in the frequency range
$15-62$\,MHz.  To mitigate the larger dispersive smearing of the
profile in this band, 32 channels were synthesised for each of the 240
subbands.  The sampling time was 491.52\,$\mu$s. The integration time
of these observations was increased to at least 2220\,s to
somewhat compensate for the lower sensitivity at
this frequency band (e.g. because of the higher sky temperature).

For some sources, 17-minute HBA observations with the Superterp were
insufficient to achieve acceptable signal-to-noise ($S/N$) profiles.  For these, longer
integration times (or more stations) were needed.  Hence, some of the
pulsars presented here were later observed with 1 hr pointings as part
of the LOFAR Tied-Array All-Sky Survey for pulsars and fast transients
(LOTAAS\footnote{http://www.astron.nl/lotaas/}: see also \citealt{coenen14}), which commenced after the official commissioning
period, during Cycle~0 of LOFAR scientific operations, and is
currently ongoing.  LOTAAS combines multiple tied-array beams (219 total)
per pointing to observe both a survey grid as well as known pulsars
within the primary field-of-view.

\begin{figure*}
\centering
\begin{tabular}{cc}
 \includegraphics[height=70mm]{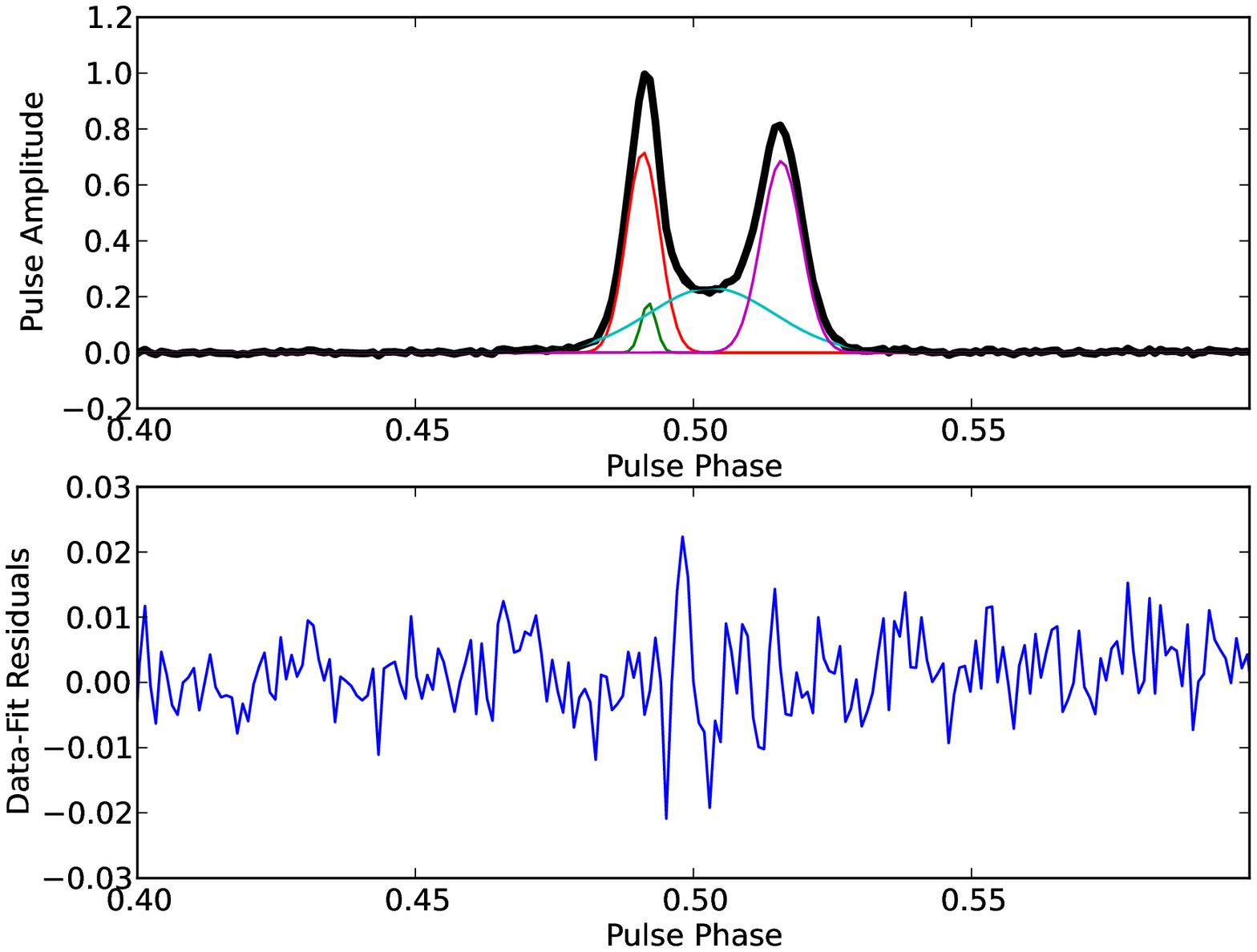} & \includegraphics[height=70mm]{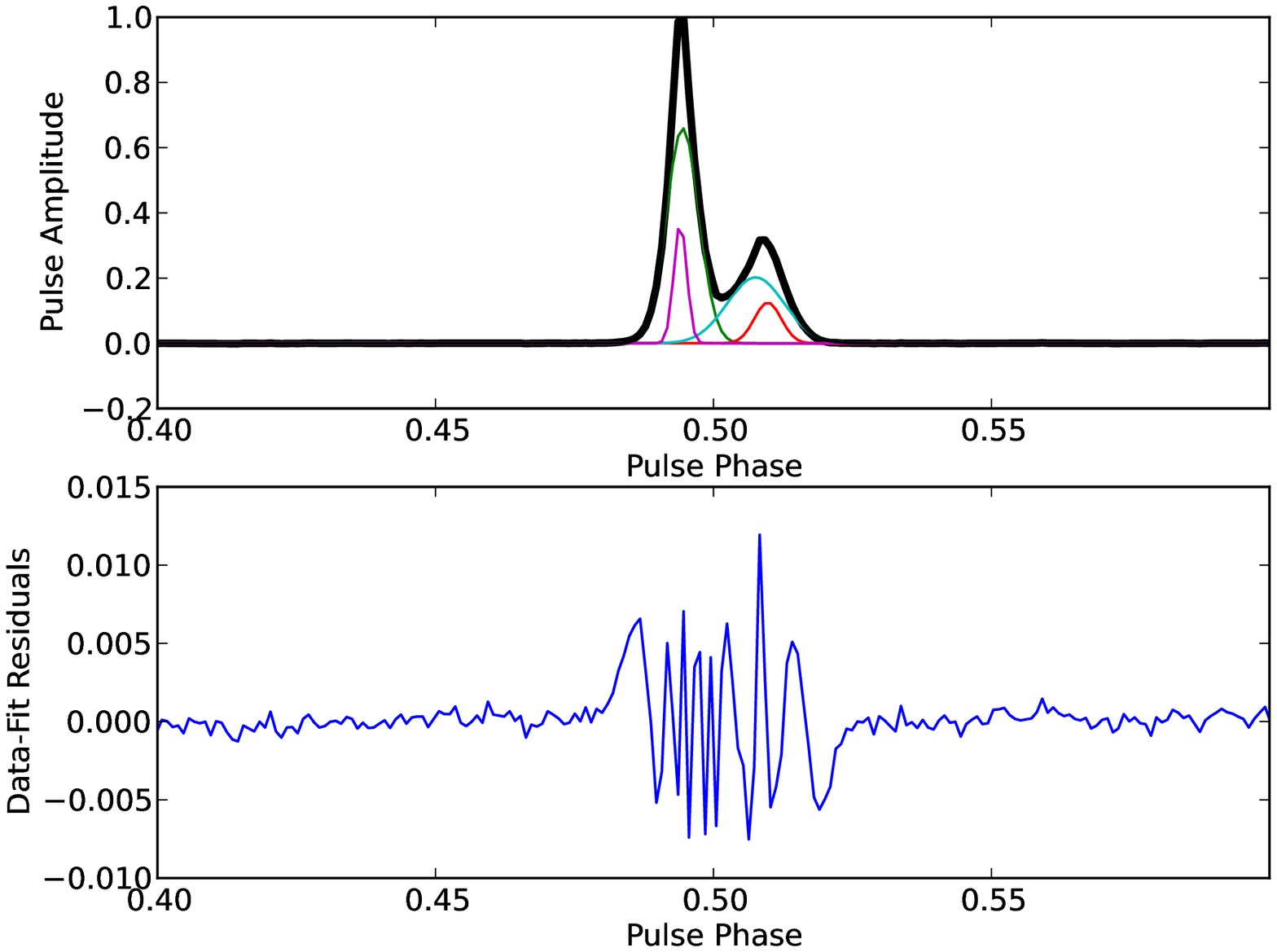}
\\
\includegraphics[height=70mm]{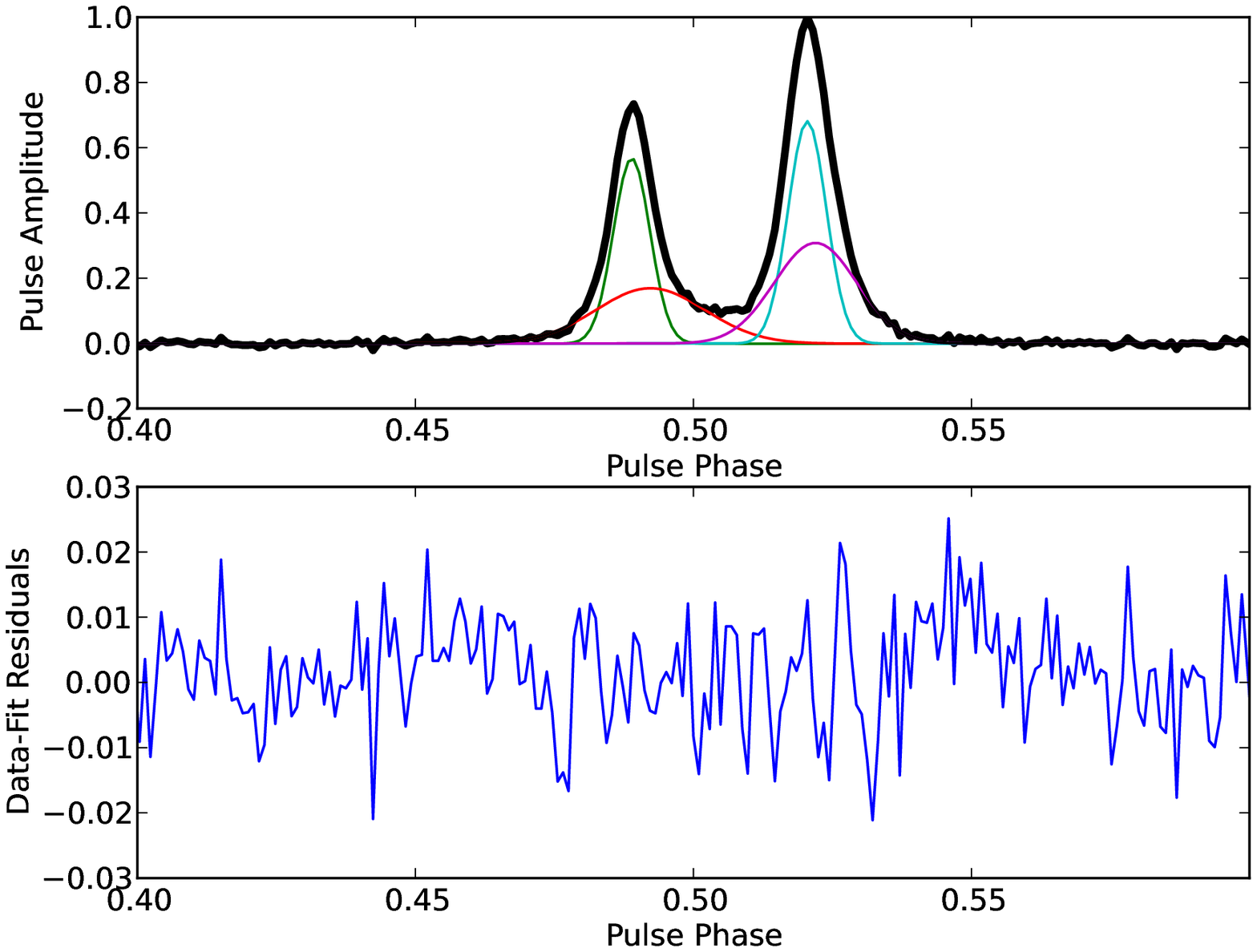} &  \includegraphics[height=70mm
]{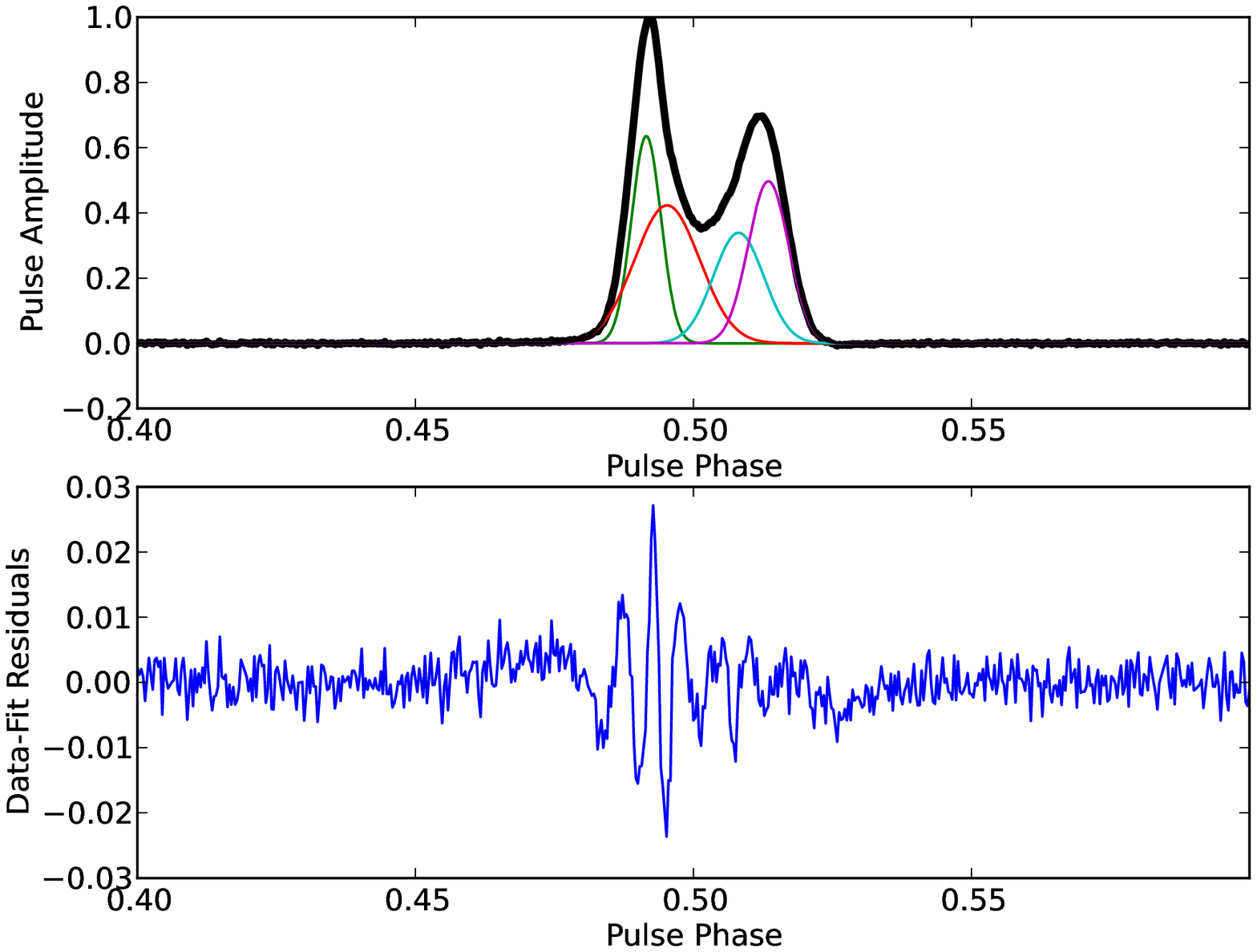}
\end{tabular}

\caption{\label{fig:example} Gaussian modelling and resulting fit residuals for the pulse profile of PSR B1133+16 at four frequencies, between 0.4--0.6 in pulse phase.  Top left: LOFAR HBA.
  Bottom left: LOFAR LBA.  Top right: 1.4\,GHz.  Bottom right:
  350\,MHz.  The Gaussian components contributing to the fit
  are shown in colours, while in black, overlapping the profile
  contours, we show the best fit obtained from them. It is evident that
  this standard `double' profile (see Sect. \ref{sec:models}) is not well fit
  by only two Gaussians. Even adopting a higher number of
  components, the scatter of the residuals is not at the same level as
  the off-pulse residuals (this is the criterion that was adopted for
  the determination of a good fit, following
  \citealt{kramer94a}). Nonetheless, when the residuals on-pulse were discrepant
  at the level of only a few percent from the ones off-pulse, we chose not to add more free
  parameters to the fit. }

\end{figure*}

\section{Analysis}
\label{sec:analysis}

The LOFAR {\tt HDF5}\footnote{http://www.hdfgroup.org/HDF5/}
  (Hierarchical Data Format5, see e.g., \citealt{alexov10}) data were
converted to {\tt PSRFITS} format \citep{hotan04} for further
processing.  Radio frequency interference (RFI) was excised by
removing affected time intervals and frequency channels, using the
tool {\tt rfifind} from the {\tt PRESTO}\footnote{https://github.com/scottransom/presto} software suite
\citep{ransom01}.  The data were dedispersed and folded using {\tt PRESTO}
and, in a first iteration, a rotational ephemeris from the ATNF pulsar
catalogue, using the automatic \lof\ pulsar pipeline `PulP'.  
The number of bins across each profile was chosen so that  each bin corresponds to approximately 
1.5~ms.

The profiles obtained
with the HBA and, where available, LBA bands
were compared with the profiles obtained with the WSRT at $\sim 350$\,MHz (from here onwards
`P-band') and at $\sim 1400$\,MHz, or with the Lovell Telescope at the
Jodrell Bank Observatory at $\sim 1500$\,MHz (from here onwards
`L-band').  The WSRT observations that we used were performed mostly between
2003 and 2004 (see \citealt{weltevrede06}, `WES' from here onwards, and \citealt{weltevrede07b} for details). The Lovell observations were all contemporary to
LOFAR observations, therefore in the cases where both sets of
observations were available, we chose the Lovell ones because
they are closer to or
overlap the epoch of the LOFAR observations. At L-band we used Lovell
observations for all but three pulsars: B0136$+$57, B0450$-$18, and
B0525$+$21.
In a handful of cases, where no profile at P-band was available from WES, we used the data from the European Pulsar Network (EPN) database\footnote{http://www.epta.eu.org/epndb/}.

To attempt to align the data absolutely, we generated ephemerides that spanned
the epochs of the observations that were used.
This did not include
those from the EPN database, however. Ephemerides were generated from the regular monitoring observations
made with the Lovell Telescope. The times of arrival were generated using
data from an analogue filterbank (AFB) up until January 2009 and a digital filterbank (DFB) since then, with
a typical observing cadence between 10 and 21 days. The observing bandwidth was 64 MHz at a central
frequency of 1402 MHz and approximately 380 MHz at a central frequency of 1520 MHz for the AFB and DFB,
respectively. The ephemerides were generated using a combination of {\tt PSRTIME}\footnote{http://www.jb.man.ac.uk/$\sim$pulsar/observing/progs/psrtime.html} and {\tt TEMPO,} and in the case of those pulsars demonstrating a high degree of timing noise, up to five spin-frequency
derivatives were fit to ensure white residuals and  thus good phase alignment.

The L-band profiles were generated from DFB observations by forming the sum of up to a dozen observations, aligned using the
same ephemerides used to align the multi-frequency data.
We re-folded both the \lof\ and the high frequency data sets using this ephemeris. 
In general, where Lovell data were available, the ephemeris was created using about 100 days' worth of data. For the WSRT observations, an ephemeris was created spanning, in some cases, ten years of data and ending at the time of the LOFAR observations. The timing procedure was the same as for the shorter data spans, except that astrometric parameters were fit and typically more spin-frequency derivatives were required.  
The epoch of the WSRT observations is specified in Table \ref{tab:100} in the {\it Notes} column.
Given the method we used to align the profiles,
the timing solution is less accurate over these longer time spans than those constructed to align the Lovell data, but this too represents a good model,
with a standard deviation (r.m.s.)  of the timing residuals $\lesssim 1$\,ms. 
We aligned the profiles in
absolute phase by calculating the phase shift between the reference epoch of the observations and the reference epoch of the ephemeris and applying this phase shift to each data set. . 

Some of the pulsar parameters derived from these ephemerides are presented in Table \ref{tab:100}. The first column lists the observed pulsars, the second and third columns list the spin period and period derivative of each pulsar, the fourth column is the reference epoch of the rotational ephemeris that was used to fold the data, and Cols. 5 and 6 list the epochs of the LOFAR HBA and LBA observations. In Cols. 7 and 8 two measurements for the DM are given: the first as originally used to dedisperse the observations at higher frequencies, and the second as the best DM obtained from the fit of the HBA \lof\ observations using PRESTO's {\tt prepfold} \citep{ransom01}.
The next three columns provide the pulsar's spin-down age, magnetic field strength, and spin-down luminosity as derived from the rotational parameters according to standard approximations (see e.g. \citealt{lorimer04}):
\begin{equation}
\tau[\mathrm{s}] =  0.5   P / \dot P,
\end{equation}
\begin{equation}
B [\mathrm{G}] = 3.219 \cdot 10^{19} \sqrt{P \dot P},
\end{equation}
\begin{equation}
\dot E[\mathrm{erg/s}] = 4\pi ^2 \cdot 10^{45}  \dot P /P^3,
\end{equation}
where $P$ is measured in s and $\dot P$ in s\,s$^{-1}$.
The resulting aligned profiles for the 100 pulsars are shown in Fig. \ref{fig:100} in the appendix. The star next to the name of the band (P-band in most cases, with the exception of B0136$+$57, where we used the P-band for absolute alignment) indicates that the corresponding band was aligned by eye based on the absolute alignment between the \lof\ data and the other high-frequency band. 
The alignment was made based only on the choice of a specific point along the rotational phase of the pulsar, at the reference epoch of the ephemeris, but unmodelled DM variations can be responsible for extra, albeit small, phase shifts (up to a few percent, see  Table \ref{tab:align}). 
Indeed, observations performed at different times, quite far apart, and at different frequencies, can possess quite different apparent DMs (up to some tenth of a percent, see  Table \ref{tab:align}). DMs that are due to the ISM have, as expected, a time dependence (see \citealt{you07}, \citealt{keith13}), and these differences can become quite relevant especially at the lowest LOFAR frequencies.
We chose to re-dedisperse  all the profiles, LOFAR and high-frequency ones, using the DM obtained as the best DM with {\tt prepfold} for the HBA LOFAR observations. {\tt prepfold} determines an optimum DM by sliding frequency subbands with respect to each other to maximise the $S/N$ of the cumulative profile.
The intra-channel smearing caused by DM over the bandwidth at the centre frequency is indicated by the filled rectangle next to each profile in Fig. \ref{fig:100}.

\begin{table}
\caption{Pulsars for which the absolute alignment was not achieved with the refolding using the same ephemeris (see text for details). The extra DM shift (in cm$^{-3}$pc) and corresponding phase shift ($\Delta\phi$) needed to align the profiles are indicated, or possibly other reasons for the observed shift, e.g., {\it S} for scattering, which notably alters the shape of the profile, {\it g} in case a glitch occurred during the range of the ephemeris, number of spin-frequency derivatives fitted to obtain a good ephemeris (i.e. {\it F\#}), or final {\it r.m.s.} (in ms) of the best timing solution.}
\label{tab:align}
\begin{tabular}{l|l}
\hline
\\
PSR Name & extra DM shift/causes for misalignment \\
\\
\hline\hline
\\
B0114+58 & S \\ 
B0525+21 & F4, S, g \\ 
B1633+24 & F4, rms=1.3, DM=$-$0.11, $\Delta\phi$=0.043 \\
B1818$-$04 & S \\ 
B1839+09 & rms=1.4, DM=$-$0.13, $\Delta\phi=0.051$\\
B1848+13 & DM=$-$0.04, $\Delta\phi=0.017$  \\ 
B1907+10 & F3, S \\ 
B1915+13 & S \\ 
B2148+63 & S \\ 
\hline
\end{tabular}
\end{table}

Only in a few cases, documented in Table \ref{tab:align}, was the DM obtained from the LOFAR observations not used for the alignment. Those are the cases for which, also evident from Fig. \ref{fig:100}, the intra-channel smearing caused by DM is a substantial fraction of the profile width (similar to or higher than the on-pulse region), and therefore the quality of the measurement is lower than that obtained at a higher observing frequency.
On the other hand, in some other cases (although rare, see Table \ref{tab:100}), even the LBA measurement was good enough to provide a DM measurement, and in these cases we were able to use that for the alignment.
In this way we obtain the best alignments, in general, even though some residual offsets could still be observed in a handful of cases. 

\begin{table}
\centering
\caption{Pulsars for which one or more
  glitches occurred during the range of the ephemerides used in this
  paper (above the horizontal line) or near the range of validity of
  our ephemerides (below the horizontal line). The uncertainty of the glitch epoch is quoted in parentheses and it is expressed in days, except for the case of PSR B1907+10, where it is quoted in seconds. Sources: Jodrell Bank and ATNF glitch archives.}
\label{tab:glitch}
\tablebib{[1] \citet{espinoza11glitch}; [2] \citet{js06}; [3] \citet{ywml10}.}
\begin{tabular}{llll}
\hline
\\
PSR Name & Start & End & Glitch Epoch \\
& {\footnotesize [MJD]}  & {\footnotesize [MJD]}  &{\footnotesize [MJD]}  \\
\\
\hline\hline
\\
B0355+54 & 51364.6 & 56262.2 & 51679(15)$^{[1]}$  \\
&&&51969(1)$^{[1]}$ \\
&&&52943(3)$^{[1]}$ \\
&&&53209(2)$^{[1]}$ \\ 
 B0525+21 & 52274.9 & 56641.1 & 52296(1)$^{[2,3]}$ \\
&&&53379$^{[2]}$ \\
&&&53980(12)$^{[3]}$ \\  
\hline
 B0919+06 & 55555.2 & 56557.5 & 55152(6)$^{[1]}$ \\ 
 B1530+27 & 51607.0 & 56535.6 & 49732(3)$^{[1]}$ \\ 
 B1822$-$09 & 54876.5 & 56571.8 & 54114.96(3)$^{[1]}$ \\ 
 B1907+00 & 54984.1 & 56556.9 & 53546(2)$^{[1]}$ \\ 
 B1907+10 & 54924.5 & 56535.1 & 54050(350 [s])$^{[3]}$ \\ 
 B2224+65 & 55359.7 & 56570.1 & 54266(14)$^{[2,3]}$ \\ 
 B2334+61 & 54635.4 & 56507.3 & 53642(13)$^{[1,3]}$ \\ 
\hline
\end{tabular}
\end{table}

For those cases (listed in Table \ref{tab:align}) where the remaining offset was noticeable by eye, we investigated the possible causes
after refolding and applying the new DM. 
We checked whether the pulsars in our sample had undergone any glitch activity during the time spanned by our ephemerides. Sixteen out of our 100 pulsars have shown glitch activity at some time. Seven of them have experienced glitches relatively close to the epoch of our observations, but only two of them during the time spanned by our ephemerides. The relevant glitch epochs of these pulsars are
  presented in Table \ref{tab:glitch} and were taken from the Jodrell Bank glitch archive\footnote{http://www.jb.man.ac.uk/pulsar/glitches/gTable.html} \citep{espinoza11glitch,espinoza12glitch}, integrated with the ATNF pulsar archive\footnote{http://www.atnf.csiro.au/people/pulsar/psrcat/glitchTbl.html}. We note that while the glitch activity could have had an influence on the shift of PSR~B0525$+$21, the recurrent activity of PSR B0355$+$54 did not cause as notable an impact on the alignment.
In some cases the
  profile is scattered in the LOFAR HBA band and rapidly becomes more
  scattered towards lower frequencies.  This will affect the accuracy
  of the DM measurement, potentially causing an extra profile shift
  between bands. In some cases the ephemeris had a large r.m.s. timing residual, or it included higher order spin-frequency derivatives beyond the second, which is typical of unstable, `noisy' pulsars. All these cases are indicated in Table \ref{tab:align}.
In the other cases, we calculated that a DM difference $\lesssim 0.05$\,cm$^{-3}$pc, compatible with the measurement uncertainties, would compensate for that shift. For this reason, we applied an extra shift to align these pulsars' profiles by eye in Fig. \ref{fig:100} and referenced this shift in Table \ref{tab:align}. 

Only a small number of pulsars in our sample show interpulses: B0823$+$26, B0950$+$08, B1822--09, B1929$+$10, and B2022$+$50. 
For these pulsars the profile longitude  is shown in its entire phase range instead of in the phase interval $0.25-0.75$, as was chosen for the other profiles. 
The phase-aligned profiles, which normally have their reference point at 0.5, have been shifted by $+0.25$ in these cases, to help the visualisation of the interpulse. 
A zoom-in of the
  interpulse region is shown in Fig. \ref{fig:ip} (LBA data were excluded because none of the interpulses were detected in that band).
Our sample also contains a few moding pulsars
  (notably B0823$+$26 and B0943$+$10).  For these pulsars we caution that the profile reflects only the mode observed in the
  particular observations presented here.  A more detailed treatment
  of the low-frequency profiles of B0823+26 and B0943+10 can be found
  in \cite{sobey15} and \cite{bilous14}, respectively.
In yet other
  cases, for instance B1857--26, no moding behaviour has previously
been   documented, but notable changes in the profile are seen from high to
  low frequencies.  These might reflect different modes of emission
  when the various observations were taken.
A more detailed discussion of some cases of peculiar profile evolution can be found in Sect. \ref{sec:profevol}. 

We fit the multi-band profiles of each pulsar using Gaussian components (see an example in Fig. \ref{fig:example}),
which are a good representation for the profiles of slow pulsars (see \citealt{kramer94a} and references therein). 
We used the program {\tt pygaussfit.py}, of the {\tt PRESTO} suite, which has the advantage of providing an interactive basis for the input parameters to the Gaussian fits. This program can be used to apply the same method as in \cite{kramer94a} (see e.g. Fig. 3 of their paper), as it provides post-fit residuals (for the discrepancy between the model and the data, see bottom half-plots of Fig. \ref{fig:example}) that we required to have approximately the same distribution in the on-pulse as in the off-pulse region.
The full rotational period, and not only the on-pulse region, was taken into account by the fit, also allowing for distinguishing interpulses or small peaks at different phases from the noise. The Gaussian components derived using this method were chosen to satisfy the condition of best fit with minimal redundancy, and no physical significance should be attributed to them. \citet{hassall12} have shown that it is possible to accurately model the
  evolution of the profile with frequency using Gaussian components,
  but a specific model has to be applied individually to each pulsar,
  requiring much careful consideration.  Such an analysis is
  beyond the scope of this paper.
In the absence of a more comprehensive treatment of scattering for LOFAR profiles, which is envisaged for subsequent papers, the most evident scattering tails of the low-frequency profiles have not been modelled and the corresponding profiles components were not included in the table and are not used in any further analysis.

\begin{figure*}
\centering
\includegraphics[angle=0,width=0.92\textwidth]{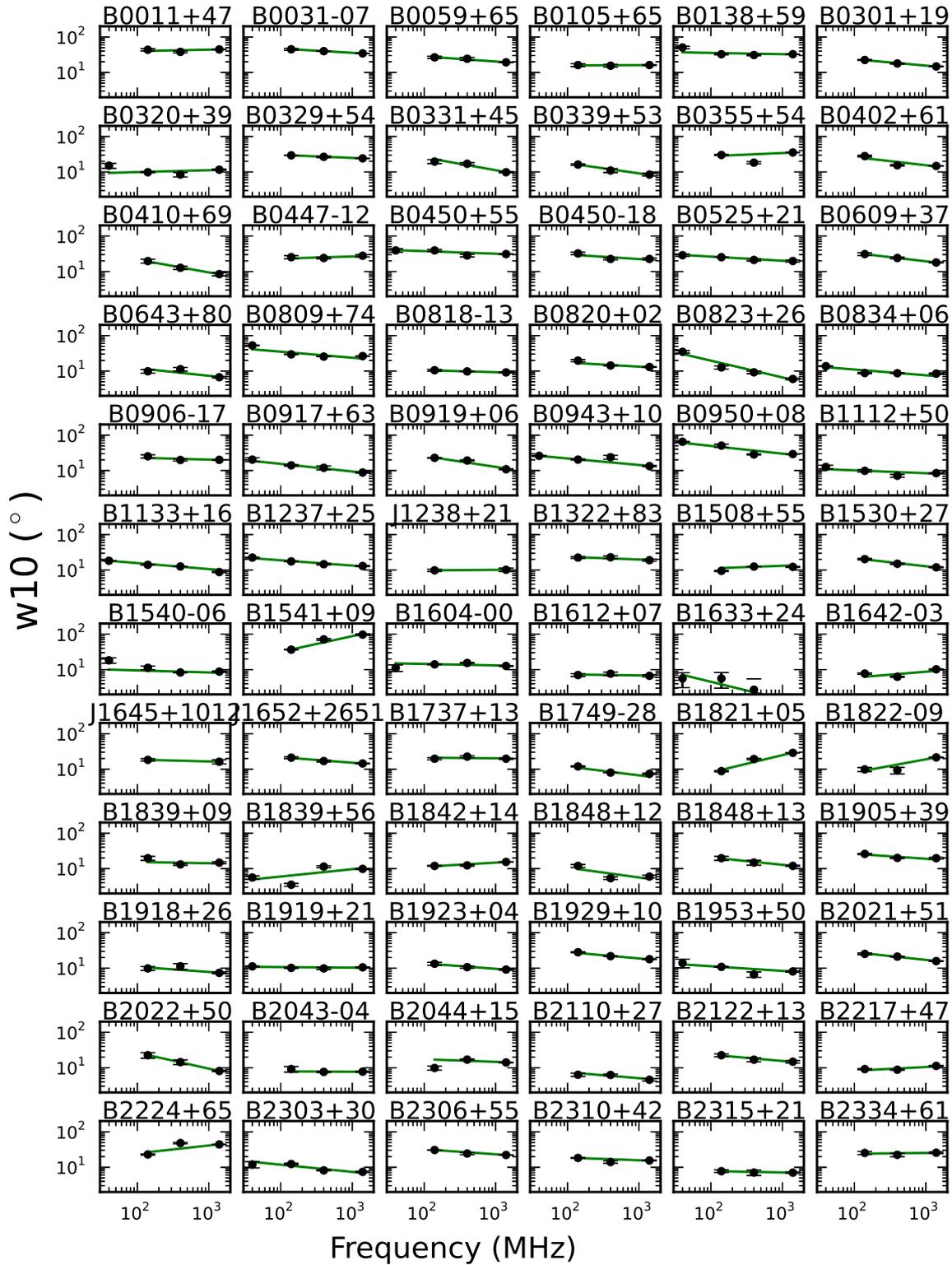} 
\caption{\label{fig:w10a}
Width at 10\% of the maximum ($w_{10}$) encompassing the outer components of the profile, where present. 
The plot shows the evolution of $w_{10}$ as a function of observing frequency for each pulsar.}
\end{figure*}

 We used the mathematical description of the profiles in terms of the Gaussian components to calculate the
widths and amplitudes of the observed peaks. For each profile we obtained the full width at half maximum ($w_{50}$) and the full width at 10\% of the maximum, $w_{10}$. 
To calculate the errors on the widths, we simulated 1000 realisations of each profile, using the noiseless Gaussian-based template and adding noise with a standard deviation equal to that measured in the off-pulse region of the observed profile. By measuring the widths in these realisations, we obtained a distribution of allowed widths from which we determined their standard deviation. We note that these errors are statistical only and do not take into account the validity of our assumptions, that is, the reliability of the template used.
To cross-check the width of the full profile, we tried different methods. An example of how these widths were calculated is described in Sect. \ref{sec:altwidths}  of the Appendix and is shown in Fig. \ref{fig:examplew}.
{In Table \ref{tab:w10}, Cols. 2 and 3 show $w_{50}$ and  $w_{10}$ in degrees at all frequencies, calculated using the Gaussian profiles and cross-checked using the on-pulse visual inspection. The last two columns represent the calculated spectral index $\delta$ of the evolution of these widths ($w_{50}$ and  $w_{10}$) with frequency as $w \propto \nu ^{\delta}$.
The table shows that this fit can be highly uncertain. We also
note that for more data, the linear fit is not always the best representation of the real trend of the profile evolution (see Fig. \ref{fig:w10a} and a more detailed discussion in Sect. \ref{sec:rfm}).
Column 4 of Table \ref{tab:w10} lists the duty cycle of each pulsar, calculated as $w_{10}/P,$ where $P$ is the pulsar period.}

\section{Results}

Here we present the
  results of LOFAR observations of 100 pulsars, considering their
  profile evolution with frequency and comparing them with observations at
  higher frequencies.
In particular, we study how the
  number of profile peaks, their widths, and the relative pulse phases vary
  with frequency.

\subsection{Pulse widths}
\label{sec:width}
We calculated the evolution of the profile width across the frequency range covered by our observations. We chose  $w_{10}$ for our calculations and cross-checked using $w_{op}$, as it is better suited for multi-peaked profiles than $w_{\rm eff}$ and less affected by low $S/N$  than $w_{\rm pow}$ (see Sect. \ref{sec:altwidths}, Fig. \ref{fig:examplew} and Table \ref{tab:w10}). When the measurements
disagreed, the Gaussian fit was refined after visual inspection of the obtained width.

\begin{figure*}
\centering
\begin{tabular}{cc}
\includegraphics[angle=0,width=0.51\textwidth]{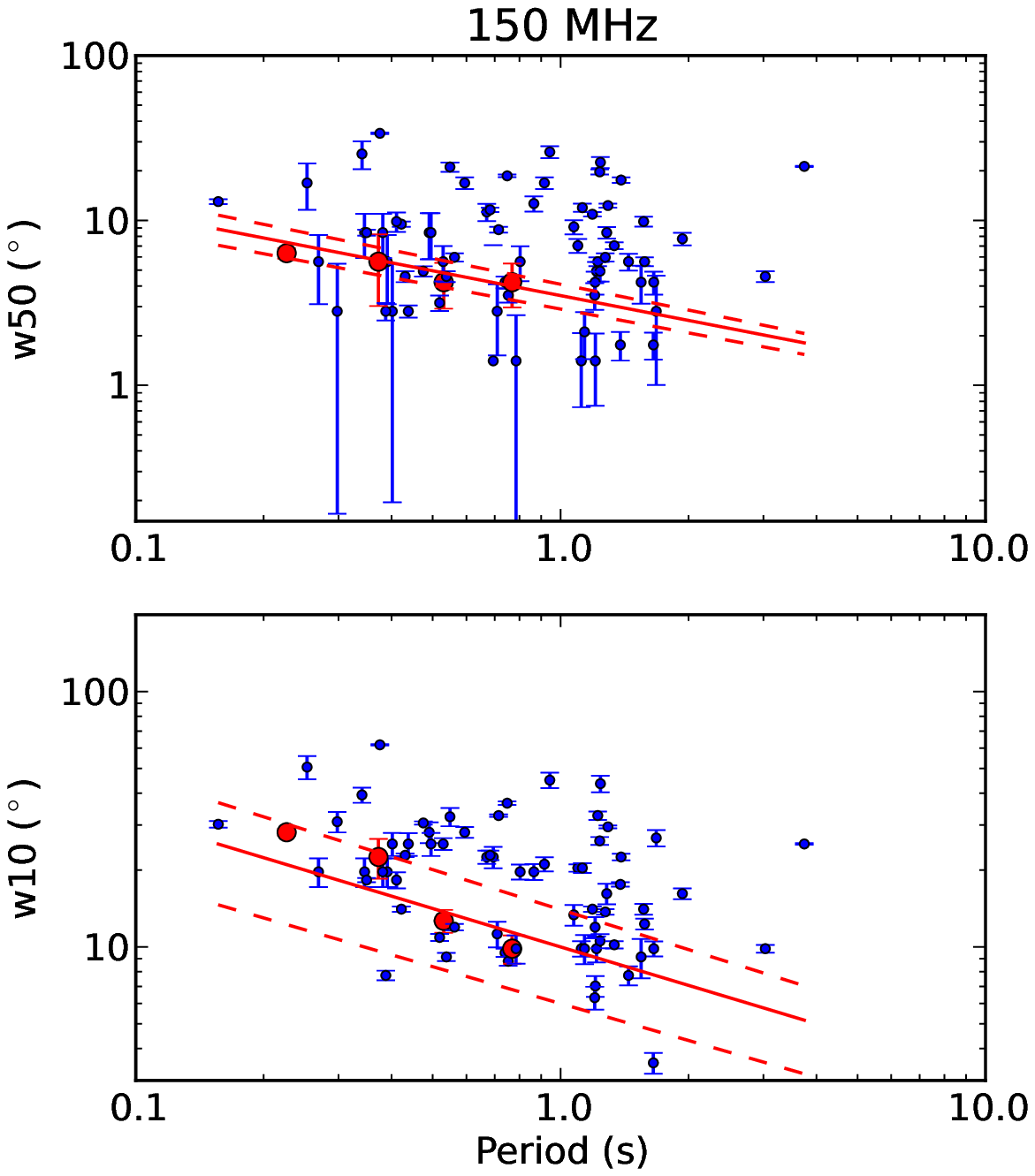} & \includegraphics[angle=0,width=0.5\textwidth]{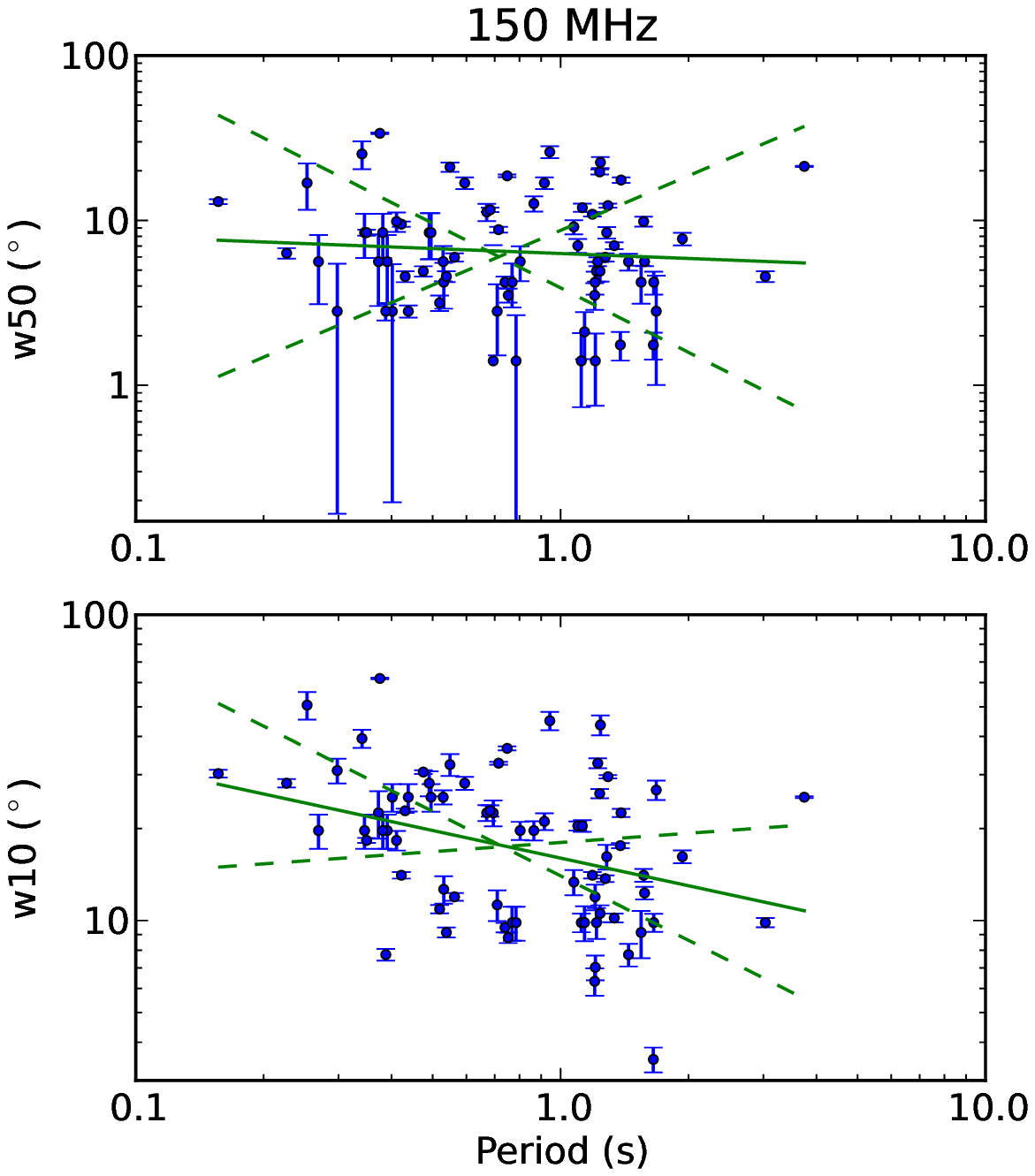} \\
\end{tabular}
\caption{\label{fig:p-w}
HBA profile widths $w_{50}$ (top) and $w_{10}$ (bottom) as a function of spin period.
{\it Left side:} 
The blue and red dots represent the data (red: interpulse pulsars, blue: other pulsars), while the red solid line represents the best fit (non-weighted)  and the red dashed lines represent its $1\sigma$ dispersion for  $w = A \cdot P^{-0.5}$. The fit is calculated using our interpulse pulsars, following \cite{rankin90} and \cite{maciesiakI} (see discussion in the text).
{\it Right side:}  The blue dots represent the data, while the green solid line represents the best (non-weighted) fit for all the pulsars: $w = A \cdot P^{-n}$ , and the green dashed lines represent its $1\sigma$ dispersion (see discussion in the text).}

\end{figure*}

We calculated the dependence of the width of the profiles on the pulse period, considering the different frequencies separately.
We note
that the pulse {\em \textup{width}} is not a direct reflection of the {\em \textup{beam
size}} or diameter (i.e. $2 \rho,$ where $\rho$ is beam radius). For a visual representation of the geometry see for instance \cite{maciesiakI} and \cite{bilous14}.
In fact, only if the observer's line of sight cuts the emission centrally for
  magnetic inclination angles, $\alpha$, that are not too small
  (i.e.~$\alpha > \sim 60^{\circ}$), $w \approx 2 \rho$. In such 
a case, when the emission beam is confined by dipolar
  open field lines, we would expect a $P^{-1/2}$ dependence, which 
has indeed been observed when correcting for geometrical effects by
transforming the pulse width into a beam radius measurement (see
  \citealt{rankin93,gil97,maciesiakIII}). For circular beams, profile width and beam radius are
  related by the relation first derived by \cite{gil84}:

\begin{equation}
{\rho_{10}} = 2\sin^{-1}{\Bigg{[}{\sin{\alpha}\sin{(\alpha+\beta)}\sin^2\bigg{(}{\frac{w_{10}}{4}}\bigg{)}+\sin^2\bigg{(}{\frac{\beta}{2}}}\bigg{)}\Bigg{]}^{1/2}}.
\end{equation}
The angle $\beta$ is the impact angle, measured at the fiducial phase,
$\phi$, which describes the closest approach of our
  line of sight to the magnetic axis.  This equation is derived under the assumption that
the beam is symmetric relative to the fiducial phase. Typically, widths are
measured at a certain intensity level (e.g.~50\% or 10\%, as here), and $\rho$ values
are derived accordingly. In many cases, profiles are indeed
often asymmetric relative to the chosen midpoint, or become
so as they evolve with frequency. 
We note that for a central cut of the beam ($\beta=0$) and for an orthogonal rotator ($\alpha \sim 90 ^\circ $) the equation reduces to
$\rho = w/2$ as expected, while in a more general case, where $\beta=0$ and $\alpha\gg\rho$ the
relation reduces to $\rho = (w/2) \sin \alpha$. In principle, it is
possible to determine 
$\alpha$ and $\beta$ with polarisation measurements. However,
in reality the duty cycle of the pulse is often too small to obtain reliable
estimates (see \citealt{lorimer04}). 
Alternatively, at least for $\alpha$, the relation reported
by \cite{rankin93} can be used:
\begin{equation}
\label{eqn:w50}
w_{50,{ \rm core}}(1 {\rm GHz}) = 2.45^\circ \cdot P^{-0.5\pm 0.2}/\sin(\alpha),
\end{equation}
calculated from the observed width dependency on period for the core
components of pulsars (see Sect. \ref{sec:models}), which is
intrinsically related to the polar cap geometry.  Equation
\ref{eqn:w50} is valid at 1 GHz, 
but can be applied at LOFAR frequencies, maintaining the same dependence, 
if the impact
angle $\beta \ll \rho_{\rm core}$; $\sin \alpha$ should be ignored for orthogonal
rotators.  
Additionally, \cite{rankin93,gks93,kramer94a,gouldlyne98} suggested that `parallel'
$\rho-P$ relations are found if the radio emission of the pulsar can be 
classified and separated into emission from `inner' and `outer'
cones, which seem to show different spectral properties (see Sect. \ref{sec:models} for
details).

Figure \ref{fig:p-w}  represents the 50\% and 10\% widths of the profiles as a function of the pulsar period. We show the results for LOFAR data, using only the HBA data, for which we have the largest sample.
In the left panel of Fig.  \ref{fig:p-w}, we adopted the assumptions from \cite{rankin90}, later followed by \cite{maciesiakI}, and used our interpulse pulsars (overlapping their samples of `core-single' pulsars that show interpulses) as our orthogonal rotators to calculate a minimum estimated width using a fixed dependence on the period as $P^{-1/2}$. The interpulse pulsars in our sample are plotted in red in Fig.  \ref{fig:p-w} and are shown in Fig. \ref{fig:ip} and labelled {\it IP} in Table \ref{tab:w10}.
The red solid and dashed lines represent the best fit of the dependency of $w_{50}$ and $w_{10}$ on $P^{-1/2}$, which should constitute a lower limit to the distribution of pulse widths. We obtain
\begin{equation}
w_{50}(150 {\rm MHz}) = (3.5\pm 0.6)^\circ \cdot P^{-0.50\pm 0.02}
\end{equation}
\begin{equation}
w_{10}(150 {\rm MHz}) = (10\pm 4)^\circ \cdot P^{-0.50\pm 0.02}
,\end{equation}
where $P$ is in seconds, the error is quoted at $1\sigma$ for the amplitude,  and the error on the power-law index ${-0.50}$ was taken to be $0.02,$ following \cite{maciesiakI}.

In the right panel of Fig. \ref{fig:p-w} we present a fit to the widths of our LOFAR sample as a function of pulse period. Because the scatter is much larger than the individual error
bars,  we performed a non-weighted fit. The lines represent the best fit to the data (solid line) and its  $1\sigma$ dispersion (dashed lines). 
Here we calculated 
\begin{equation}
w_{50}(150 {\rm MHz}) = (6.2\pm 0.8)^\circ \cdot P^{-0.1\pm 0.4}
\end{equation}
\begin{equation}
w_{10}(150 {\rm MHz}) = (16\pm 2)^\circ \cdot P^{-0.3\pm 0.4}
,\end{equation}
where the errors are $1\sigma$.

The widths follow an inverse dependency with the pulsar period, consistent with previous analyses at higher frequencies (e.g. \citealt{rankin90,maciesiakI}, but also \citealt{gks93,arzoumanian02}) and at these frequencies \citep{kuzlos99}.
 In general, broadening by external effects may also be
expected, even though not dominant: while we were careful to avoid
evidently scattered profiles in our sample, DM smearing can also
contaminate it.  Finally, because our data are
chosen according to detectability of the pulsars at low frequencies,
a different bias in the observed sample compared with high frequencies cannot be
excluded.  In conclusion, our determination of the relationship
between $w_{50}$ or $w_{10}$ and $P$ is only a first step to determining the relation for the model-independent beam shape, which is to be determined when more polarisation measurements are available.

\begin{figure}
\includegraphics[angle=0,width=0.5\textwidth]{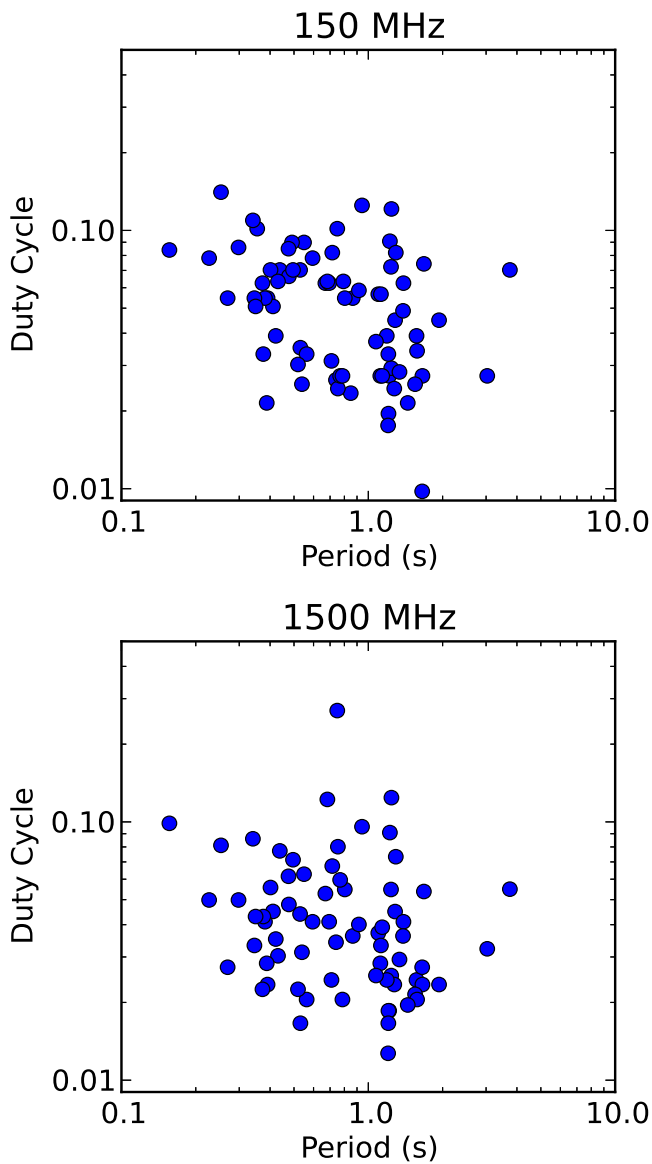}
\centering
\caption{\label{fig:dcycle}
Duty cycle ($w_{10}/P$) vs period of the pulse longitude at the two frequencies for which we have data for all the pulsars in our sample: HBA band from LOFAR and L-band, for comparison. The error bars are omitted, as in \cite{lorimer95}, to more easily guide the eye on the trend. }
\end{figure}

Figure \ref{fig:dcycle} presents the duty cycle ($w_{10}/P$) of the pulsars in two bands: LOFAR HBA and L-band, for comparison,  plotted against the period. The values of the duty cycle are reported in Col. 5 of Table \ref{tab:w10}. The inverse correlation that is observed, implying that shorter period pulsars have larger beams, is also evident at \lof\ frequencies. 
It can be used to characterise pulsar beams and help create accurate beaming models for pulsars in the Milky Way, which would in turn constrain the Galactic population and its birth rates \citep{lorimer95}.

\subsection{Spectral evolution of individual components}

\begin{figure*}
\includegraphics[angle=0,width=1.\textwidth]{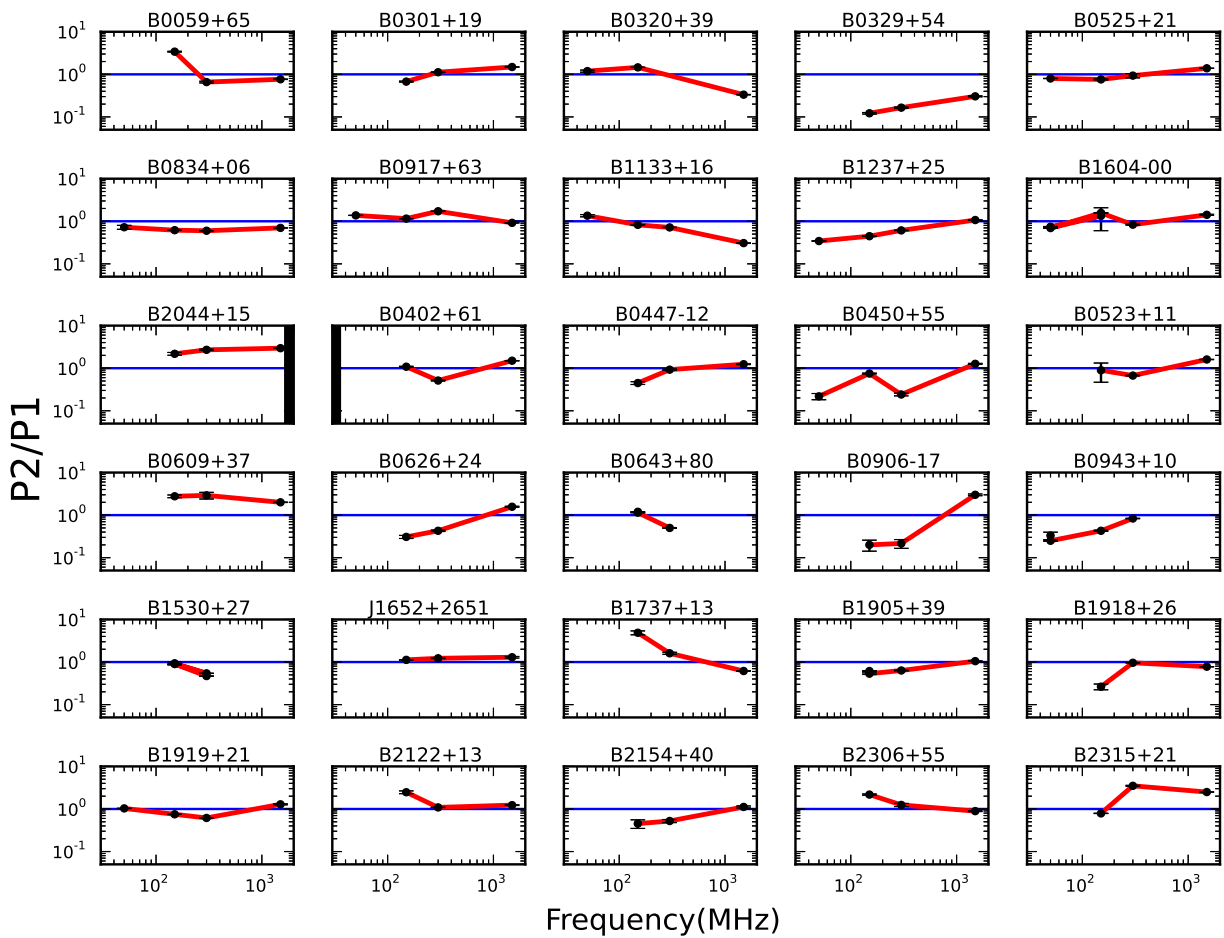}
\caption{\label{fig:PRPL}
Ratio of peak amplitudes for pulsars with multiple peaks as a function of observing frequency. The red line connects the dots (it is not representative of the
power law used to fit an exponential dependence: $\rm{P}2/\rm{P}1(\nu) \propto \nu^{\gamma}$. It is evident that in most cases a power law does not represent the best fit for the frequency evolution of the peaks' ratio, although care should be taken as it is hard to reliably track the
      peak amplitudes in some cases and to follow the same P2 and P1 (see Sect. \ref{sec:rfm} for details). The blue line corresponds to the inversion point where the peaks have equal amplitude. It is evident that in a number of cases the relative amplitudes of the peaks invert as a function of frequency (see Sect. \ref{sec:rfm} for details).
The black vertical lines in between plots 11 and 12 mark the change between previously studied cases and new (see text for details).}
\end{figure*}

No absolute flux calibration of beam-formed LOFAR data was possible with the observing setup used for these observations.
Therefore no spectral characterisation could be attempted yet.
Nonetheless, we attempted a characterisation of the relative amplitudes of  pulse profile components for pulsars with multiple peaks.
In Table \ref{tab:peaks} we list the pulsars for which double or multiple components can be observed and separated in at least two frequency bands. For these pulsars we selected the two most prominent peaks and calculated the evolution of the relative heights with observing frequency. We chose to select the peaks as the two most prominent maxima of the smoothed Gaussian profiles and verified by eye that we were consistently following the same peak at all frequencies. We note that as a result of subtle profile evolution (see e.g. \citealt{hassall12}), shifts of the peaks in profile longitude cannot
be excluded, which we did not track here.
The profile evolution is in some
  cases quite complex and the profiles are sometimes noisy, therefore
the profiles as presented in Fig. \ref{fig:100} need to be reviewed
before
  drawing any strong conclusions based on Fig. \ref{fig:PRPL}.
We first attempted to apply a power-law fit to calculate how the ratio between the two peak amplitudes changes with observing frequency for all the pulsars: $[P2/P1](\nu) \propto \nu^{\gamma}$, where $P1$ is the peak at the earlier phase and $P2$ is the peak at the later phase.
This fit presented large errors, and the distribution of the spectral indices was peaked close to 0, with an average of $0\pm 2$, indicating no systematic evolution despite the large scatter.
A similar finding was obtained by \cite{wang01}, who only selected
conal double pulsars.
They concluded that a steeper spectral index for the leading or trailing component are equally likely, arguing in favour of a same origin of the peaks in the magnetosphere, as expected if both components correspond to two sides of a conal beam. 
They also found a dominance of small spectral indices, with a quasi-Gaussian distribution, indicating no systematic evolutive trends.
The different evolution of the peaks would then be due to geometric beaming effects.
While in their case the pulsars were carefully selected so as to include only the conal doubles, in our case no such distinction was followed, so that the different relative spectral indices could also depend on a different origin of the emission regions (see also Sect. \ref{sec:profevol}).

Given the large scatter
  of this result, and because our sample is heterogeneous (with
  double and multi-peaked pulsars), we investigated the ratios more
  closely.
Figure \ref{fig:PRPL} shows the evolution of the ratios with observing frequency for each of the pulsars used for this calculation. 
We ordered the pulsars into two groups, taking first the pulsars for which
the profiles were already studied in previous works, and sorted by right ascension
within the groups.
In most cases it is apparent that the simple power law is not a good fit to the data and can be misleading if  measurements are only possible at two frequencies.

Table \ref{tab:peaks} provides a detailed summary of these measurements. We observe that the prominence of the peaks seems to shift from low to high frequencies with, in most cases, a net inversion in the dominance of $P1$ from \lof\ HBAs to L-band. The inversion point is also indicated in Fig. \ref{fig:PRPL} by the blue horizontal line.
In general, starting from LOFAR frequencies,
there seems to be a trend that the peaks change from being more dissimilar amplitudes at low frequencies to becoming more similar at high frequencies.
We note that Fig. 1 of \cite{wang01} shows that a linear trend of the peaks' ratio with frequency fits the data well in most cases, meaning that at higher frequencies the relative amplitude of the peaks will again depart from equality.
Notable changes in the observed pulse
  profile properties at low frequencies with respect to high
  frequencies have previously been observed for instance by \cite{hassall12}, \cite{hanran10}, and \cite{izvekova93}.

An observed feature that can contribute to this behaviour was discussed by 
\cite{hassall12}: they modelled the profile evolution with Gaussian components that were free to evolve longitudinally in a dynamic template. The examples presented there (two of which are also in our sample: B0329$+$54 and B1133$+$16) showed that the components change amplitudes and move relative to one another. 

Hardening of the spectrum of the second peak is observed in gamma-rays in the typical case of two prominent caustic peaks \citep{abdo13} and is explained with the different paths that curvature-emitting photons follow in the leading and trailing side of the profile (see e.g. \citealt{hirotani11}), but it is not obvious that this should also follow for the radio emission.

\section{Discussion}

\subsection{Phenomenological models for radio emission}
\label{sec:models}

Based on the findings discussed above, we drew some conclusions on the models that have been proposed to explain the observed properties of  pulsar profiles and on some predicted effects such as radius-to-frequency mapping (RFM).
These models have largely been developed based on observations performed at $> 200$\,MHz.

Rankin's model (\citealt{rankin83b,rankin83,rankin86,rankin90,radranV_90,rankin93,mrVII_02}) proposed that the emission comes from the field lines originating at the polar caps of the pulsar, forming two concentric hollow emission cones and a central, filled, core. There is a one-to-one relation between the emission height and the observing frequency, so that
at different frequencies the profile evolves, as more components come into view or disappear.
Rankin based her classification on the number of peaks and polarisation of the pulse profiles.
Profiles with up to five components are observed (although see \citealt{kramer94b} for a more detailed classification). 
The profiles can be single ($S_t$ or $S_d$ based on whether the profile will become triple or double at higher or lower frequencies, respectively), double ($D$), triple ($T$ or $cT$), tentative quadruple ($cQ$), and quintuple (indicated as multiple $M$), where $c$ represents the presumed core origin.

\cite{lyma88} found that their data agreed with the hollow cone model, and they also observed a distinction in spectral properties between core and cone emission, or at least inner and outer emission. However, based on asymmetries of the components relative to the midpoint of the profile and the presence of so-called partial-cone profiles, they proposed that a window function defines the profile shape, and within this, the locations of emission components can be randomly distributed.

A further step in this direction was made by \cite{kj07},
who assumed a single hollow cone structure without core emission but instead with patches of emission from the cone rims.
Emission could come from different heights at the same observing frequency, but still following RFM; the number of patches changes with the age of the pulsar: up to $\sim 10$ patches, but only at one (large) height for the young pulsars, and up to $\sim 4-5$ at $\sim 4$ different (lower) heights for the older pulsars. This also explains the narrowing of the profiles as the period increases (see Sect. \ref{sec:width}), and their simulations successfully reproduce the observed number of profile components (i.e. typically $N_{comp} \lesssim 5$). 
The central components are then simply more internal and surrounded by the external ones coming from higher up in the magnetosphere: this is why they show single peak profiles and steeper spectra. 
Younger pulsars have been observed to have simpler profiles, but typically with longer duty cycles than those of older ones.
 \cite{kj07} predicted that there should be a maximum height of emission of $\sim 1000$ km. The minimum height, on the other hand, is quite varied but is close to the maximum allowed for young pulsars, which then emit only from one or two patches. Because of the width depends on the period, the opening angle of the cone would then be comparatively larger at the same height for younger than for older pulsars.

\subsubsection{Radius-to-frequency mapping}

\label{sec:rfm}

In the framework of the standard models for pulsar emission, where the radio emission is predicted to come from the polar caps of the pulsar, it has been hypothesised (e.g. \citealt{komesaroff70} and \citealt{rusu75}) and in some cases observed \citep{cordes78} 
that the emission cone widens as we observe it at lower frequencies
because we
  are probing regions further away from the stellar surface where the
  opening angle of the closed magnetic field lines is broader.
The phenomenon is more apparent at low frequencies ($< 200$\,MHz) and is therefore ideal to study using \lof.

A limited observational sample has always biased the conclusions about RFM. Originally (e.g. \citealt{komesaroff70}) it was thought that the RFM behaviour could be observed as a power-law dependence of the increase in peak separation with decreasing 
observing frequency, and asymptotically approaching a constant
  separation at high frequencies ($> 1$\,GHz).
It was therefore proposed that two power laws (i.e. two different mechanisms) regulated the evolution of the pulse profile, with a break frequency at approximately 1\,GHz.

\cite{thorsett91} analysed this dependency using a sample of pulsars observed at various frequencies. He concluded that no break frequency seemed to be necessary to model the evolution of the pulse profile. On the other hand, a simple power law (or a quadratic dependence, indiscernible with his data) and the additional constraint of a minimal emission width (or peak separation) could fit the data at all frequencies.
He obtained the following functional dependence from a phenomenological model of component separation:
\begin{equation}
\label{eqn:tho}
\Delta\theta= A\cdot\nu^{\delta}+\Delta\theta_{min}
,\end{equation}
where $\Delta\theta$ is the component separation, $\delta$ is the separation power-law index of the components, and $\Delta\theta_{min}$ the constant value at high frequency that the pulse separation tends to.
The predictions for $\delta$ are quite varied depending on the theoretical model (see Table 1 in \citealt{xilouris96}).

\cite{mrVII_02} also did extensive work on RFM. They assumed double profiles to derive from conal emission and therefore analysed a sample of ten bright pulsars showing prominent cone components.
They found that inner cones are not affected by RFM and that their component separation does not vary with observing frequency, while outer cone components show RFM and increase their separation with decreasing observing frequency. 

\cite{hassall12} have discussed that RFM does not seem to be at play for some pulsars observed with LOFAR.
Here, with a more statistically significant sample, we can discuss the matter in more detail.
Following \cite{thorsett91} and in particular \cite{mrVII_02}, we investigated double-peaked
  pulse profiles and their component separation (see peak phases in Cols. 5 and 6 of Table \ref{tab:peaks}).

\cite{mrVII_02} divided a group of double profile pulsars into three groups 
The pulsars from groups A and B are associated with outer cone emission (the difference between the two being a fit with or without the constraint $\rho_{0}=\rho_{\rm pc}$, where $\rho_0$ is the constant equivalent to  $\Delta\theta_{min}$ from Eq. \ref{eqn:tho}, relative to the beam radius, and $\rho_{\rm pc}$ is the beam radius at the polar cap edge) while the pulsars from group C are associated with inner cone emission.
Of the ten pulsars of \cite{mrVII_02}, the pulsars from our sample that fall inside each group are\\
Group A: B0301+19, B0525+21, and B1237+25\\
Group B: B0329+54 and B1133+16\\
Group C: B0834+06, B1604$-$00, and B1919+21.\\
\cite{mrVII_02} reported that pulsars from groups A and B show  RFM, while the pulsars from group C show almost no evolution at all. 
Although the profile in our sample evolves rapidly at low frequency, it seems that a similar behaviour can be observed (see single cases in Fig. \ref{fig:100}).

\begin{figure}
\includegraphics[angle=0,width=0.5\textwidth]{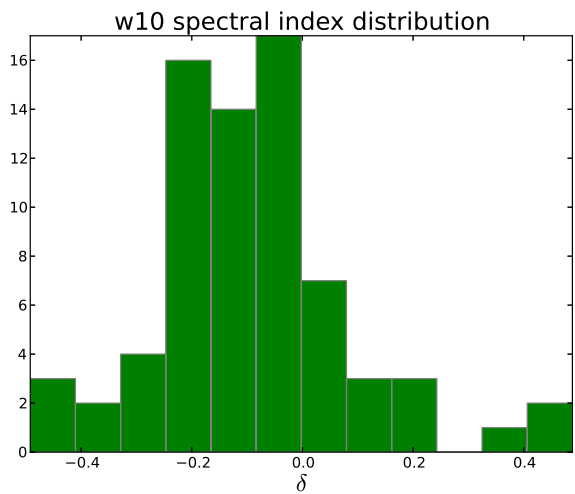} 
\caption{\label{fig:w10b}
Distribution of the indices $\delta$ of the power law used to compute the evolution of $w_{10}$ across our frequency range as  $w_{10}(\nu) \propto \nu^{\delta}$. }
\end{figure}

Figures \ref{fig:w10a} and \ref{fig:w10b} show a similar calculation using $w_{10}$. We plotted the evolution of $w_{10}$ as a function of observing frequency for the single-peaked pulsars and the histogram of the spectral indices of this evolution. We excluded the LBA and HBA profiles that showed significant scattering. 
The errors were calculated from the Gaussian fit,
taking into account both the noise contribution and any unaccounted scattering of the profile.
Although the values in Fig. \ref{fig:w10a} seem to follow the power-law, the profile width in some cases effectively behaves in a non-linear way, as can be cross checked in Fig. \ref{fig:100} for the single cases.

The weighted mean spectral index from Fig. \ref{fig:w10b} is $\delta = -0.1(2)$.
Our results are compatible at 1$\sigma$ with the predictions
made by \cite{ba86}: the component separation does not vary (no RFM, $\delta \sim 0.0$), but the distribution in Fig. \ref{fig:w10b} peaks at negative spectral indices, which is evidence  for a weak widening of the profile at low frequencies. Following the predictions of \cite{gil97}, the dependence of $w_{10}$ with observing frequency based on their calculations should be $\delta = -0.21$ for RFM and conal beams.
This is in support of their model, while the fact that we see a broad
distribution and a flatter median index might be explained by a
subdivision of pulsar behaviours according to the Rankin groups. 
For a future
complete analysis, geometry should be taken into
account to perform a study on the beam radii ($\rho$) rather than the
pulse widths. 

An alternative but complementary explanation to the observed widening of pulsar profiles with decreasing observing frequency can be found in the theory of birefringence of two different propagation modes of a magneto-active plasma \citep{mckinnon97}. These two modes of propagation follow a different path along the open field lines.
The nature of birefringence is such that the two polarisations are spatially closer together at higher frequencies, and depolarisation will occur where they overlap.
\cite{beskin88} predicted that the two modes of propagation would result in two different indices: $\delta = -0.14$ and $\delta = -0.29$ for the ordinary mode and $\delta = -0.5$ for the extraordinary mode. 
While our observations are compatible with either scenario or a combination of them, polarisation studies will help discern between the two
interpretations of this phenomenon (see \citealt{noutsos15}).

\subsubsection{Profile complexity}

\cite{kj07}  searched for a relation between the number of peaks in the profile of a pulsar and some observed or derived parameters, such as its period, age, and rotational energy loss.
As a general trend, they found that faster, younger, more energetic pulsars would typically show less complex profiles, which prompted them to assume that the regions of emission for these pulsars arise at higher altitudes in the magnetosphere and are, therefore, less numerous (see Sect. \ref{sec:models}).
Notably, an abrupt change in this respect can be observed at $P<150$\,ms, $\tau < 10^5$\,yr and $\dot E > 10^{35}$\,erg\,s$^{-1}$.

\begin{figure}
\includegraphics[angle=0,width=0.45\textwidth]{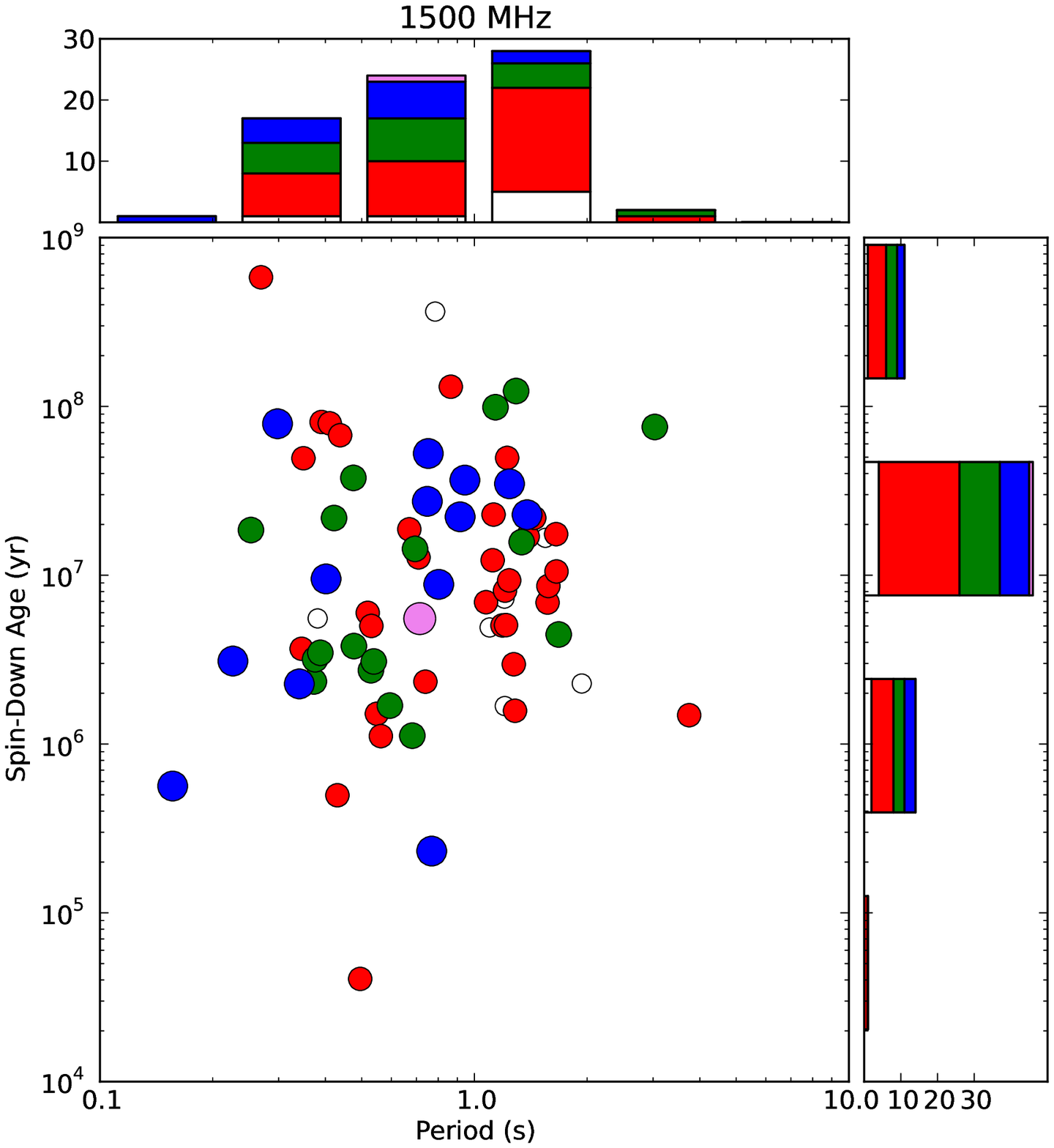}
\includegraphics[angle=0,width=0.45\textwidth]{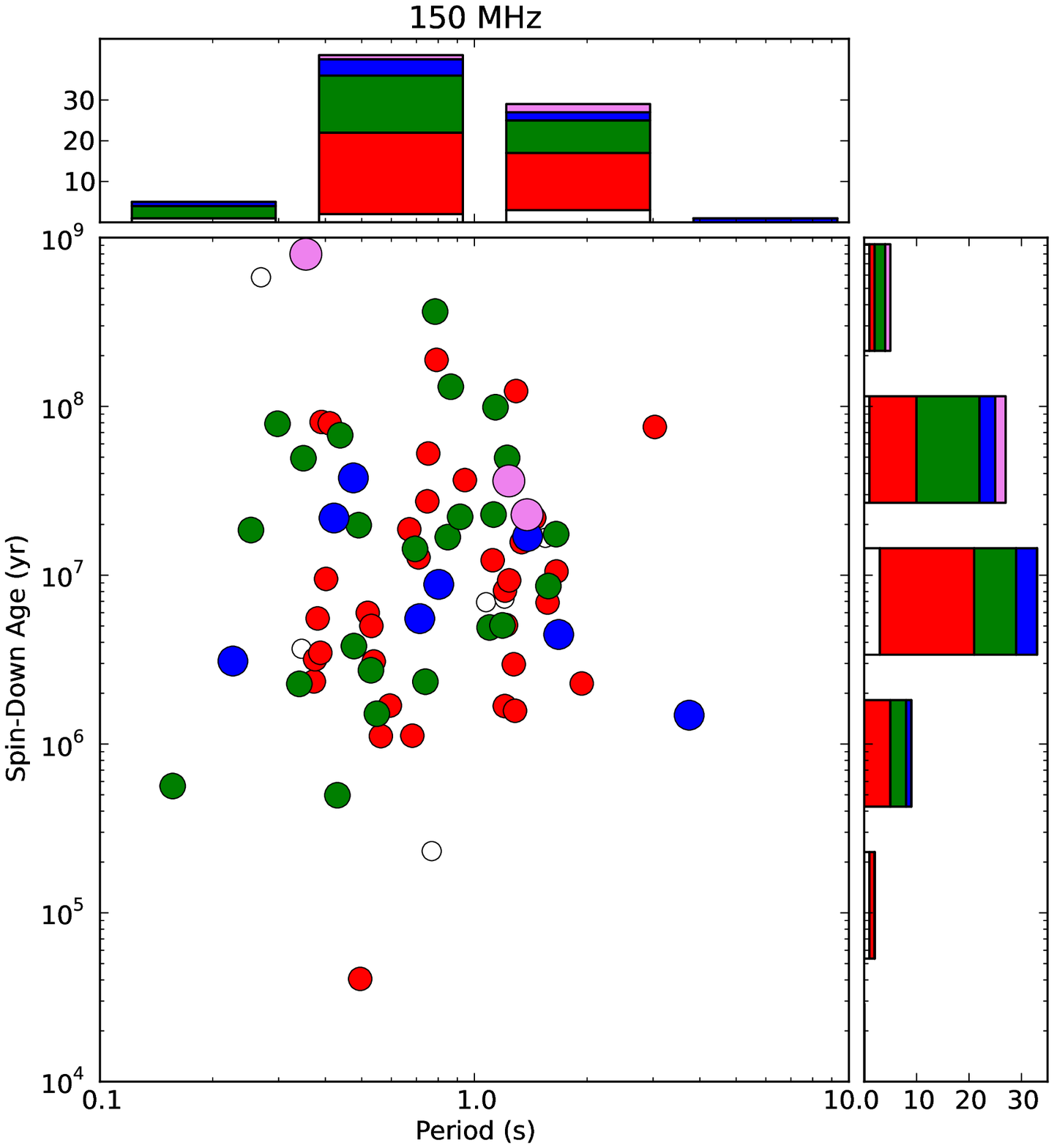}
\centering
\caption{\label{fig:np0t}
Spin-down age vs period in HBAs and at L-band  for the pulsars whose profiles are not affected by scattering in HBAs. The colour (and symbol size) code for the number of peaks is white=1, red=2,
green=3, blue=4, and violet $>4$. 
The histograms at the sides represent the distribution of these values as a function of spin-down age (right-most histogram) and of period (top-most histogram). From the plots no trend is visible in the number of peaks.}
\end{figure}

We calculated the same relations using our sample of 100 pulsars and made the comparisons using LOFAR HBA band and the L-band data.
Figure \ref{fig:np0t} shows the relation between the period of the pulsar and its spin-down age $\tau$ for the two frequency bands. Each circle represents a pulsar, and its colour and diameter represent a different number of peaks of its profile. The circles are larger for growing number of peaks with frequency, and the number of peaks is ordered by colour, in the order white, red, green, blue, and purple. 
In the histograms we summed the pulsars separated by number of peaks in their profiles to search for trends as a function of either period or $\tau$. No trends are evident in any of the histograms.
Our sample does not cover the region of young energetic pulsars in a statistically significant way so that while the few cases might confirm the predictions of \cite{kj07}, nothing in favour or disfavour can be stated in this respect.

\begin{figure}
\includegraphics[angle=0,width=0.5\textwidth]{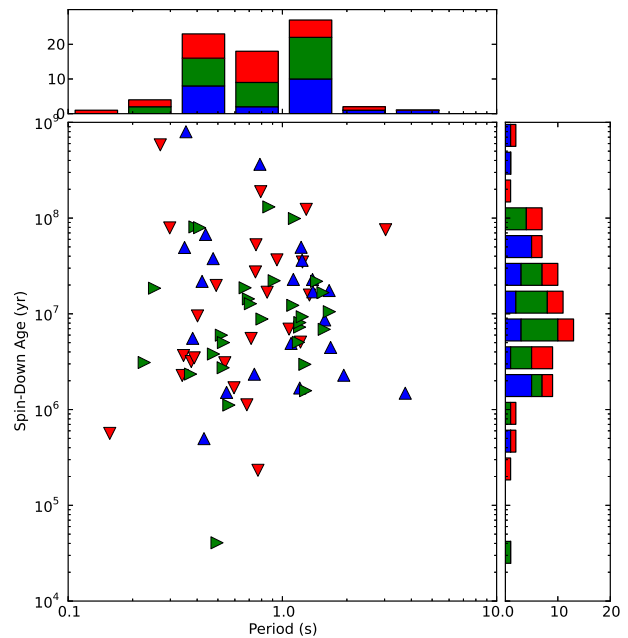}
\centering
\caption{\label{fig:np_diff}
Spin-down age vs period for the pulsars whose profiles are not affected by scattering in HBAs. Here the shape- and colour-coding represent the difference of the number of peaks for each pulsar between HBAs and L-band relative to the HBAs.
In red (triangles down) are the pulsars for which the number of peaks becomes smaller at HBAs than it is in L-band, in blue (triangles up)  are the pulsars for which the number of peaks becomes higher at HBAs than it is in L-band, and in green (triangles right) are the pulsars for which the number of peaks does not change.
At the top and on the right-hand sides the stacked histograms represent the trends of the number of peaks as a function of period and age of the pulsar, respectively.}
\end{figure}

Comparing the histograms from the two plots it appears, in general, that the number of peaks changes from one band to the other. 
For more clarity, we plotted the trend of the number of peaks from HBAs to L-band in Fig. \ref{fig:np_diff}. Here the triangles pointing down (red) represent the pulsars that show fewer peaks at HBA frequencies than at L-band: a total of 30 pulsars; and the ones pointing up (blue) are those for which the number of peaks is higher at HBA frequencies than at L-band: 16 pulsars. The green triangles pointing right represent the pulsars whose number of peaks does not change between the two frequencies: 30 pulsars.
There is a slight indication 
that faster-spinning pulsars have a higher number of peaks at
  L-band than at HBA frequencies while the slower spinners have more
  peaks at HBA frequencies than at L-band.
The younger pulsars in general have more peaks at L-band than HBAs; the older have fewer peaks at L-band than at HBAs.
In both cases the null hypothesis probability that the two distributions are the same based on the Kolmogorov-Smirnoff statistics cannot be rejected at the 20\% confidence level.
This, as pointed out by \cite{lyma88}, might be a selection effect or
a resolution effect (as fast pulsars tend to have wider pulses
  that can be resolved more easily), but given our statistics, we can
also tentatively assume that our sample includes pulsars with different behaviours.

Additionally, a number of
pulsars show an increase in  the number of components at HBA compared to L-band, which
 may also be explained at HBA frequencies by the general expectation that a higher portion of
the beam can come into view at lower frequencies, according to RFM.
On the other hand, the fastest and youngest pulsars show an opposite
trend, similar to what is also observed in millisecond pulsars (see
\citealt{kondratiev15}), which are even faster and therefore have a
wider duty-cycle (see Fig. \ref{fig:dcycle}).  A similar finding was
also reported by \cite{hankins86}, who analysed the
frequency dependence of pulsar profiles for 12 pulsars in the
135--2380 MHz frequency range.  They argued that the occurrence of
single profiles at low frequencies that become multiple at high
frequencies can be explained in the framework of the `core
and cones' models (in the formulation of \citealt{rankin93}).  In
particular, this is expected to occur if we can only observe the core
emission at low frequencies, which is observed to typically have a
steeper spectrum, while the outriding conal components only emerge at
higher observing frequency (see \citealt{kramer94a} for arguments
why this is caused by geometrical reasons and applies to inner
and outer components regardless of their `nature' as core or
cones). 
 A word of caution is needed here, related to
how the number of peaks was determined: it is possible that we achieved a good fit to the profile using a smaller number of Gaussians in HBAs relative to L-band
because the quality of the profile is lower and so fewer components need be fit
(for details on the method see Sect. \ref{sec:analysis},
and the single cases can be studied by comparing Fig. \ref{fig:100} and
Table \ref{tab:w10}).

\subsection{External effects on profile evolution}
\label{sec:external}
The interstellar medium affects the pulse signal while it travels towards the observer, and  it strongly depends on observing frequency, with observations at low frequencies being more strongly affected by scattering and dispersion delay (see also \citealt{zak13}).
We here did not correct the profiles presented for scattering effects that can smear out the signal especially at LBA frequencies, but we considered the effect of intra-channel dispersive smearing.

The DM represents the integrated column density of free electrons between the pulsar and the observer. It produces a time delay in the signal, between the observing frequency and infinite frequency, that can be approximated as
\begin{equation}
{\Delta t_{DM}} = {[{{DM}\over{\rm{cm^{-3}pc}}}]\over{2.41 \cdot 10^{-4} [{{\nu}\over{\rm{MHz}}}]^2}}~\rm{s}.\end{equation}
This approximation is valid if the plasma is tenuous and thus collisionless and if the observing frequency is much greater than the plasma frequency and the electron gyrofrequency.
Signals at different frequencies will be delayed, with the lowest frequencies being delayed the most.
These delays can change on timescales of some years, up to $10^{-3}$~cm$^{-3}$\,pc (see e.g. \citealt{keith13}). 

When aligning the profiles absolutely as we did, connecting the reference point of the profile to the reference epoch of the ephemeris, the DM delays had to be taken into account and all reference times converted to the corresponding times at infinite frequency to correct for dispersive delay.
\cite{hankins86} described a way to measure DM variations based on the alignment of pulsar profiles at different frequencies. \cite{hankins91} and \cite{hanran10} followed this method using increasingly higher resolution profiles. 
They absolutely aligned the profiles spanning, where possible, all the octaves of radio frequency. Leaving the DM as the only free parameter, they identified a reference point in the profile and adjusted the DM value to compensate for the remaining misalignment of the profiles. The value of the best DM was obtained this way, its accuracy strongly depending on the lower frequency that can be used and on the precision with which the time difference of the misalignment can be measured.

All the pulsars from our sample were aligned by refolding their profiles at all frequencies using the same ephemeris. The DM that was adopted for the alignment was the one obtained from \lof\ HBA data: a first folding was performed with {\tt prepfold} using the new ephemeris created from the Lovell data (see Sect. \ref{sec:analysis}), but allowing for a search over DM values, and then the best DM obtained from this search was included in the ephemeris and the profile was dedeispersed once more, without any search option.  
The same was done for the LBA and the P- and L-band observations.
Figure B1 shows that the multi-frequency profiles are aligned  in most cases.
There were some cases, discussed in Sect. \ref{sec:analysis}, where the alignment is visibly incorrect or where we had to apply an extra adjustment to DM to compensate for a visible offset between the profiles (see Table \ref{tab:align}).

One possible cause for the misalignment is the evolution of the profile across the frequencies, so that it is not possible to easily identify a fiducial point in the profile. In addition, pulsars strongly affected by scattering will not only have an exponential scattering tail, but also an absolute delay.
Our profile alignment seemed to also be affected by the
  significantly different observing epochs (e.g. the WSRT
  observations were performed more than ten years before LOFAR
  observations).
On one hand, this meant that we had to adopt a timing solution spanning a long period of time, where timing noise and other effects can become substantial. On the other hand, with observations so far apart, we might also be probing gradients in electron density.

In our case it is not yet possible to perform a systematic study of these variations, as was done by \cite{keith13}. Nonetheless, \lof\ data can provide a new wealth of DM measurements to be compared with previous observations to map the evolution of the interstellar dispersion with time. 
A first conclusion that can be drawn from
  this, simply by comparing the DM values in Table \ref{tab:100}, is that there
  is no significant indication of DMs being systematically higher at
  low frequencies (at least at our measurement precision).
Some authors (e.g. \citealt{shitov83}) 
have postulated “superdispersion” due to
  the sweepback of field lines in the pulsar magnetosphere, which
  would be responsible for lower dispersion delays at low frequencies
  and could create an observed profile misalignment over a wide
  observing band.  
We find that the ratio DM$_{\rm HBA}/ \rm{DM}_{eph}$ ranges from 0.97 to 1.06 (see Table \ref{tab:100}), thus differing by a significant percentage in some cases, but in both directions, 
thus not favouring superdispersion. This agrees with previous findings using \lof\ data \citep{hassall12} and previous measurements \citep{hankins91}. 

\subsection{Some examples of unexpected profile evolution}
\label{sec:profevol}

It is not in
  the scope of this initial paper to enter into much detail about the
  profile evolution of specific pulsars.
These will be the
subject of future dedicated work. There are, however, several interesting
cases of pulse evolution that are worth pointing out at this time.

Figures \ref{fig:w10a} and \ref{fig:w10b} showed that
there are some cases (e.g. B1541$+$09, B1821$+$05, and B1822--09,
B2224$+$65) where the width of the profile is observed to increase
with increasing observing frequency. If we compare these results
with the single profiles in Fig. \ref{fig:100}, we notice that in
these cases this is caused by new peaks appearing in the
profile at higher frequencies (e.g. the well-known
  `precursor' in the case of B1822--09). This is not common in the standard
core and cones geometries, even though it is predicted that new
components can come into view if our viewing angle changes with
increasing observing frequency, thus allowing us to see deeper into the
beam.  While this would explain some of the cases, in some others the
profile evolution can hardly be ascribed to a symmetric core and
conal structure (see e.g. B0355$+$54, B0450$+$55, B1831--04,
and B1857--26).

These narrowing profiles at low frequencies
  might be interpreted as evidence for fan beam models \citep{michel87}.
In fan beam models by\ \citet[][]{drd10,DR12,dyks15}, for instance, the emission comes from elongated
broad-band streams that follow the magnetic field lines.  The model,
based on the cut angle at which the line of sight crosses the beam,
can explain the lack of RFM, for example, if the stream is very narrow (also
the case for millisecond pulsars), and it can also explain the
`inverse' RFM if there is spectral non-uniformity along the
azimuthal direction of the beam through which our line of sight cuts.
The fan-beam formulation proposed by
  \cite{wang14} can even explain `regular' RFM by assuming that
a fan beam composed of a (small) number of
  sub-beams will produce a so-called `limb-darkening pattern'
that is  caused by the decrease in intensity with beam radius
of the emission at higher altitudes because the emission is
  farther from the magnetic pole. Their model, based on observations and simulations, predicts
  that the non-circularly bound beam (different in this from the beam
  predicted by the narrow-band models) can depart from the
  relation $w\propto P^{-1/2}$ (where $w$ is the measured width of the profile, see Sect. \ref{sec:width}).

\cite{chenwang14}, who recently analysed the pulse width evolution
with frequency of 150 pulsars from the EPN database, reached a similar
conclusion: the emission must be broad-band, and the observed
behaviour of width at different frequencies is caused by spectral
changes along the flux tube. They found that if the spectral index
variation along pulse phase is symmetric, there can be  either canonical RFM or anti-RFM,
while in cases where there are substantial deviations from the
symmetric case, then there can be the non-monotonic trends of $w_{10}$
with $\nu$, which we also observed. This is supported theoretically
  by the particle-in-cell simulations of pair production in the vacuum
  gaps \citep{timokhin10}, which predict that the secondary plasma does
  not necessarily have a monotonic momentum spectrum.

These results, and in particular the fact that a wide stream is
expected to produce spectral variations longitudinally in the
profiles, could also explain the observed behaviour of the peak
ratios shown in Fig. \ref{fig:PRPL}.  Alternatively, the peak
ratio changing with observing frequency and, in particular, its
changing sign, might be related to the frequency dependance of the two modes of polarisation
(ordinary, `O', and extraordinary, `X') 
(e.g., \citealt{smits06}) and to that they might be differently dominant in
different peaks.  While it is not possible to give a comprehensive
analysis of the phenomenon here, our studies on the polarised emission
from pulsars with \lof\ (see \citealt{noutsos15}) address the questions
related to the orthogonally polarised modes and the related jumps in
the polarisation angle.

While in this section we have discussed some unexpected profile
evolution, it remains the case that most pulsar profiles (cf. Sects. \ref{sec:width} and \ref{sec:rfm}) are well
described by RFM.

\section{Summary and future work}

We have presented the profiles of the first 100 pulsars observed by \lof\ in the frequency range 119--167 MHz. Twenty-six of them were also detected with 57\,min integrations using \lof\  in the interval 15--63 MHz.
All the \lof\ profiles presented in this work will be made available through the EPN database\footnote{http://www.epta.eu.org/epndb/}.
\lof\ observations were compared with archival WSRT or Lovell observations in P- and L-band, after first folding and aligning all profiles using an ephemeris spanning the full range of the observations. The rotational and derived parameters are presented in Table \ref{tab:100}. Two values of DM were presented as well: one obtained by the best timing fit and one from the best fit of \lof\ data. For each pulsar we aligned the profiles at different frequencies in absolute phase, using the latter DM value. The 100 profiles are presented in Fig. \ref{fig:100}.

Each profile of every pulsar was described using a multi-Gaussian fit following the approach of \cite{kramer94a}, so that in general more components were needed to fit the profiles than evident at first glance or traditionally considered (e.g. \citealt{rankin83}). The results of the Gaussian fit (the measurement of the widths at half and at 10\% of the maximum of the profile and the spectral index of their evolution with observing frequency) are reported in Table  \ref{tab:w10}.
Using the components' widths, we calculated the ratio of the peaks for pulsars with double or multiple peaks (the two most prominent, in the case of multiple peaks pulsars). We concluded that the ratio of the main peak to the second peak does not follow a unique trend, although we note that in most cases  the dominant peak alternates with changing observing frequency. Using $w_{10}$ , we followed the evolution of the width of the full profile with observing frequency. We concluded that while our average spectral index is compatible with no evolution of the pulse width, the distribution of the values is quite large and compatible with the presence of different behaviours for different pulsars, for example, based on inner or outer cone emission, as discussed by \cite{mrVII_02}, or on different propagation modes in the magnetosphere (e.g. \citealt{beskin88,besphil12}).

Future work will be needed, and is in progress, to add more elements to complete this puzzle.
Parallel to this work, a similar one is being conducted on the evolution of the profiles of millisecond pulsars at low frequencies \citep{kondratiev15}, and the spectral behaviour of the slow and recycled pulsars has been analysed (Hassall et al. in prep.).
Complementary to our work is the study of the polarisation properties of pulsars \citep{noutsos15}: the study of the polarisation properties can give constraints on the geometry of the pulsar emission and therefore on its height and on the intrinsic opening angle of the beam of the emission. Additionally, polarisation can help distinguish between orthogonal polarisation modes and therefore determine whether the observed widening of the profiles is caused by birefringence.

To better constrain the width of the pulse, it is also important to be able to deconvolve the scattering tail, which in some cases becomes dominant at low frequencies, from the intrinsic width of the profile. Studies on the characterisation of scattering at low frequencies and modelling of the scattering tail are being conducted (\citealt{archibald14}, Zagkouris et al. in prep.). 
At the same time, the effects of the ISM on \lof\ profiles are being used to create an `ISM weather' database, where the DM variations, independently measured with \lof\ (Verbiest et al. in prep.), can be used for high-precision timing measurements from pulsar timing arrays (see \citealt{keith13}).  

Finally, the observations presented in this work were from commissioning \lof\ data; as mentioned, \lof\ is currently performing the LOTAAS all-sky pulsar survey. When completed, it will provide an extraordinary database of 1.5~h coverage of the whole northern sky with 0.49\,ms time resolution and down to a flux $S_{min} \sim 6$\,mJy at 135 MHz. At present we are able to use the full \lof\ core instead of only the `Superterp' and can cover a full 80 MHz bandwidth contiguously.
Moreover, further improvements will include coherent dedispersion of the data and will enable single pulse studies to probe the `instantaneous' magnetosphere.

\begin{acknowledgements}
MP wishes to thank A. Archibald, V. Beskin and J. Dyks for useful discussion. We are grateful to an anonymous referee whose comments notably improved the quality of our work. This work was made possible by an NWO Dynamisering grant to ASTRON, with additional contributions from European Commission grant FP7-PEOPLE-2007-4-3-IRG-224838 to JVL. 
LOFAR, the Low-Frequency Array designed and constructed by ASTRON, has
facilities in several countries, that are owned by various parties
(each with their own funding sources), and that are collectively
operated by the International LOFAR Telescope (ILT) foundation under a
joint scientific policy.  MP acknowledges financial support from the RAS, Autonomous Region of Sardinia. JWTH, VIK and JVL acknowledge support from the European Research Council
under the
European Union’s Seventh Framework Programme (FP/2007-2013) / ERC Grant
Agreement nrs. 337062 (JWTH, VIK) and 617199 (JVL).
SO is supported by the Alexander von Humboldt Foundation.
C. Ferrari acknowledges financial support by the {\it “Agence  
Nationale de la Recherche”} through grant ANR-09-JCJC-0001-01.
\end{acknowledgements}

\def\pasa{Publications of the Astronomical Society of Australia}

\bibliographystyle{aa}
\bibliography{biblio}

\appendix
\Online


\onecolumn
\section{Determination of pulse widths}
\label{sec:altwidths}
As discussed in the Sects. \ref{sec:analysis} and \ref{sec:width}, the widths of the full profile that are to be used in the subsequent calculations were determined from the fit of Gaussian components to the profile shape.
However, a number of alternative methods were also tested and, in one case, used to cross check the widths obtained from the Gaussian fits. Here we give a brief description of each of them.

\begin{figure*}[h!]
\centering
\begin{tabular}{cc}
\includegraphics[height=120mm,width=0.49\textwidth]{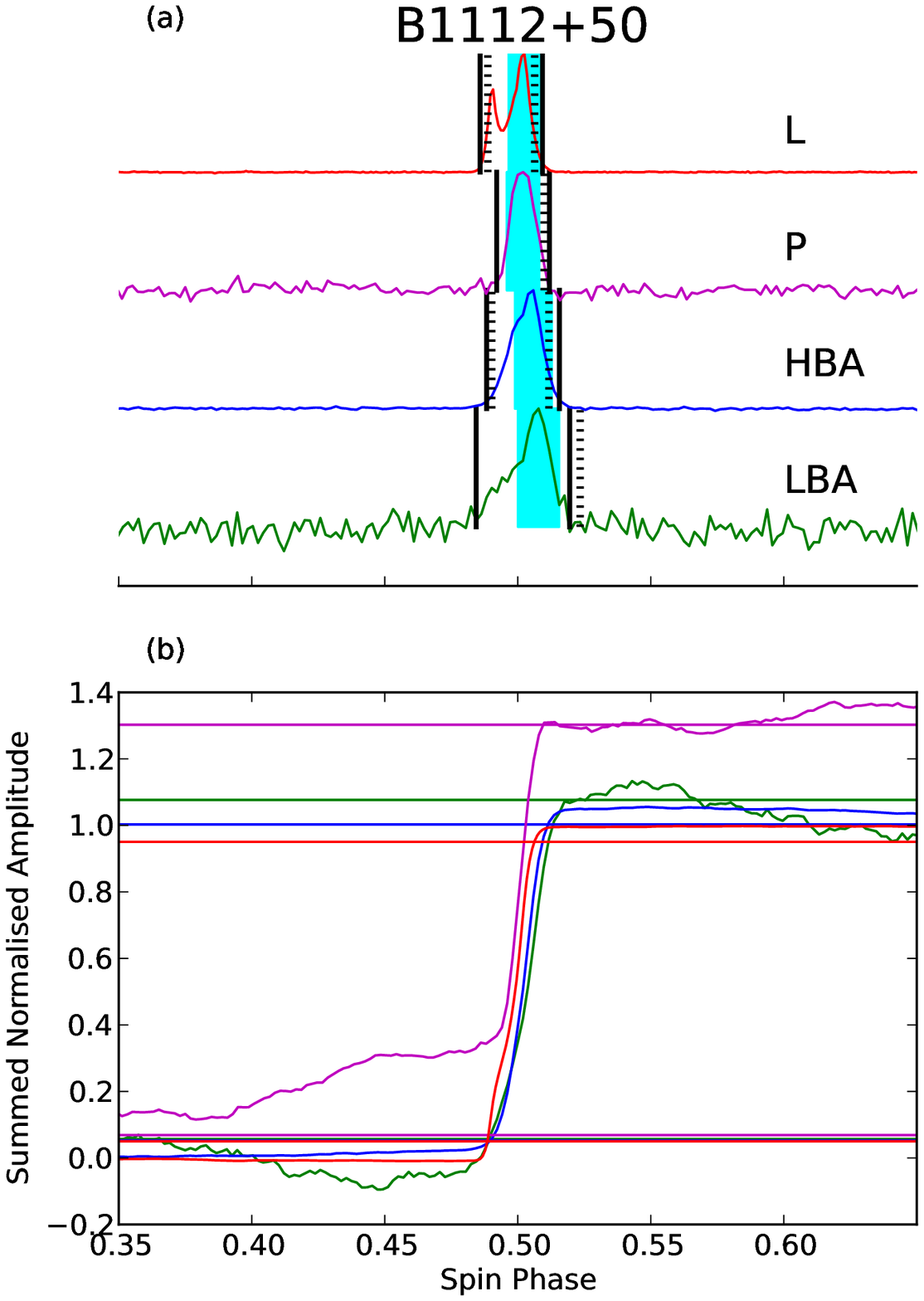} & \includegraphics[height=120mm,width=0.49\textwidth]{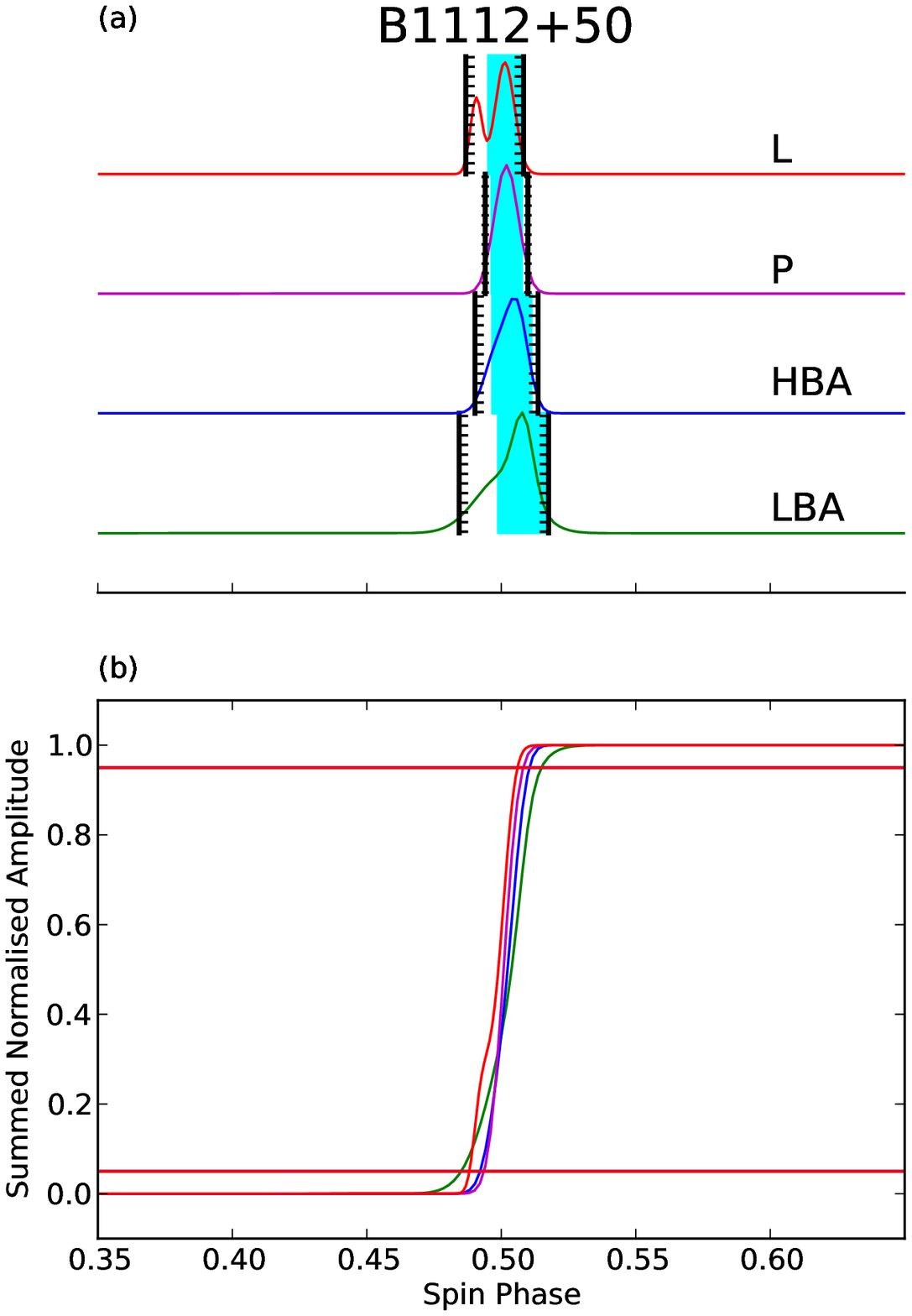} \\
\end{tabular}

\caption{\label{fig:examplew}
PSR B1112+50: example of how the widths of the full profile were calculated using the four methods described in the text. 
On the left-hand side is the real profile. On the right-hand side, for comparison, the noise-free profile obtained from the Gaussian fit.
In panel (a), the shaded cyan area represents $w_{\rm eff}$.  
The dashed vertical lines represent $w_{\rm pow}$, calculated as shown in  panel (b): the cumulative distribution of the normalised profile. 
The solid line demarcates $w_{10}$ as obtained from the Gaussian smoothed profiles (which are shown in panel (a), on the right). The measurement of $w_{10}$ is consistent and overlapping with  $w_{\rm op}$ (see text for details).
}
\end{figure*}

We calculated the effective width, $w_{\rm eff}$, as the integrated pulsed flux divided by the
  peak flux; this is represented by the cyan-shaded areas in Fig. \ref{fig:examplew} centred at the main peak. This width metric typically does not represent all the profile components
well, as is
  observed in the P– and L–band profiles. The method underestimates the width of the profile for multi-peaked profiles, which is the majority of profiles in this work, mainly characterising the width of the main peak in these cases. 

The total-power width $w_{\rm pow}$ is  calculated as the phase interval that includes 90\% of the total pulse energy in the cumulative flux distribution of the normalised profile (see lower plots in Fig. \ref{fig:examplew}). 
We selected the phase included in the interval between 5\% of the total flux and 95\% of the total flux 
(indicated by the intersection of the summed profile with the horizontal lines in panel (b) of Fig. \ref{fig:examplew}).
This distribution should start at 0 and increase monotomically to 1 if no noise is present (see the noise-free distribution on the right side of Fig. \ref{fig:examplew}). 
When noise is present (left-hand side of Fig.  \ref{fig:examplew}), the sum does not grow smoothly, and sometimes negative terms in the off-pulse region add up quite substantially.
In the real profiles (left-hand side of Fig.  \ref{fig:examplew}), the horizontal lines should all overlap at 0.05 and 0.95, like they do in the case of the noise-free profiles (right-hand side of Fig.  \ref{fig:examplew}), but they differ in some cases as they take into account that the highest and lowest value of a noisy cumulative distribution  can 
be negative or greater than 1.
As can be seen from the green (LBA) and magenta (P-band) curves, when the profile is noisier, this measurement is less stable because the cumulative distribution oscillates more, and the width can be consistently overestimated (in these cases the left-hand dashed line precedes the phase range shown in the Fig.).
This method is represented by the dotted line in the upper plot of Fig. \ref{fig:examplew}, which, in the case of the LBA and P-band profiles, starts at phase 0.

The full-width at 10\% of the maximum, $w_{\rm op}$, represents the width of the full profile (including noise) at the 10\%  level of the outer components, including the full on-pulse region: it is coincident with the solid vertical lines in  Fig. \ref{fig:examplew}. It was calculated as the on-pulse region with a flux above 10\% of the main peak in the baseline-subtracted profile.

\clearpage
\section{Multi-frequency profiles and tables of the derived profile properties.}
\begin{figure*}[h!]
\caption{\label{fig:100}
Pulse profiles of the 100 pulsars observed by LOFAR, showing from bottom to top LBA  (green) and HBA (blue) profiles from LOFAR observations, 350 MHz (magenta) and 1400 MHz (red) profiles from WSRT and Lovell observations. We caution that the moding pulsars, for which only one mode of pulsation is represented by our observation,  might not be the same for the profiles at different frequencies. The profile phase is zoomed-in on the interval 0.25 -- 0.75 in all cases except for the pulsars with interpulses, where the full pulse phase is shown, and the profile is rotated by 0.25 in phase to shift the interpulse from phase 0. The star next to the band name indicates that the alignment  for that band was made manually (see text for details).
}

\centering
\begin{tabular}{cccc}

 \includegraphics[width=32mm]{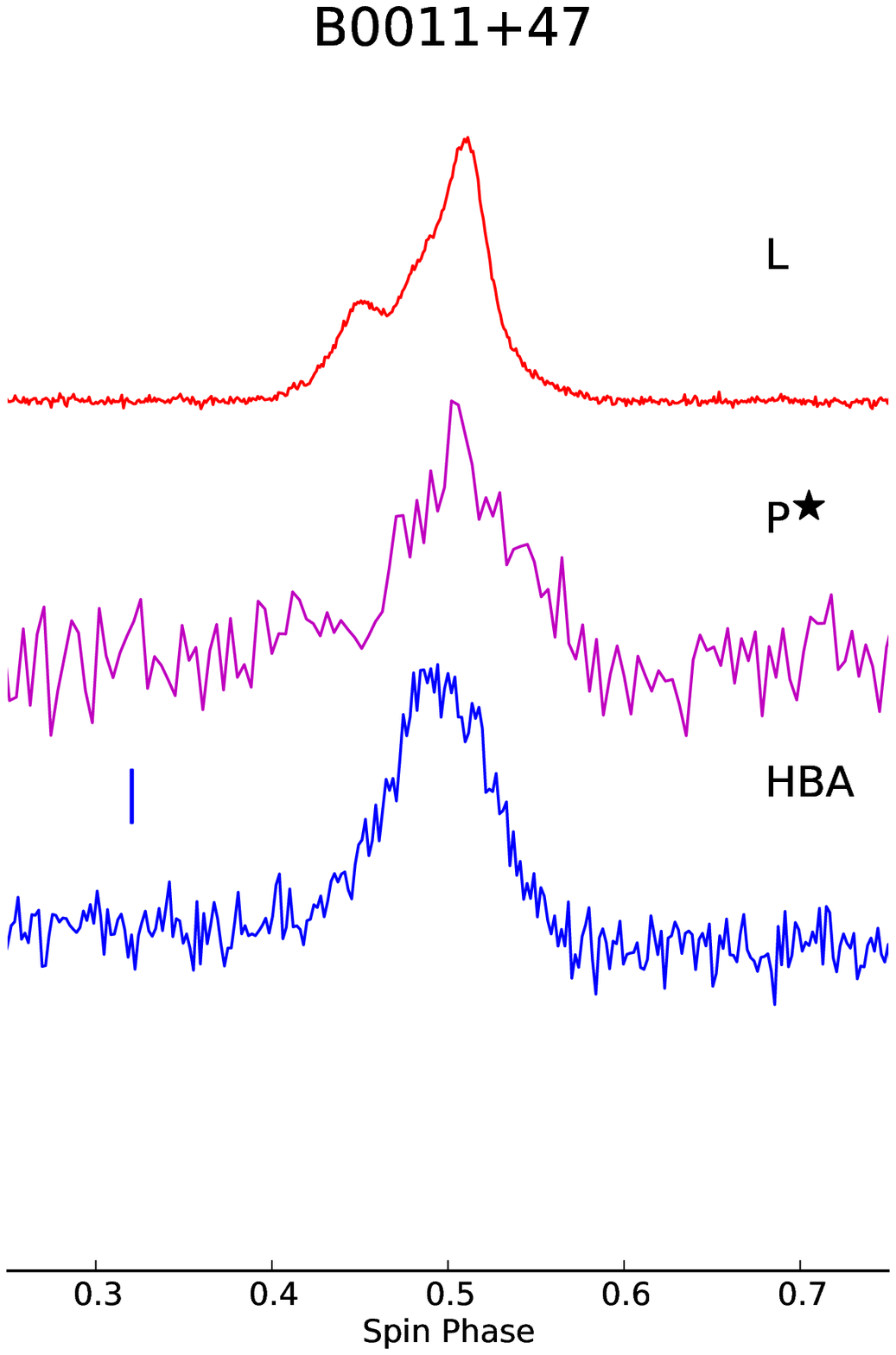} &   \includegraphics[width=32mm]{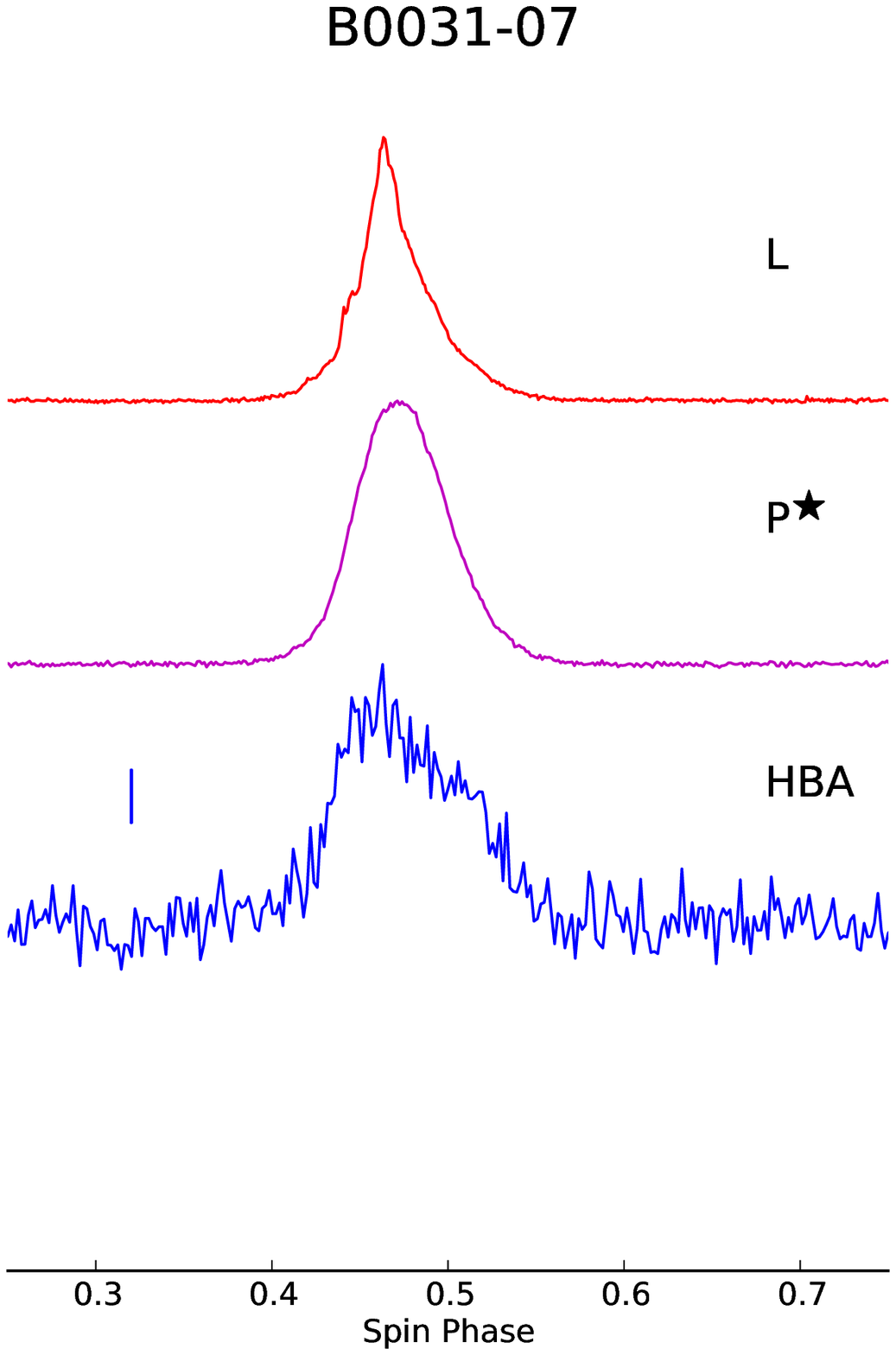} &   \includegraphics[width=32mm]{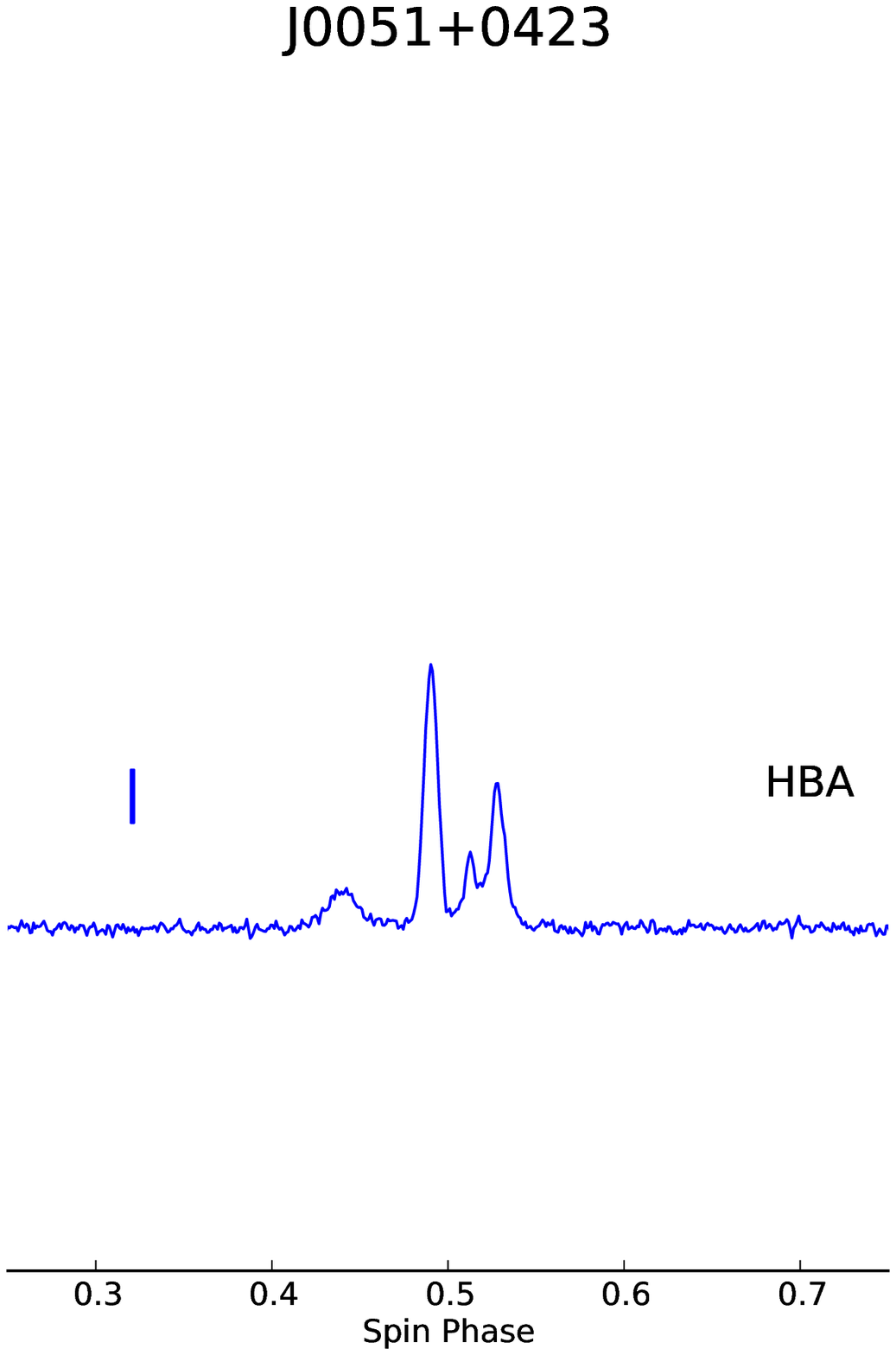} &   \includegraphics[width=32mm]{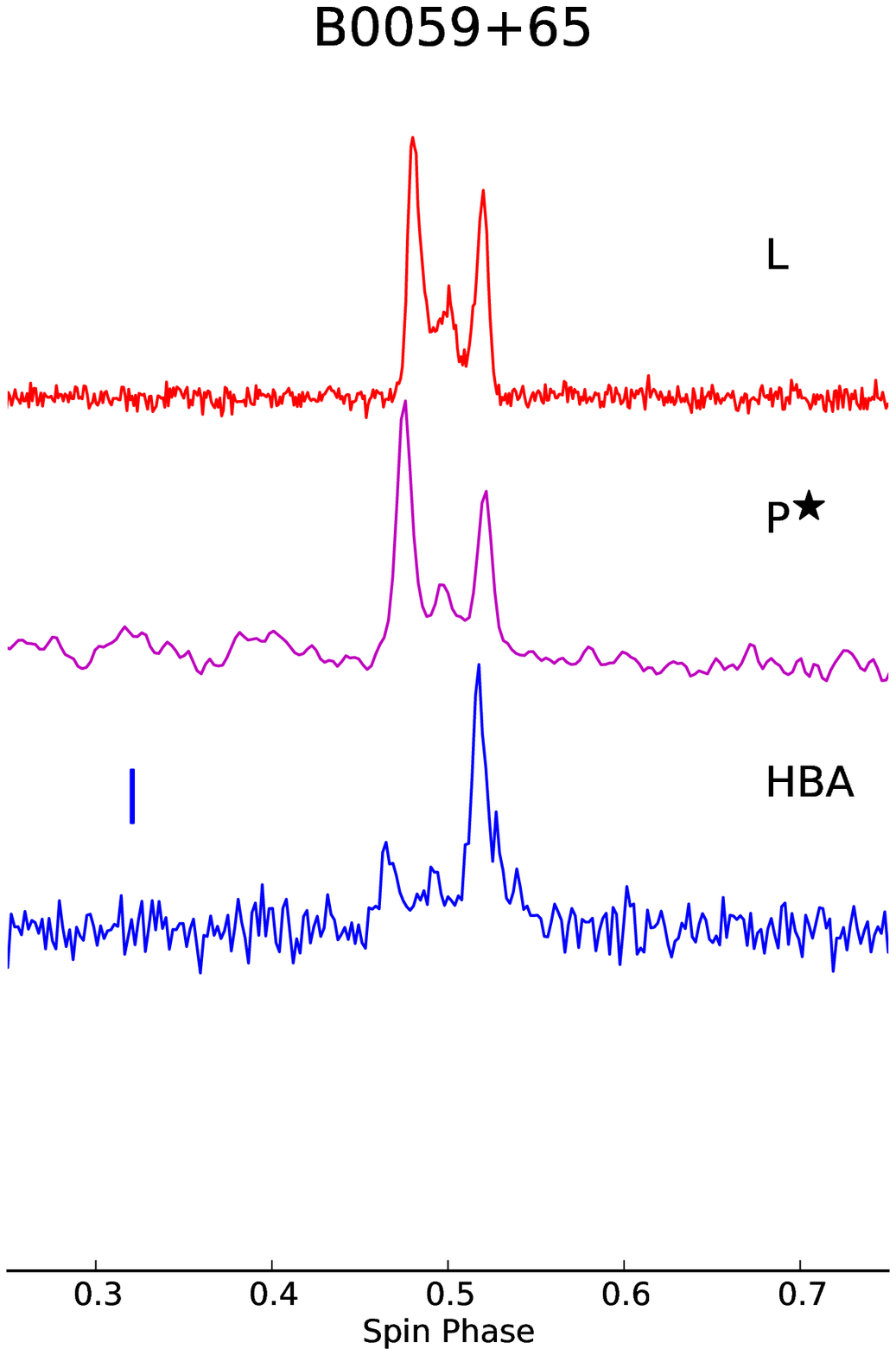}  \\

 \includegraphics[width=32mm]{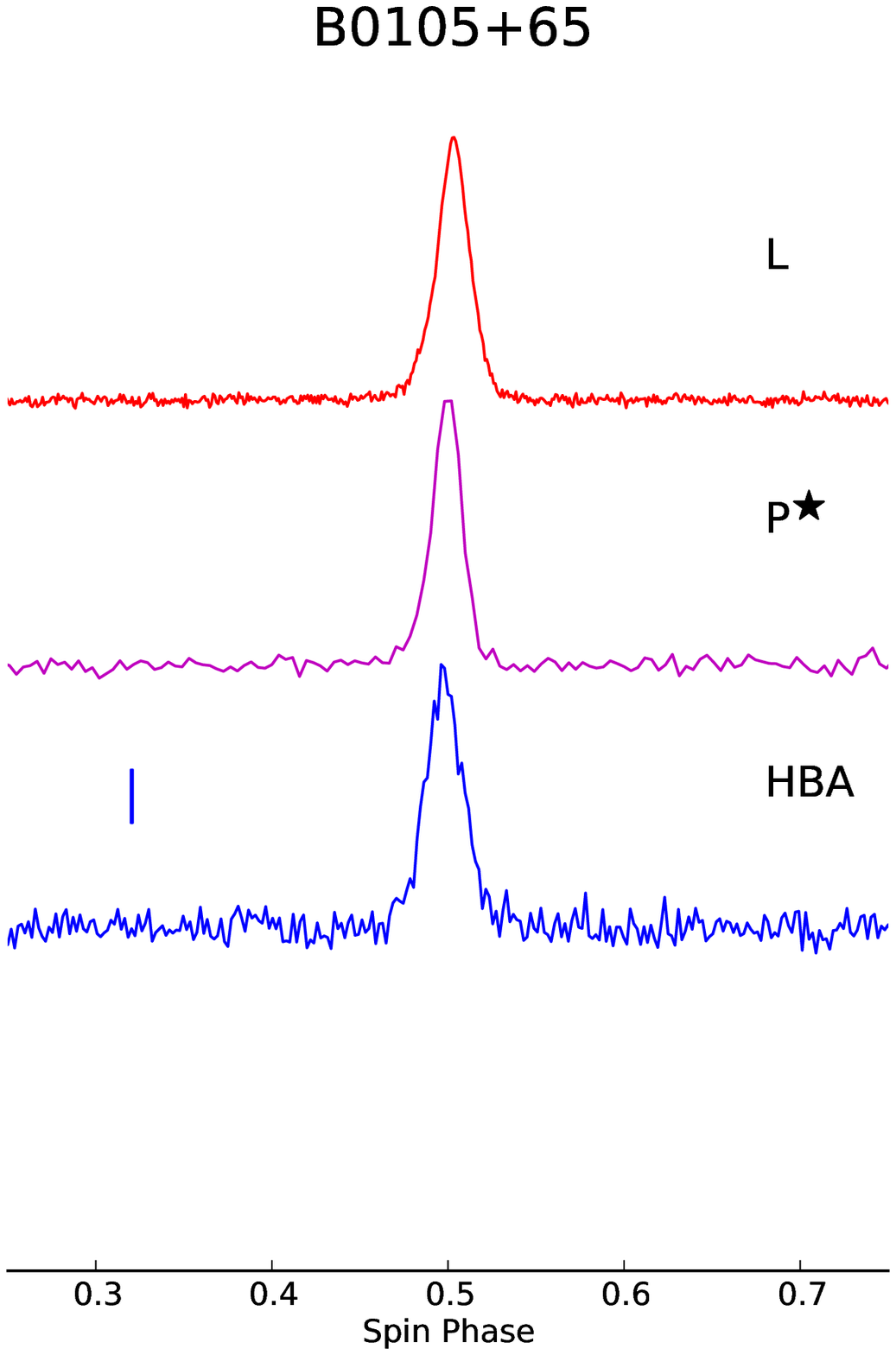} & \includegraphics[width=32mm]{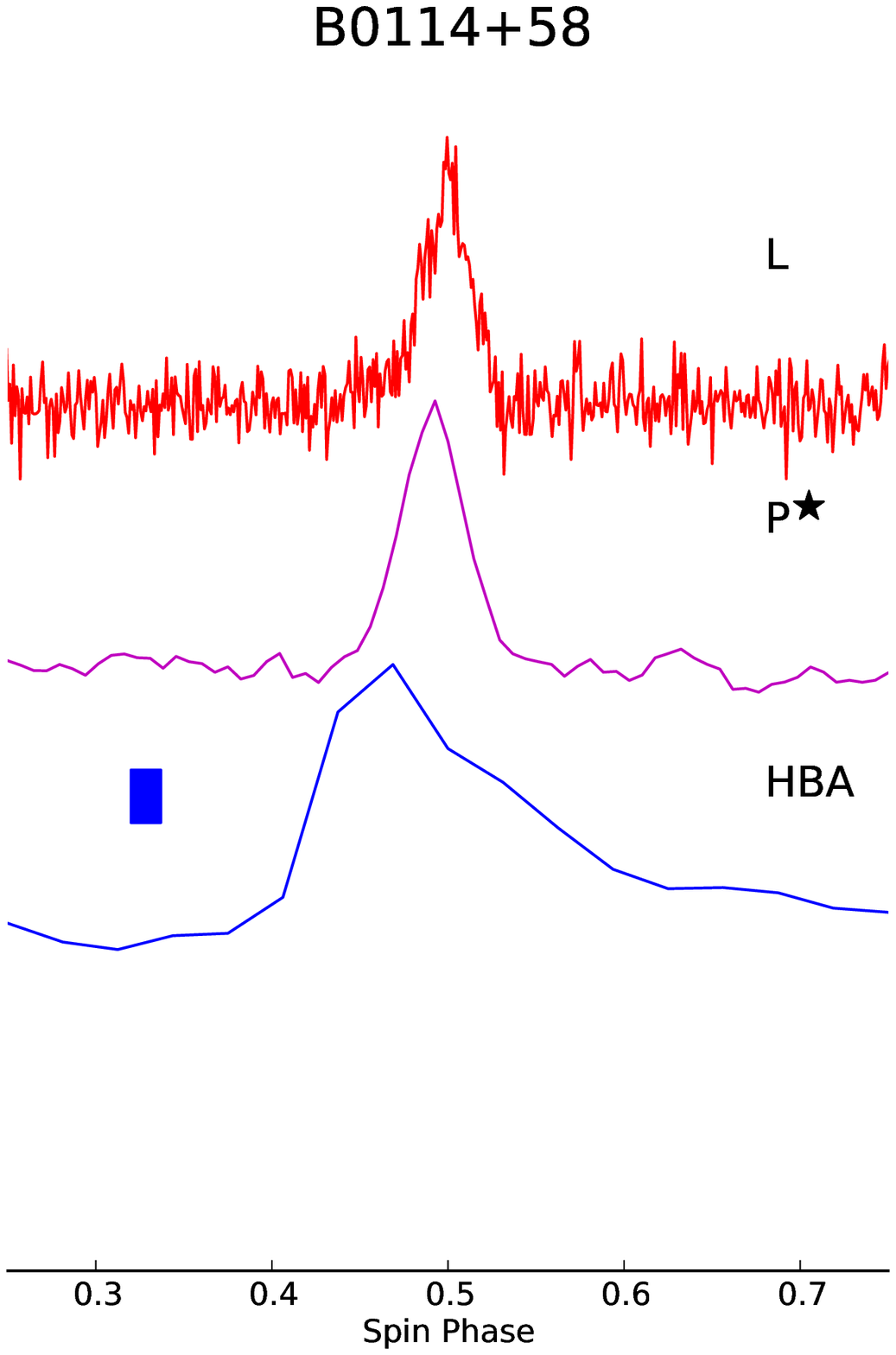} &   \includegraphics[width=32mm]{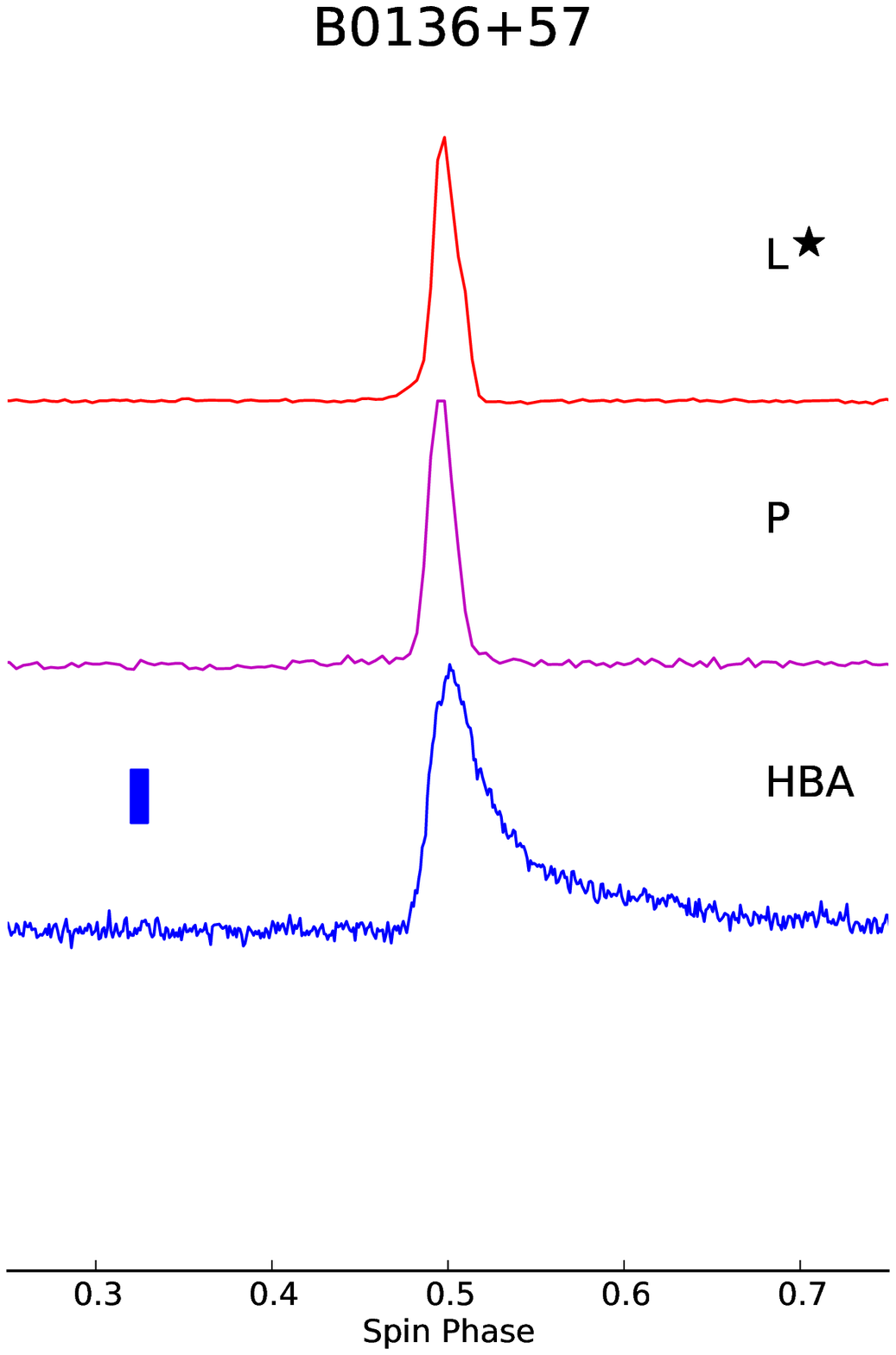} &   \includegraphics[width=32mm]{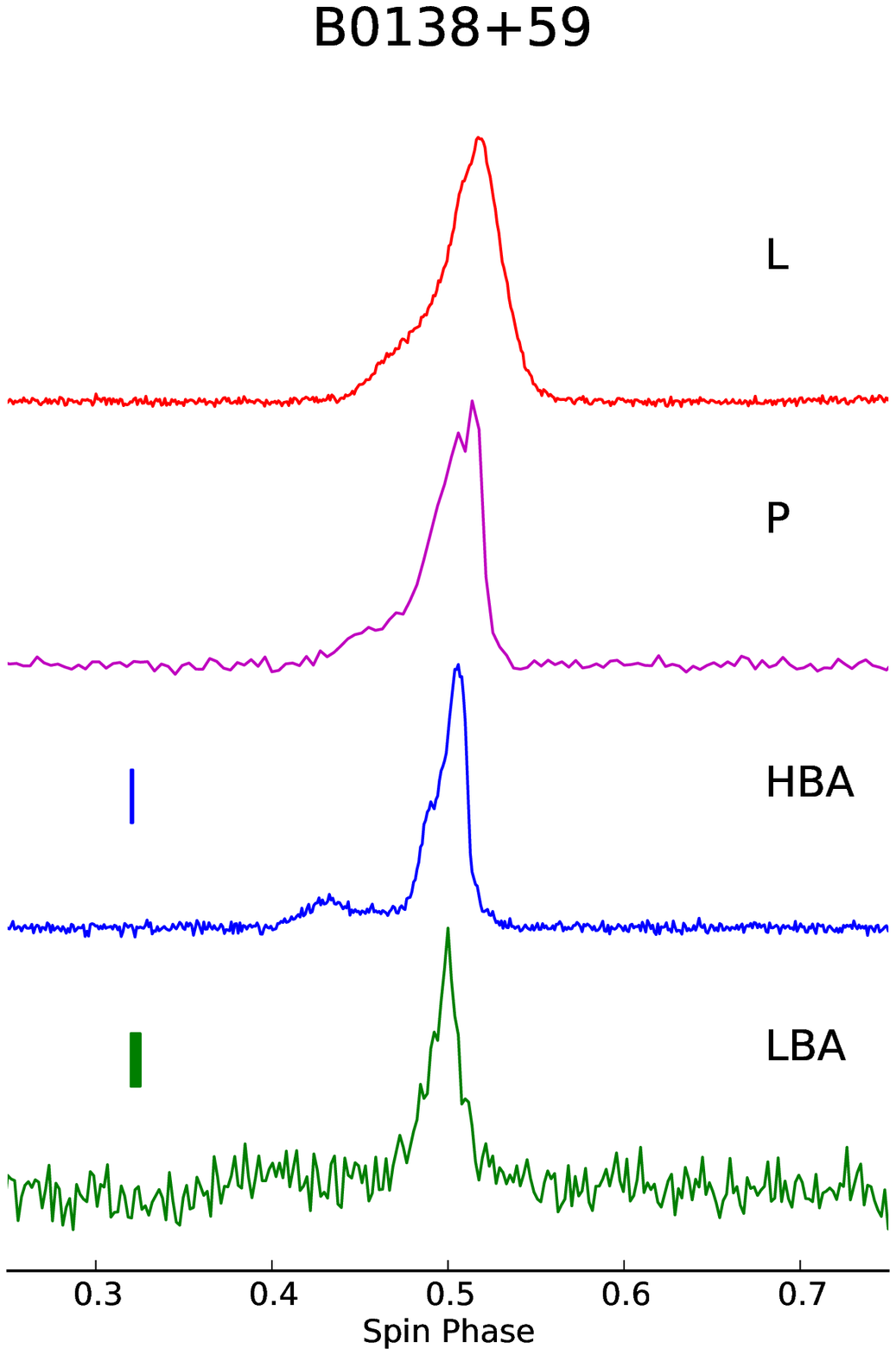} \\

\includegraphics[width=32mm]{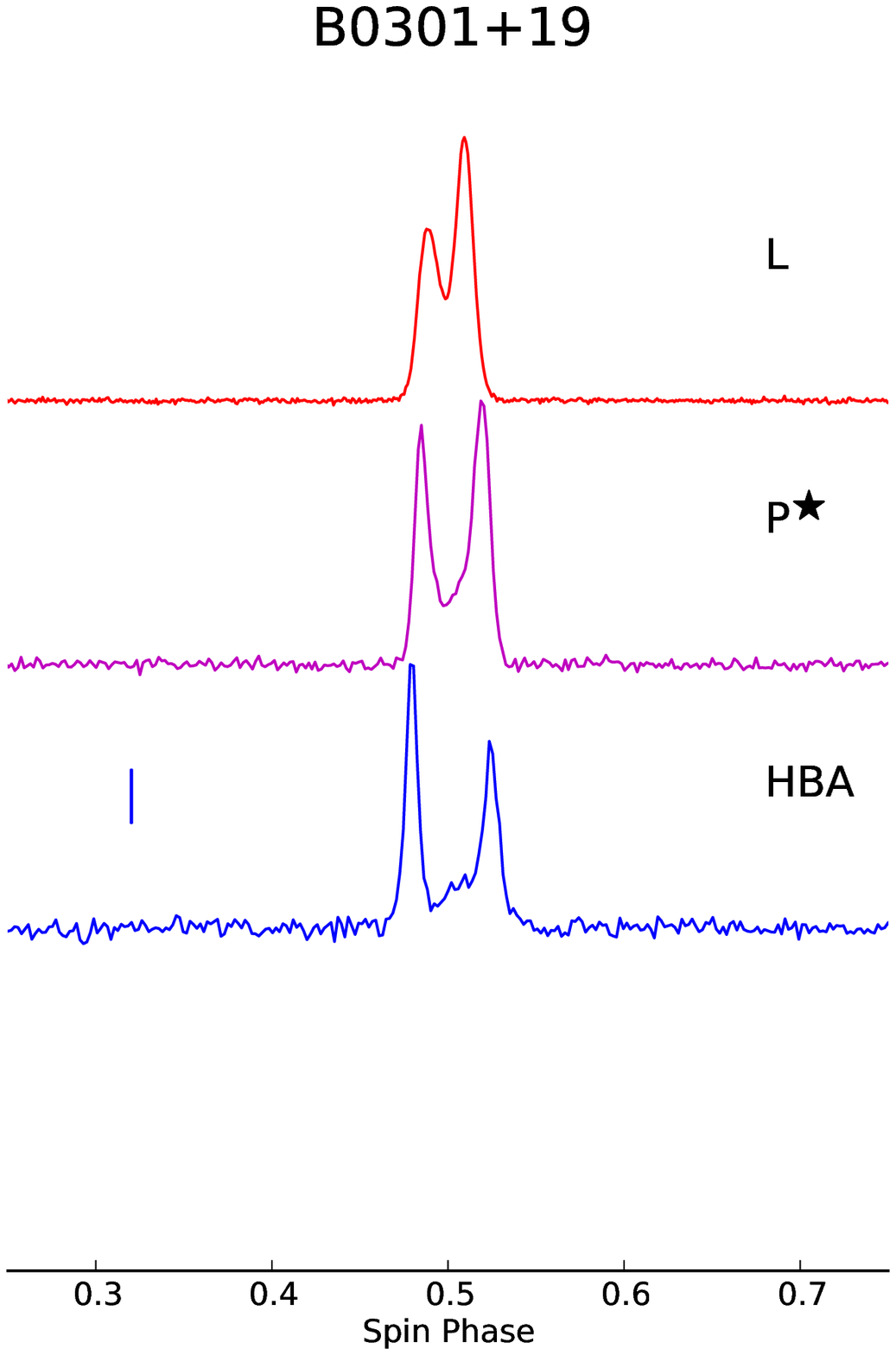} &   \includegraphics[width=32mm]{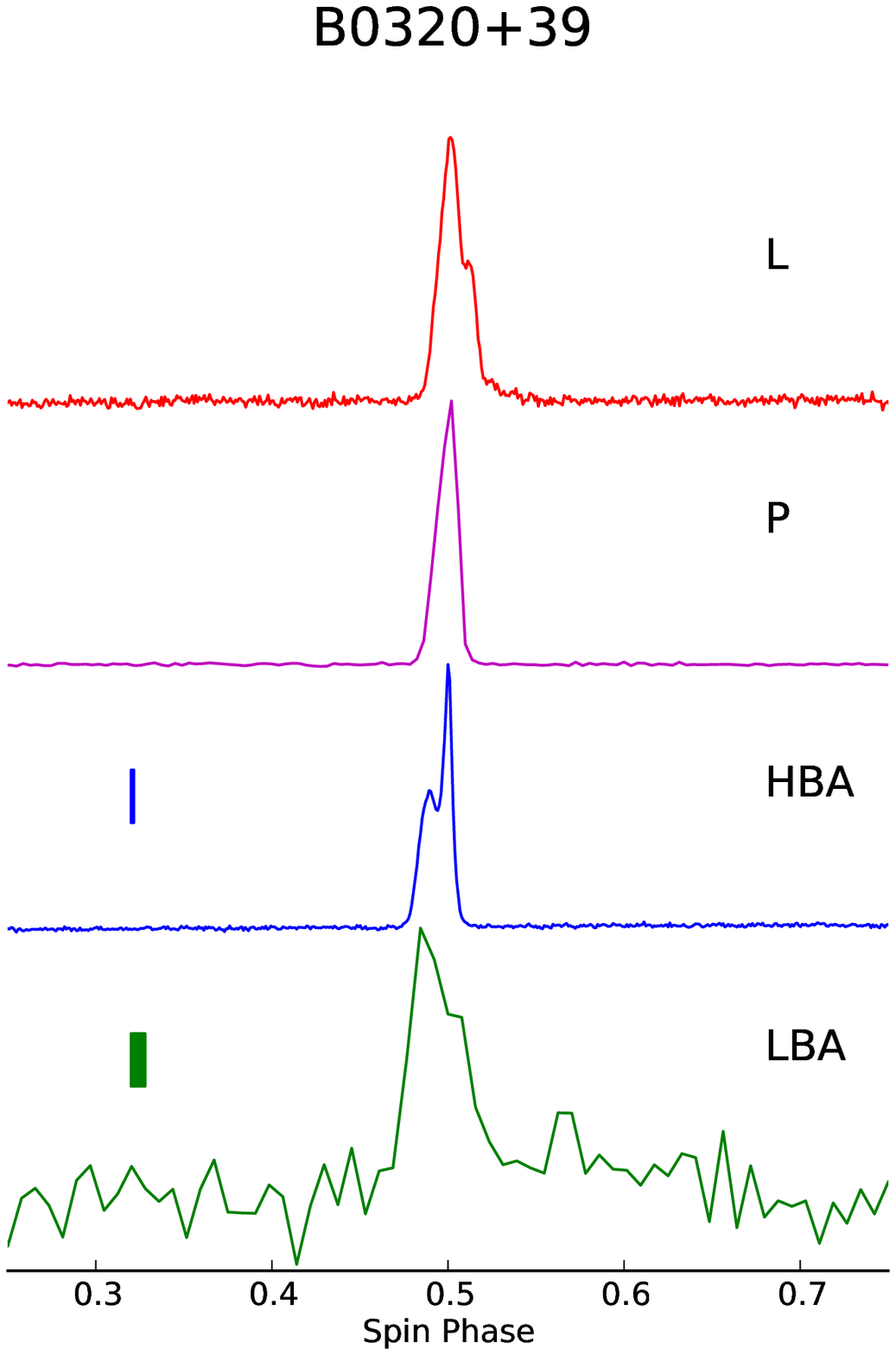} & \includegraphics[width=32mm]{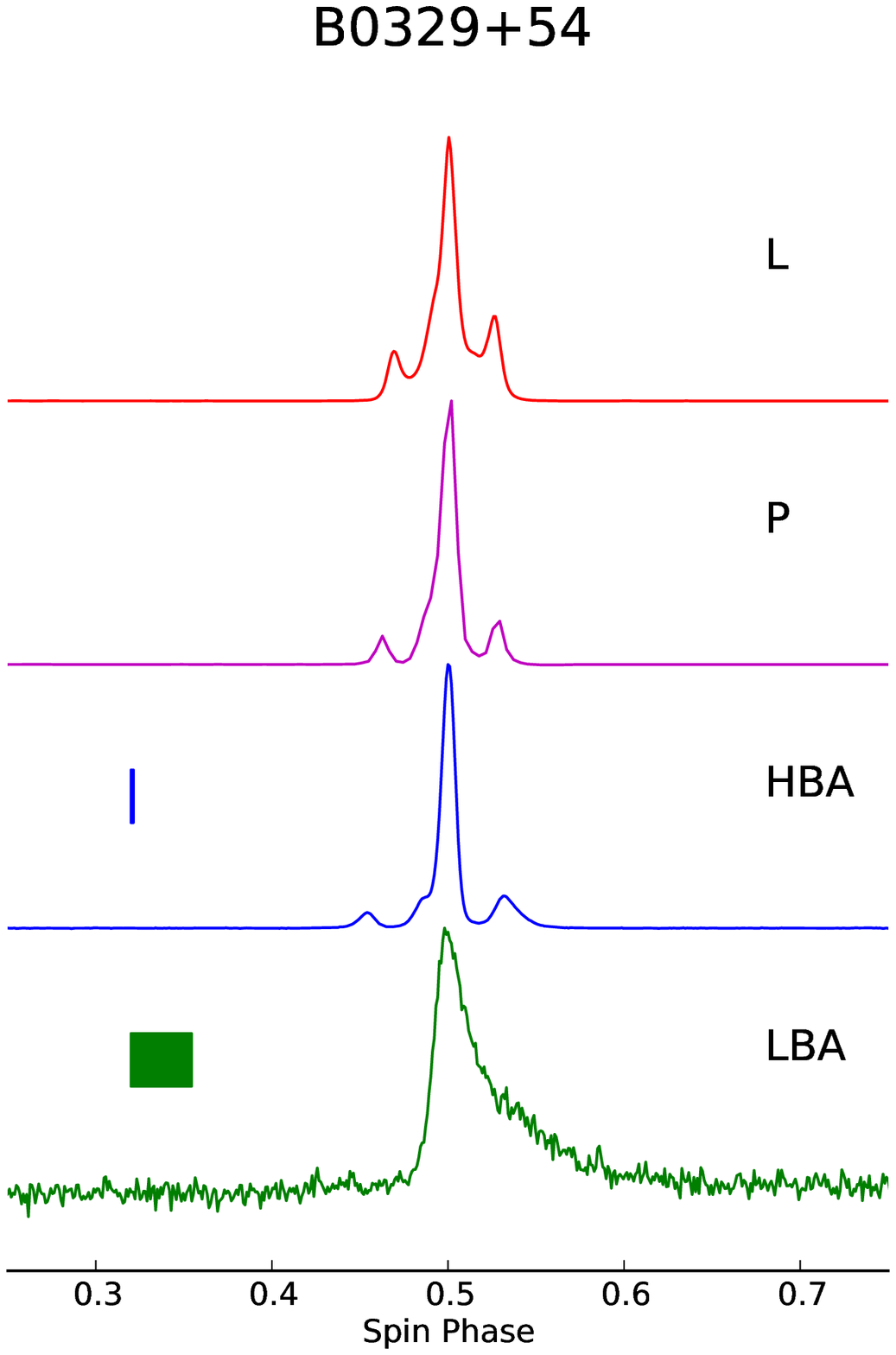} &   \includegraphics[width=32mm]{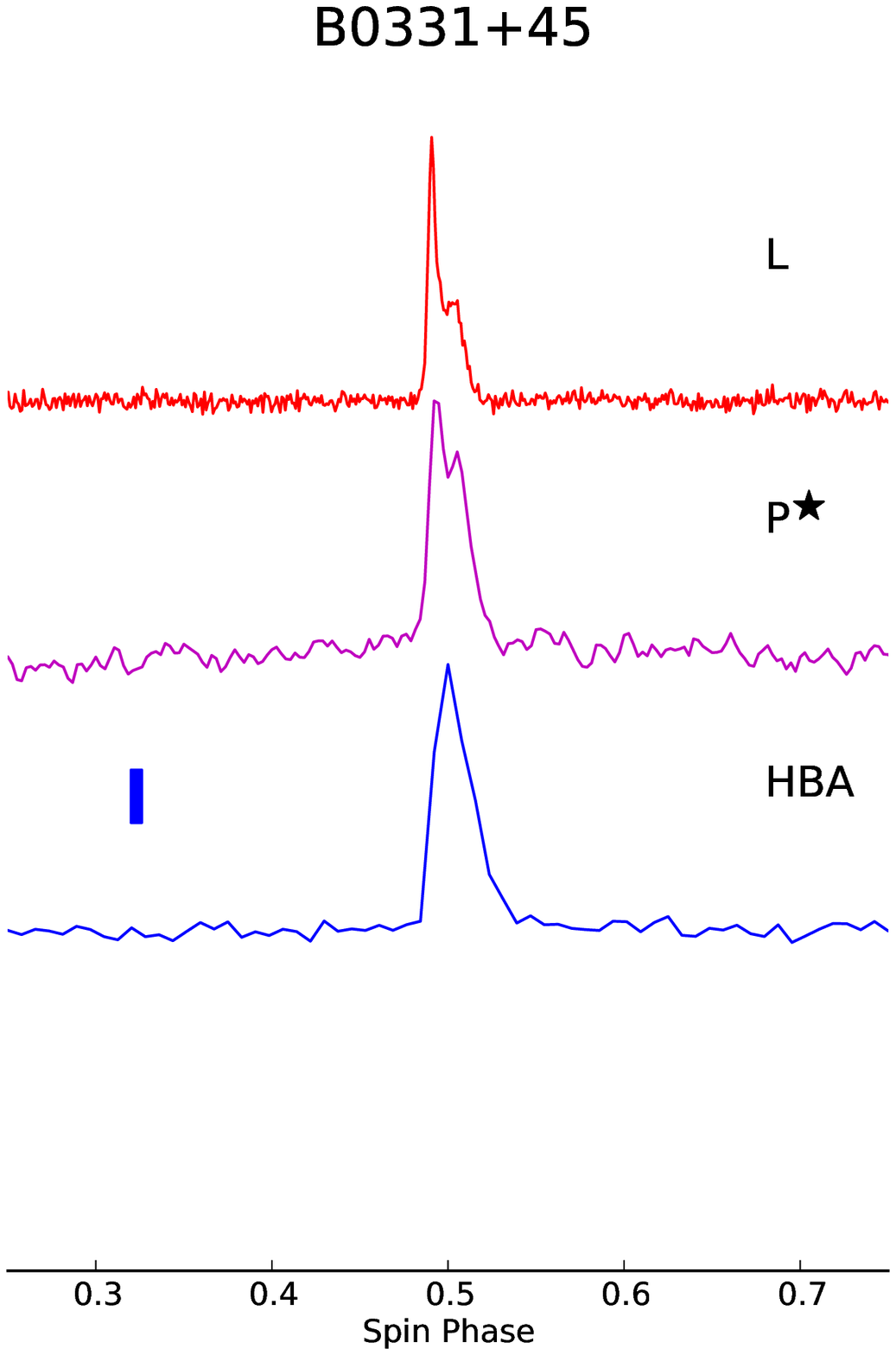} \\

\end{tabular}
\end{figure*}

\begin{figure*}
\centering
\begin{tabular}{cccc}

\includegraphics[width=32mm]{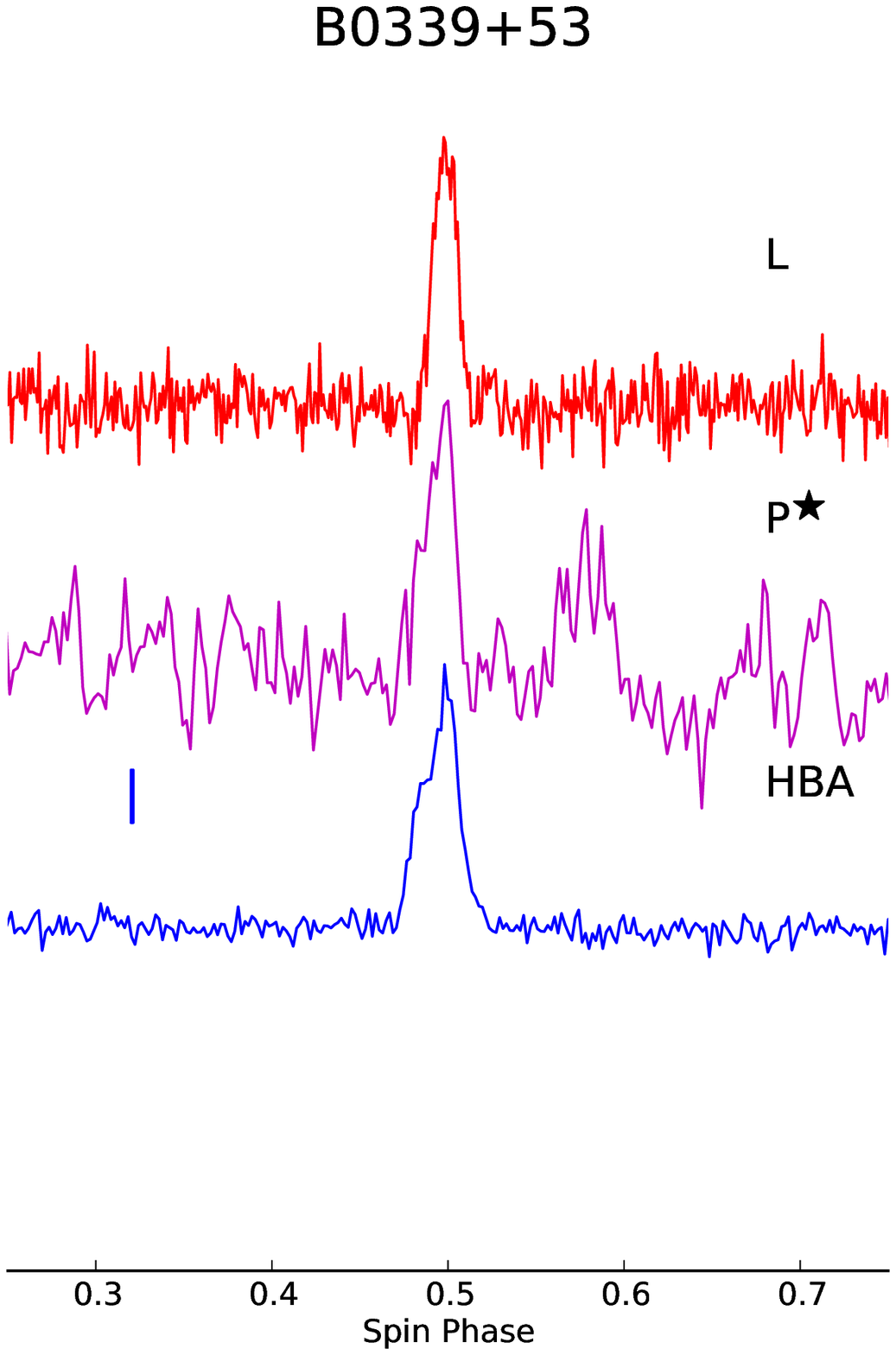} &  \includegraphics[width=32mm]{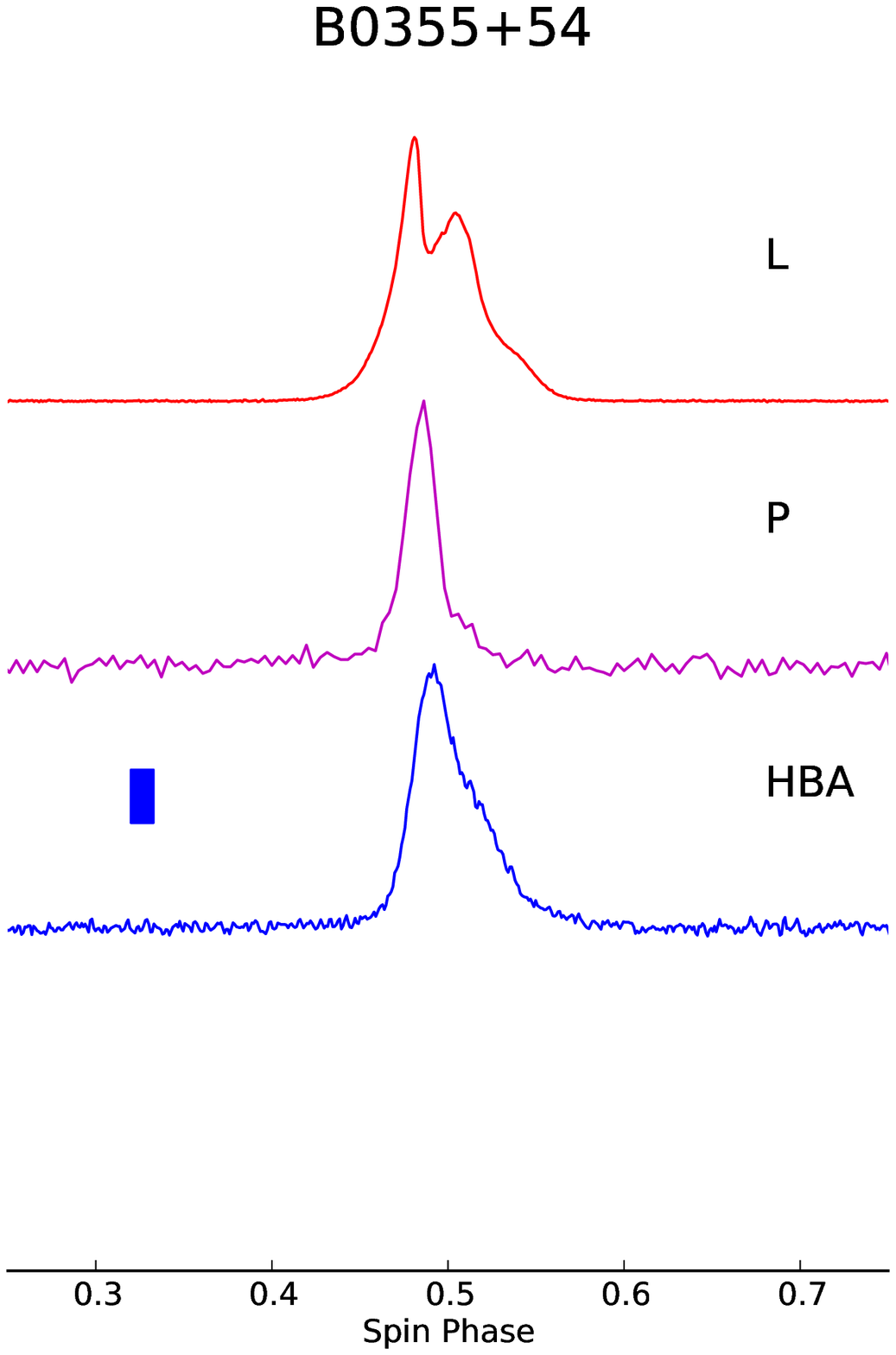} &   \includegraphics[width=32mm]{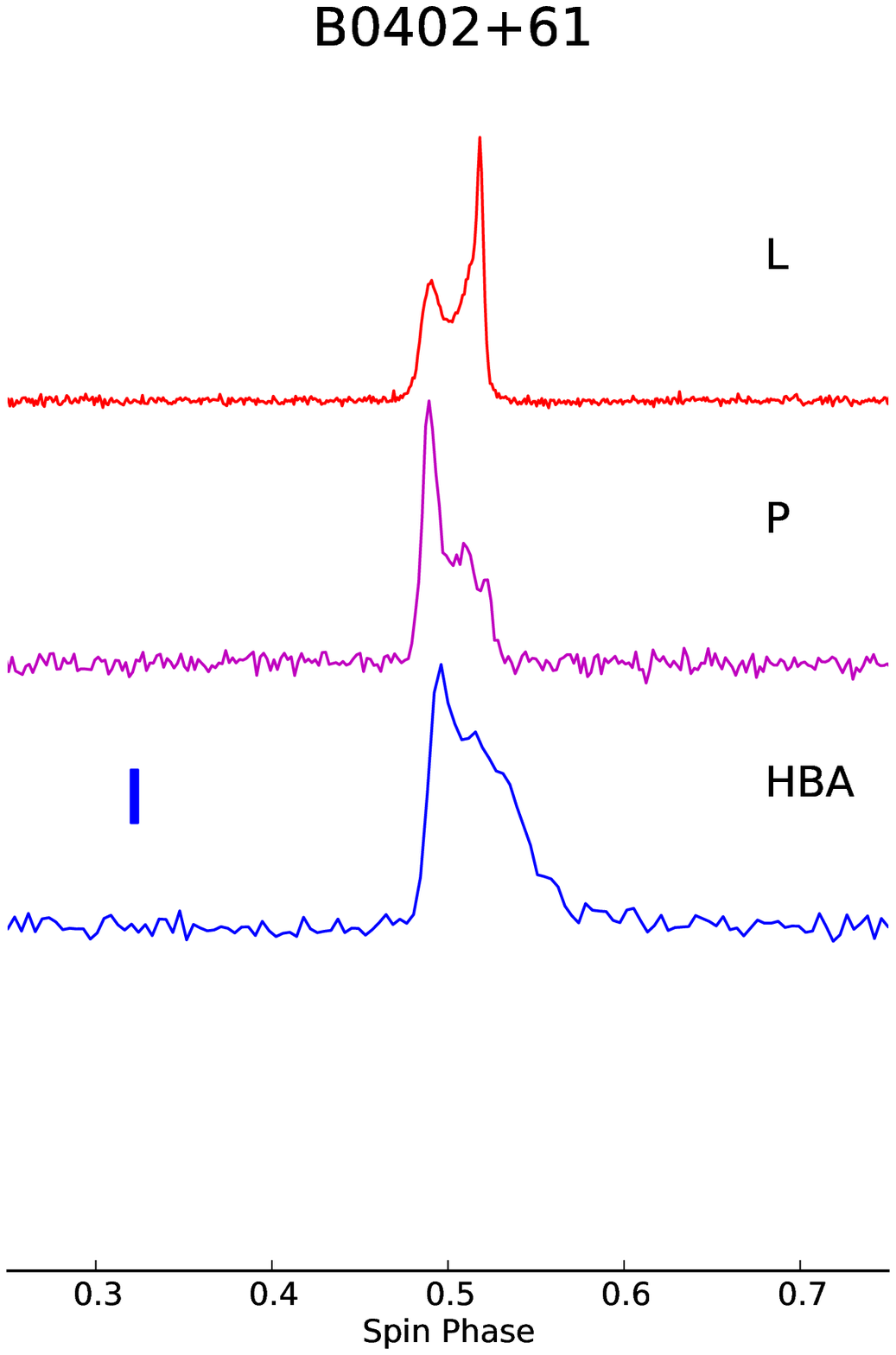} & \includegraphics[width=32mm]{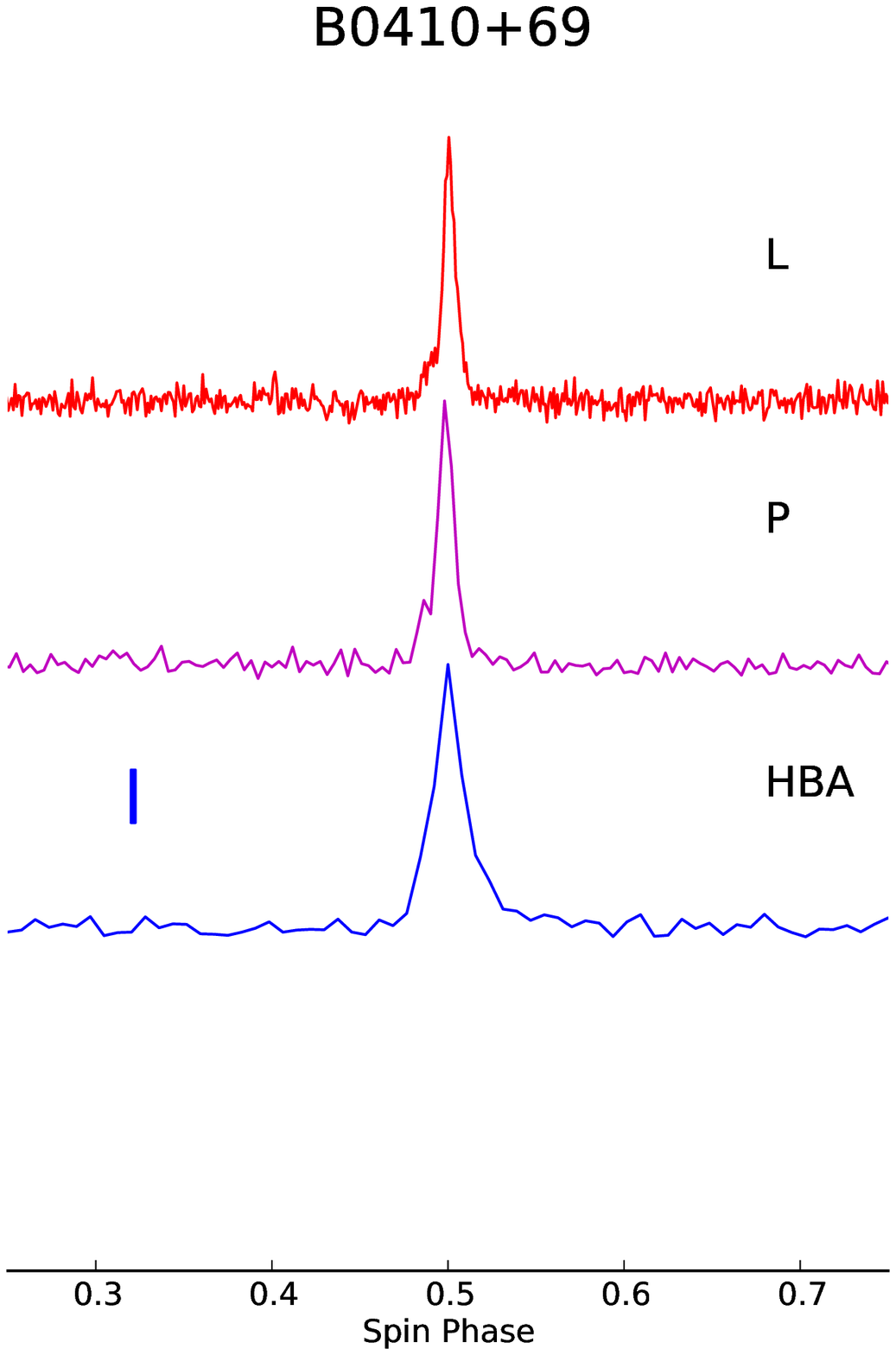} \\

 \includegraphics[width=32mm]{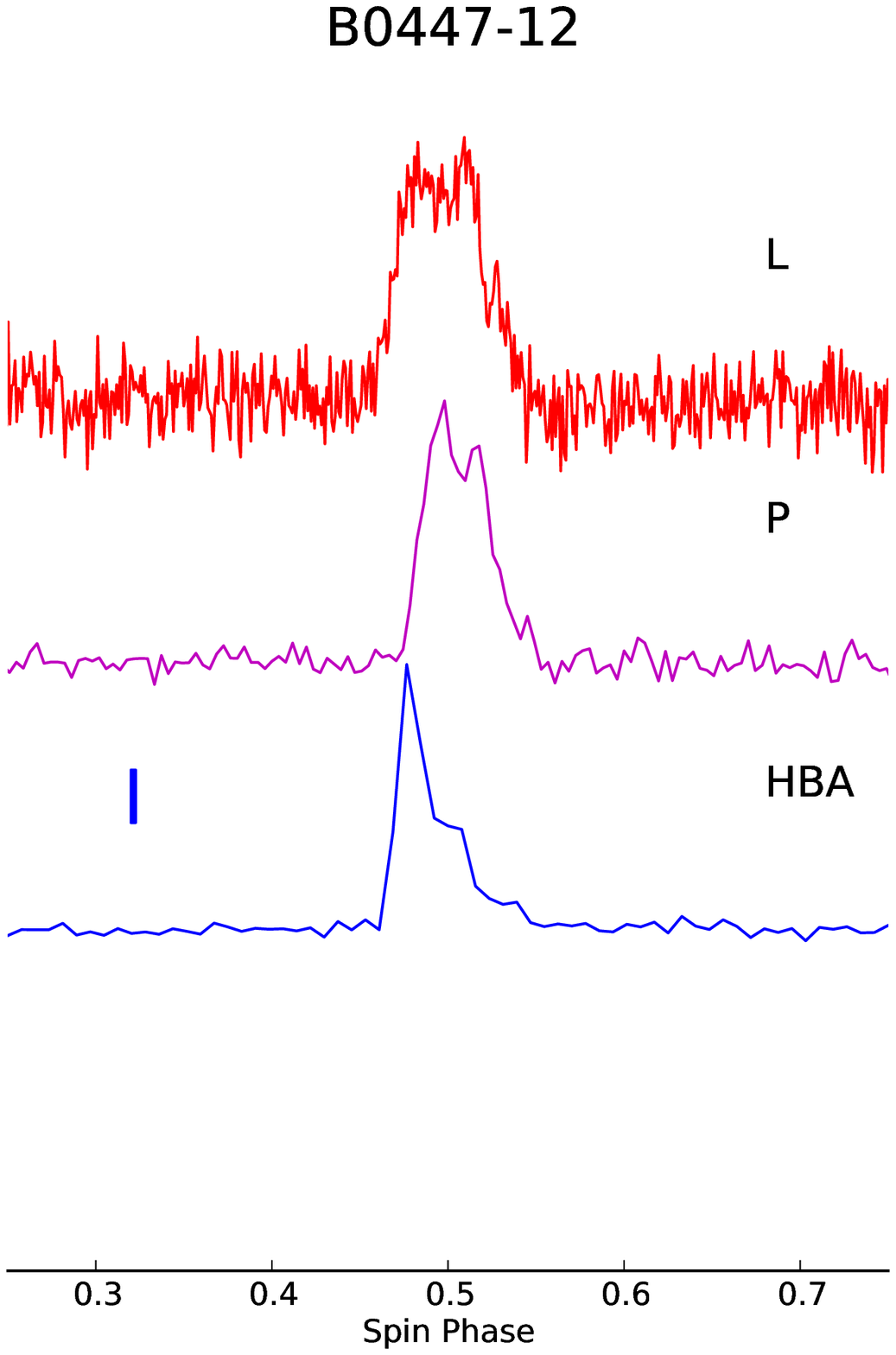} &  \includegraphics[width=32mm]{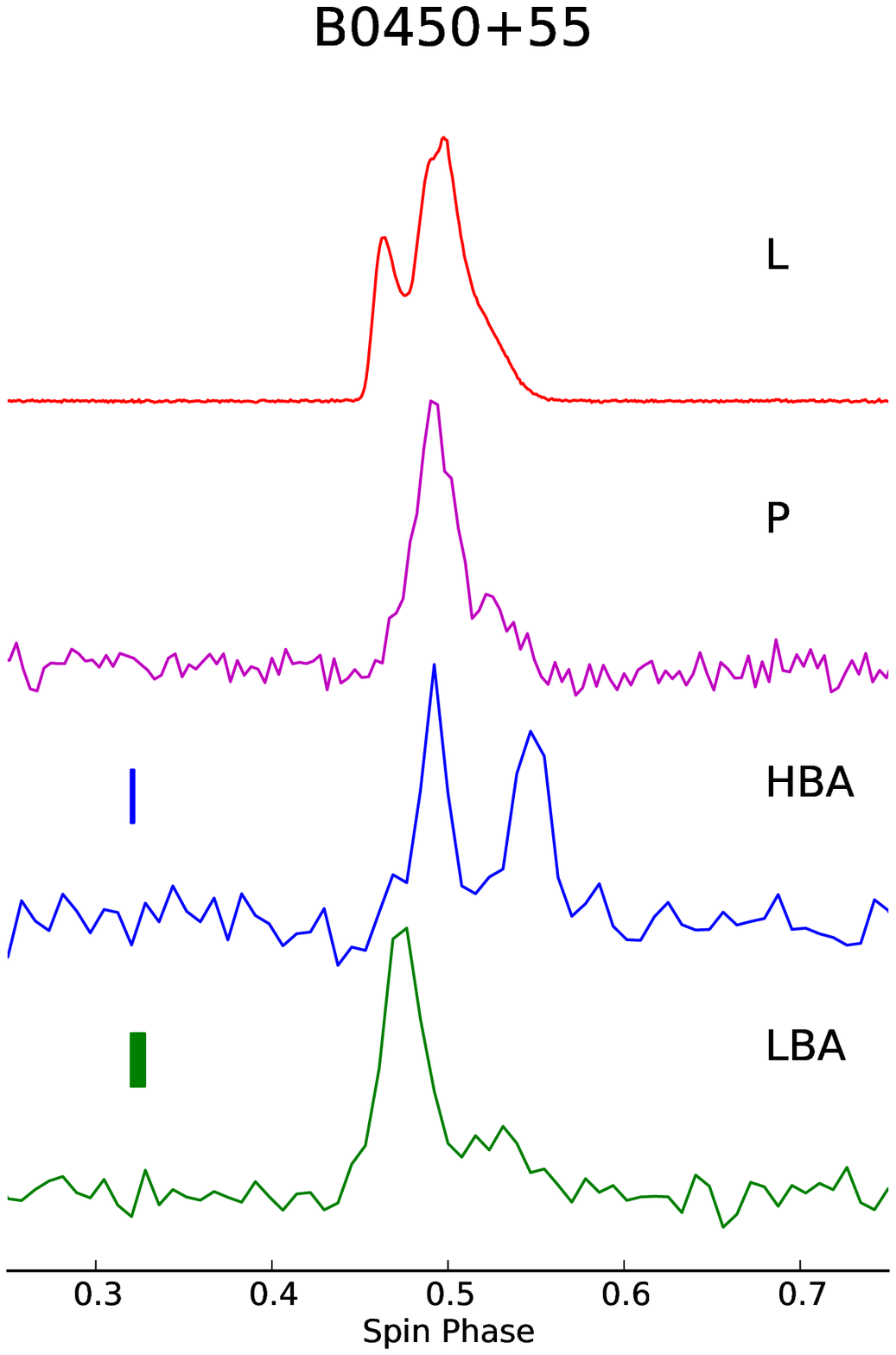} &   \includegraphics[width=32mm]{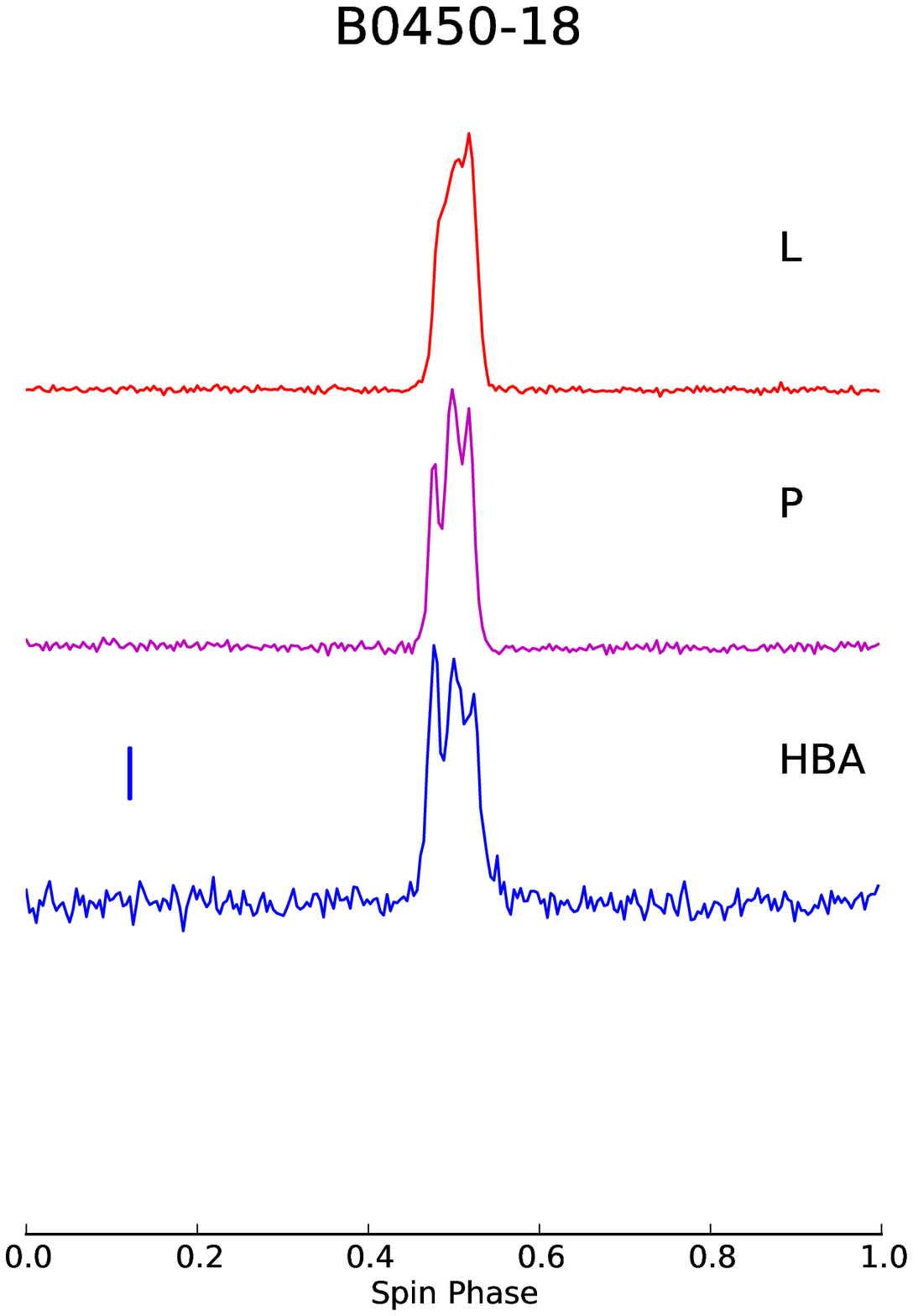} &   \includegraphics[width=32mm]{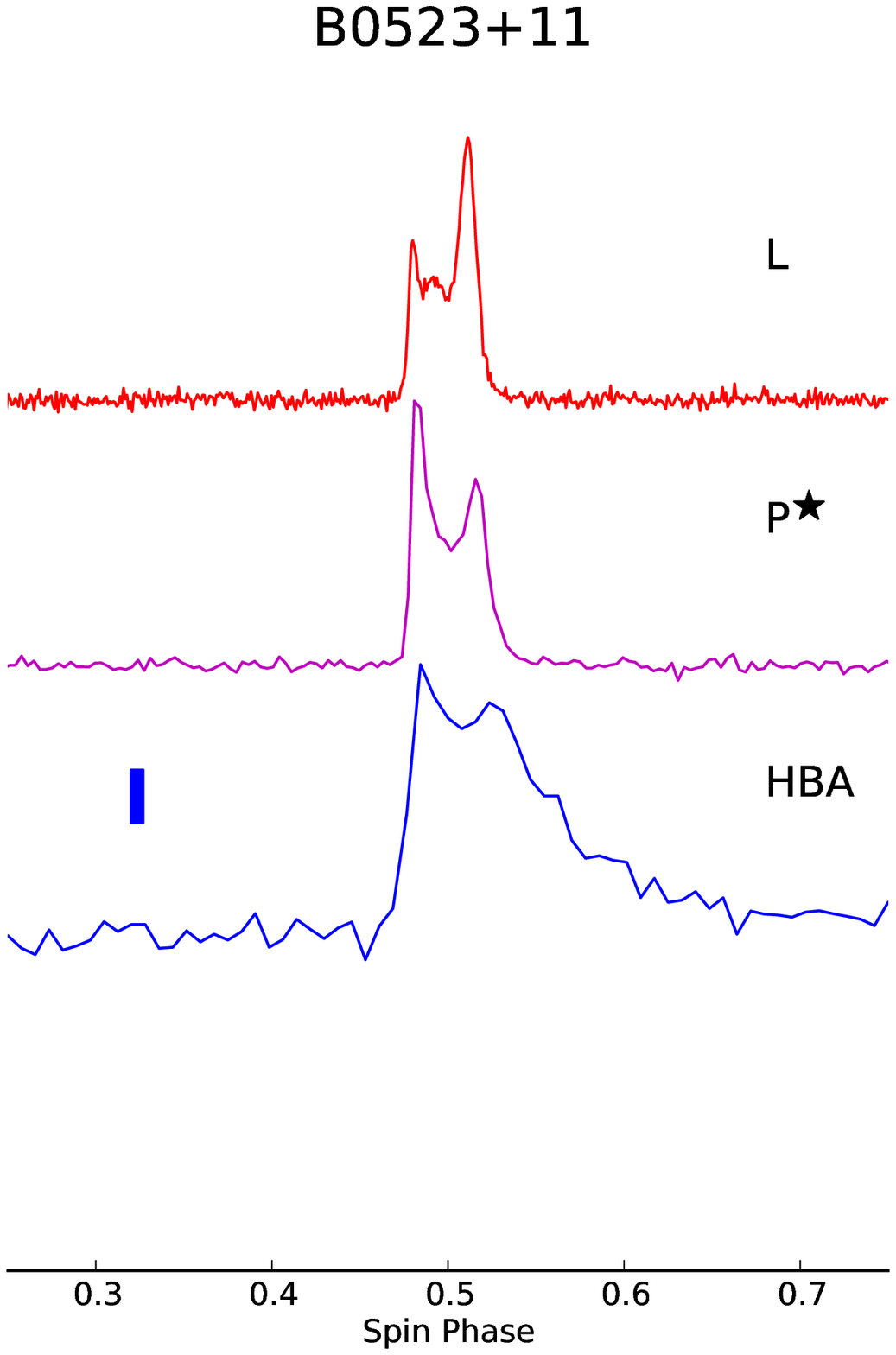}  \\

 \includegraphics[width=32mm]{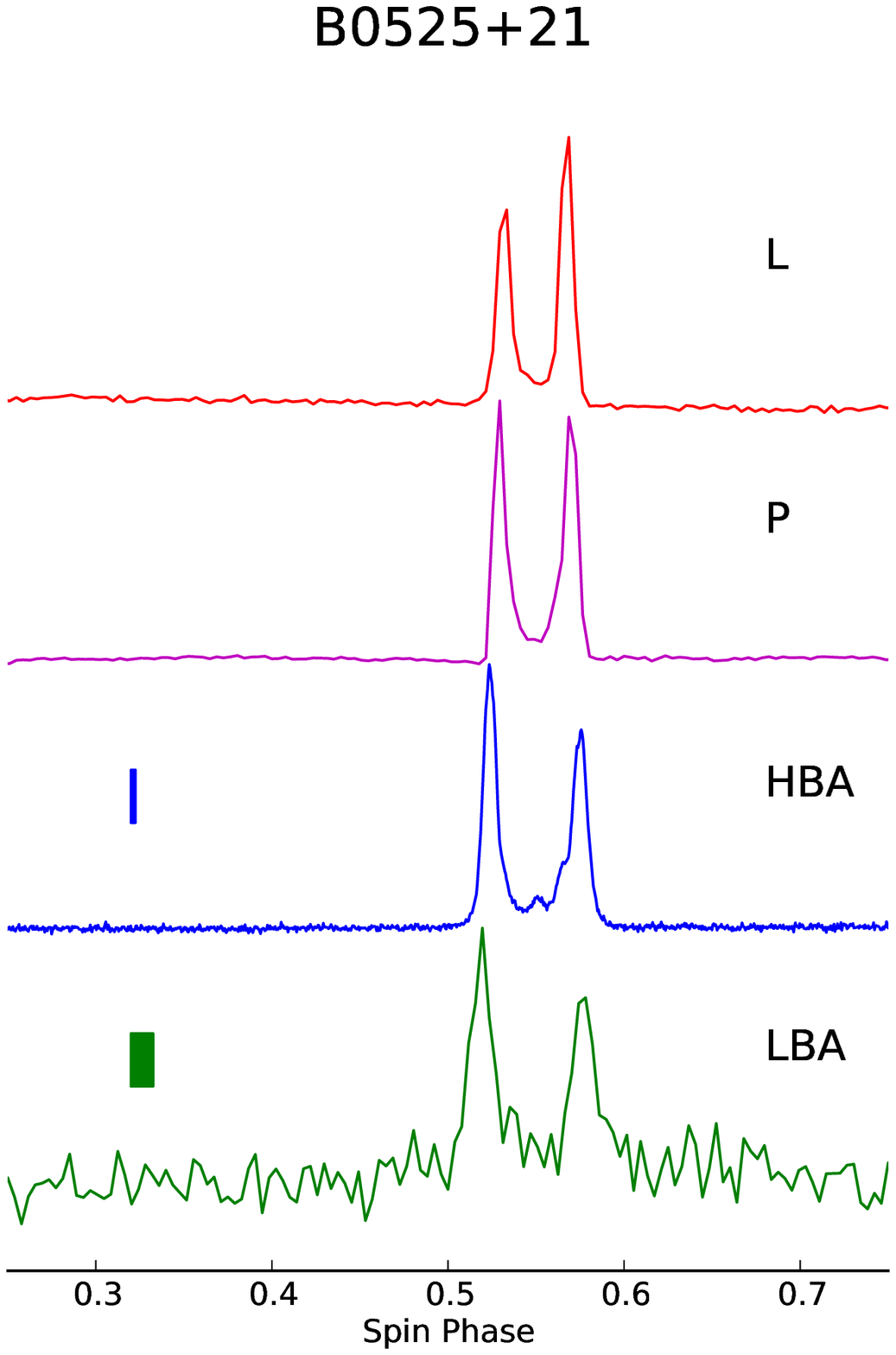} &  \includegraphics[width=32mm]{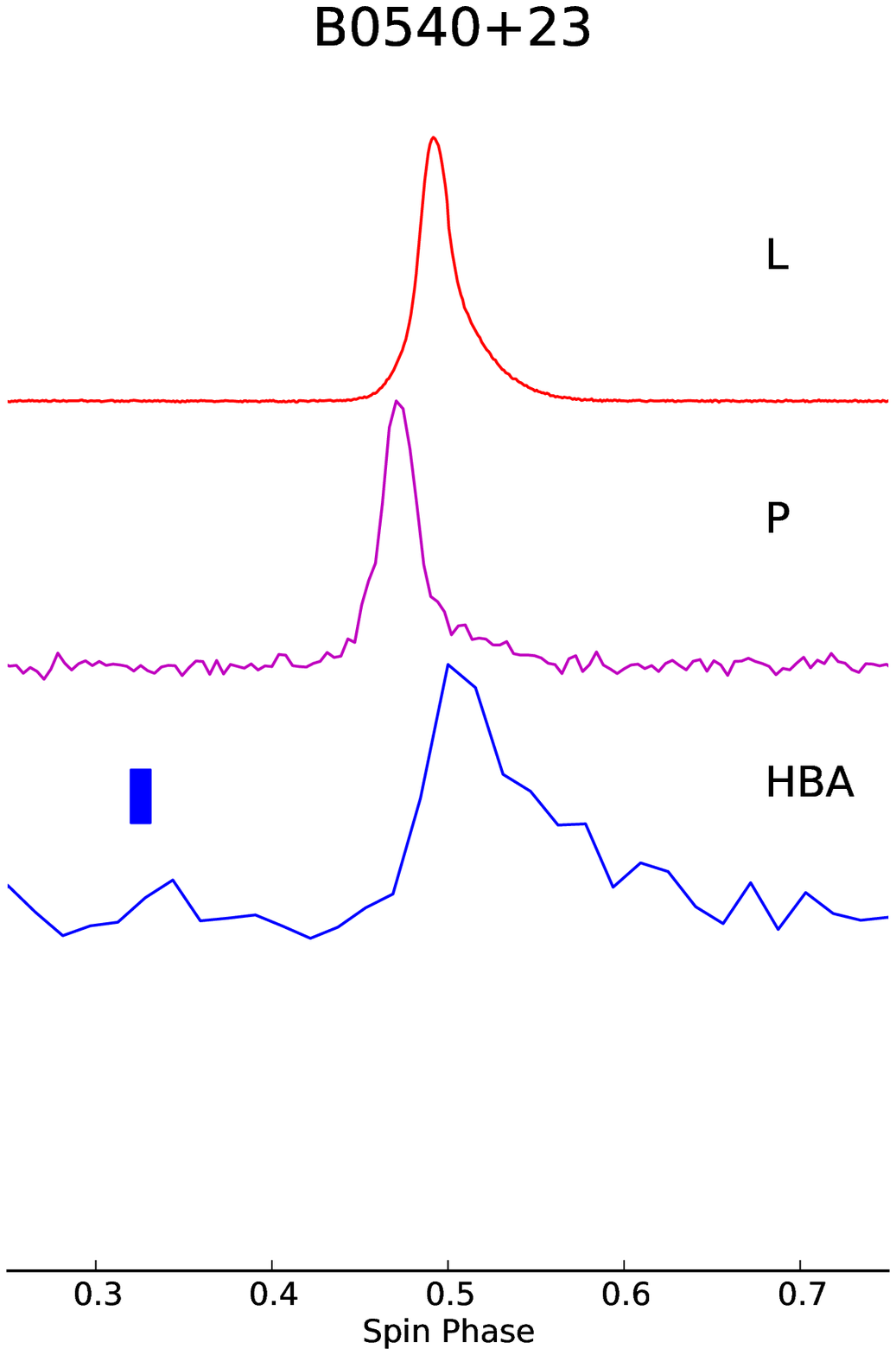} &   \includegraphics[width=32mm]{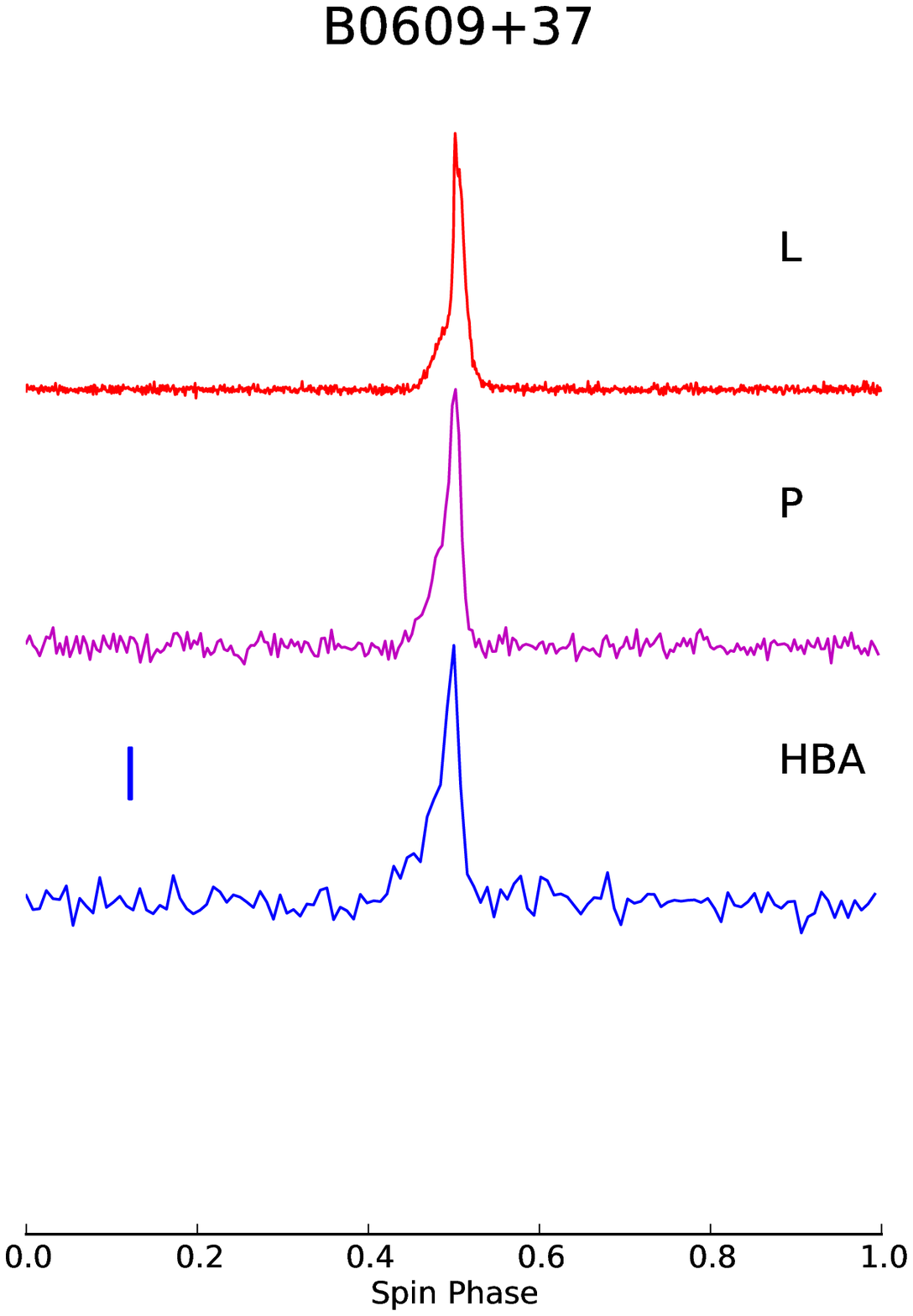} &   \includegraphics[width=32mm]{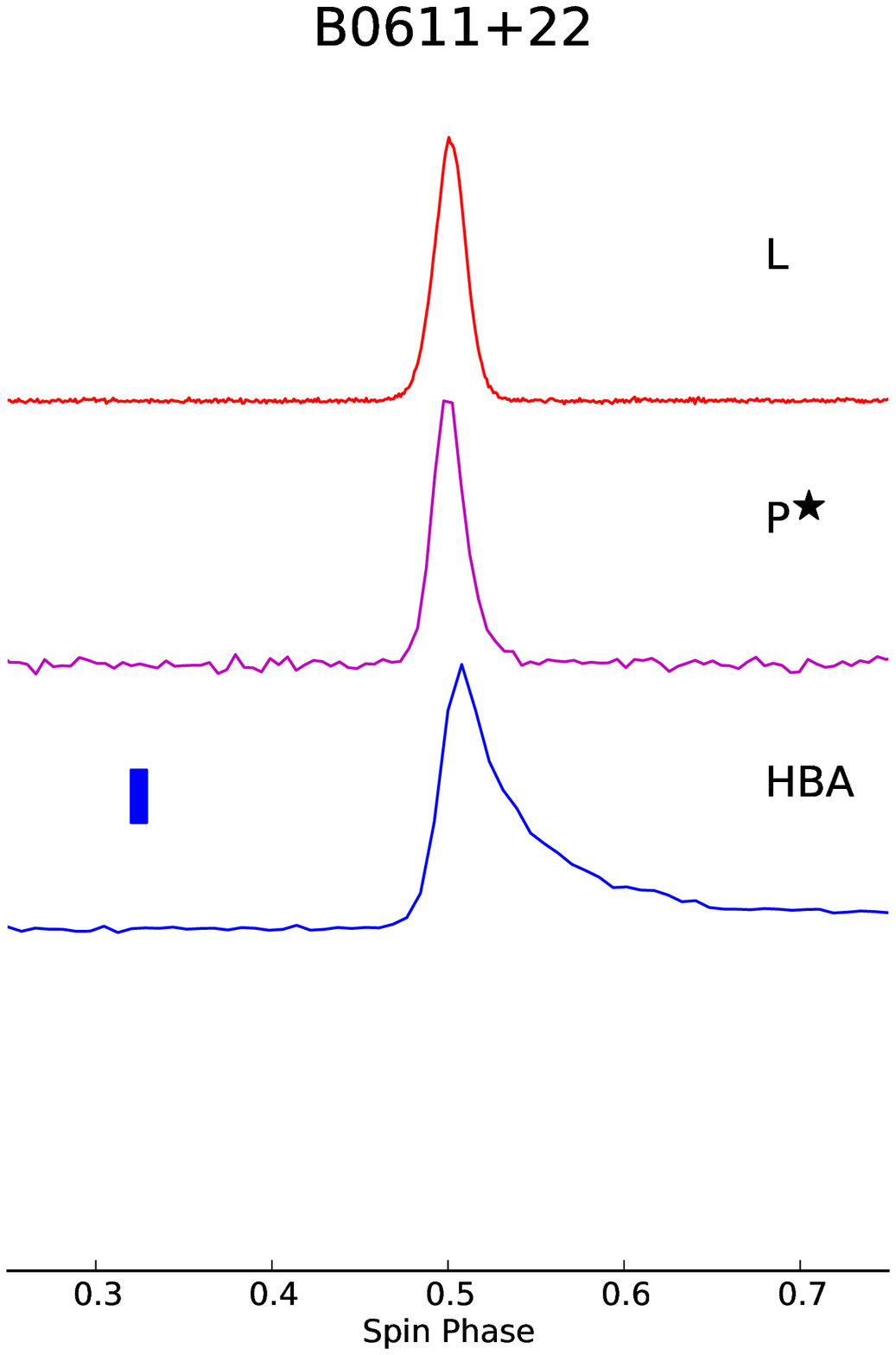} \\

 \includegraphics[width=32mm]{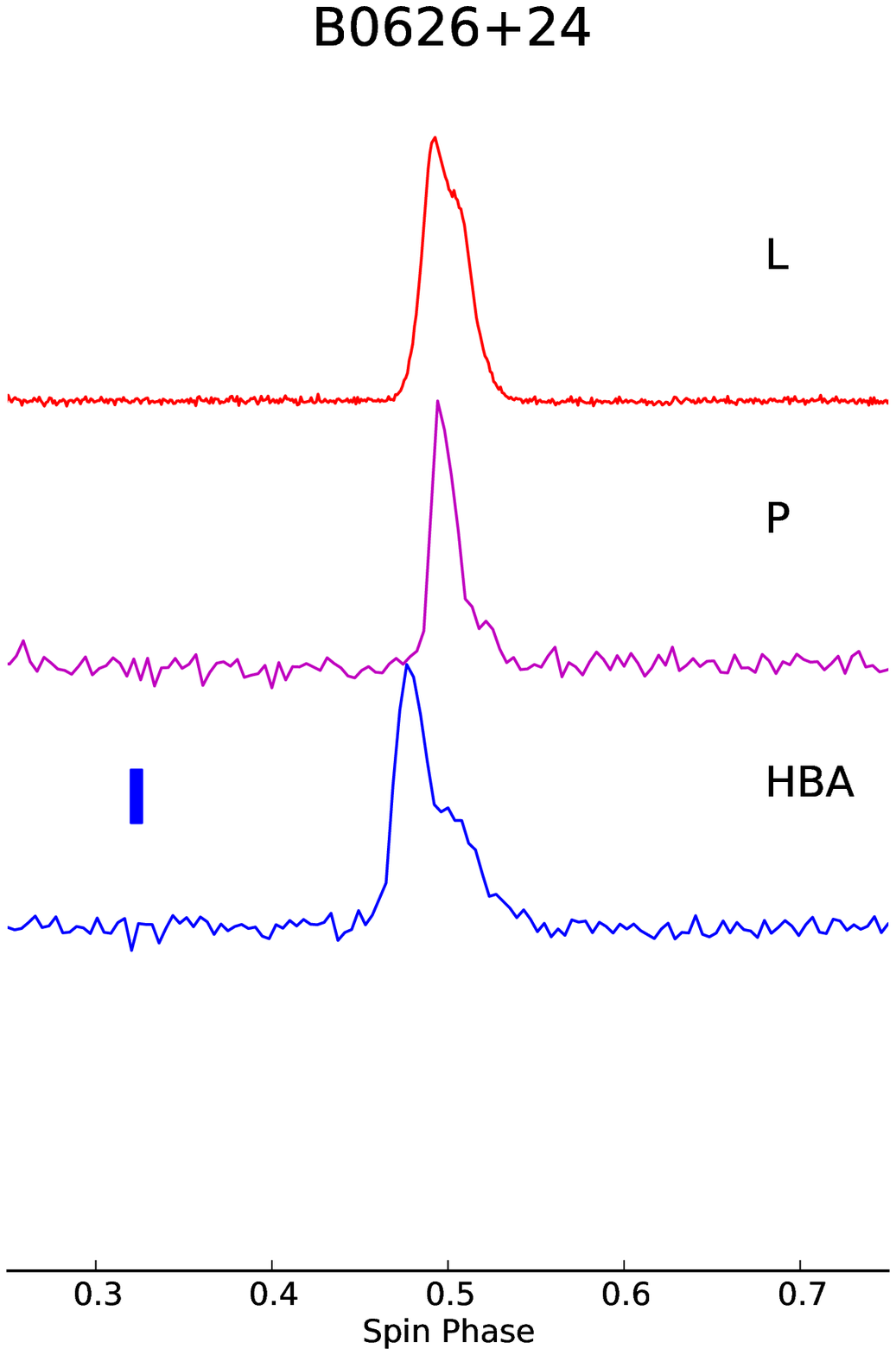} &  \includegraphics[width=32mm]{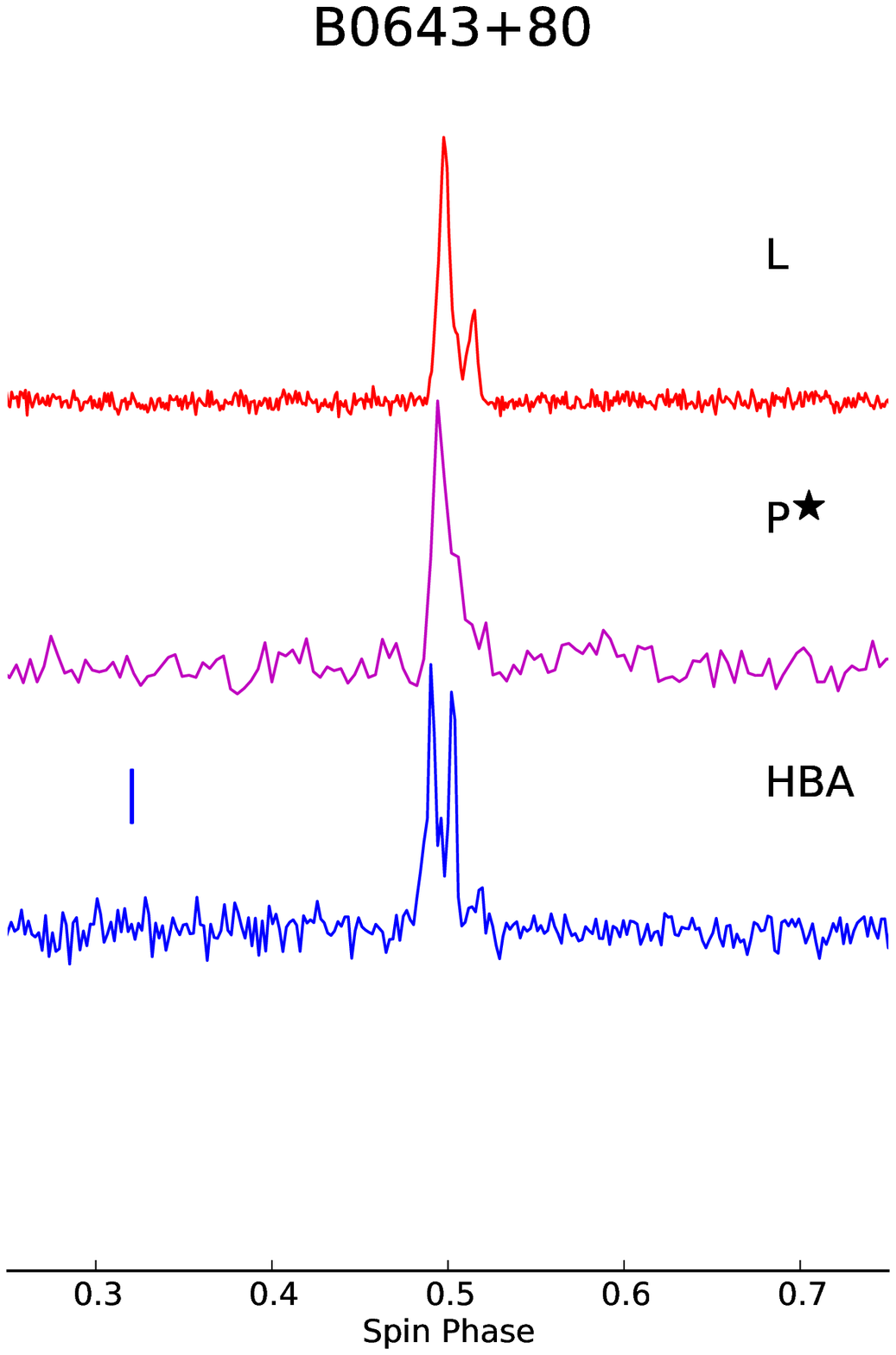} &   \includegraphics[width=32mm]{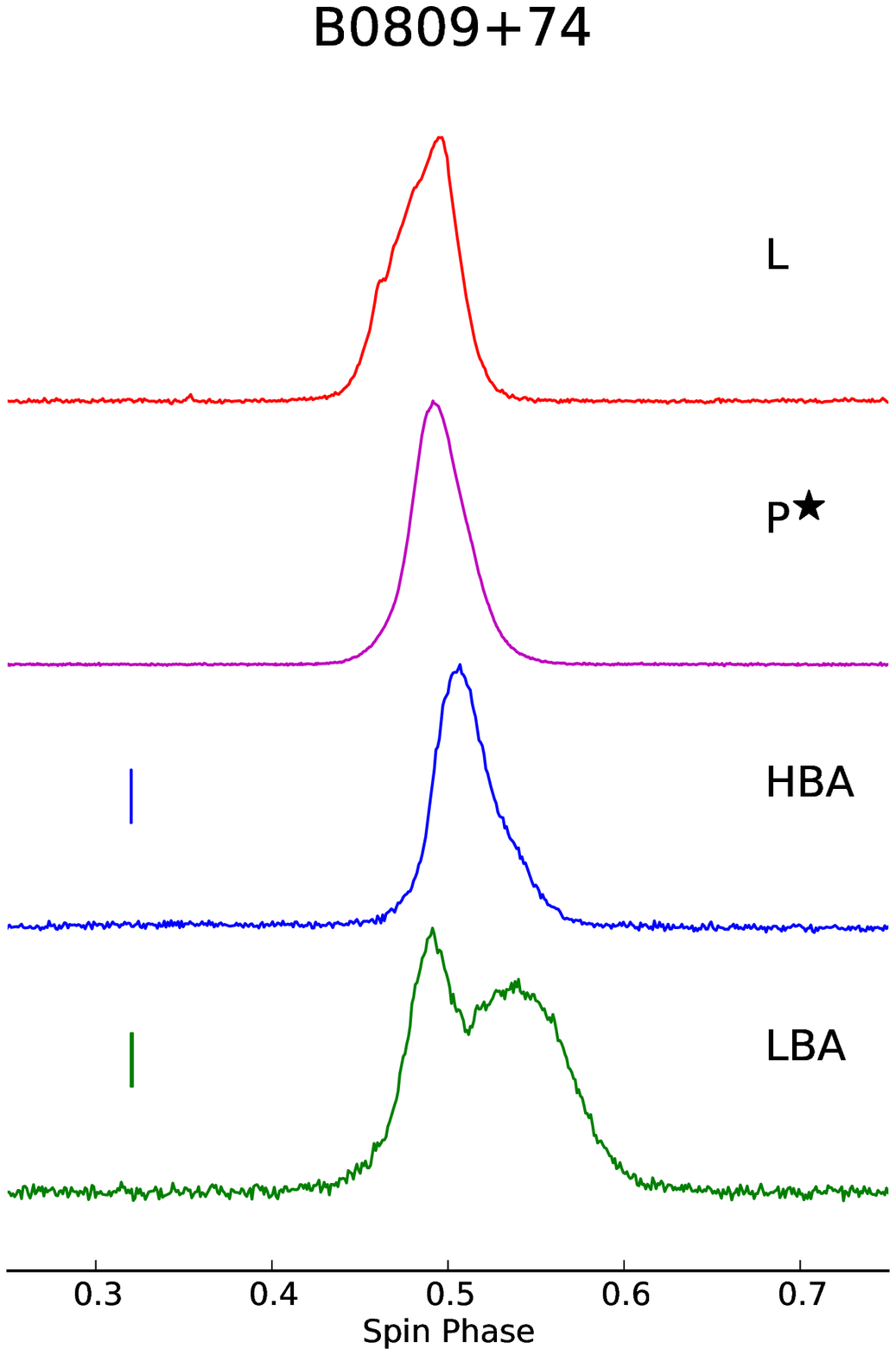} &   \includegraphics[width=32mm]{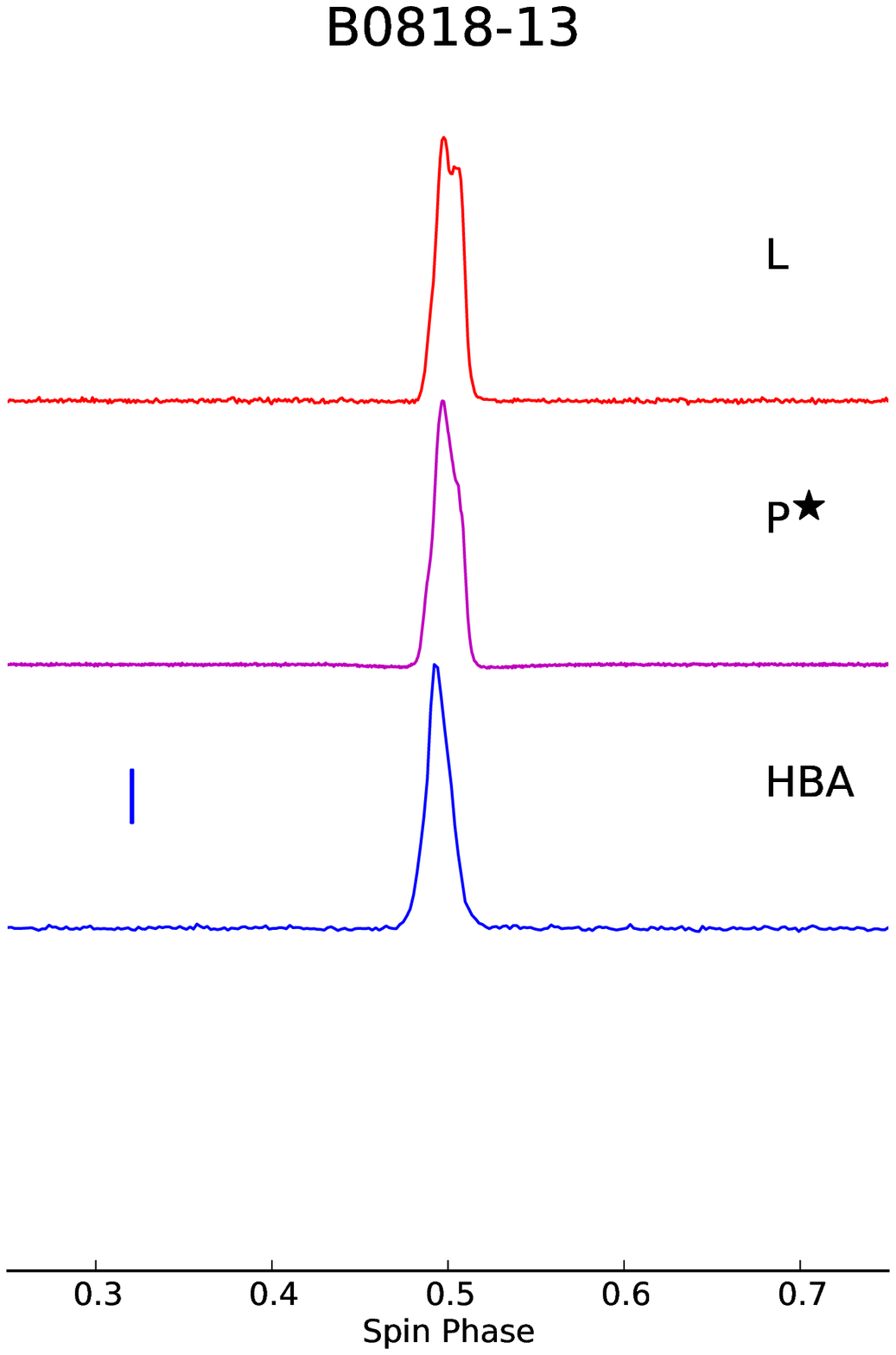} \\

 \includegraphics[width=32mm]{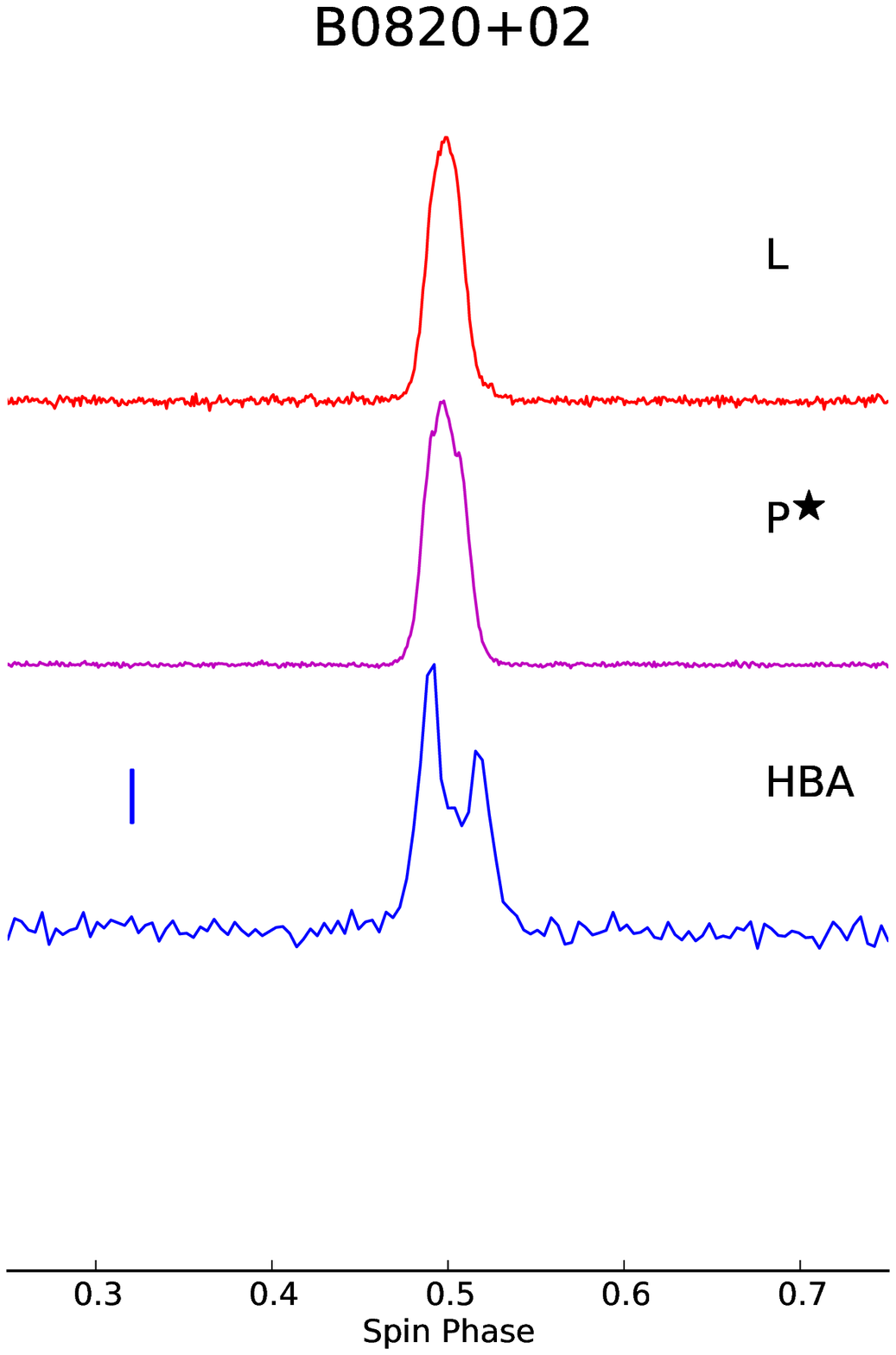} &  \includegraphics[width=32mm]{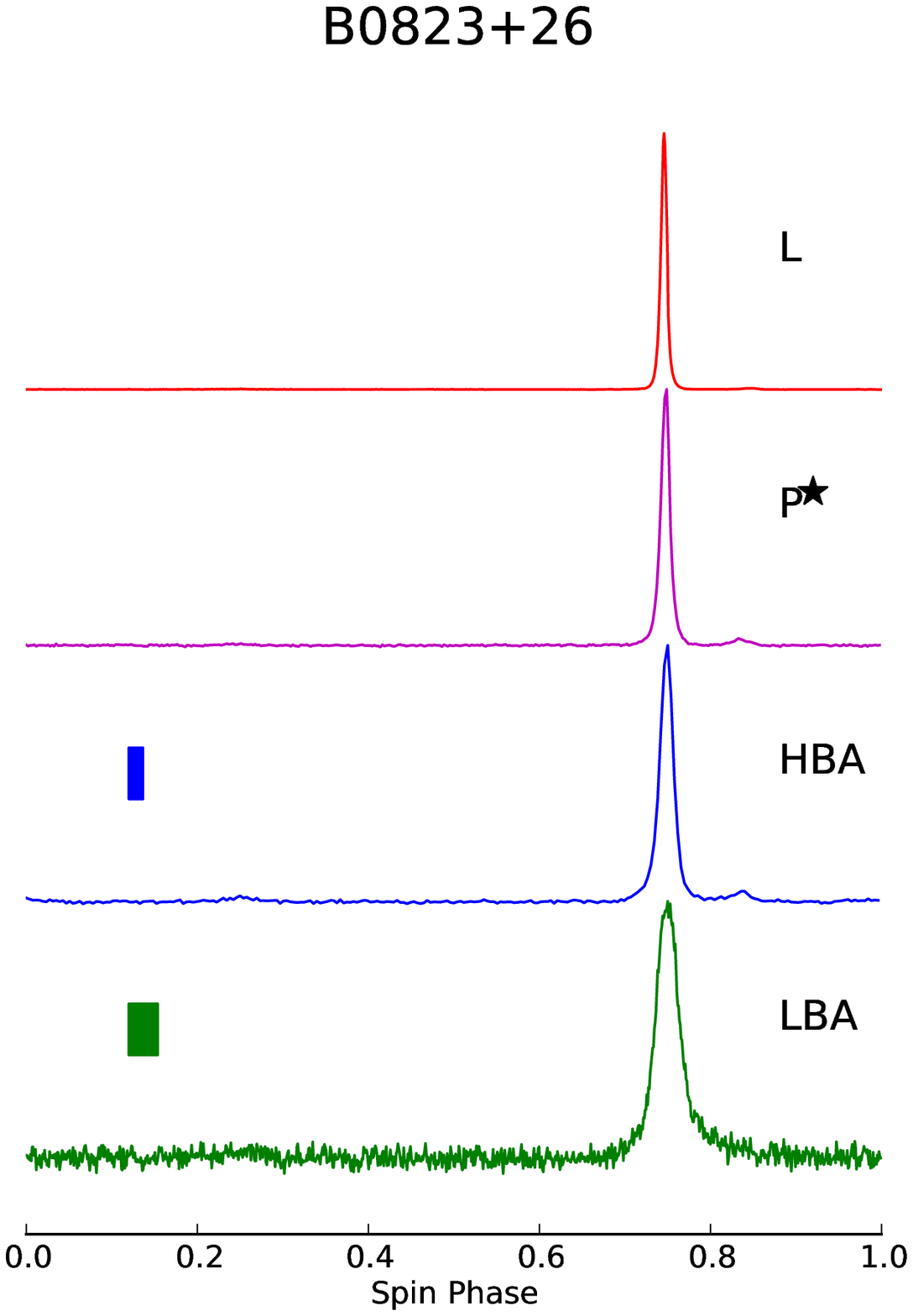} &   \includegraphics[width=32mm]{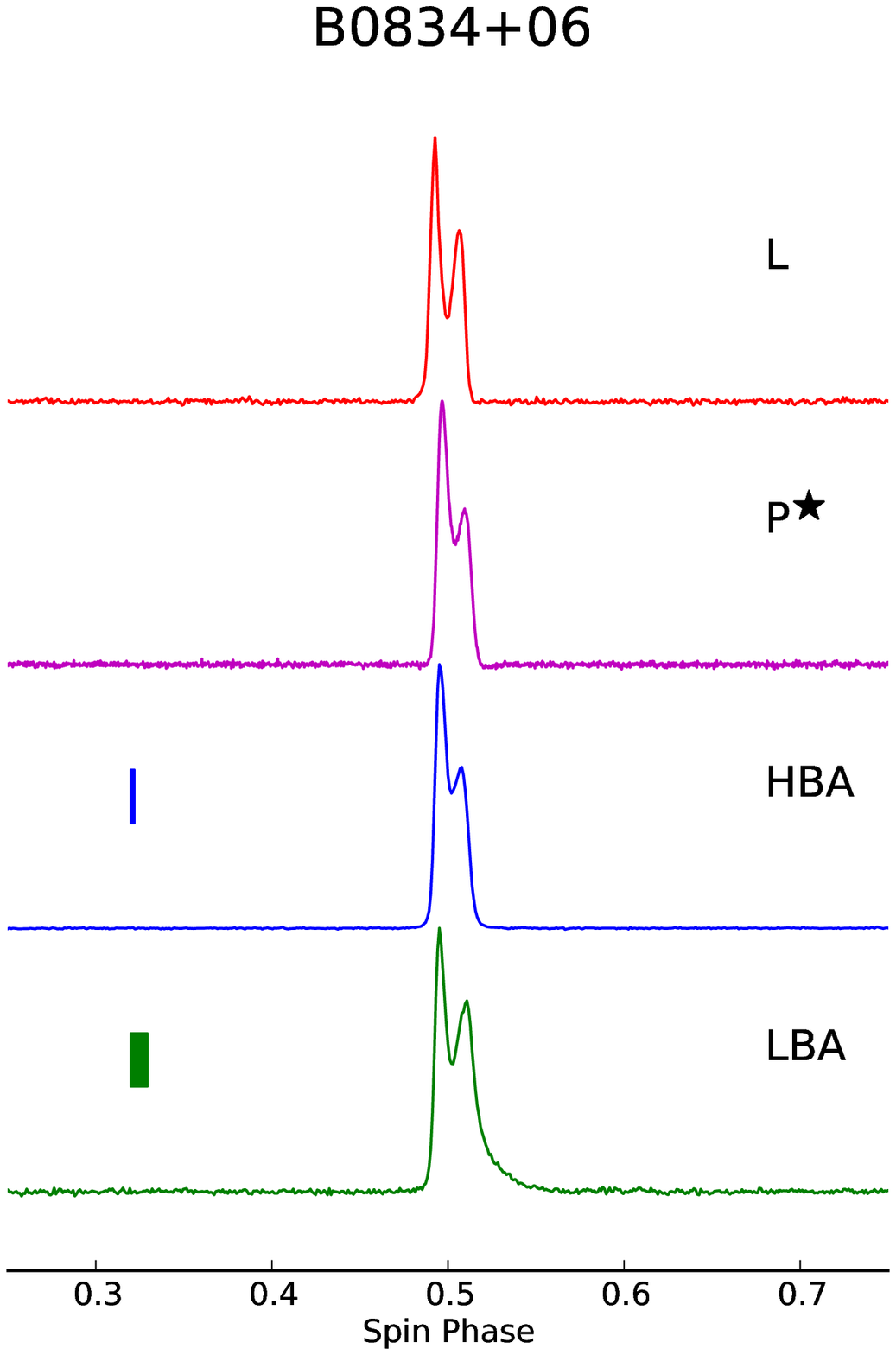} &   \includegraphics[width=32mm]{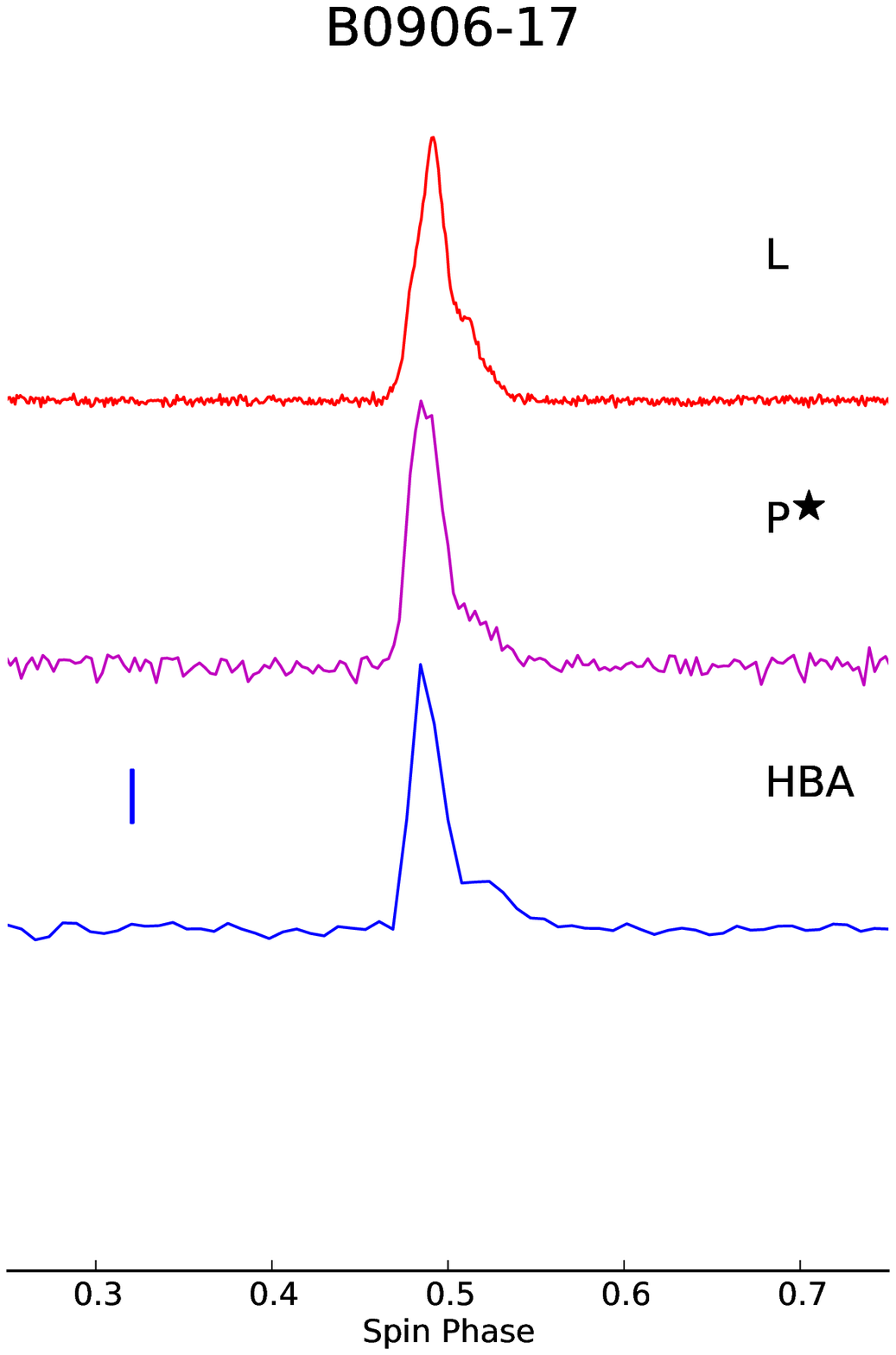} \\

\end{tabular}
\caption*{\label{fig:100_3}
Fig.\ref{fig:100}: continued
}
\end{figure*}

\begin{figure*}
\centering
\begin{tabular}{cccc}

 \includegraphics[width=32mm]{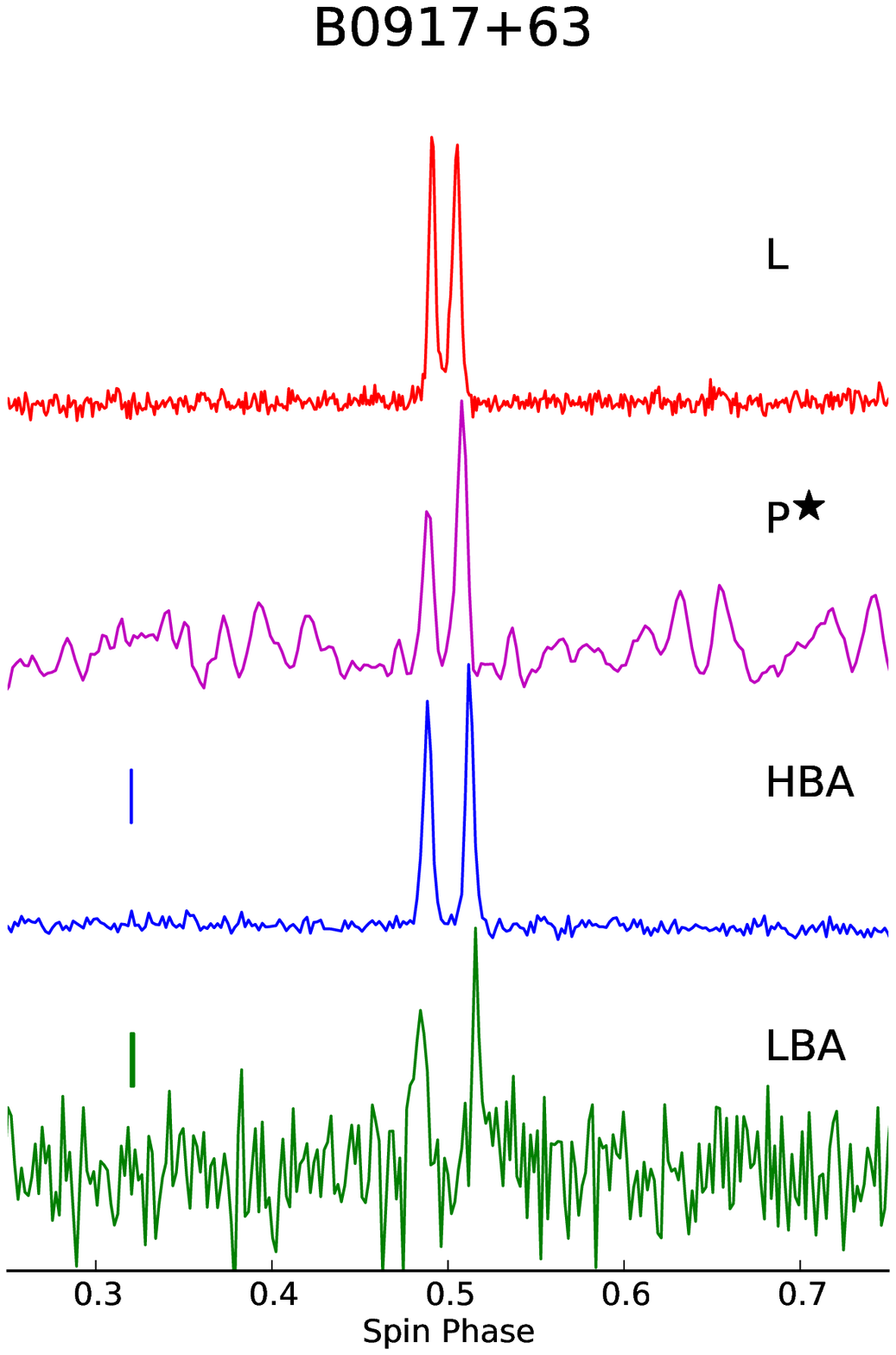} & \includegraphics[width=32mm]{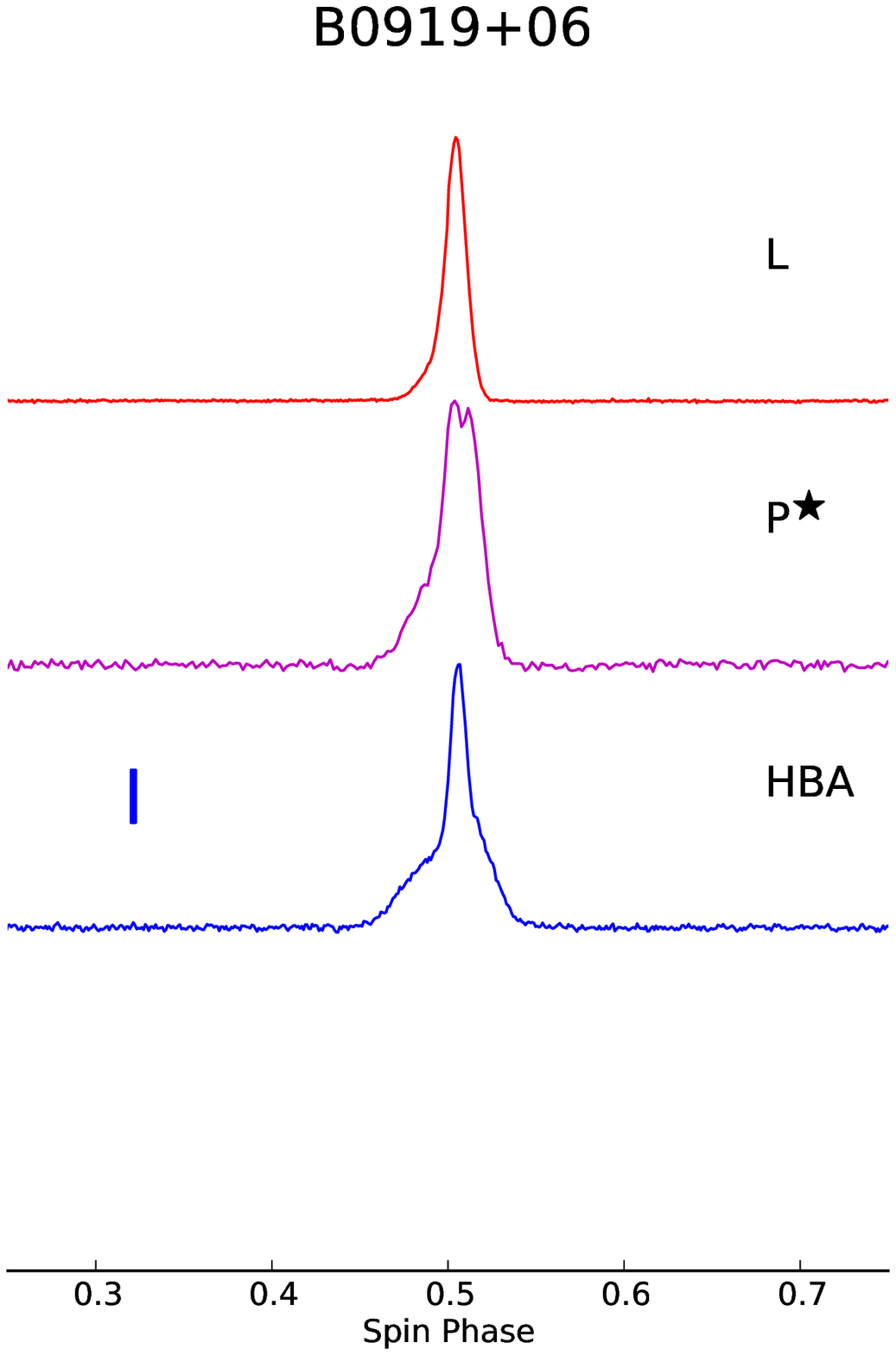} & \includegraphics[width=32mm]{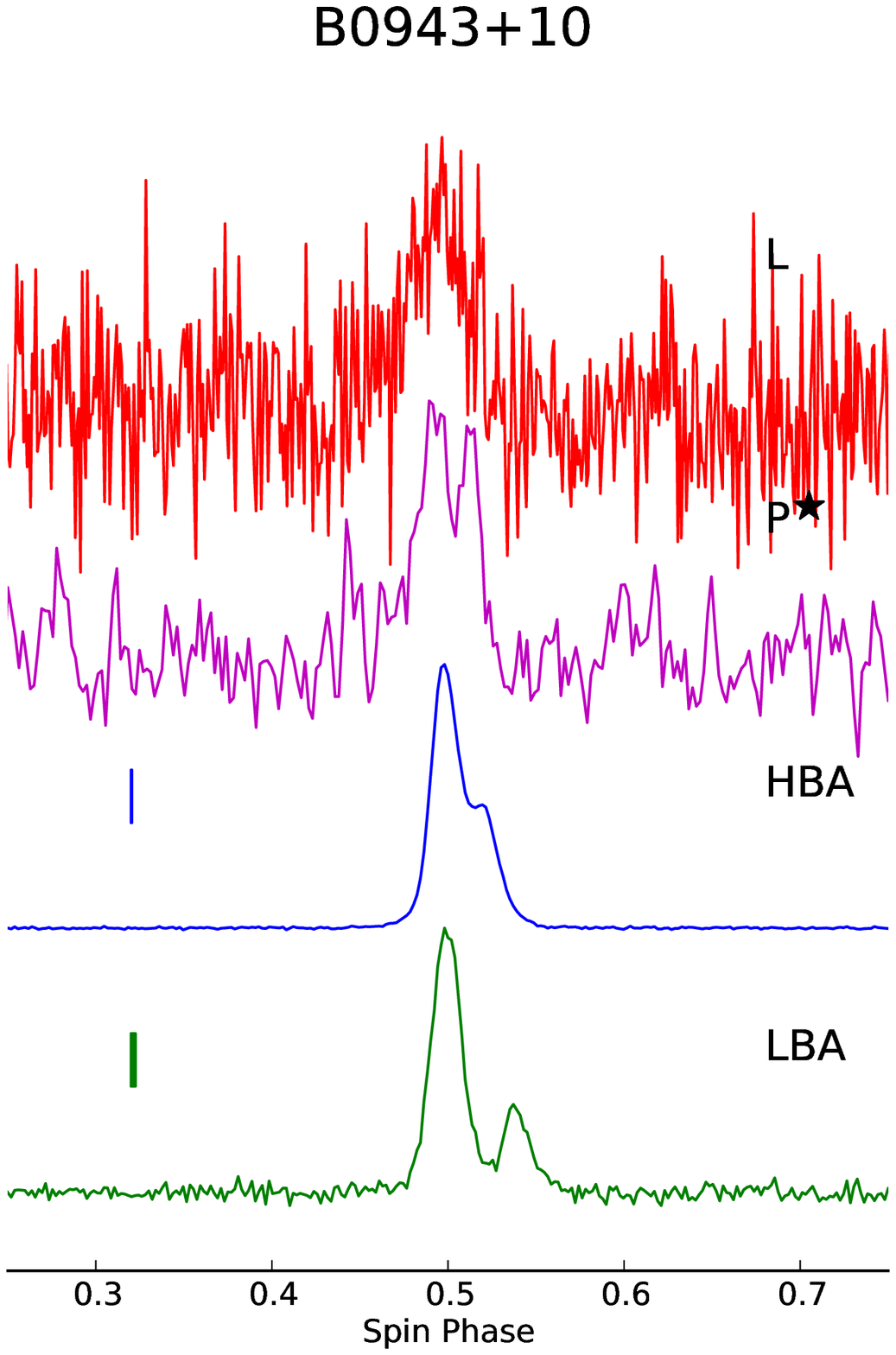} & \includegraphics[width=32mm]{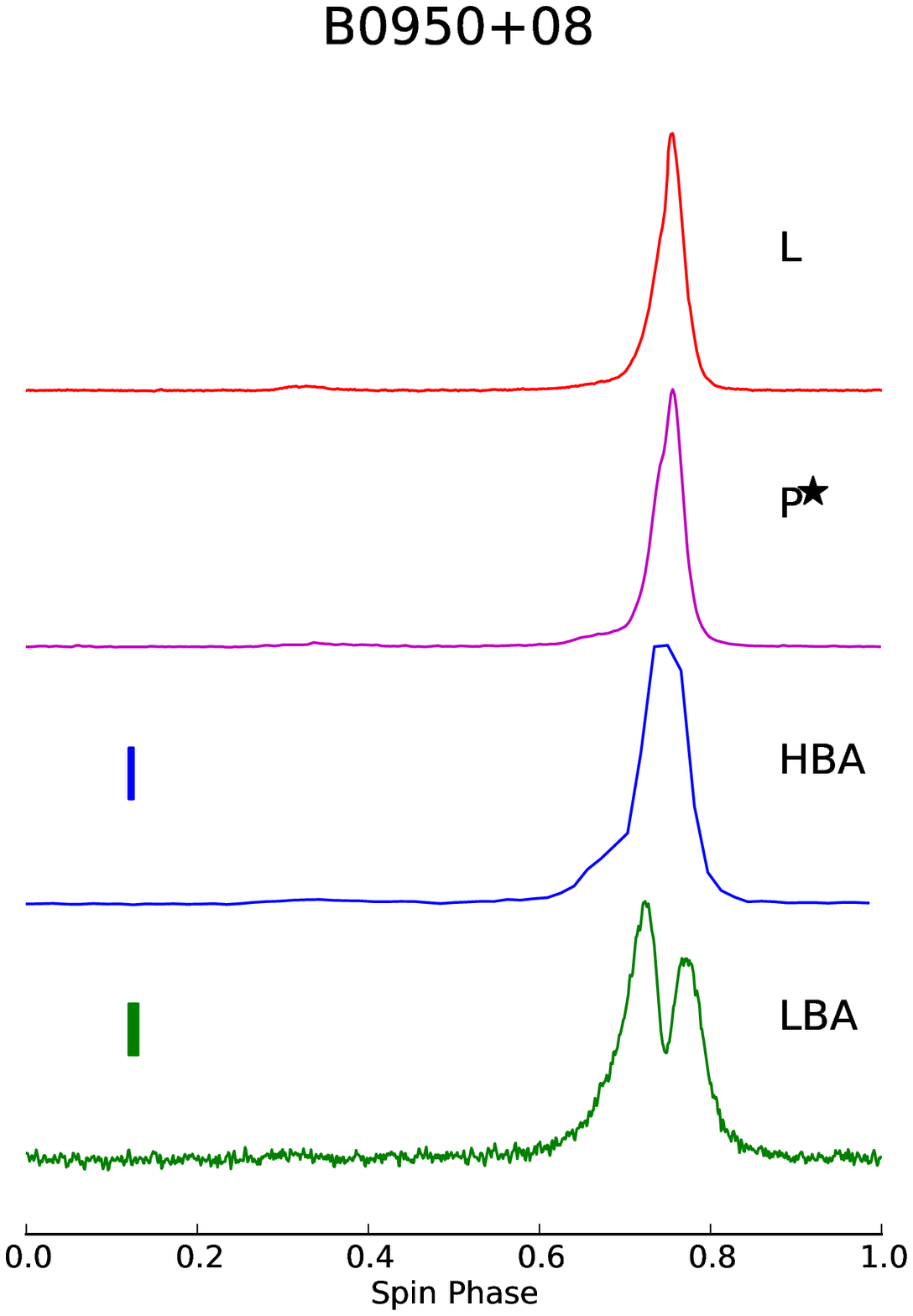} \\

 \includegraphics[width=32mm]{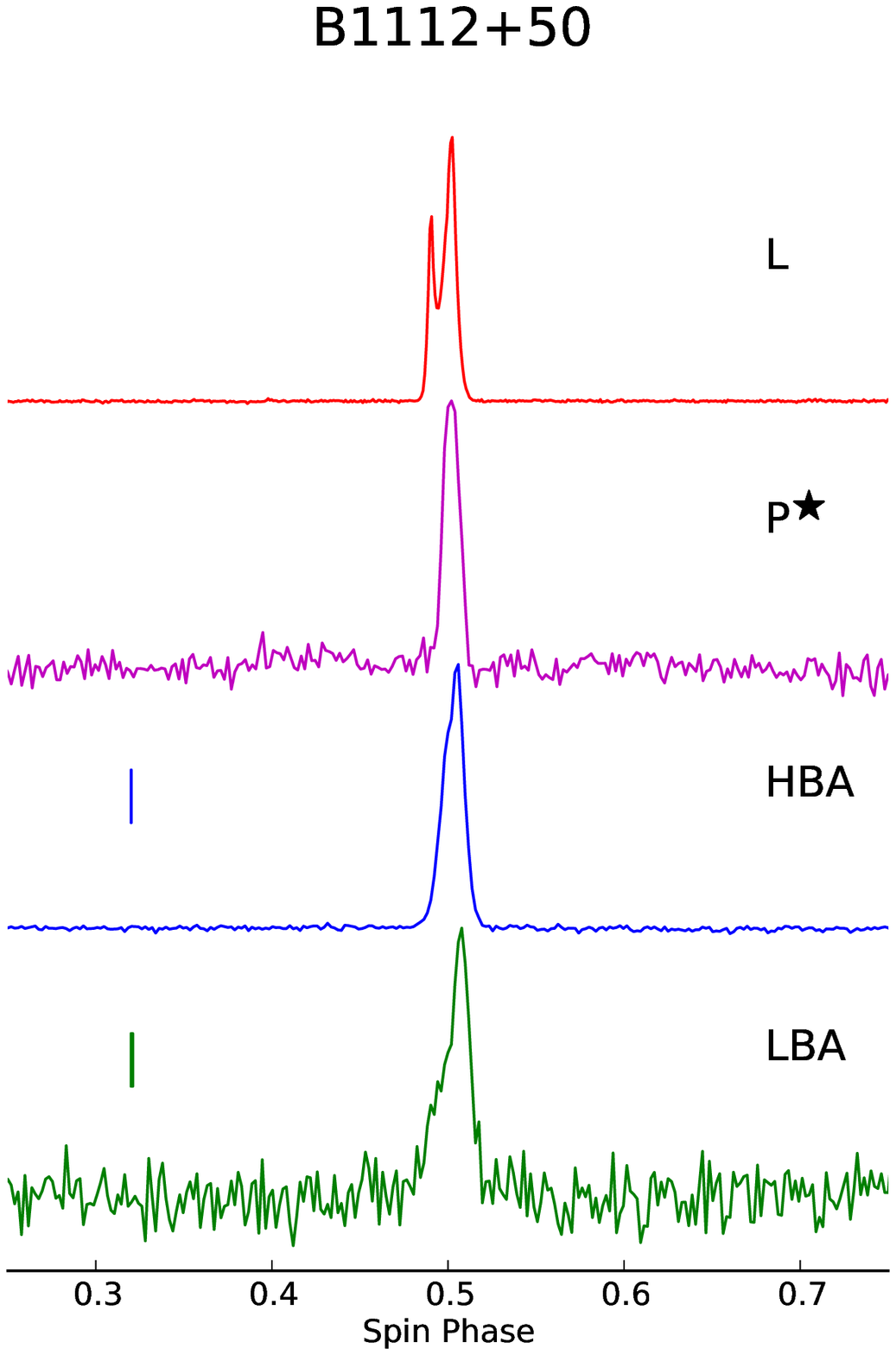} & \includegraphics[width=32mm]{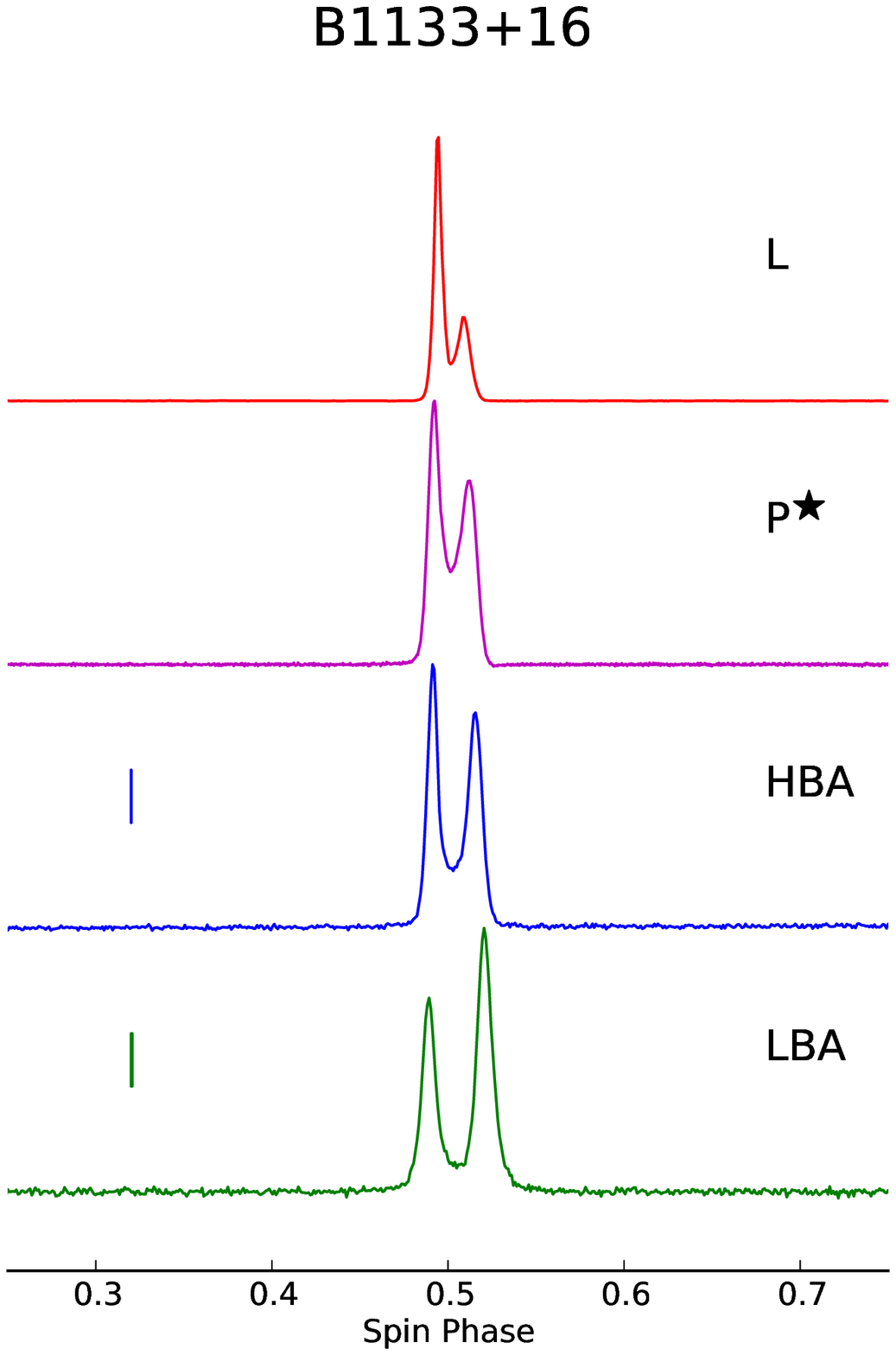} & \includegraphics[width=32mm]{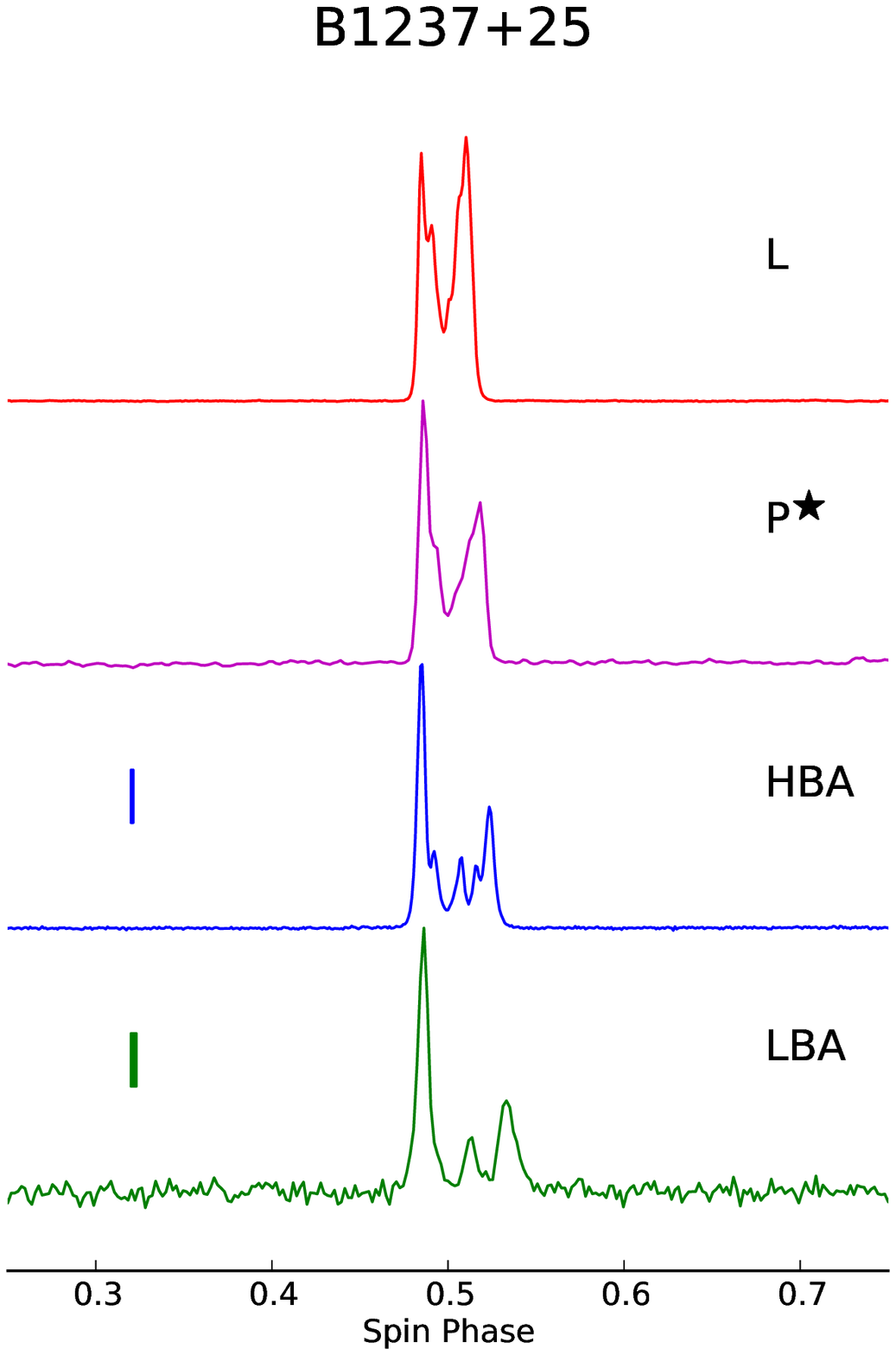} & \includegraphics[width=32mm]{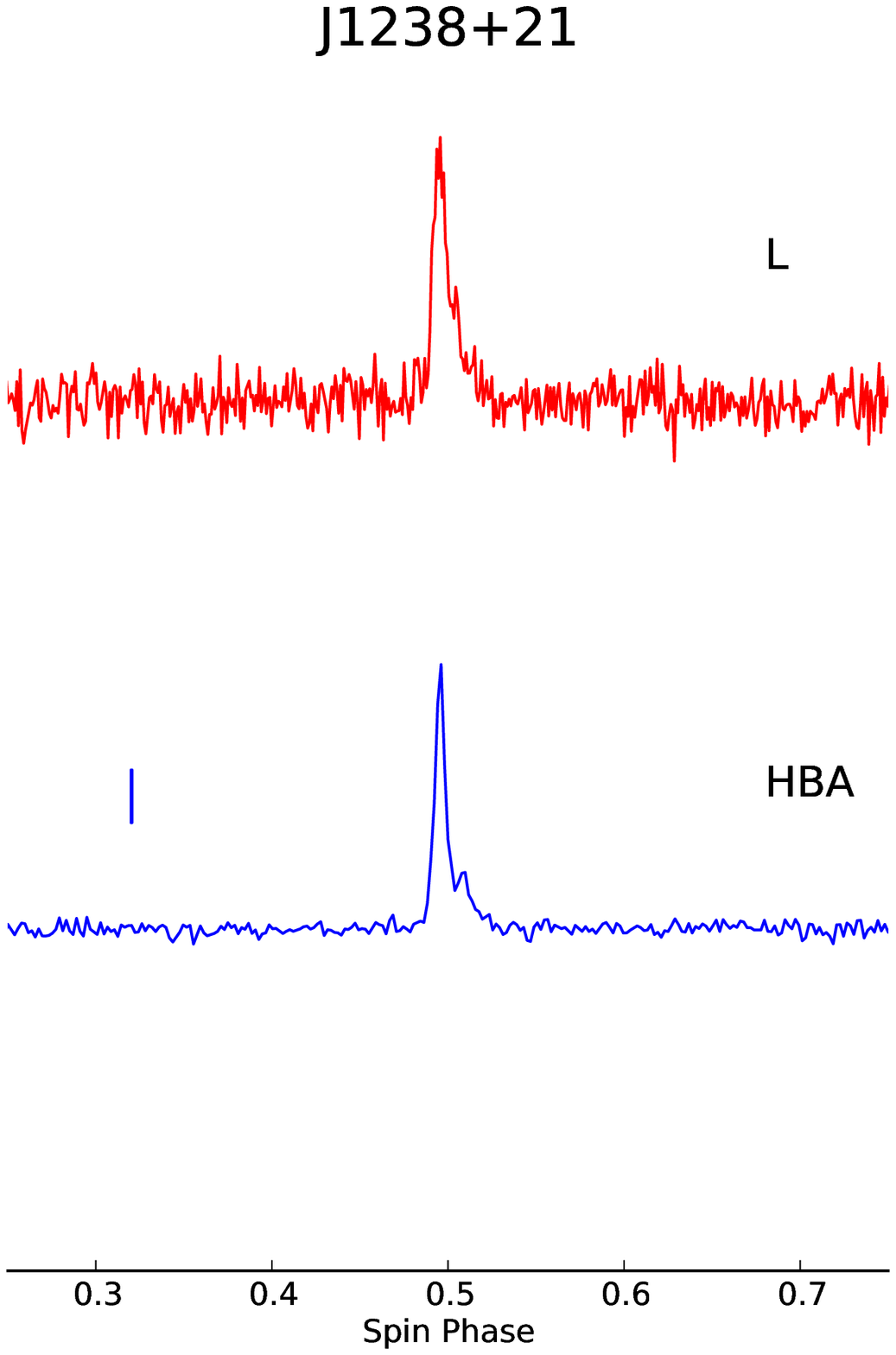} \\

 \includegraphics[width=32mm]{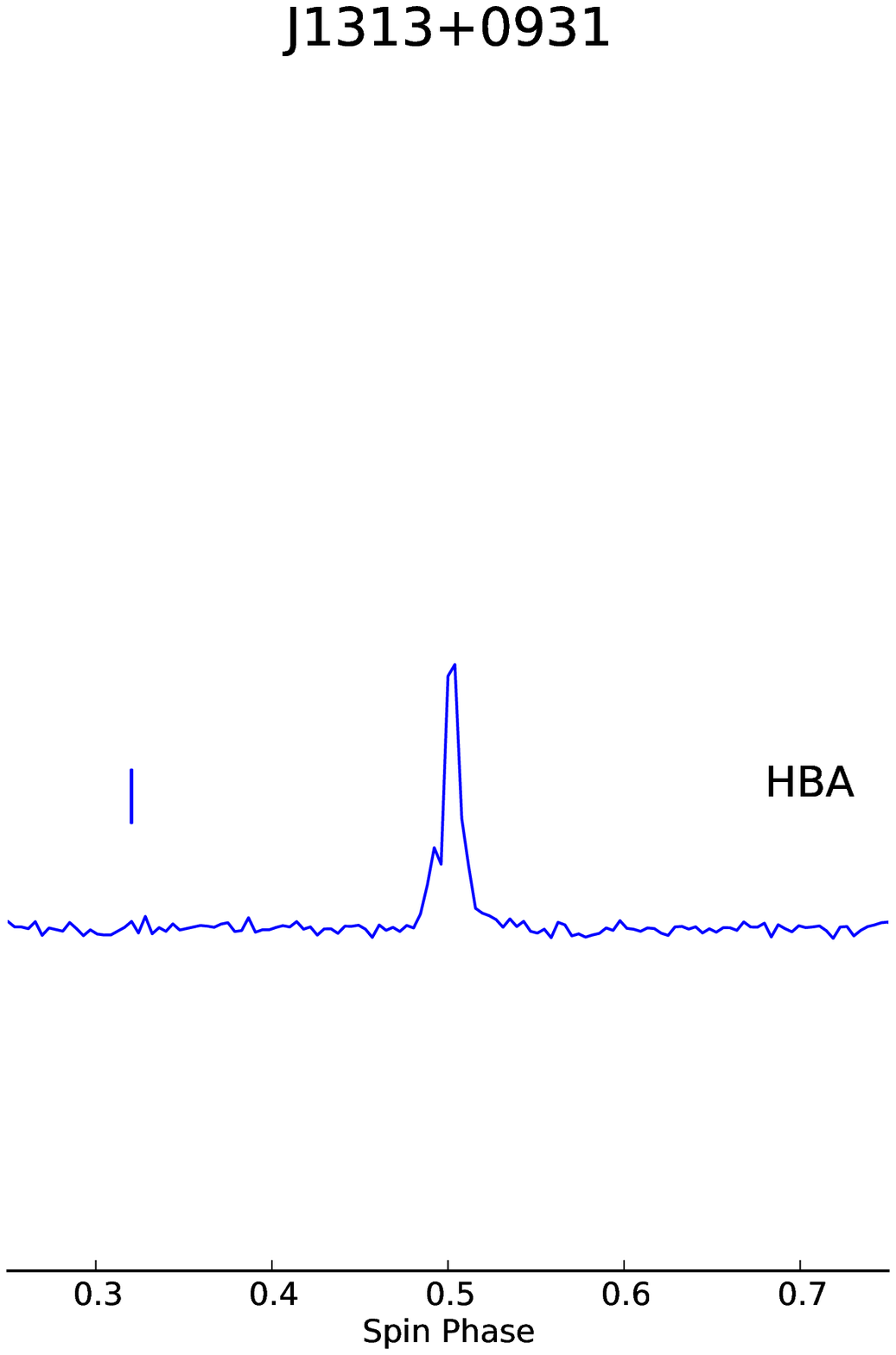} &  \includegraphics[width=32mm]{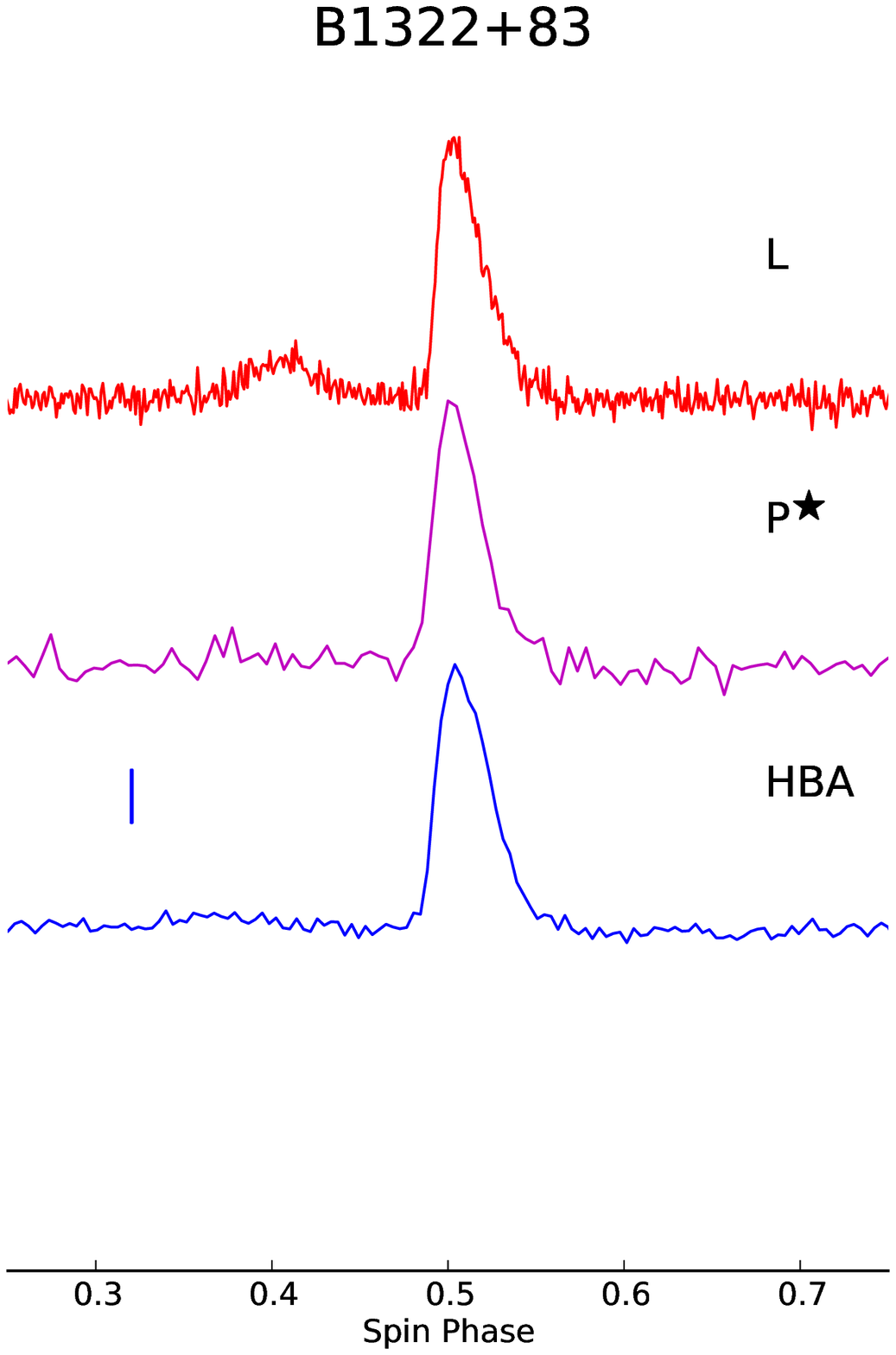} &   \includegraphics[width=32mm]{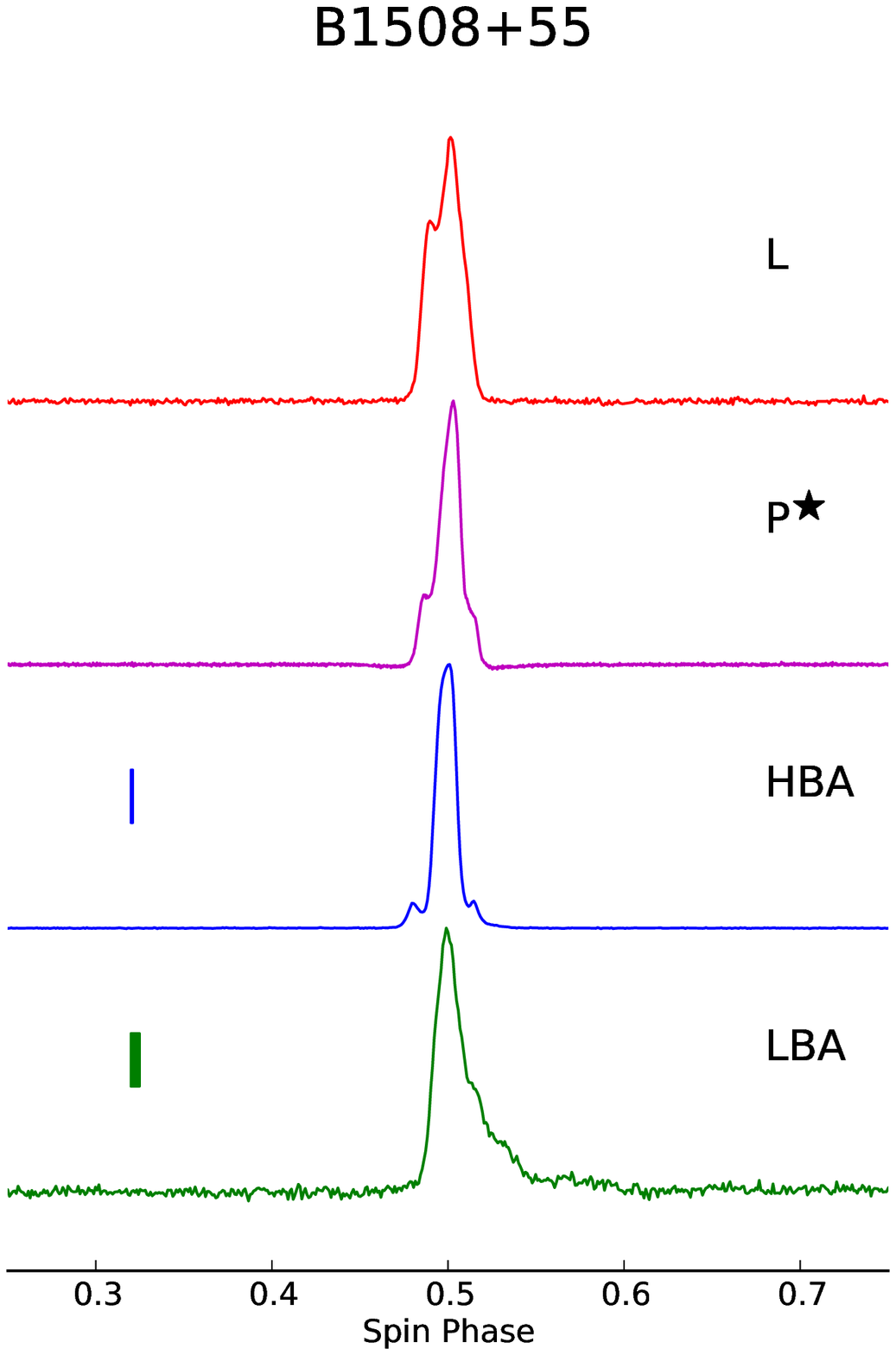} &   \includegraphics[width=32mm]{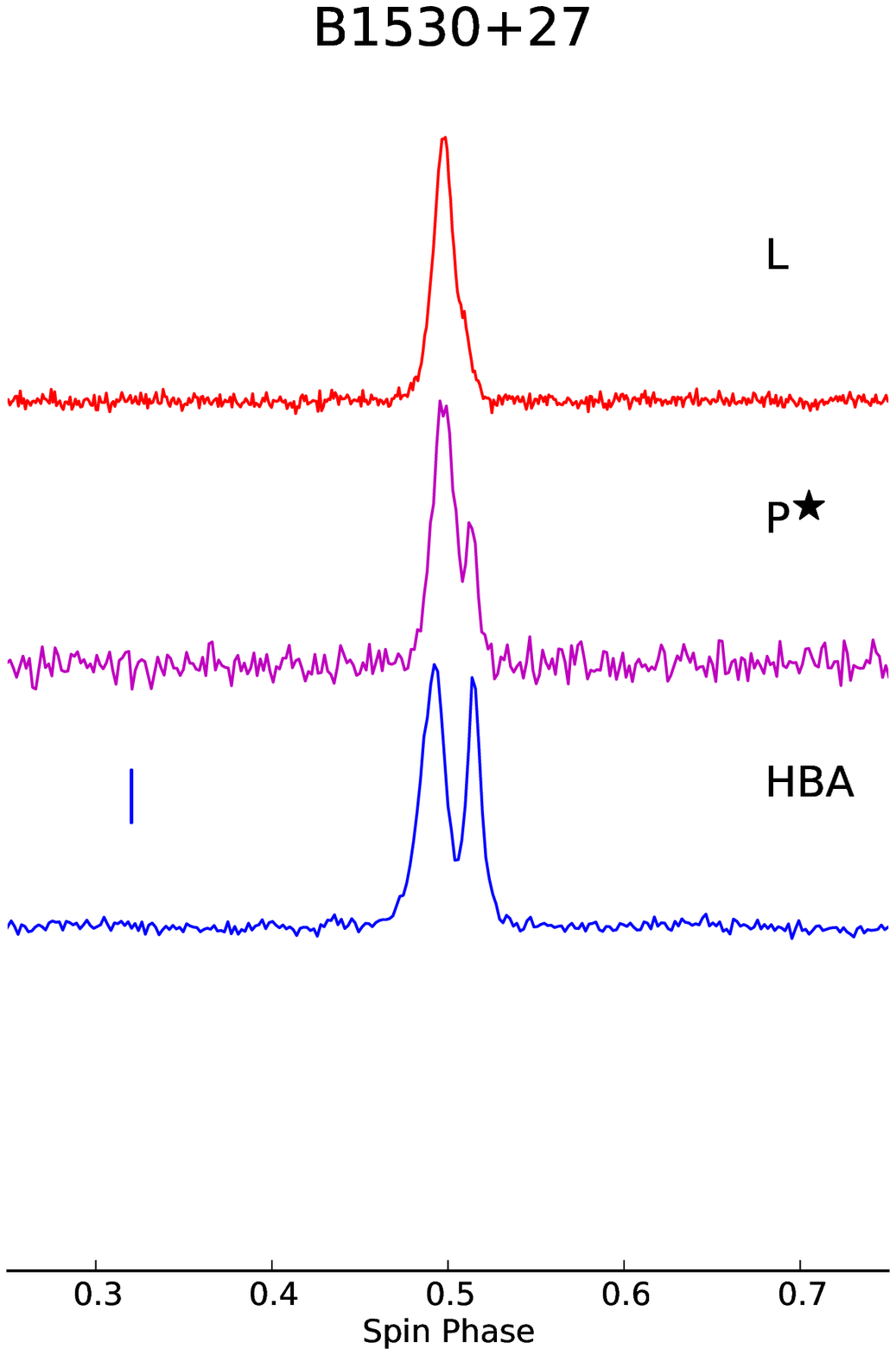} \\

 \includegraphics[width=32mm]{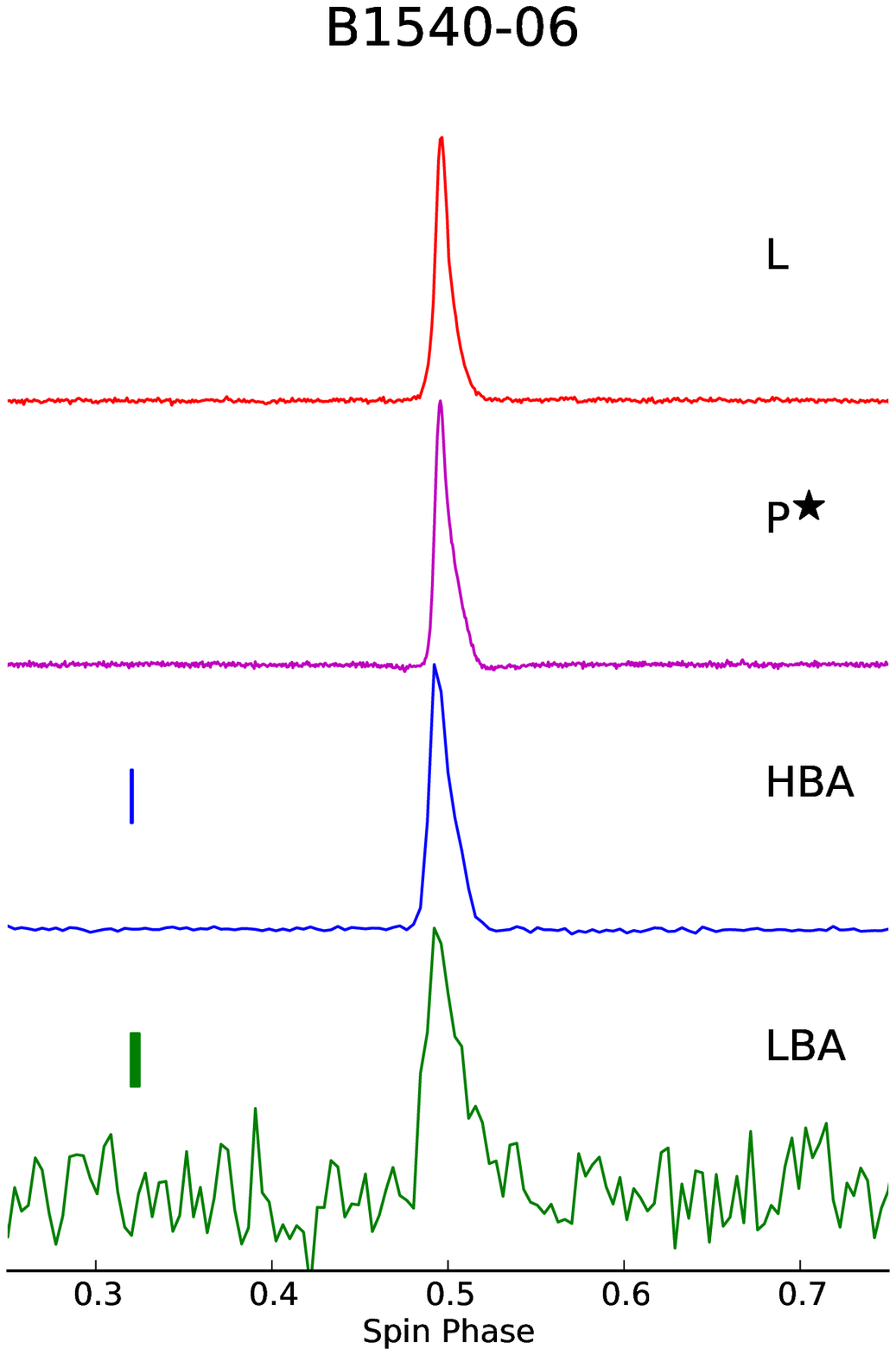} &  \includegraphics[width=32mm]{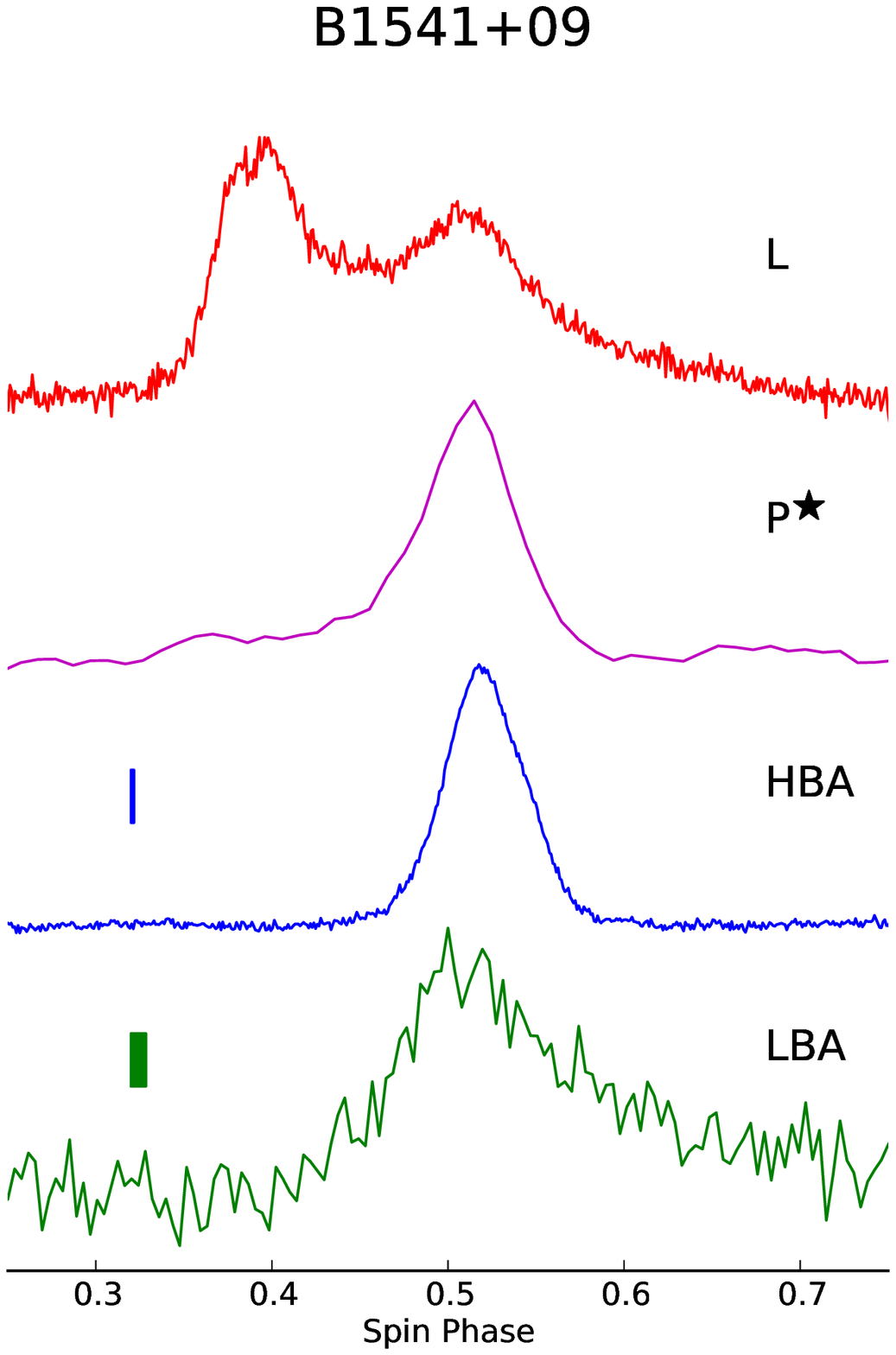} &   \includegraphics[width=32mm]{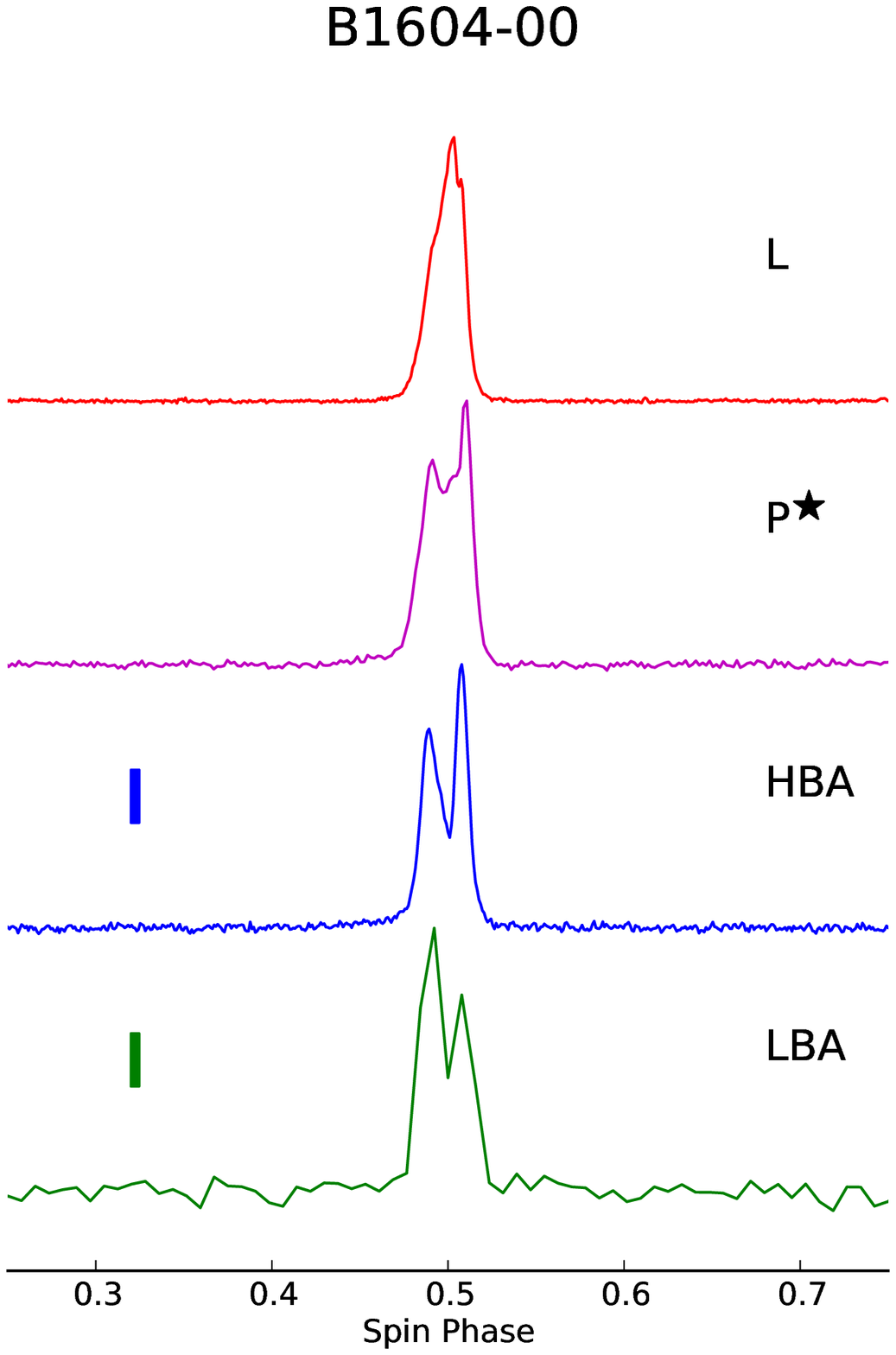} &   \includegraphics[width=32mm]{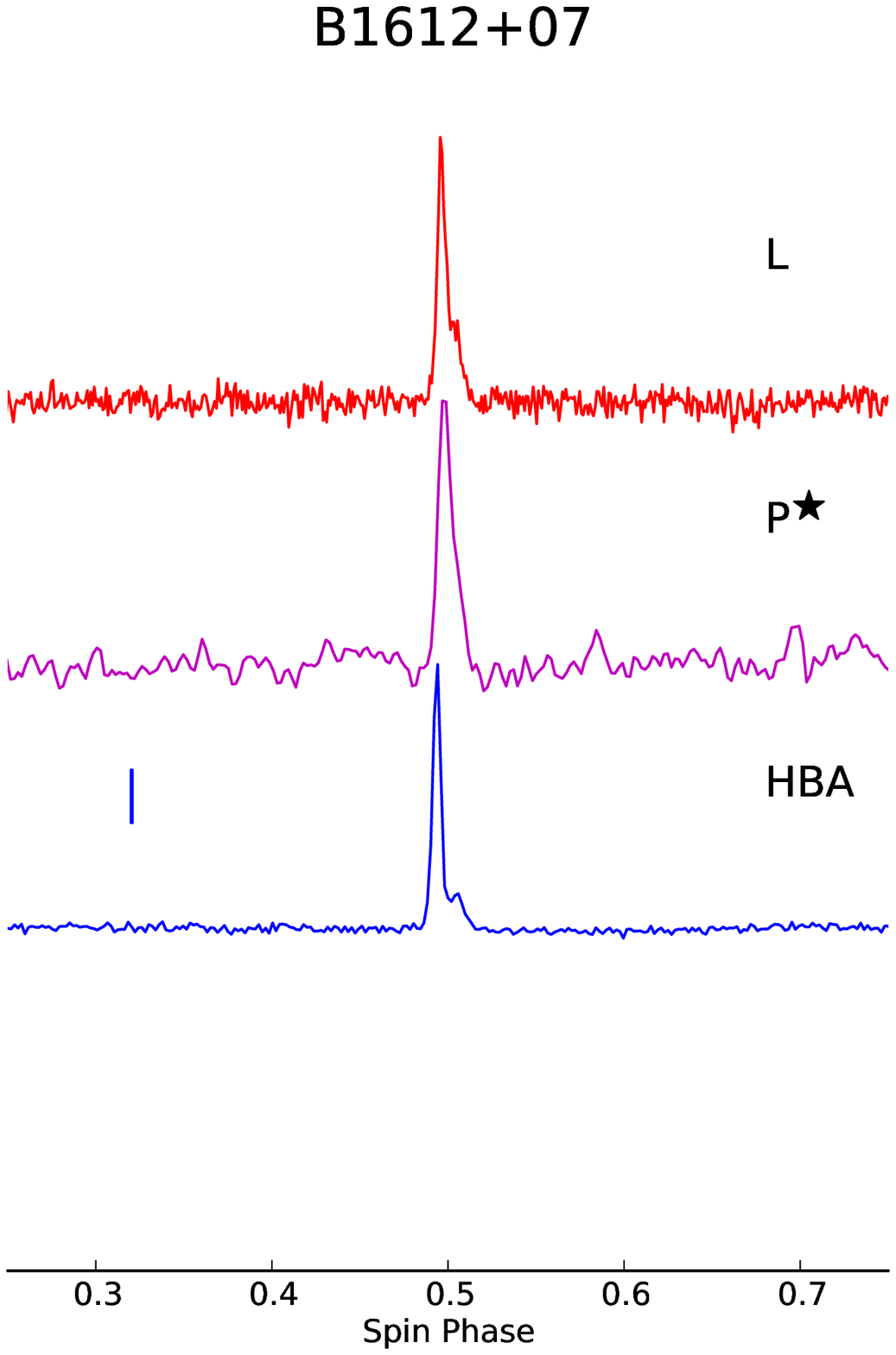} \\

 \includegraphics[width=32mm]{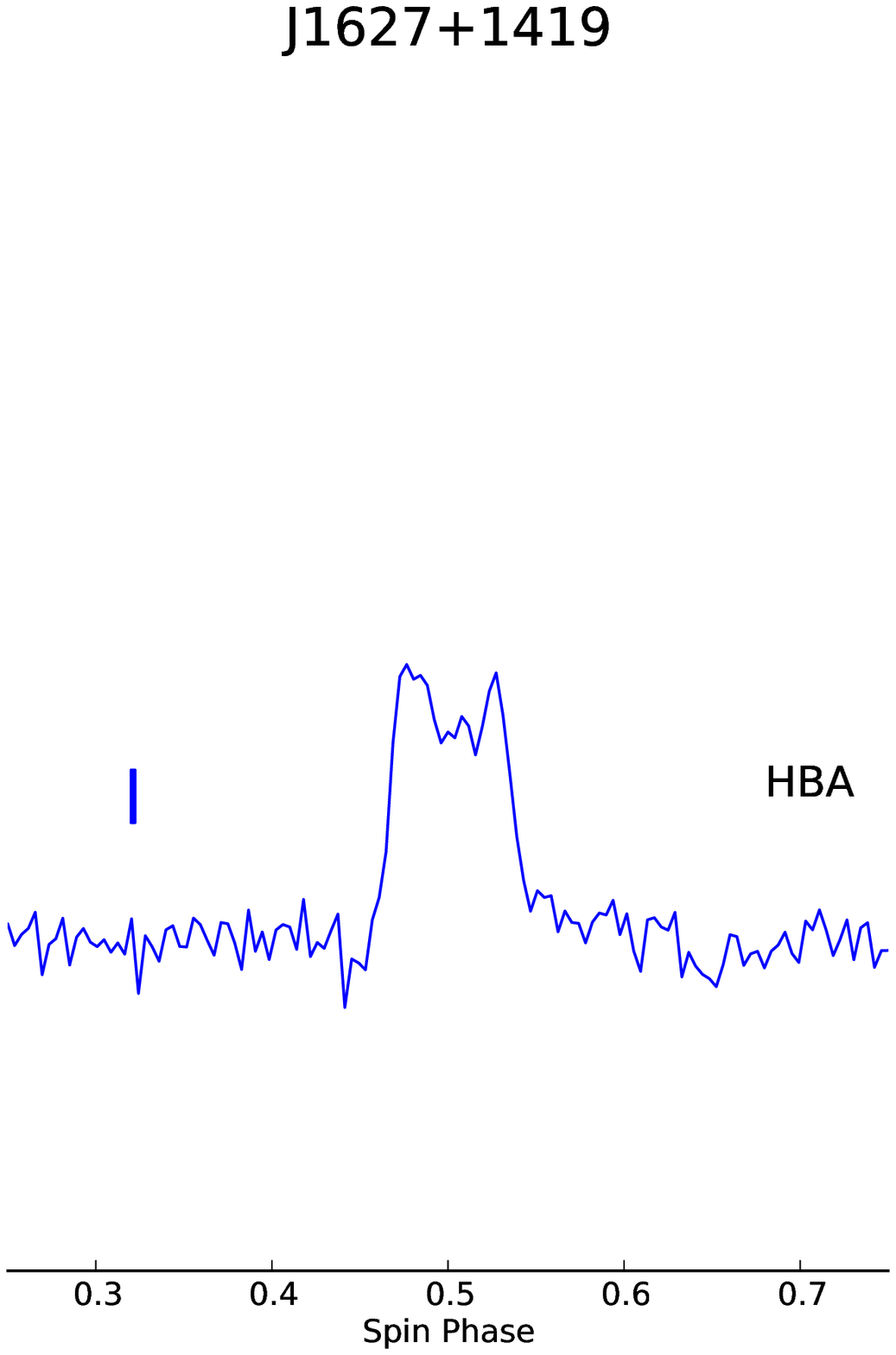} &  \includegraphics[width=32mm]{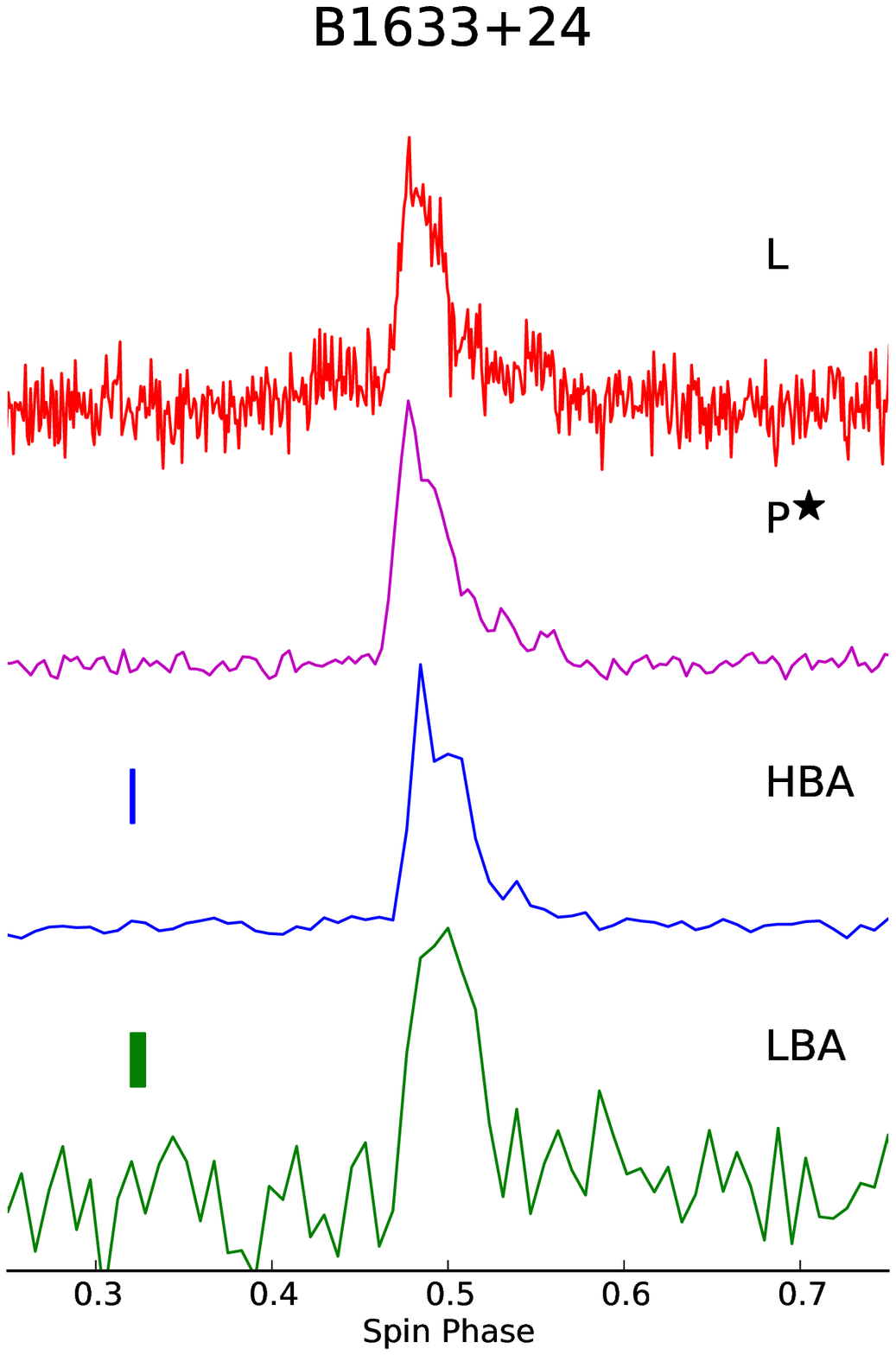} &   \includegraphics[width=32mm]{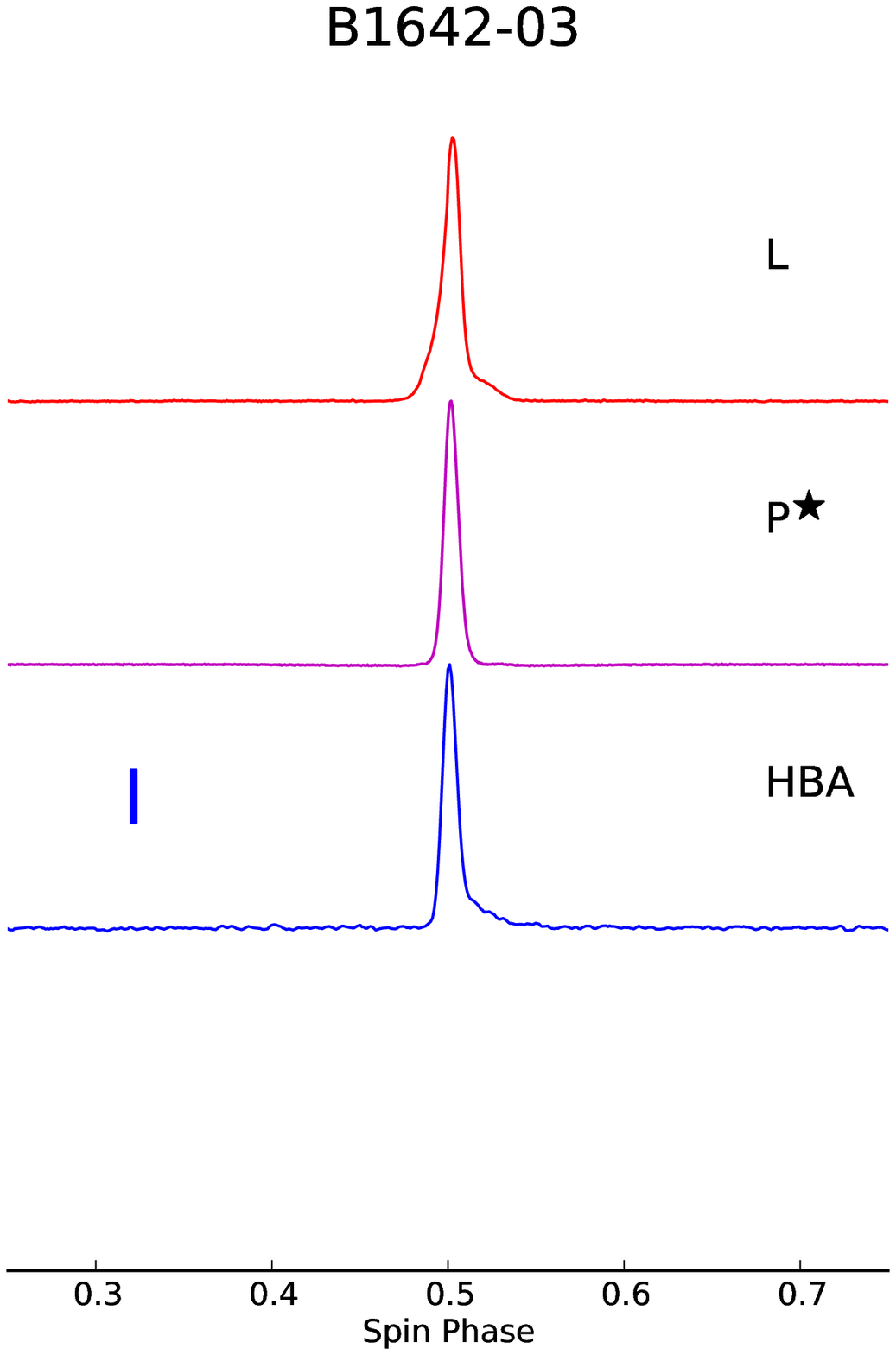} &   \includegraphics[width=32mm]{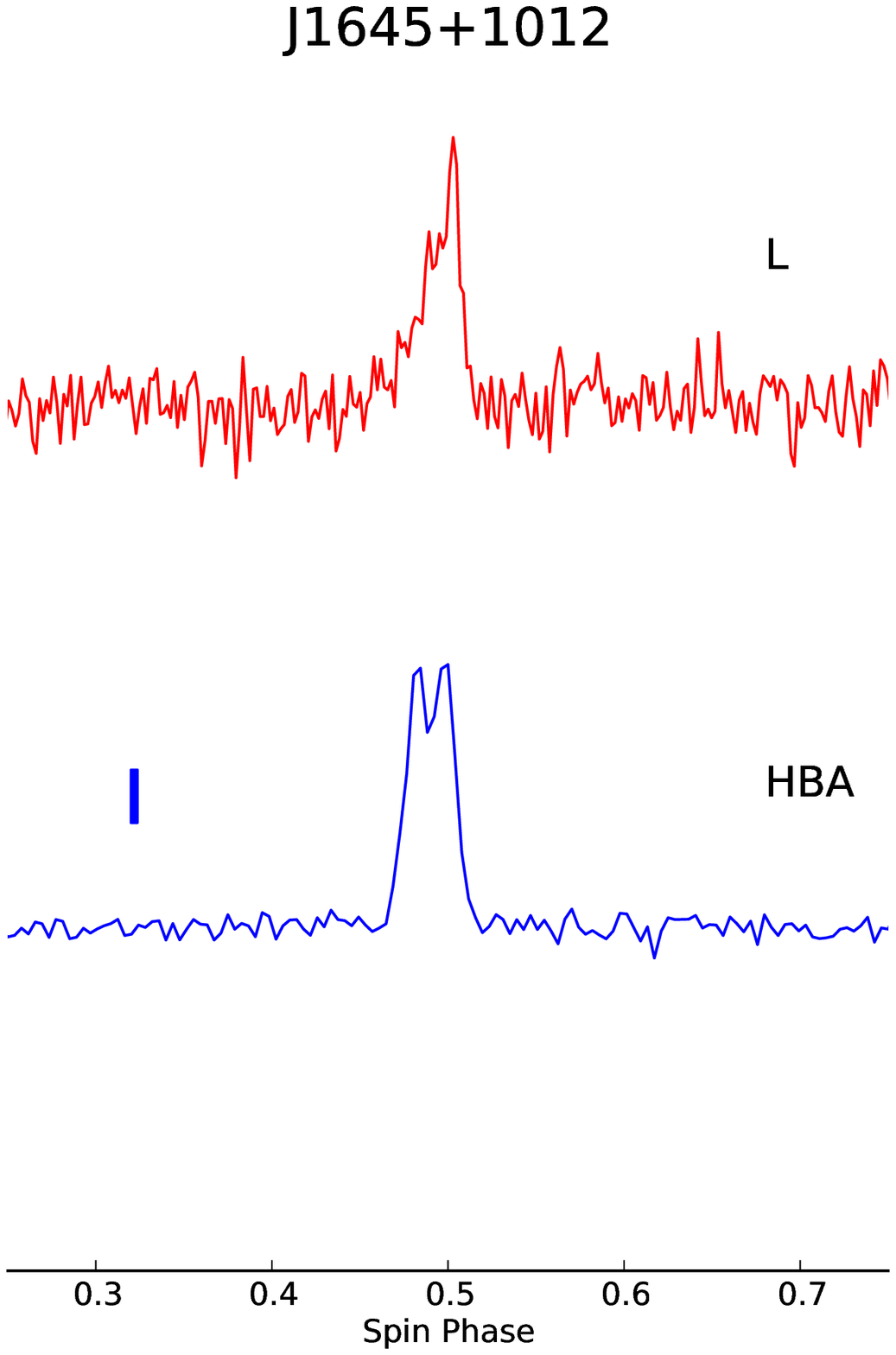} \\

\end{tabular}
\caption*{\label{fig:100_4}
Fig.\ref{fig:100}: continued
}
\end{figure*}

\begin{figure*}
\centering
\begin{tabular}{cccc}
 \includegraphics[width=32mm]{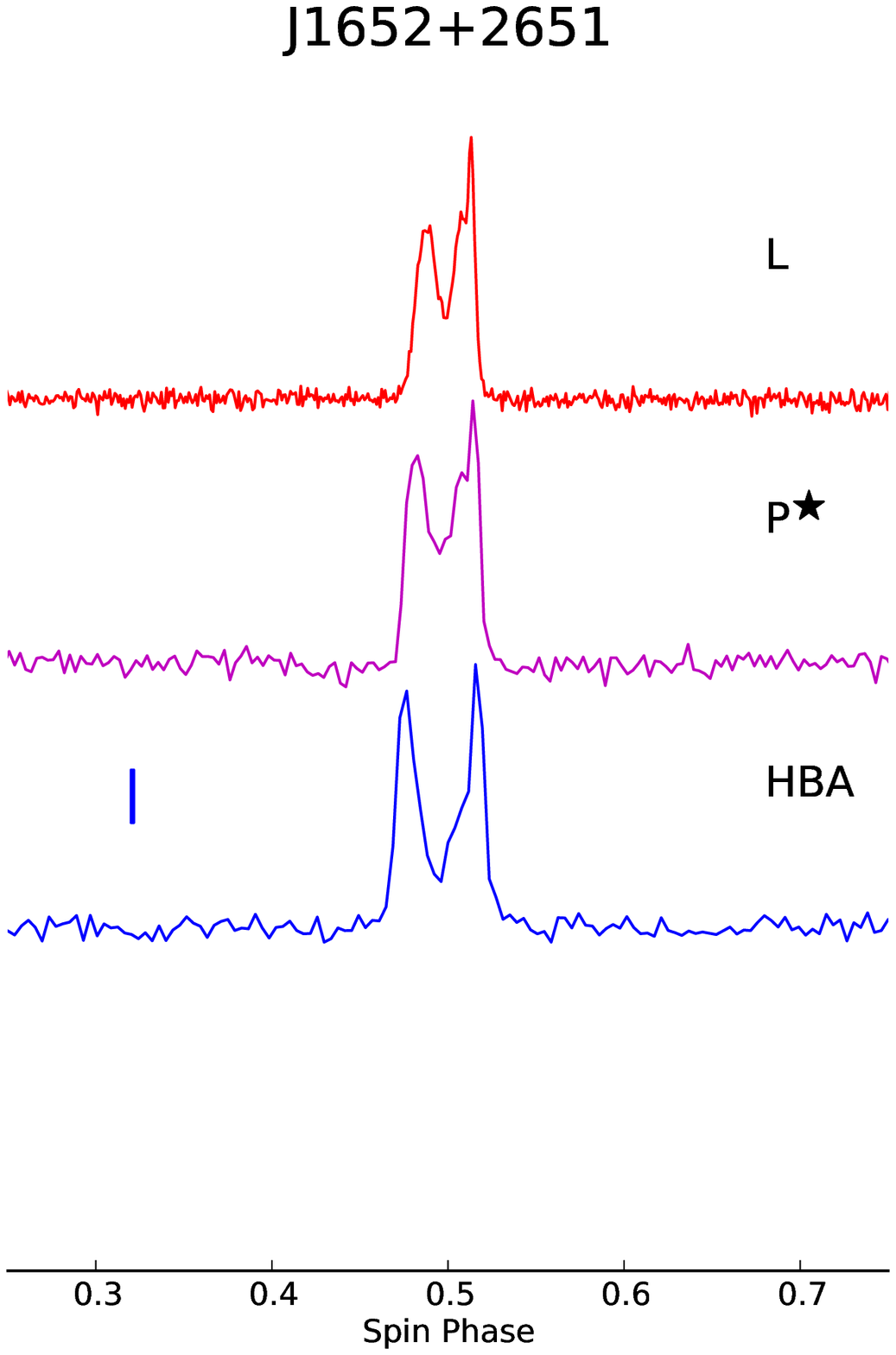} &  \includegraphics[width=32mm]{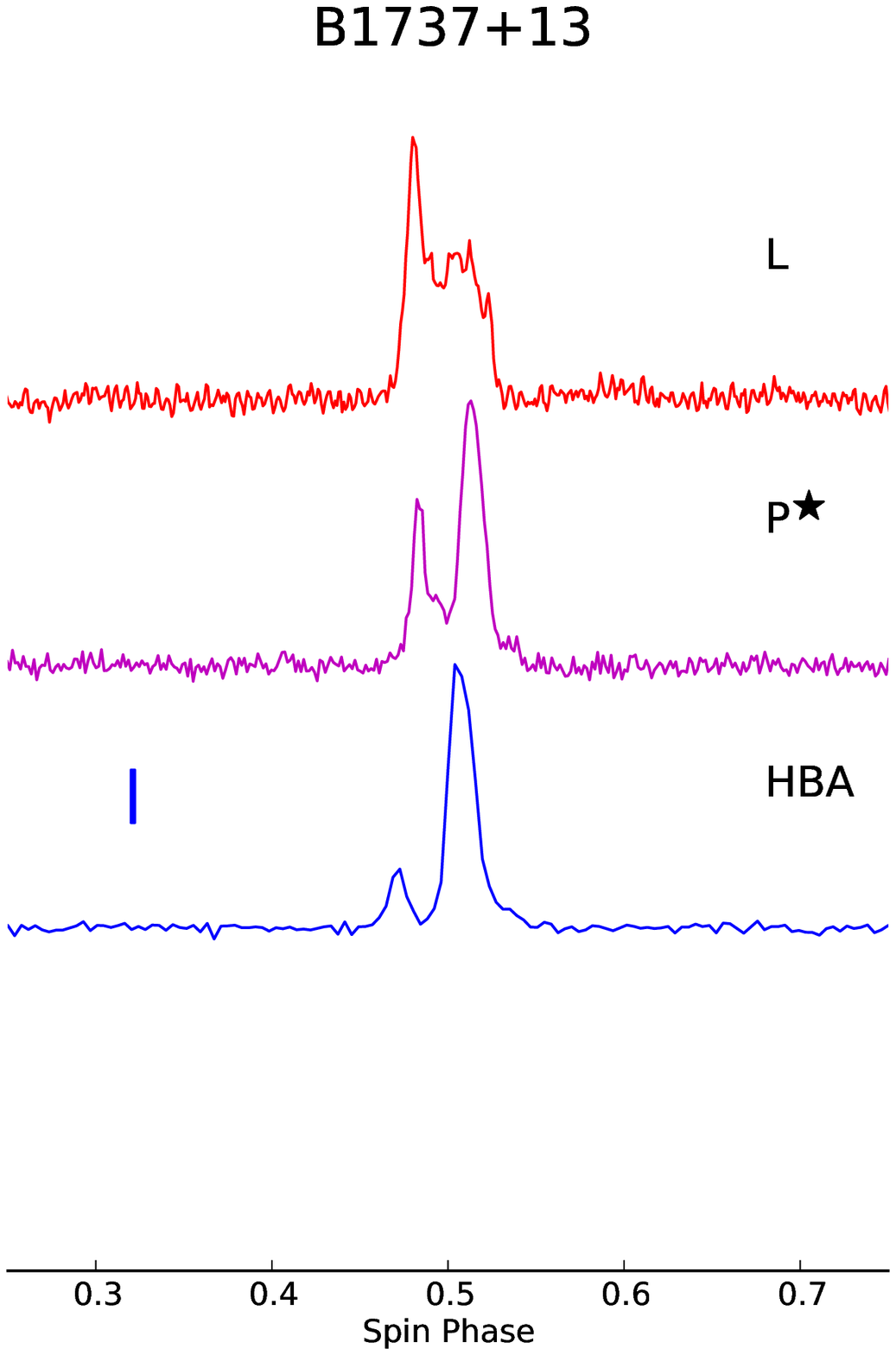} &   \includegraphics[width=32mm]{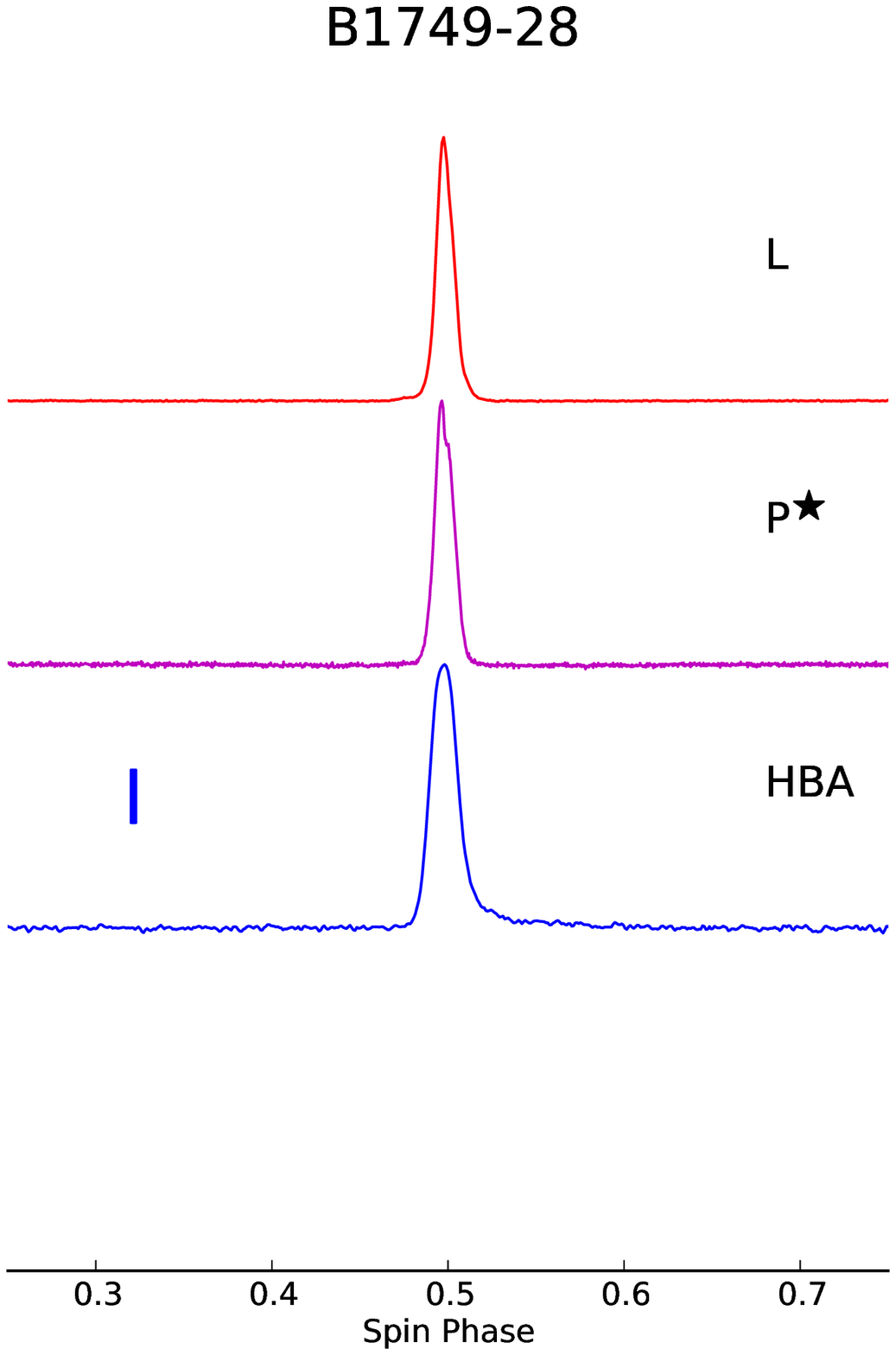} &   \includegraphics[width=32mm]{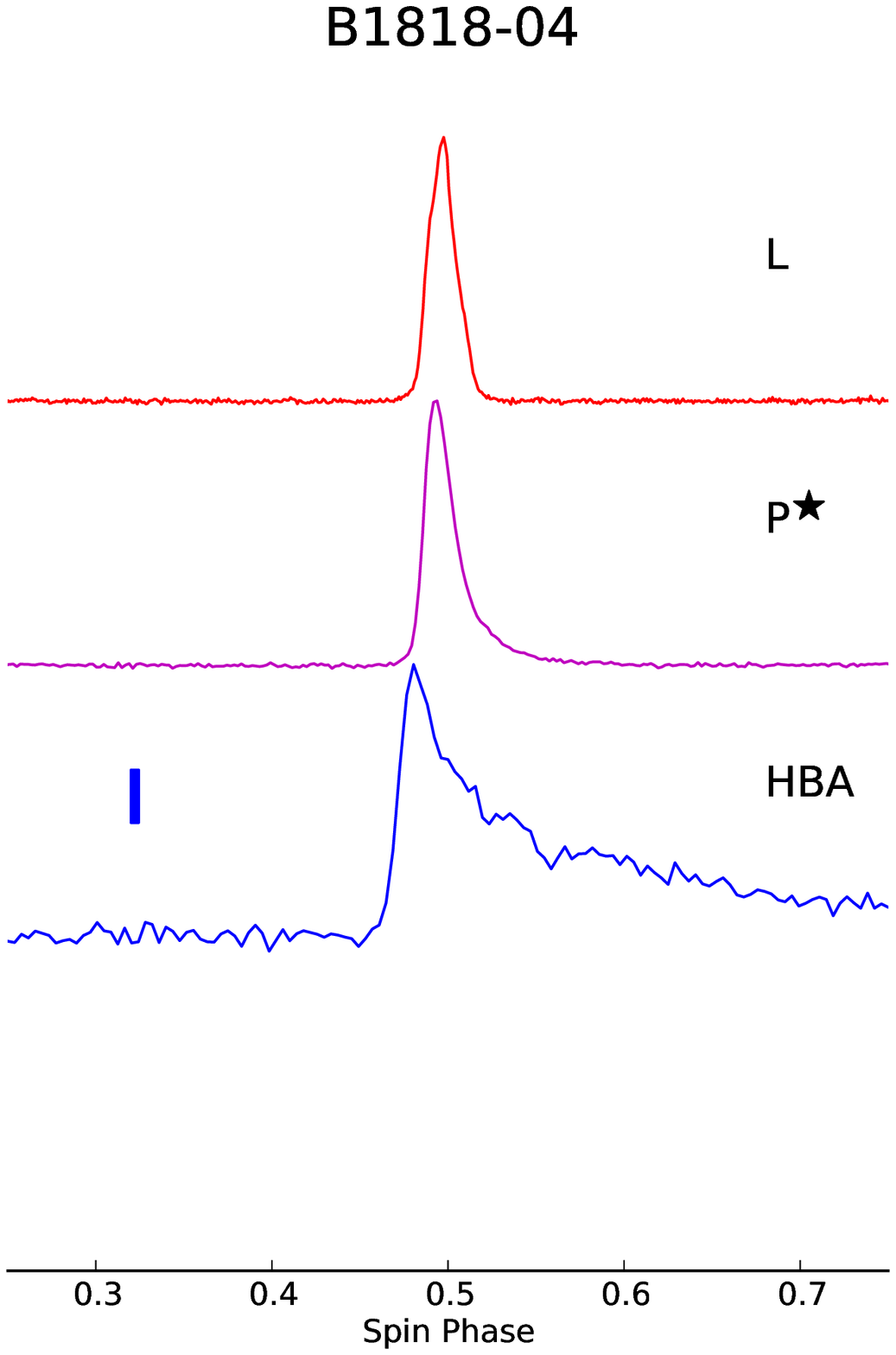} \\

 \includegraphics[width=32mm]{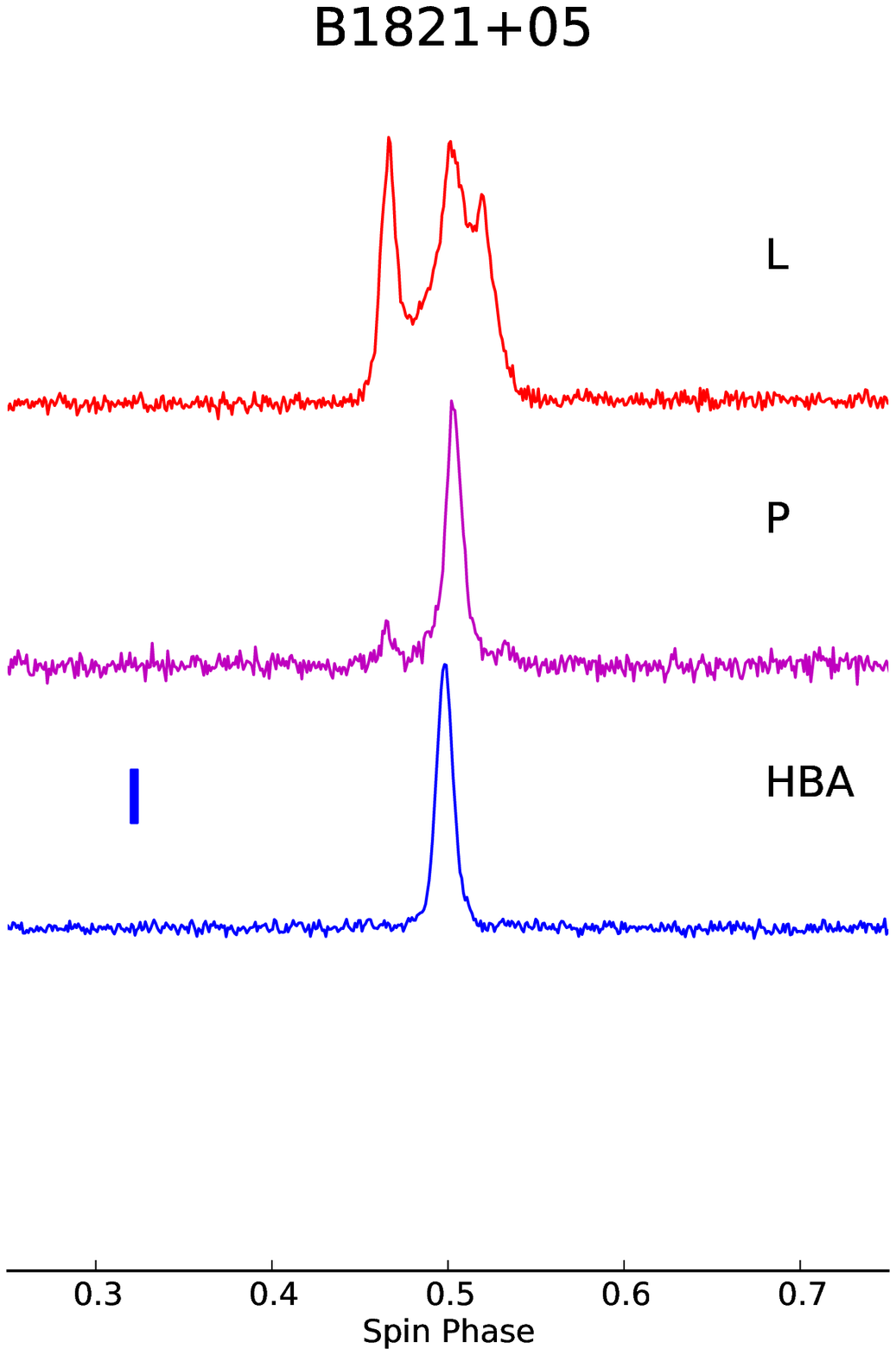} &  \includegraphics[width=32mm]{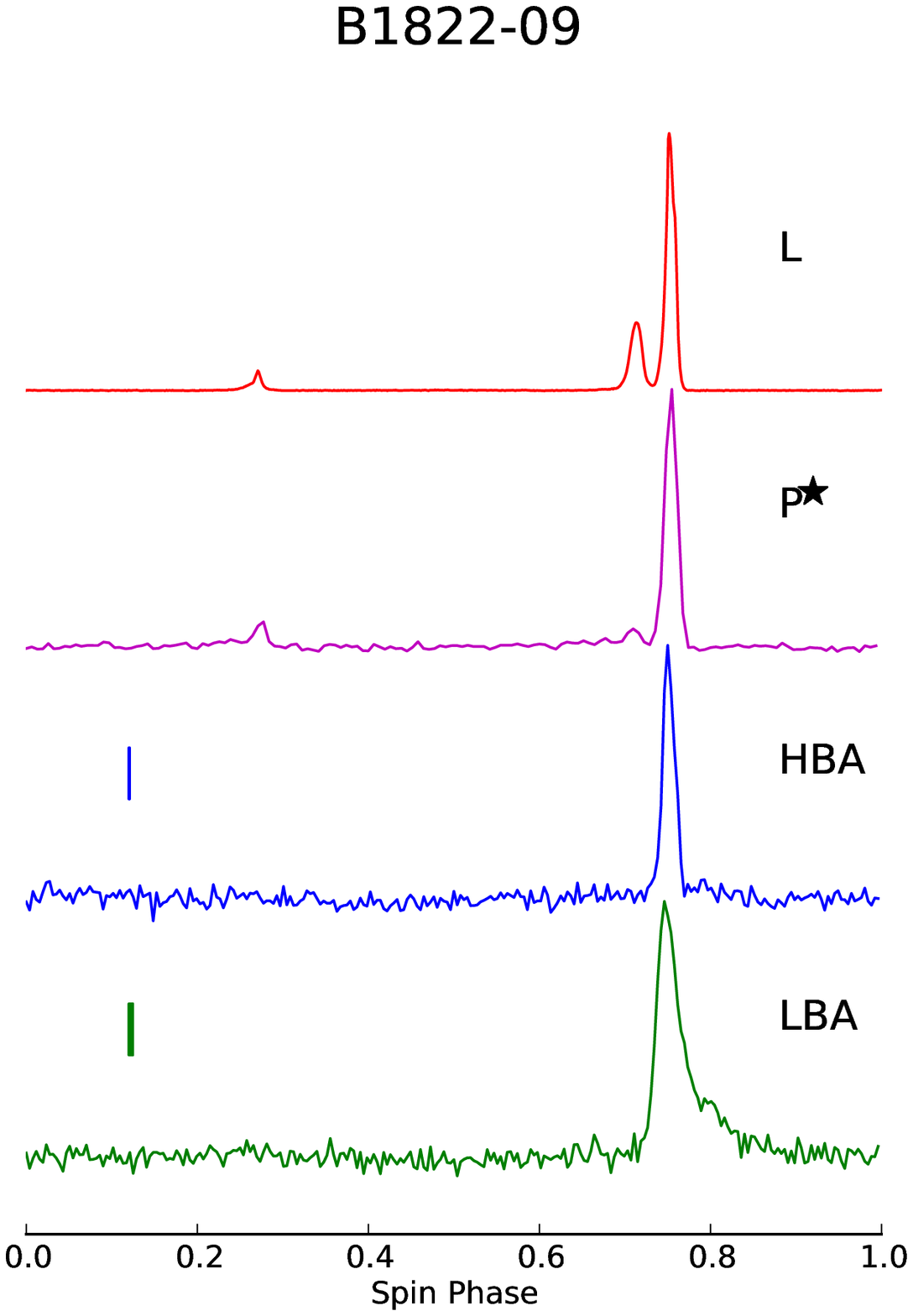} &   \includegraphics[width=32mm]{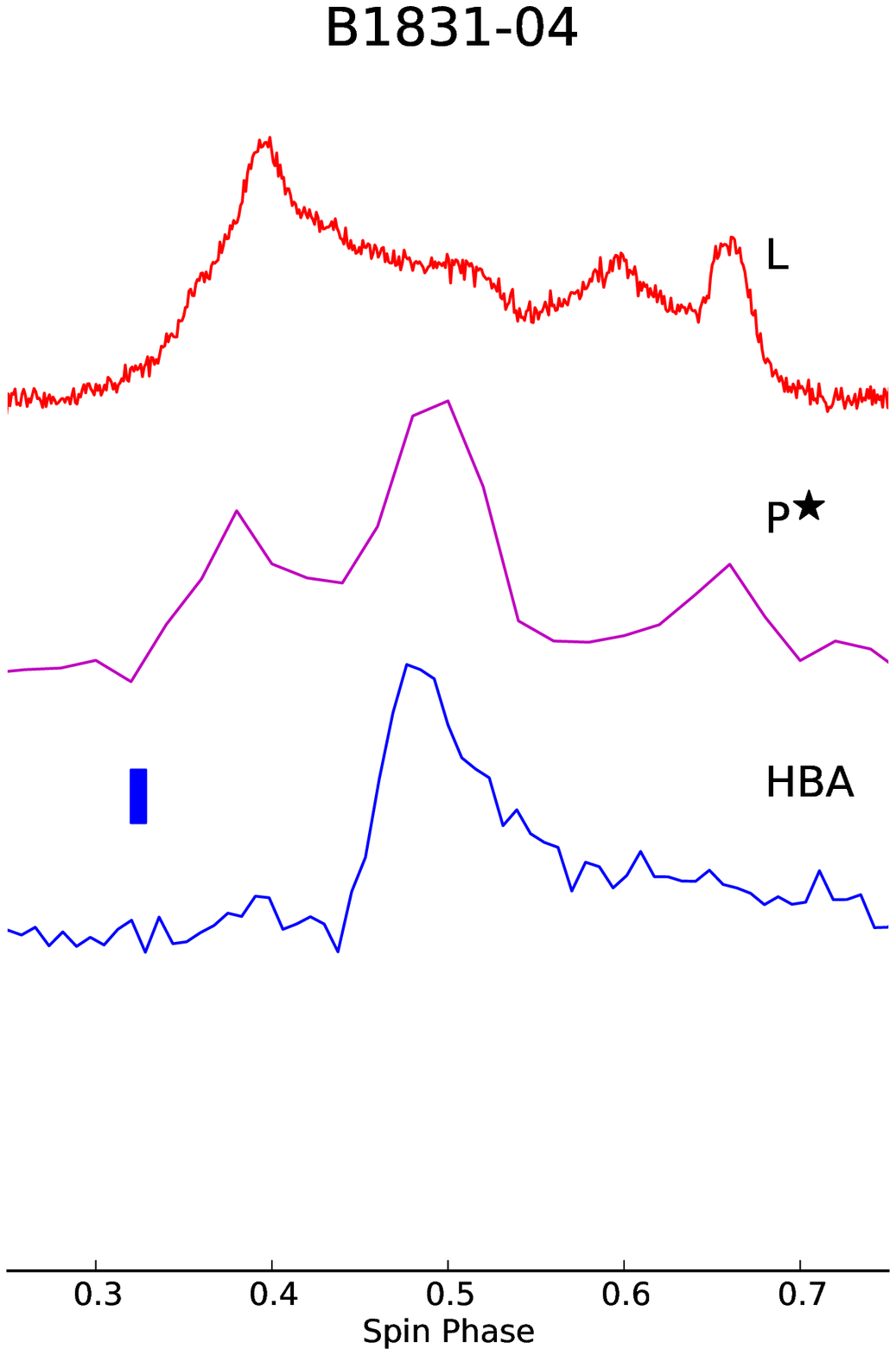} &   \includegraphics[width=32mm]{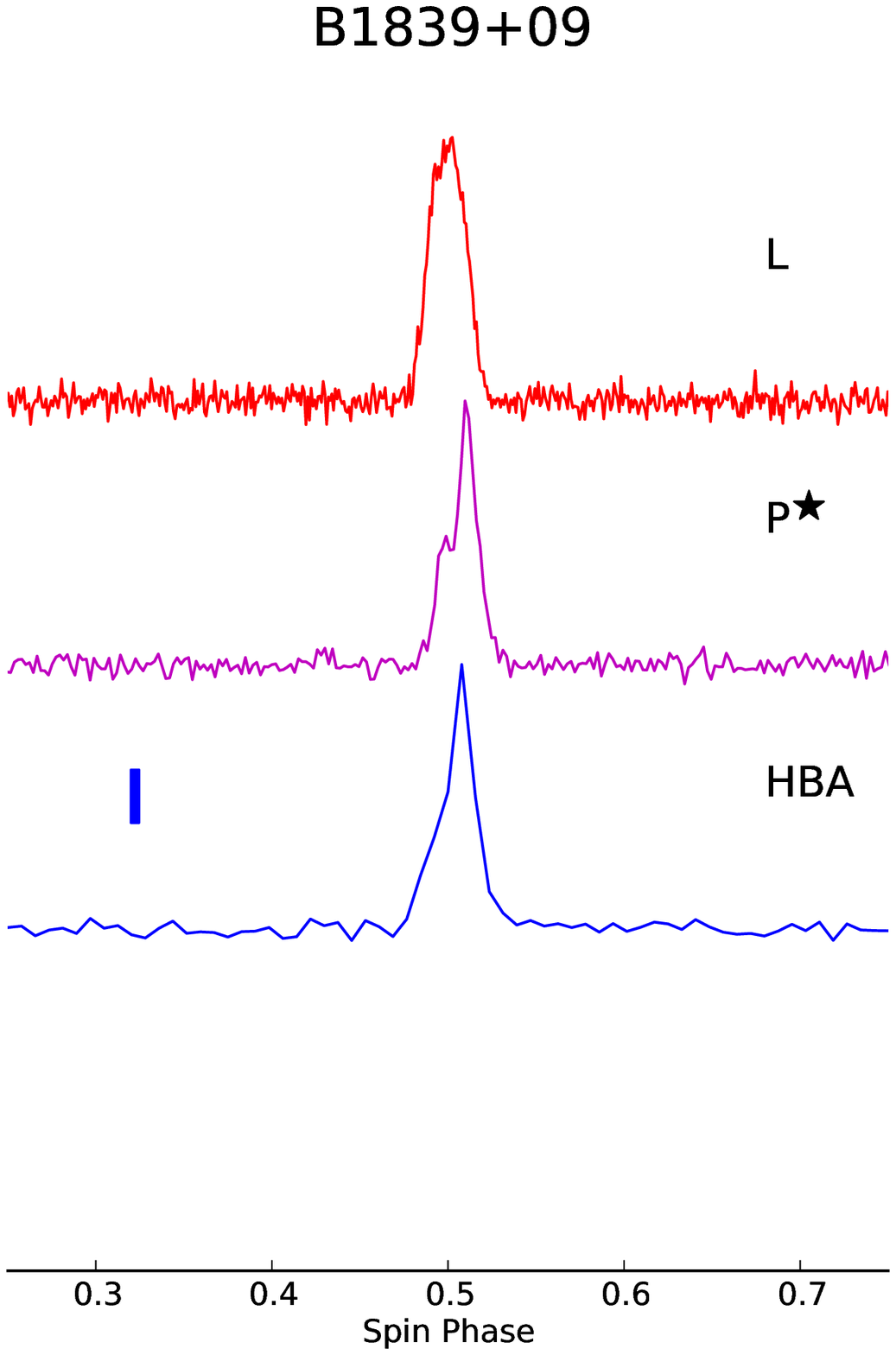} \\

 \includegraphics[width=32mm]{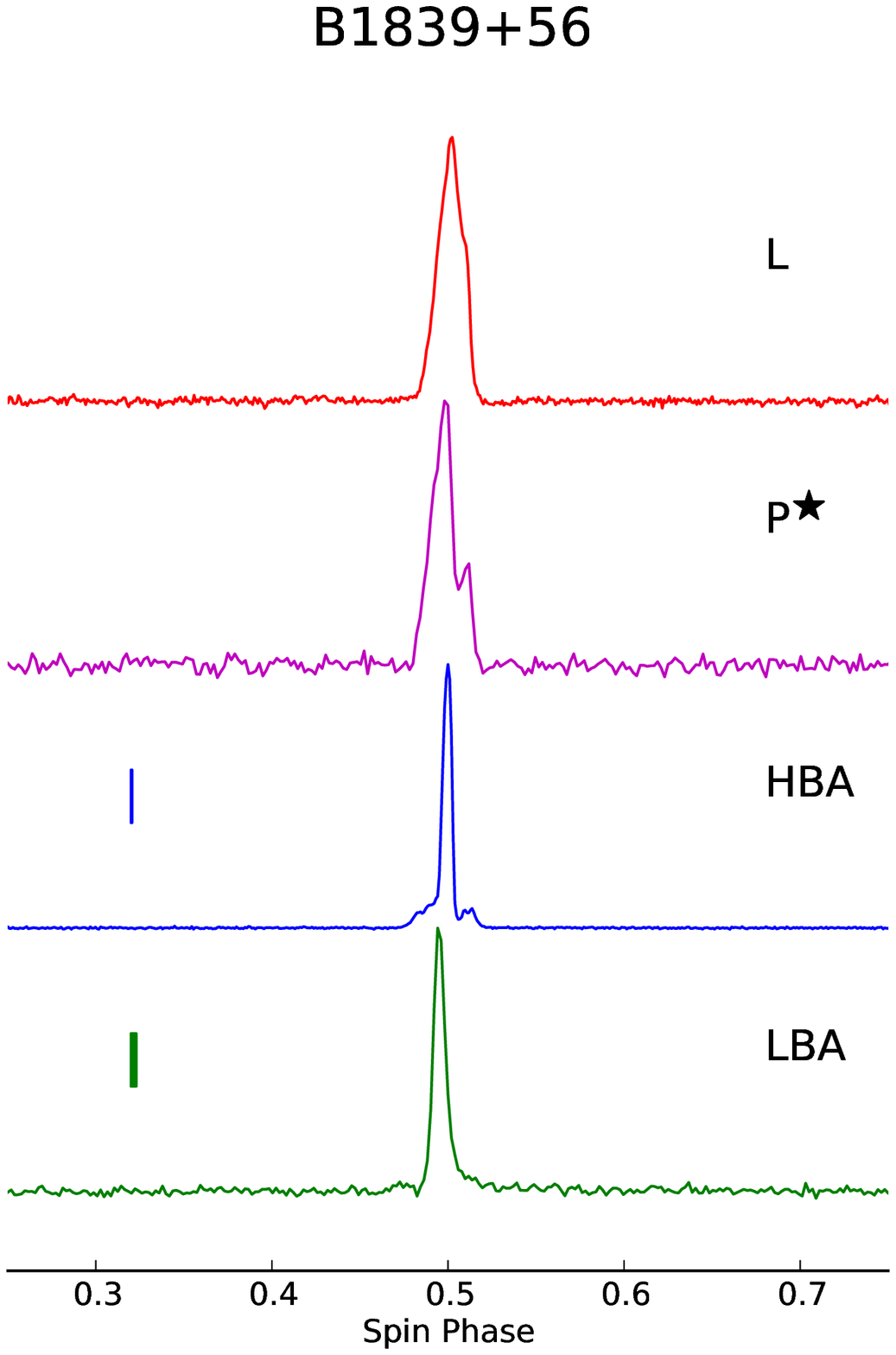} &  \includegraphics[width=32mm]{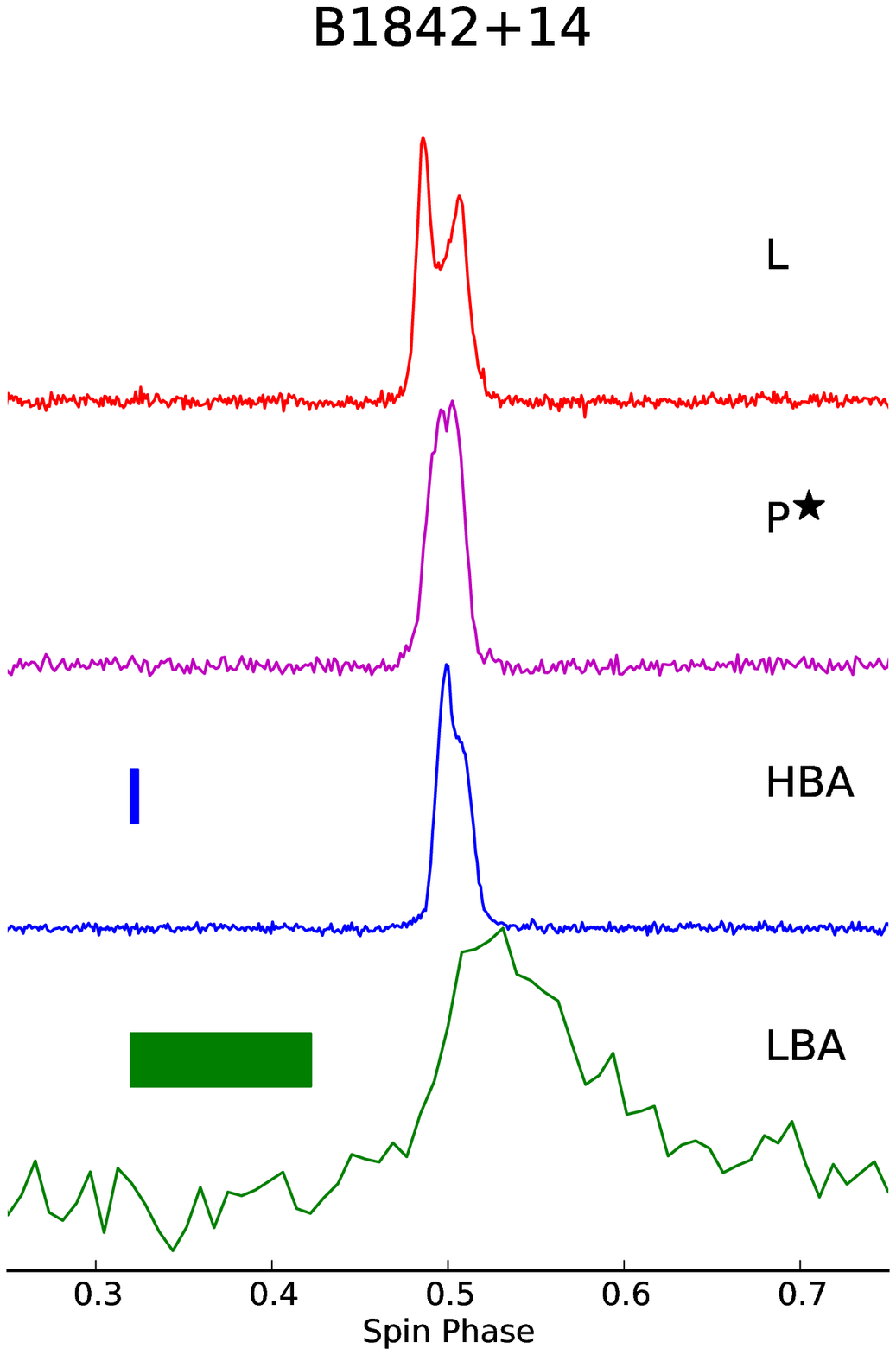} &   \includegraphics[width=32mm]{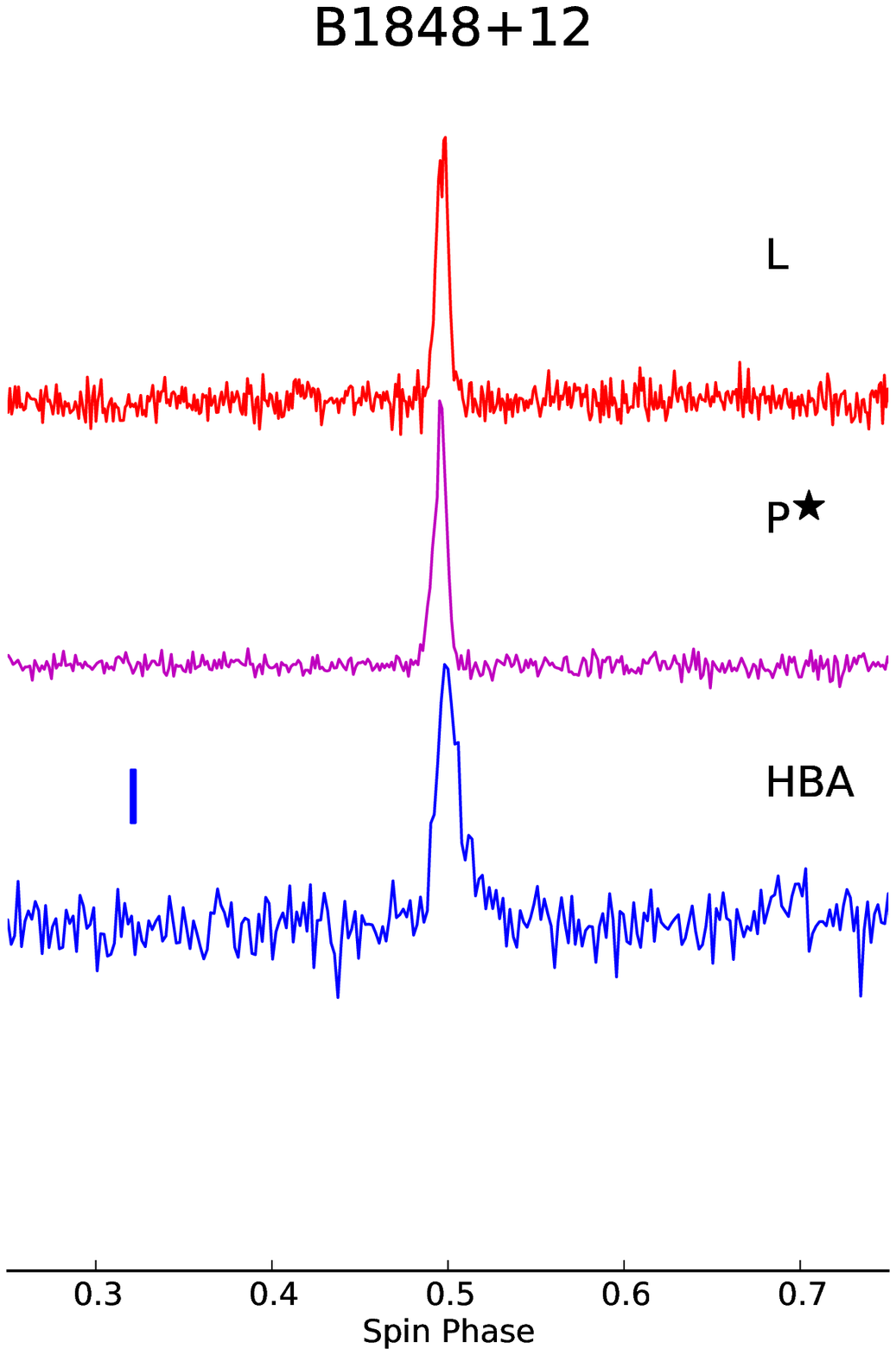} &   \includegraphics[width=32mm]{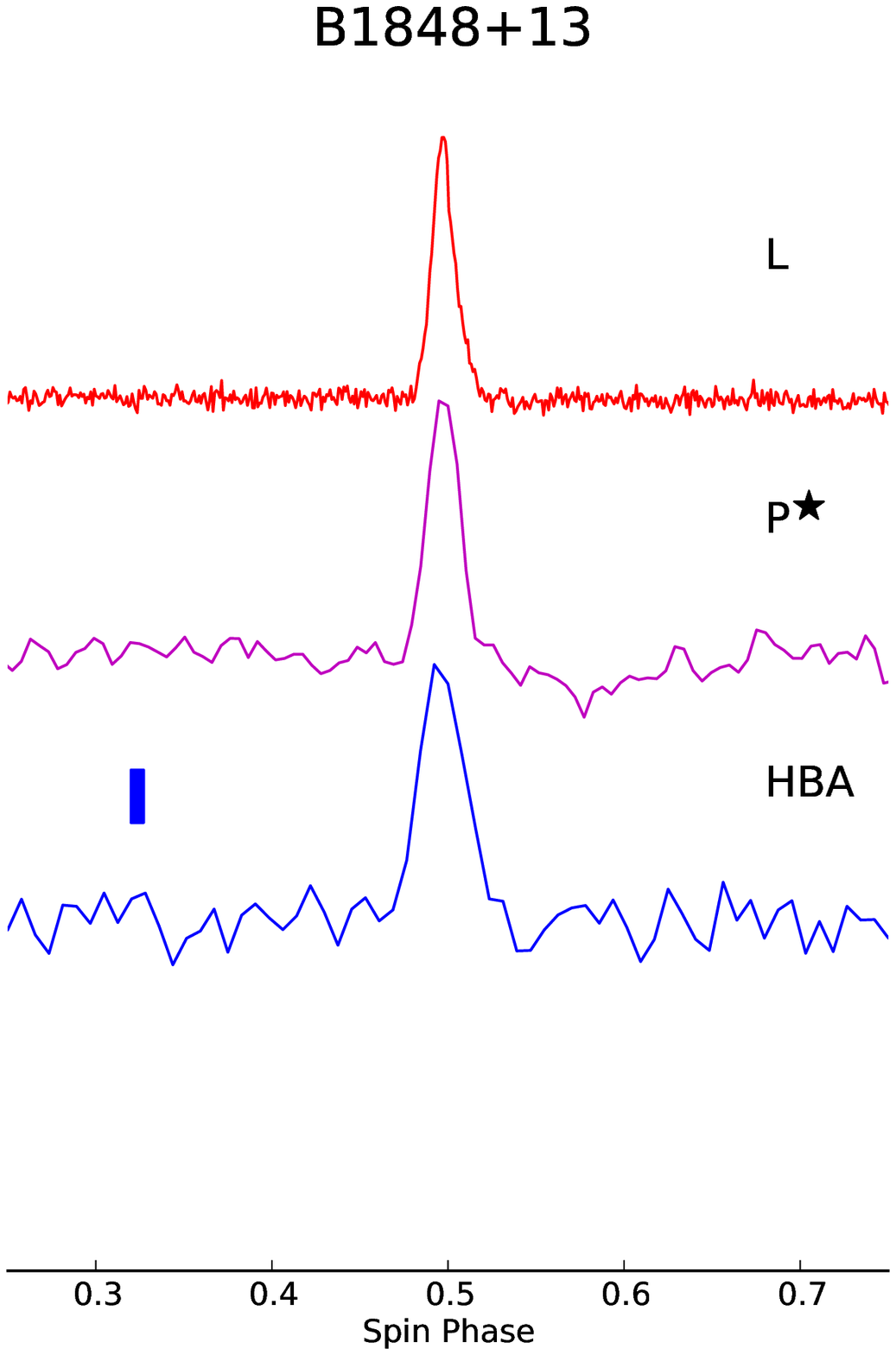} \\

 \includegraphics[width=32mm]{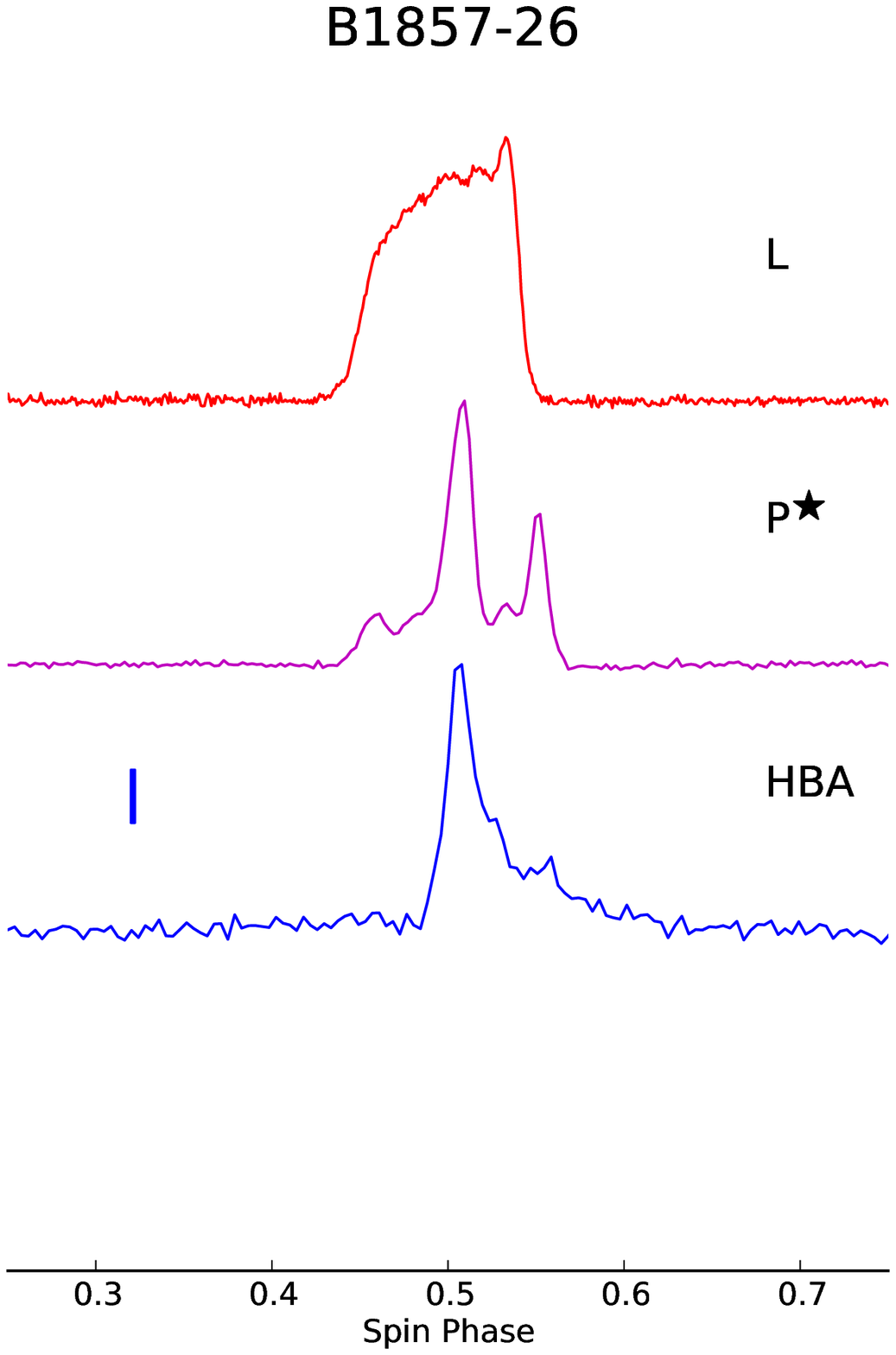} &  \includegraphics[width=32mm]{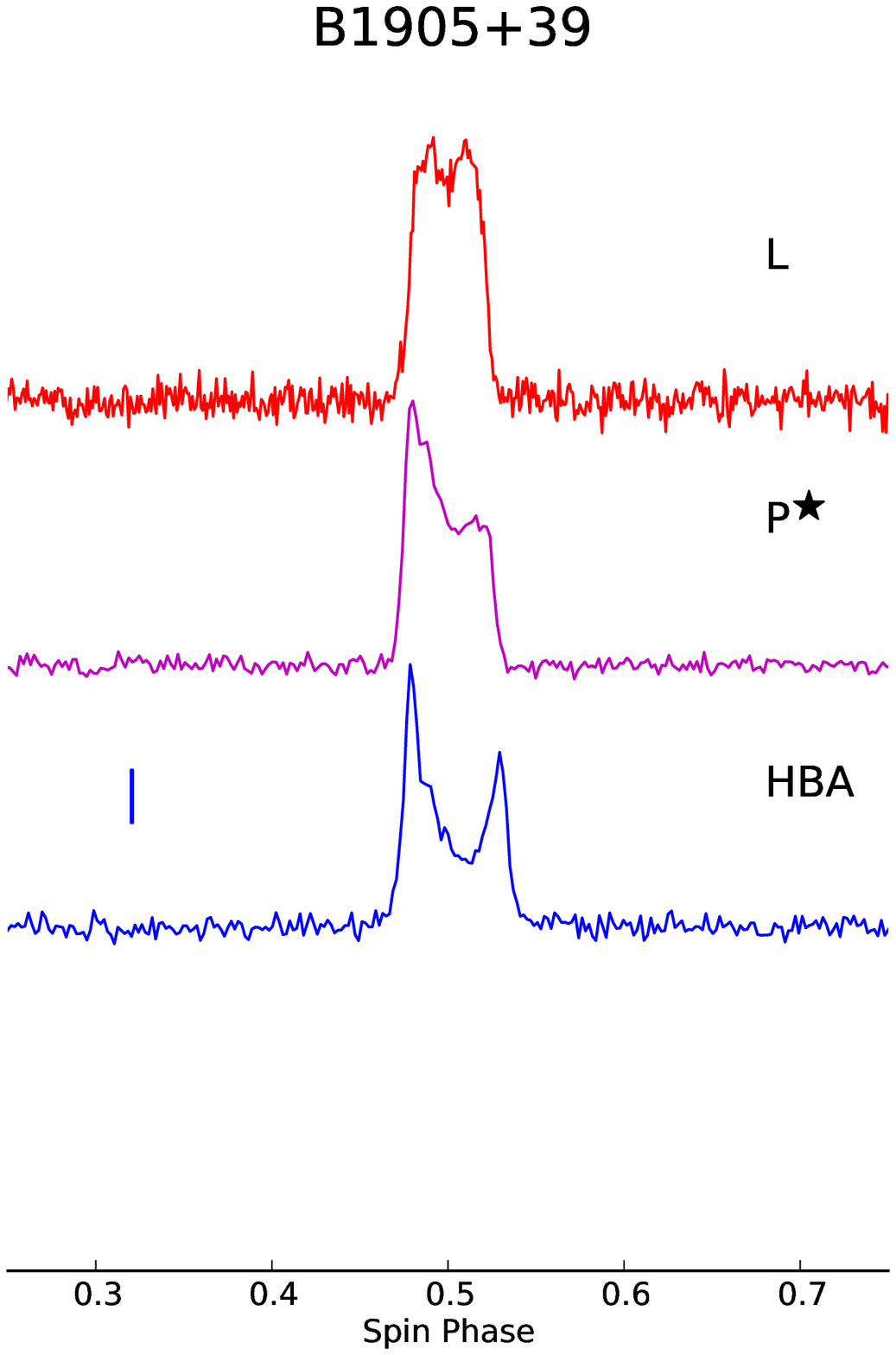} &   \includegraphics[width=32mm]{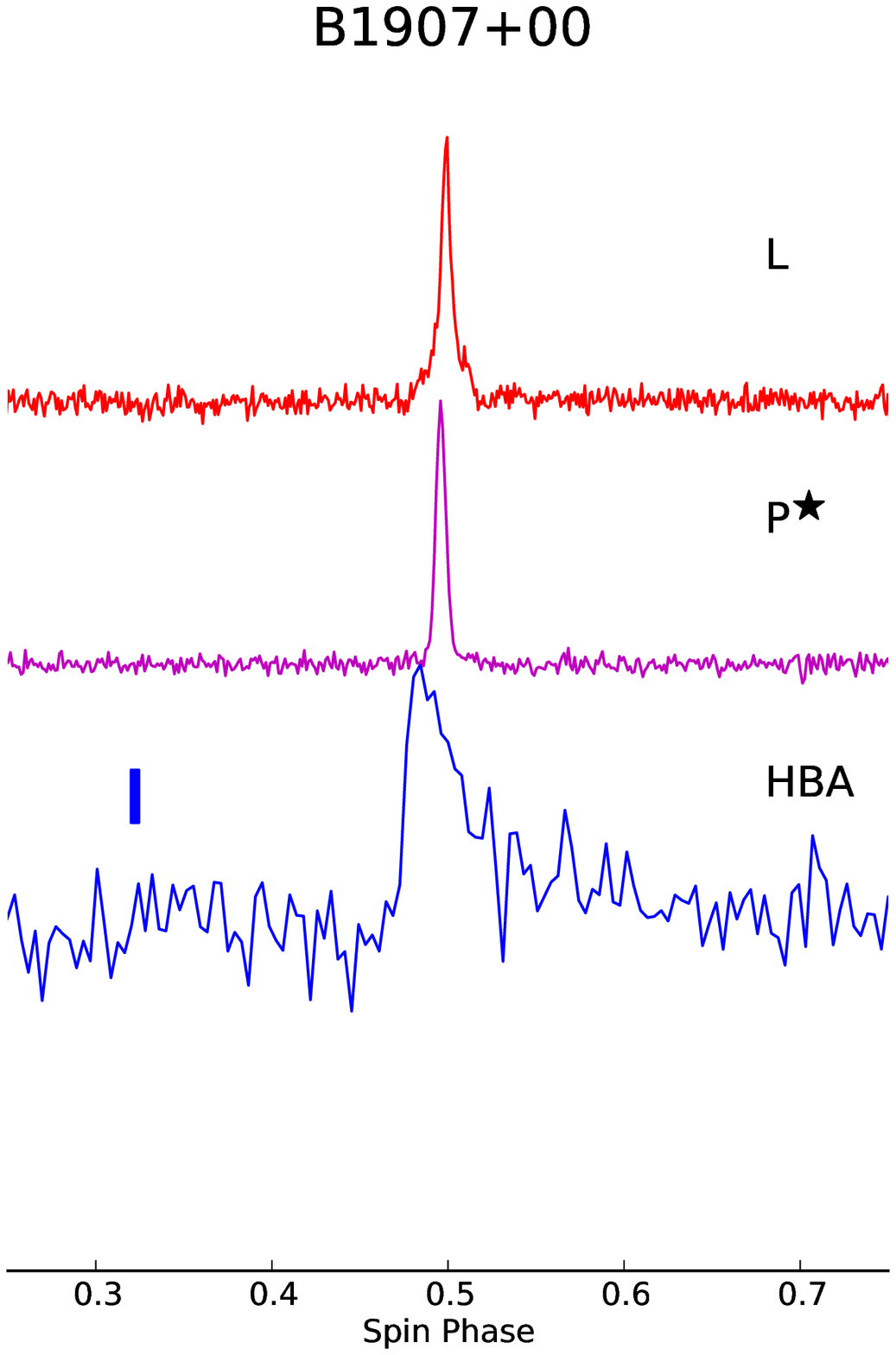} &   \includegraphics[width=32mm]{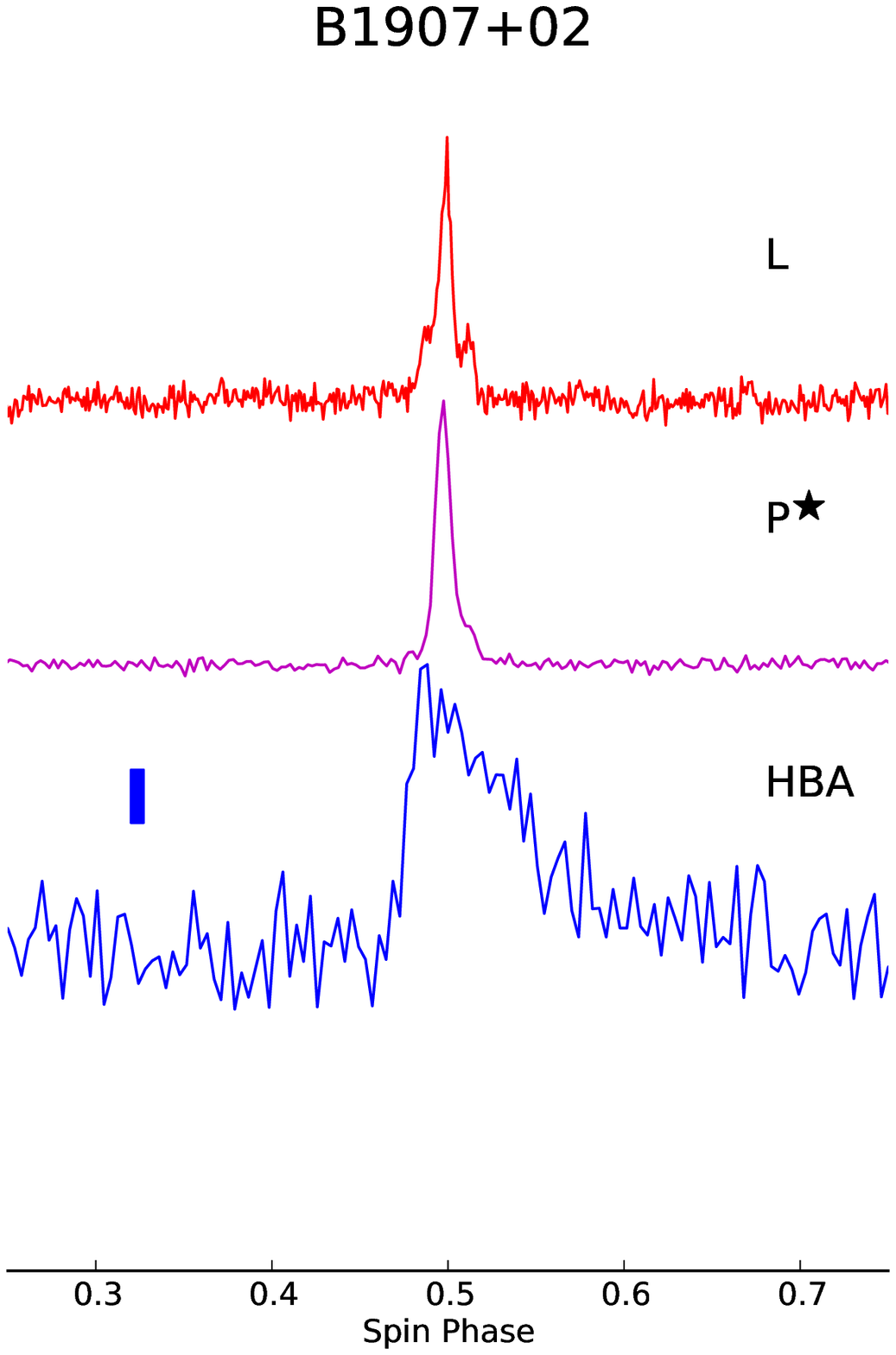} \\

 \includegraphics[width=32mm]{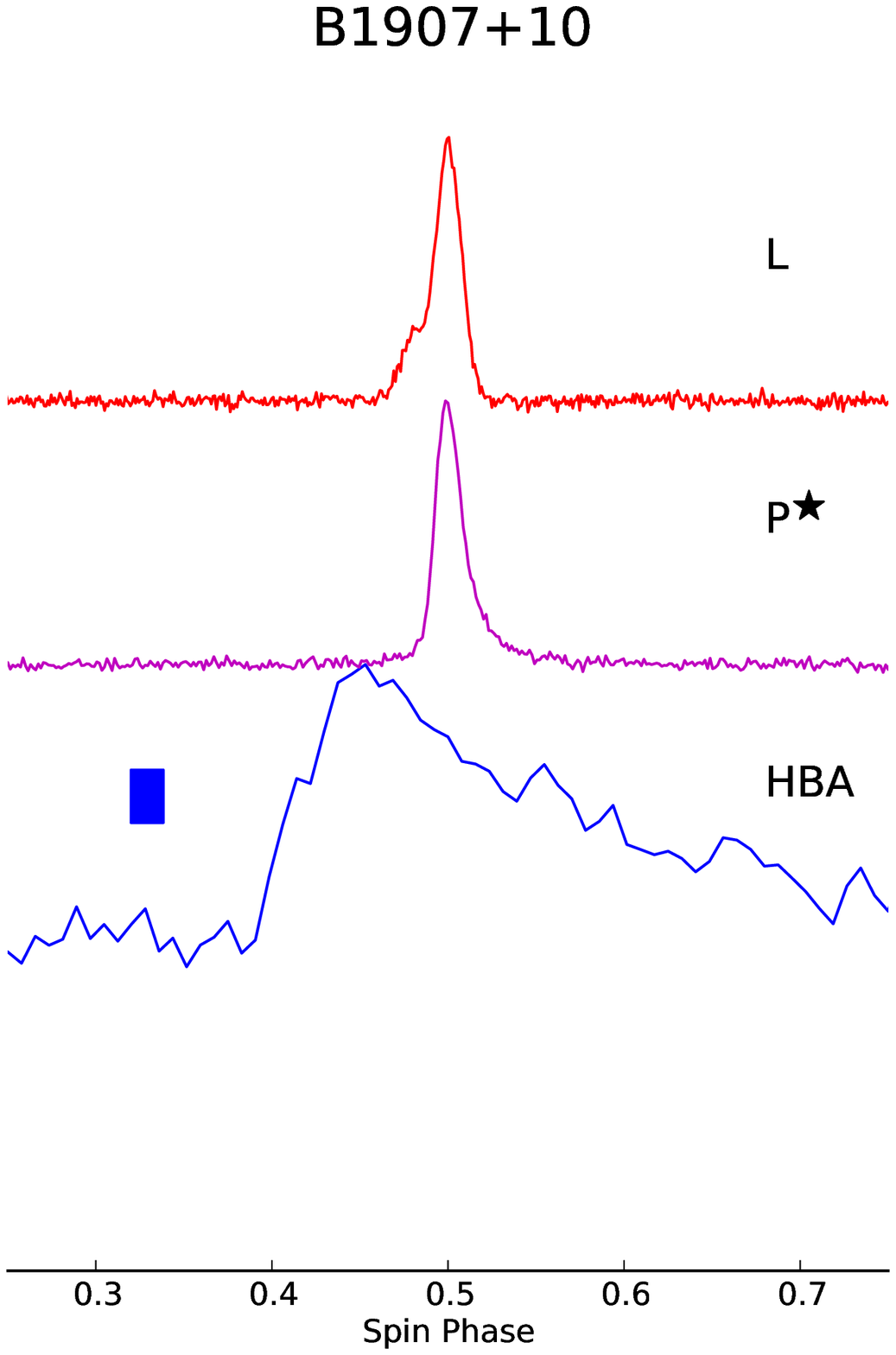} &  \includegraphics[width=32mm]{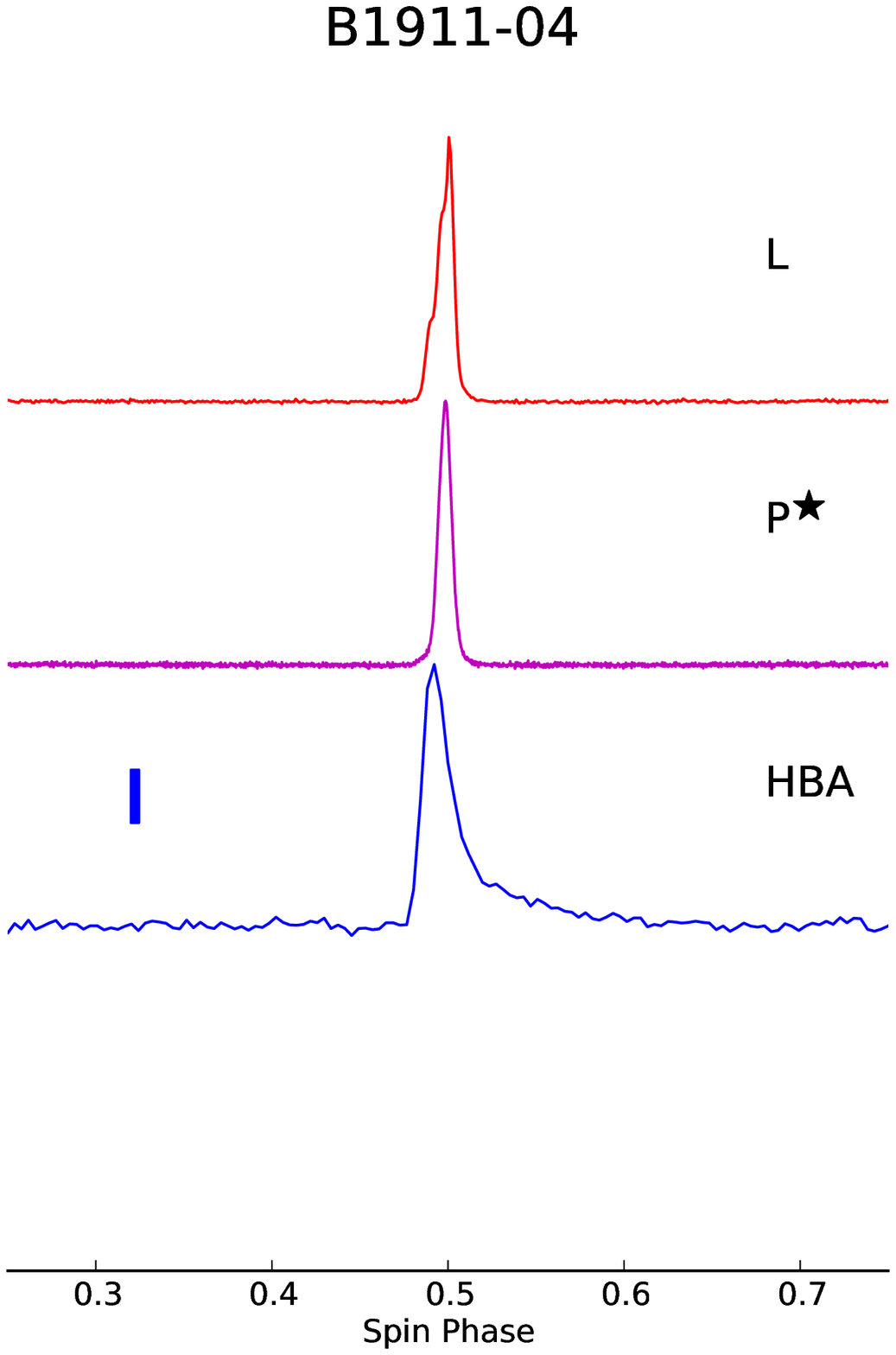} &   \includegraphics[width=32mm]{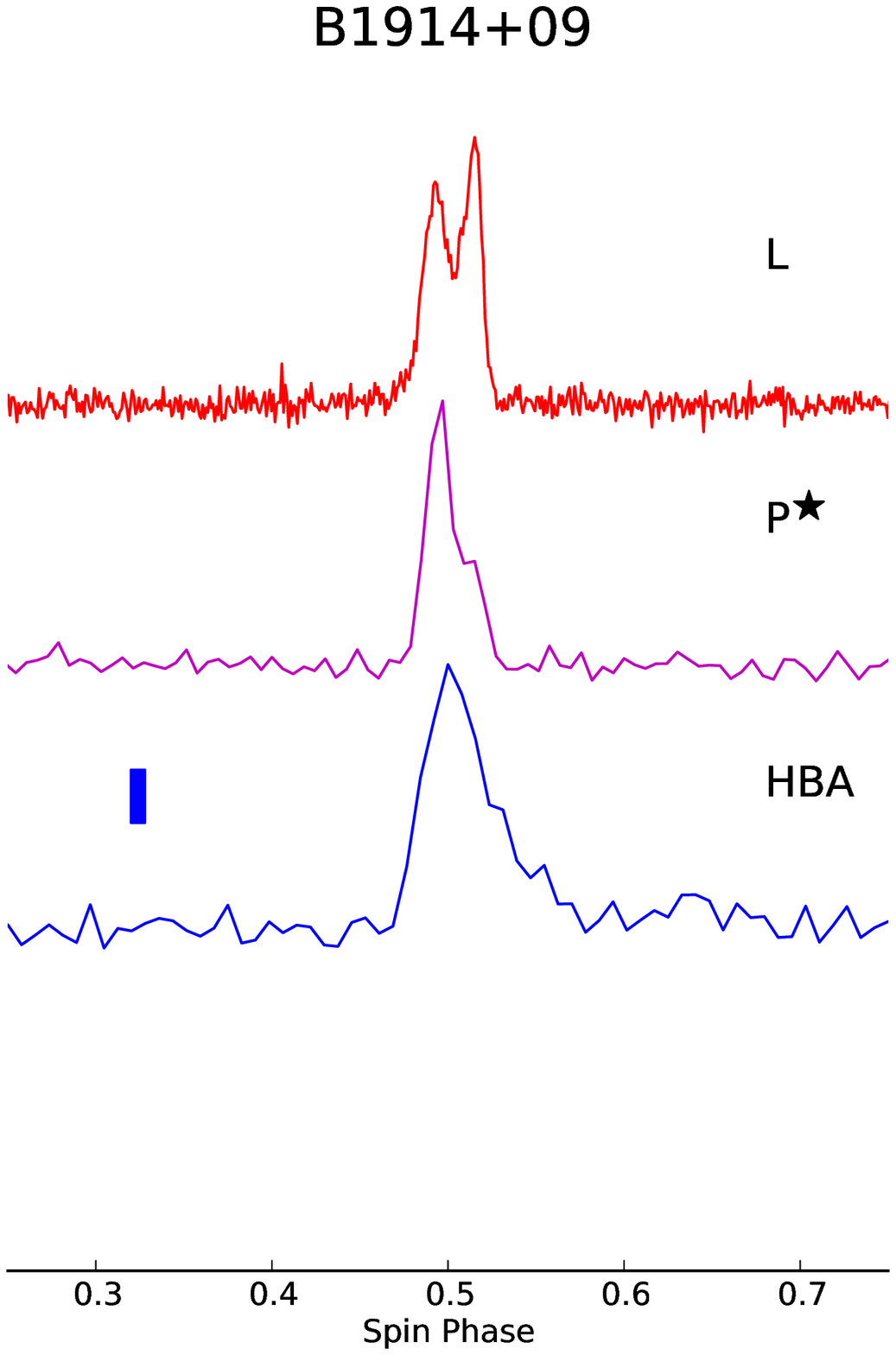} &   \includegraphics[width=32mm]{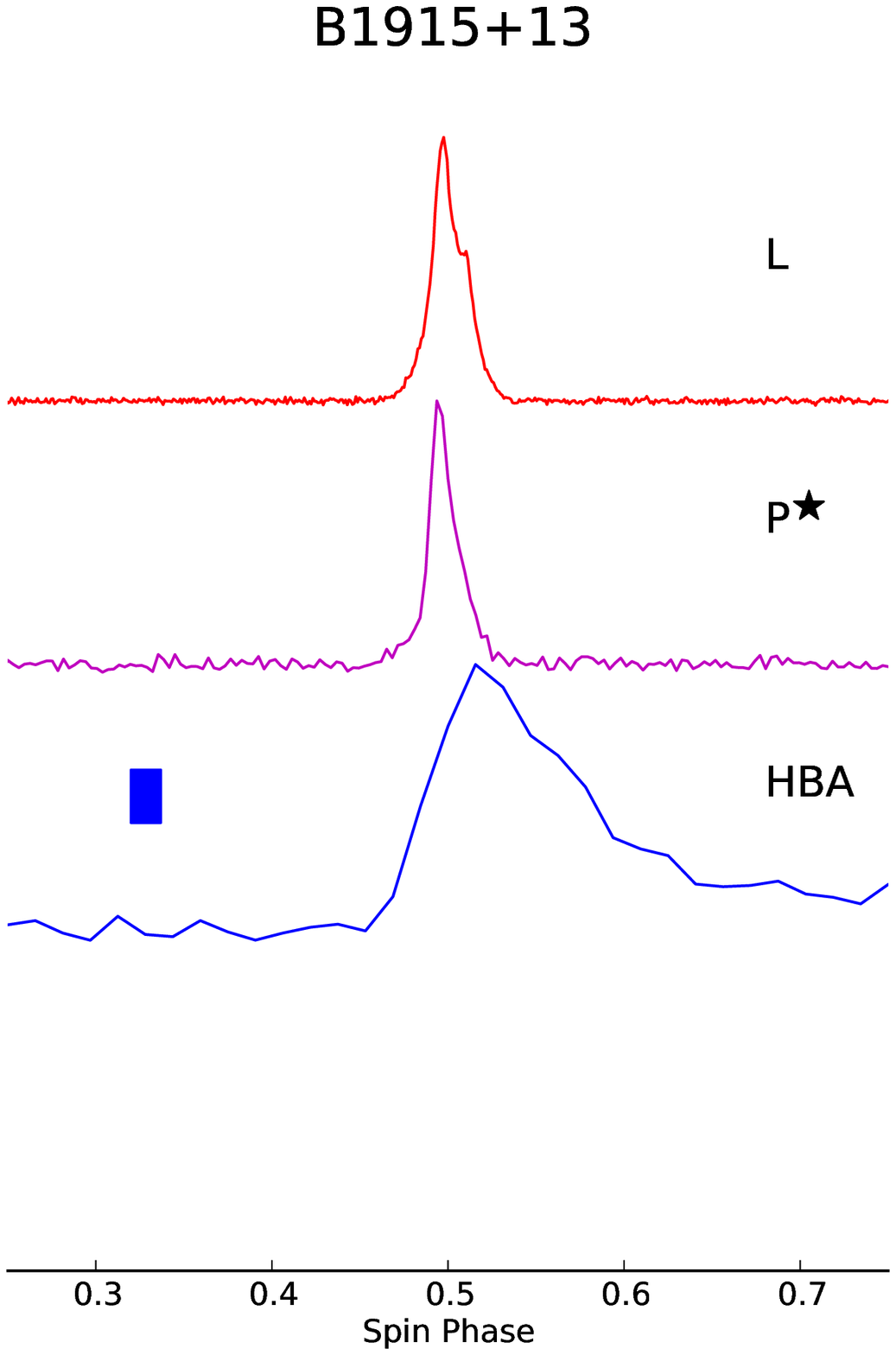} \\

\end{tabular}
\caption*{\label{fig:100_5}
Fig.\ref{fig:100}: continued
}
\end{figure*}

\begin{figure*}[b]
\centering
\begin{tabular}{cccc}

 \includegraphics[width=32mm]{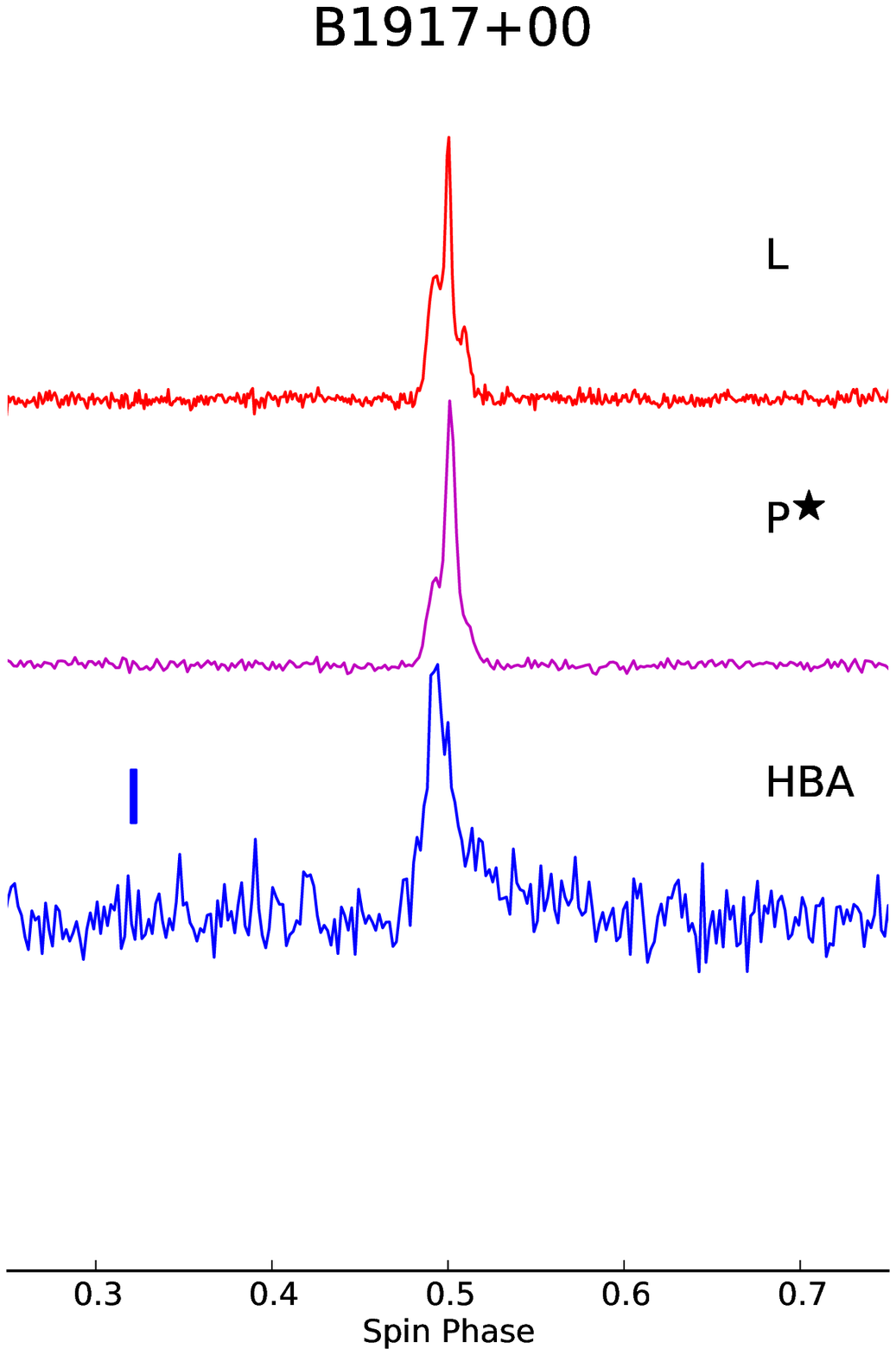} &  \includegraphics[width=32mm]{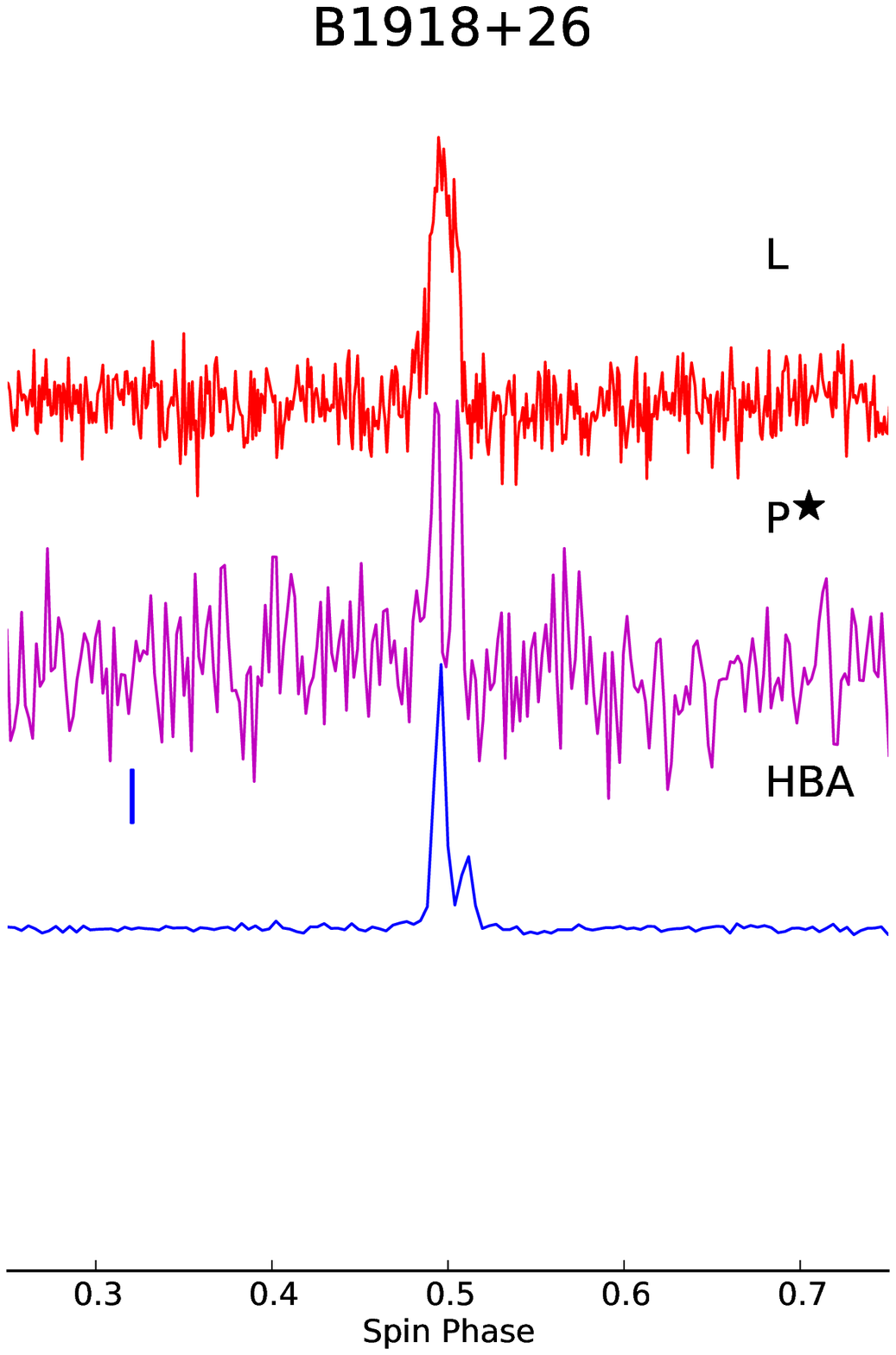} &   \includegraphics[width=32mm]{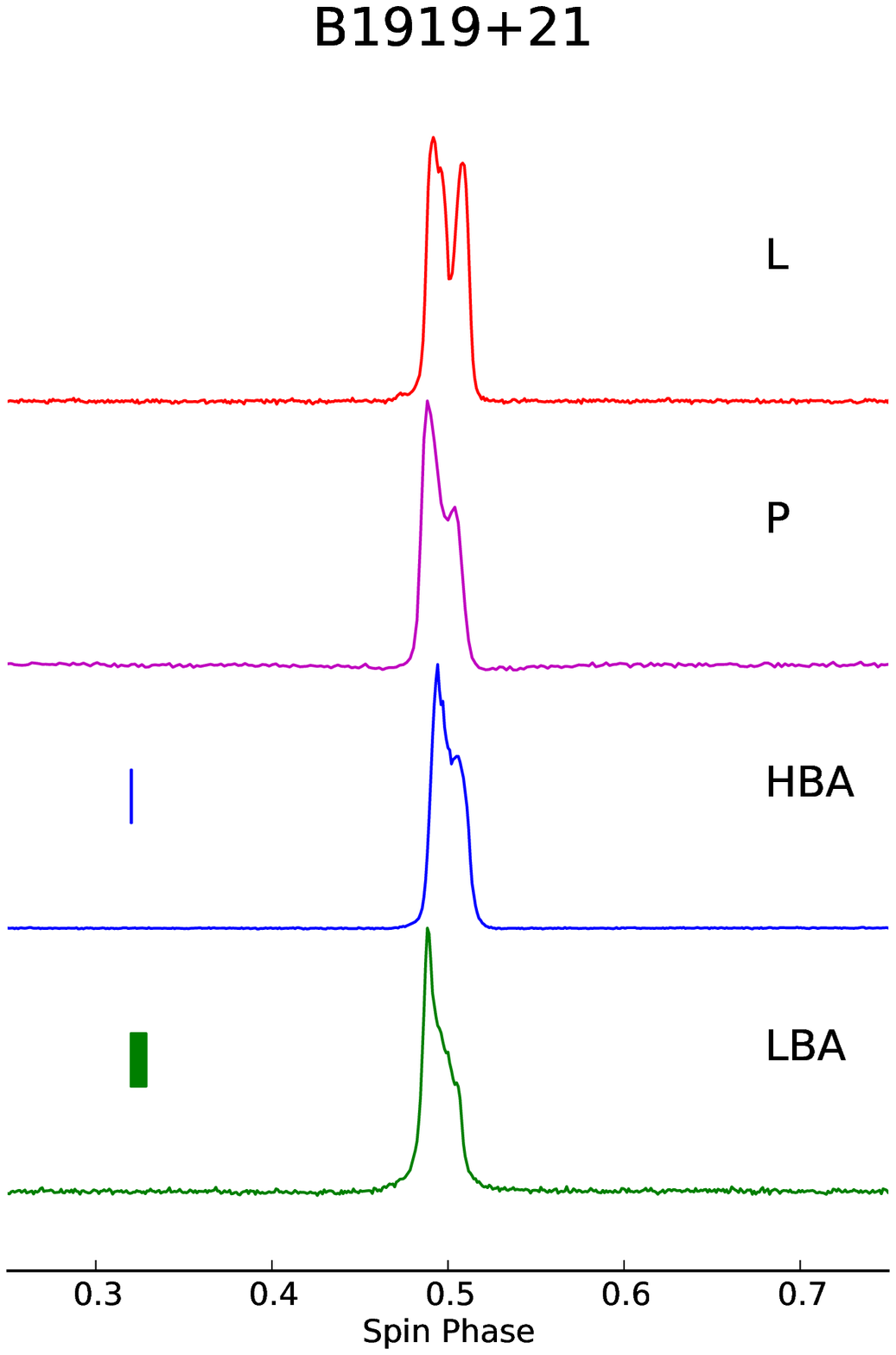} &   \includegraphics[width=32mm]{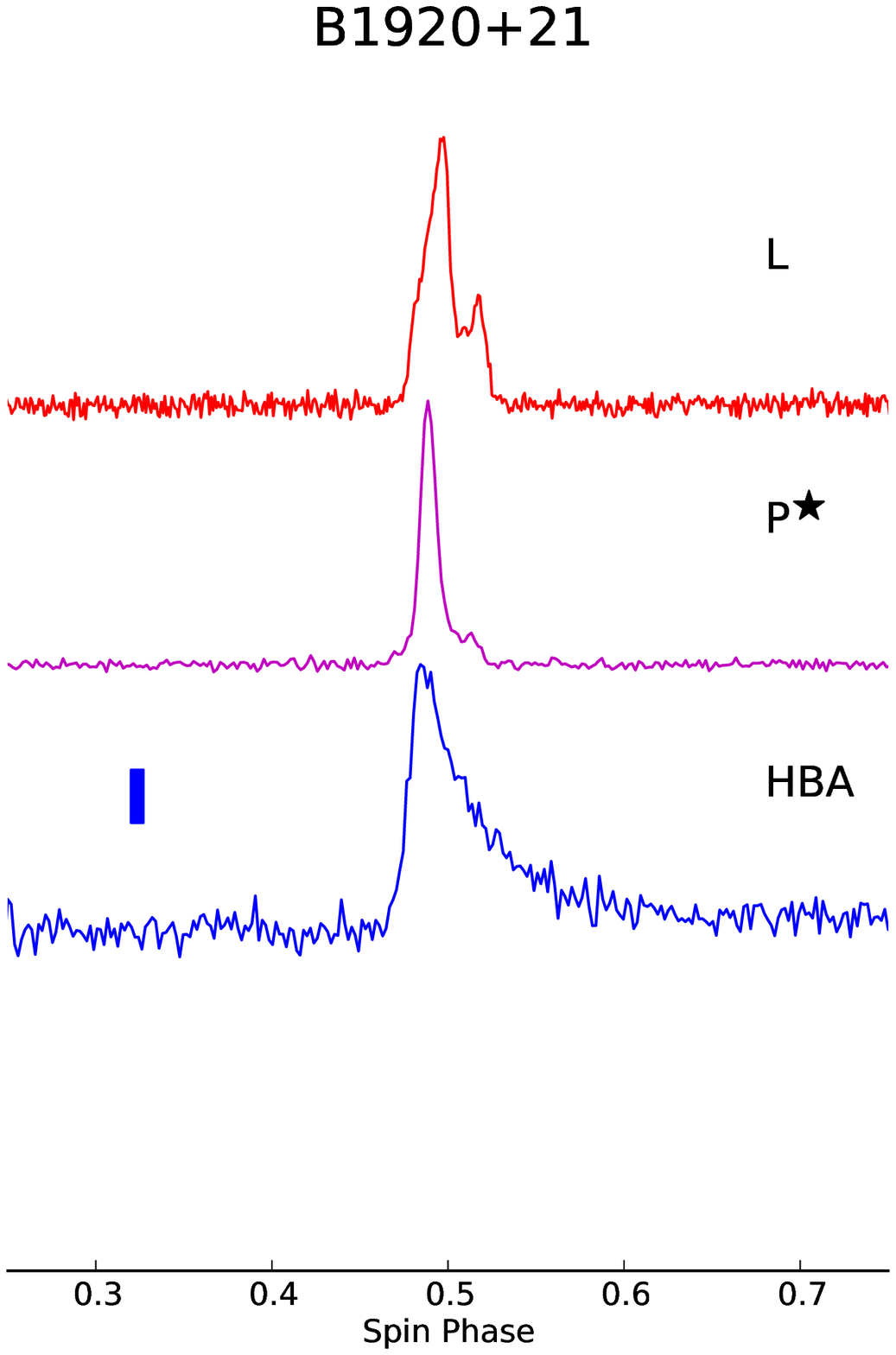} \\

 \includegraphics[width=32mm]{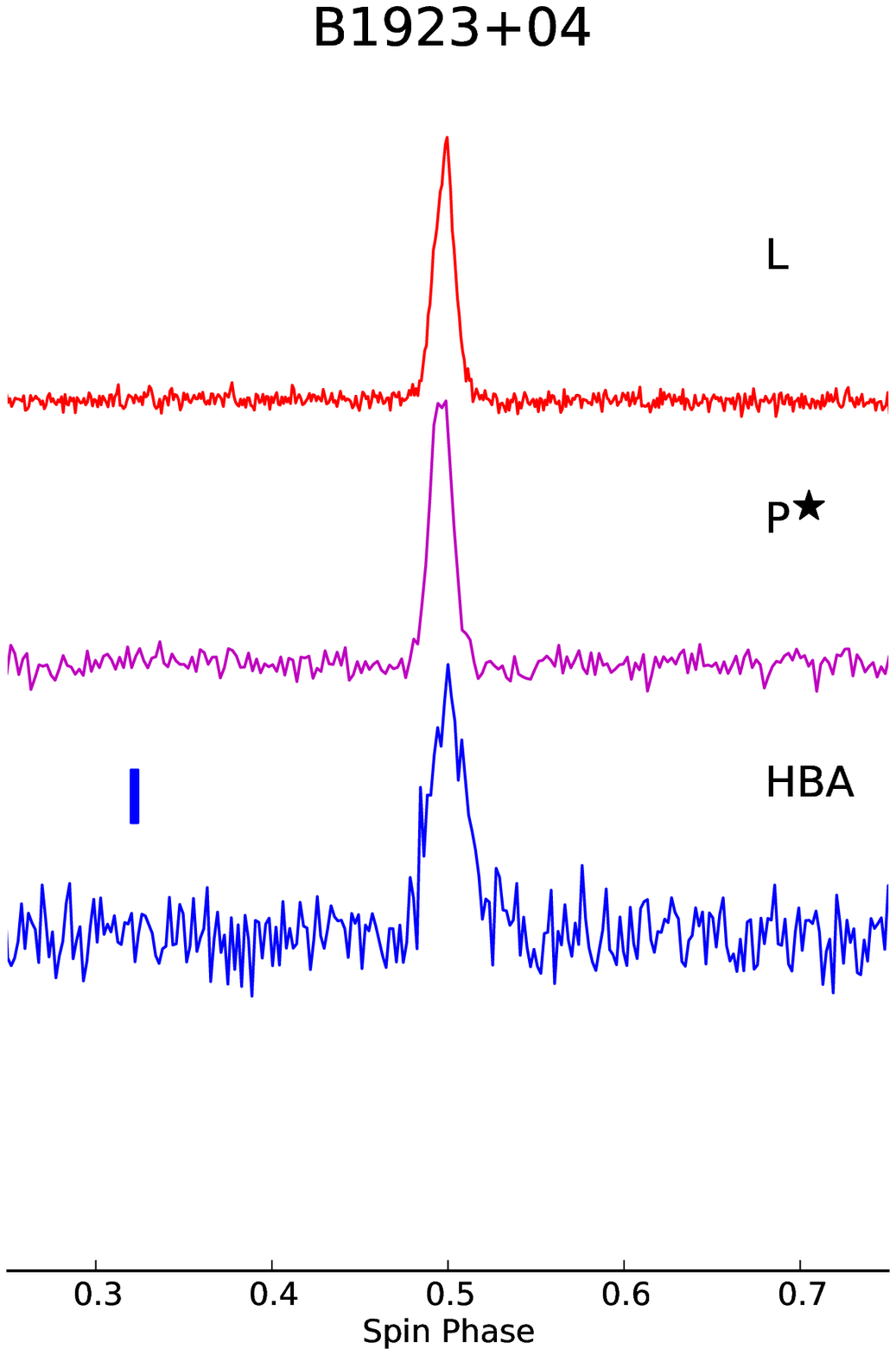} &  \includegraphics[width=32mm]{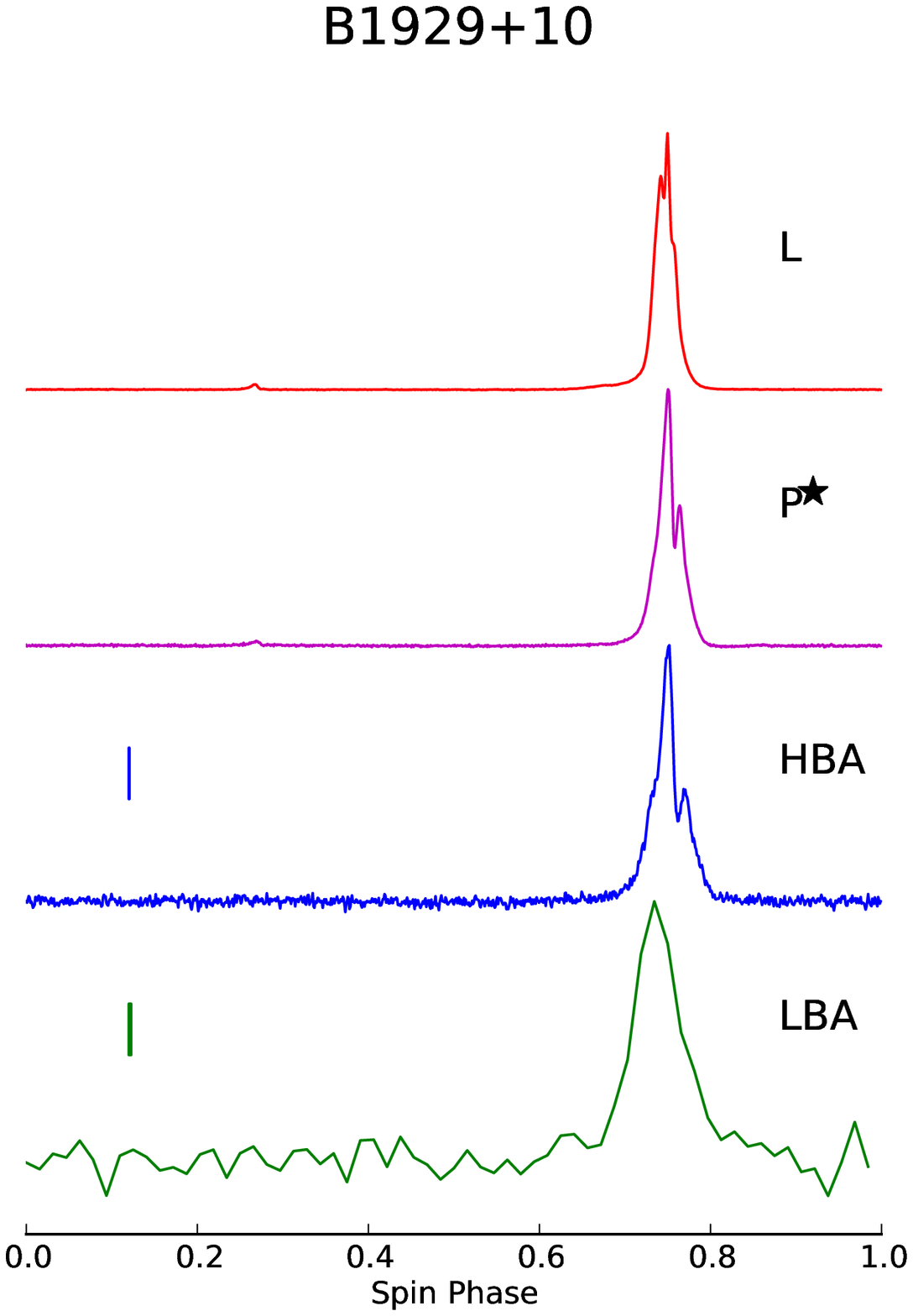} &   \includegraphics[width=32mm]{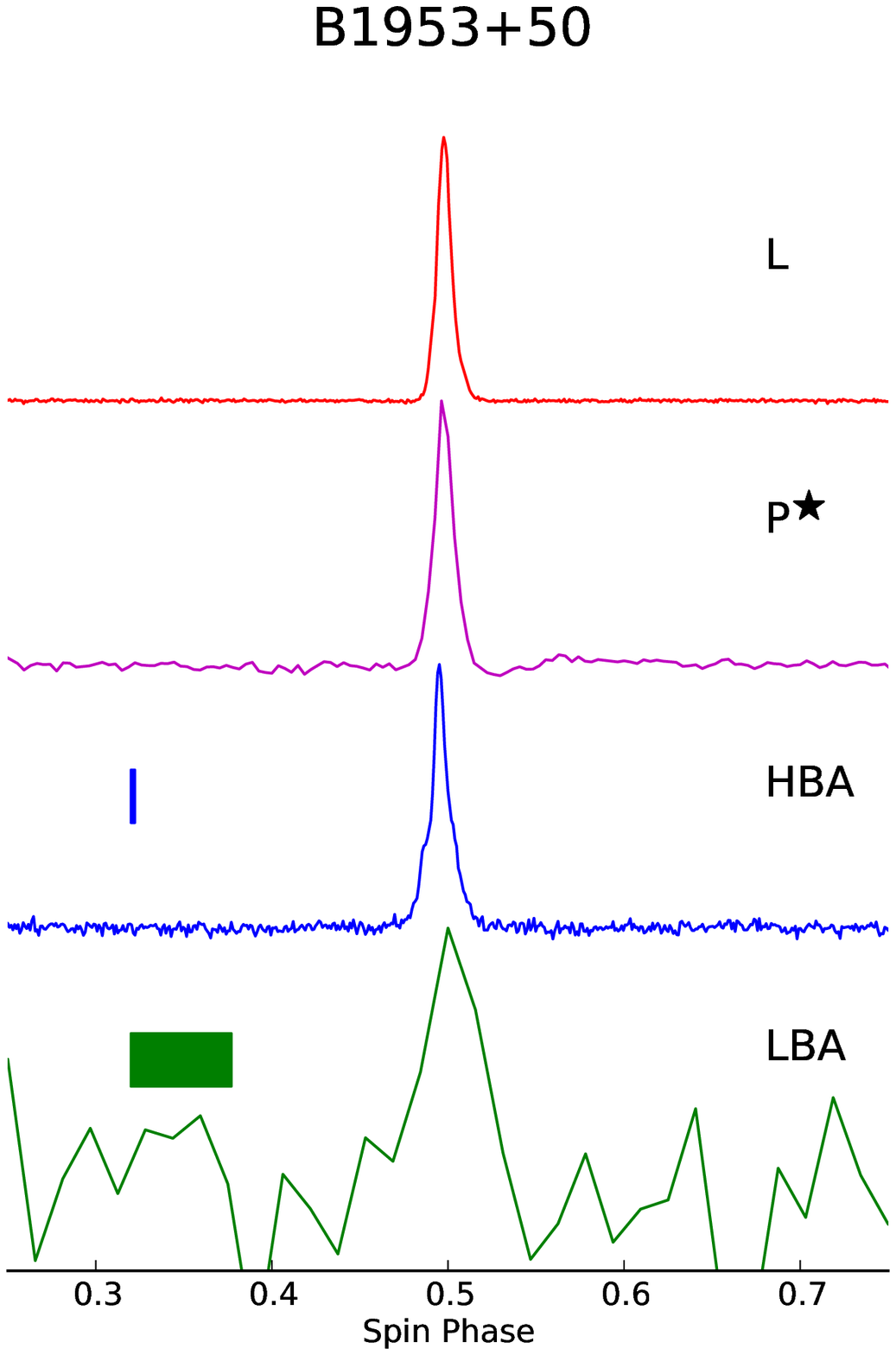} &   \includegraphics[width=32mm]{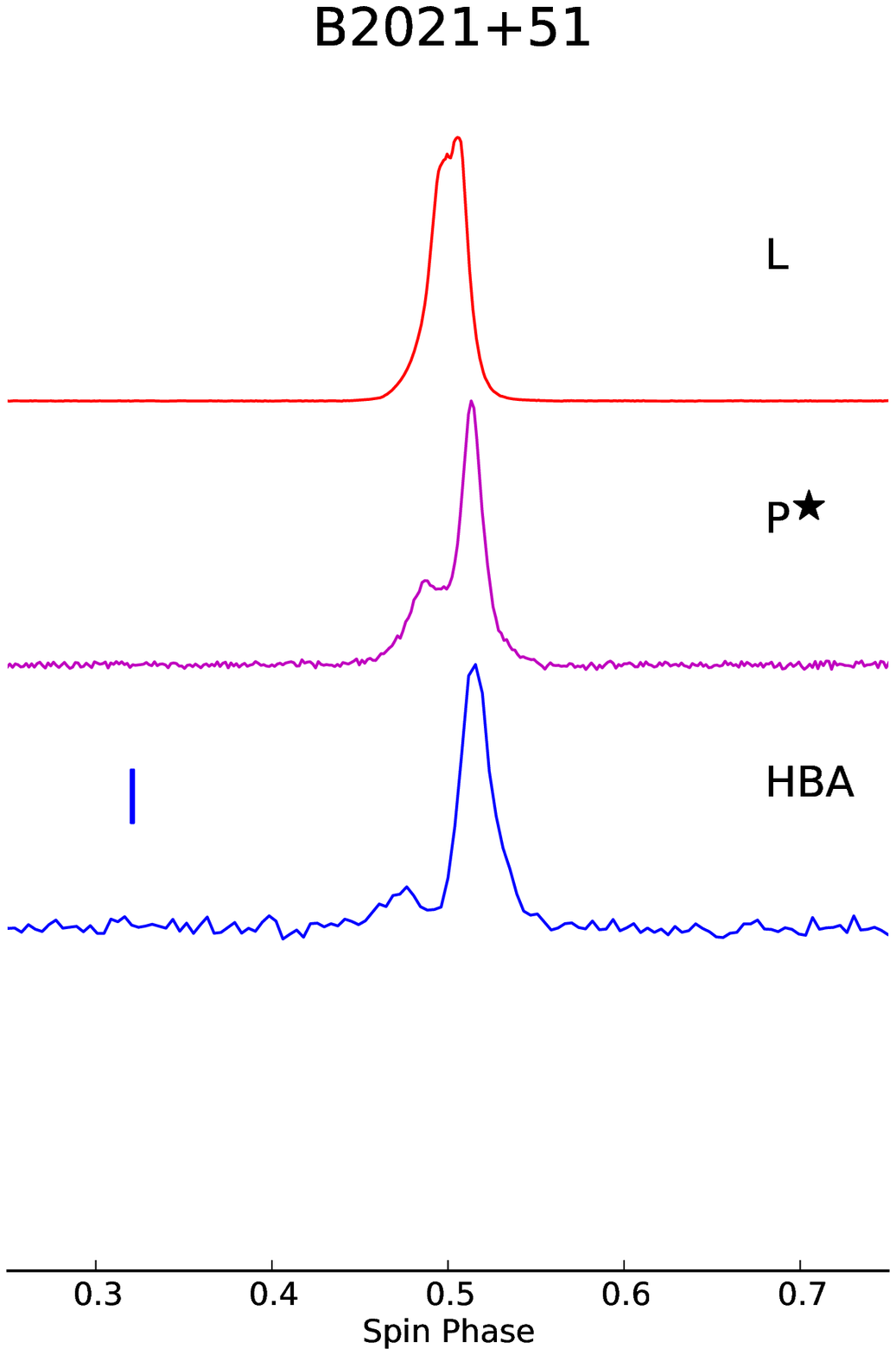} \\

 \includegraphics[width=32mm]{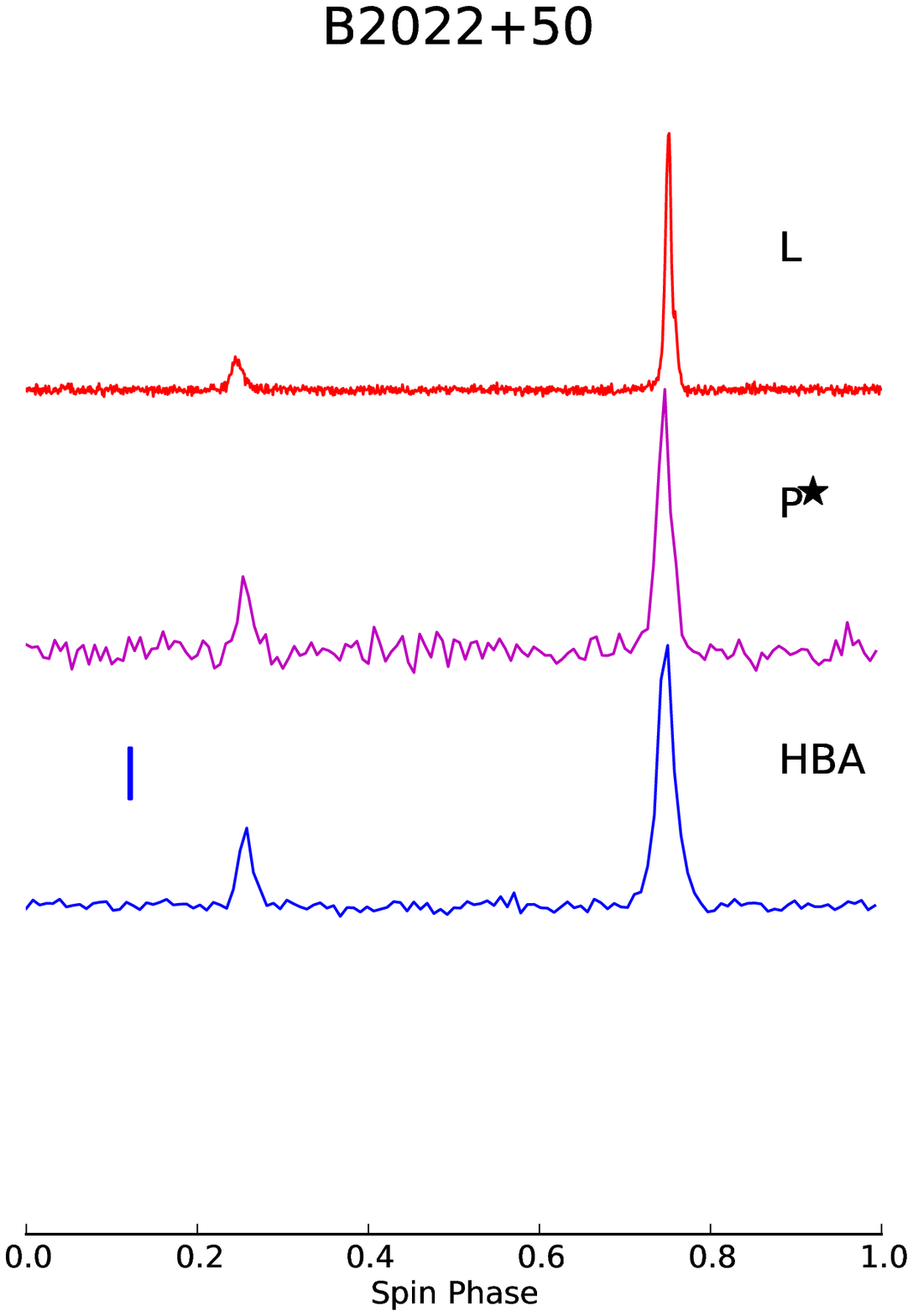} &  \includegraphics[width=32mm]{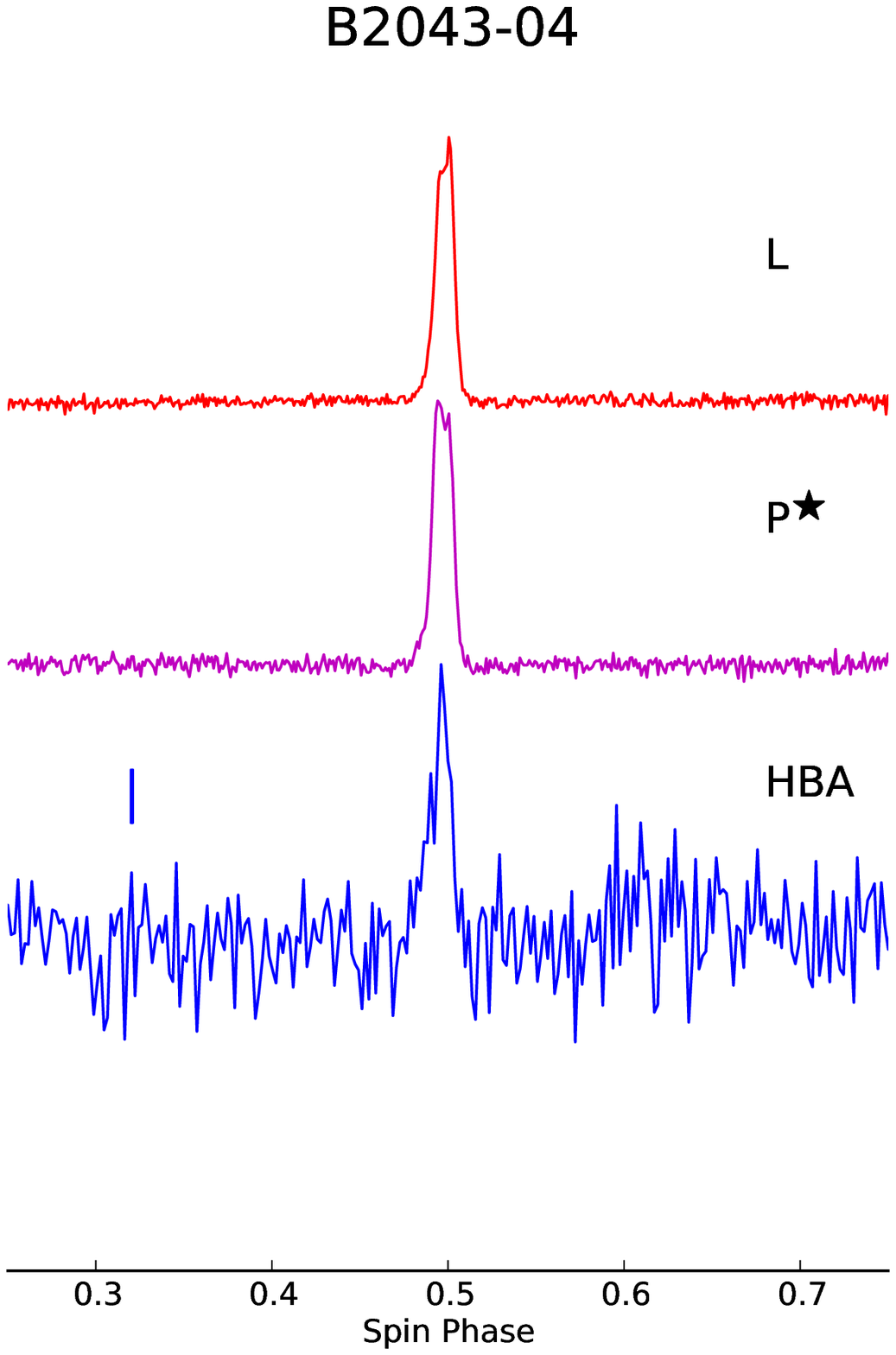} &   \includegraphics[width=32mm]{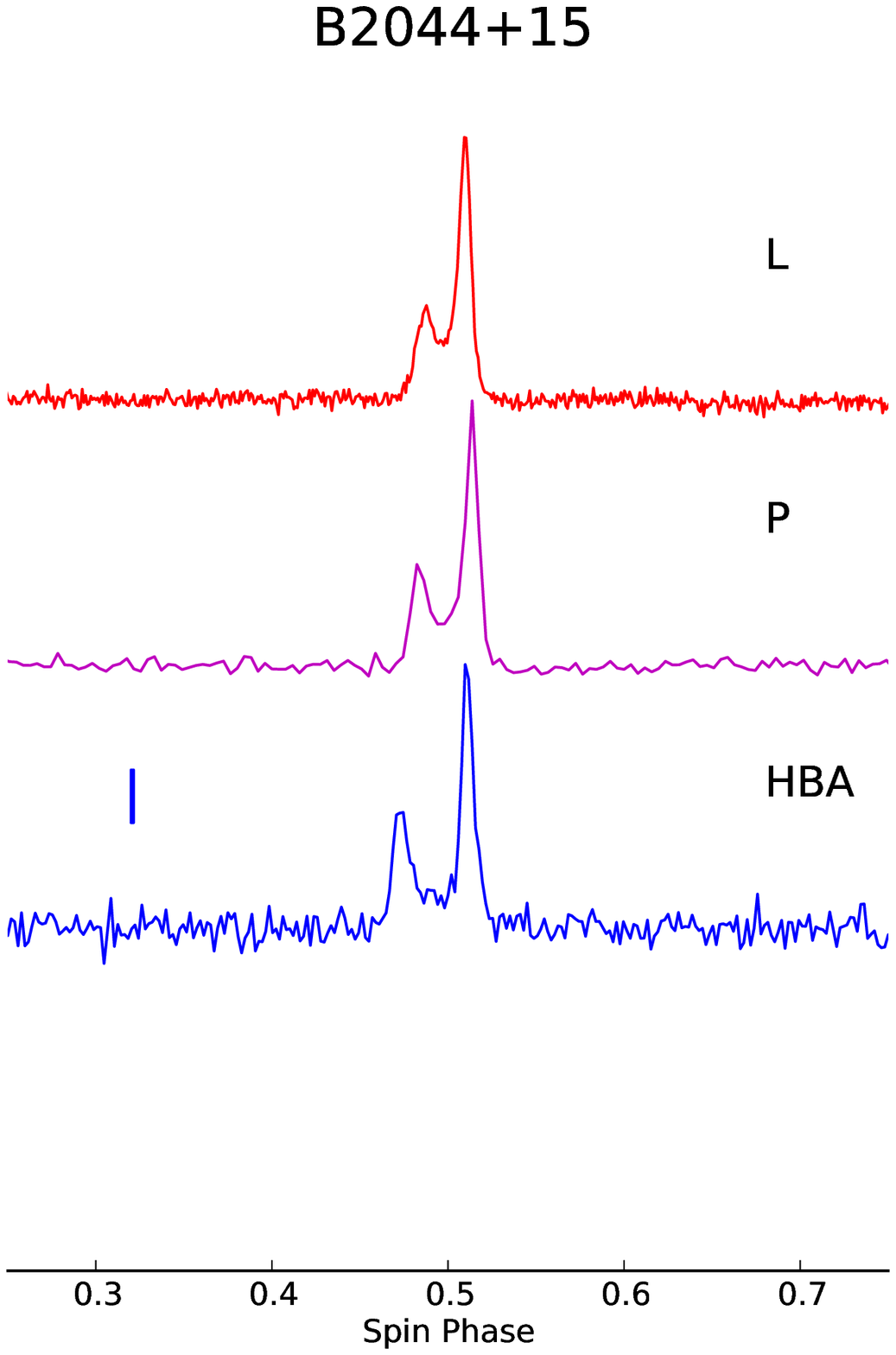} &   \includegraphics[width=32mm]{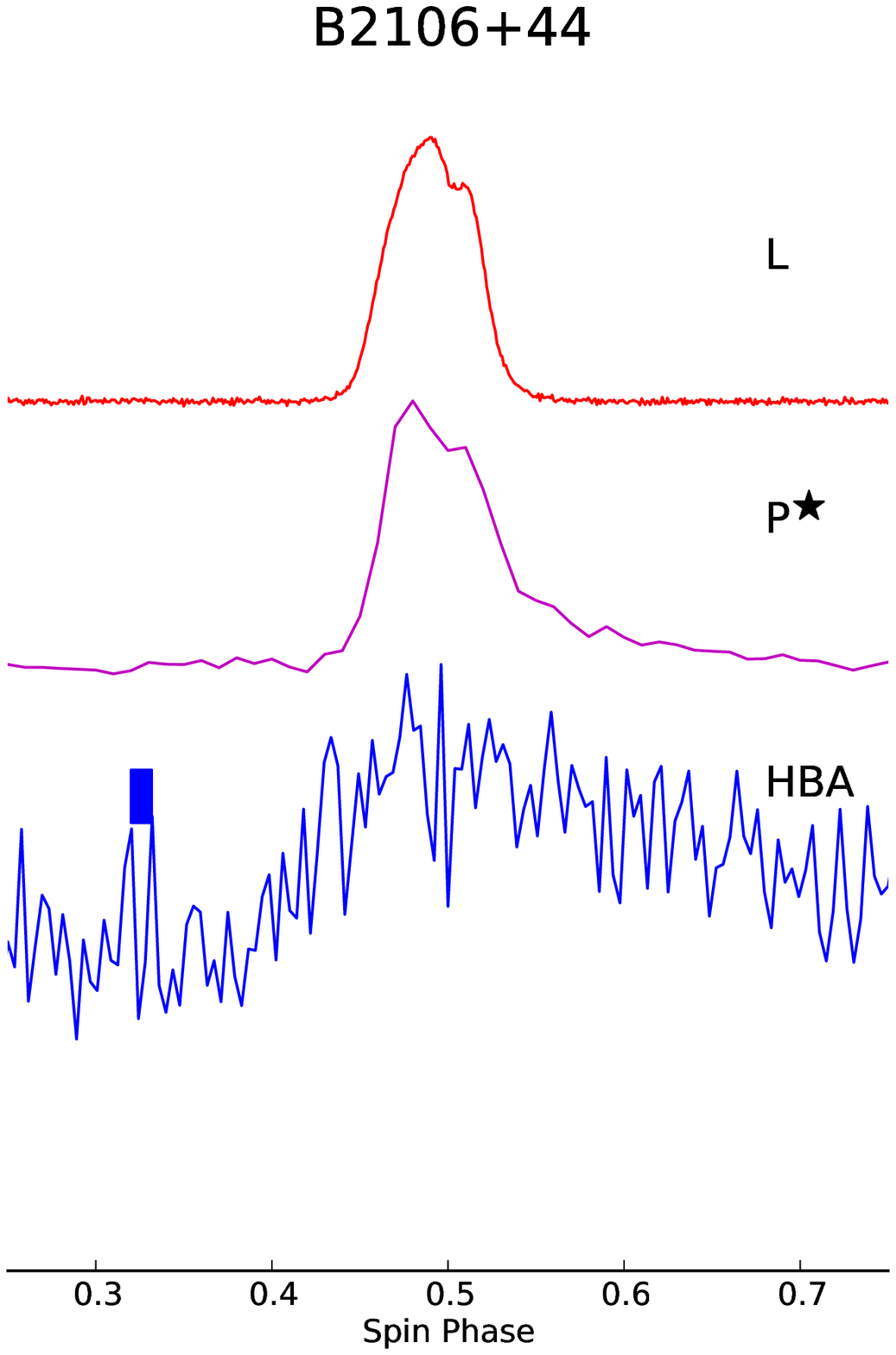} \\

 \includegraphics[width=32mm]{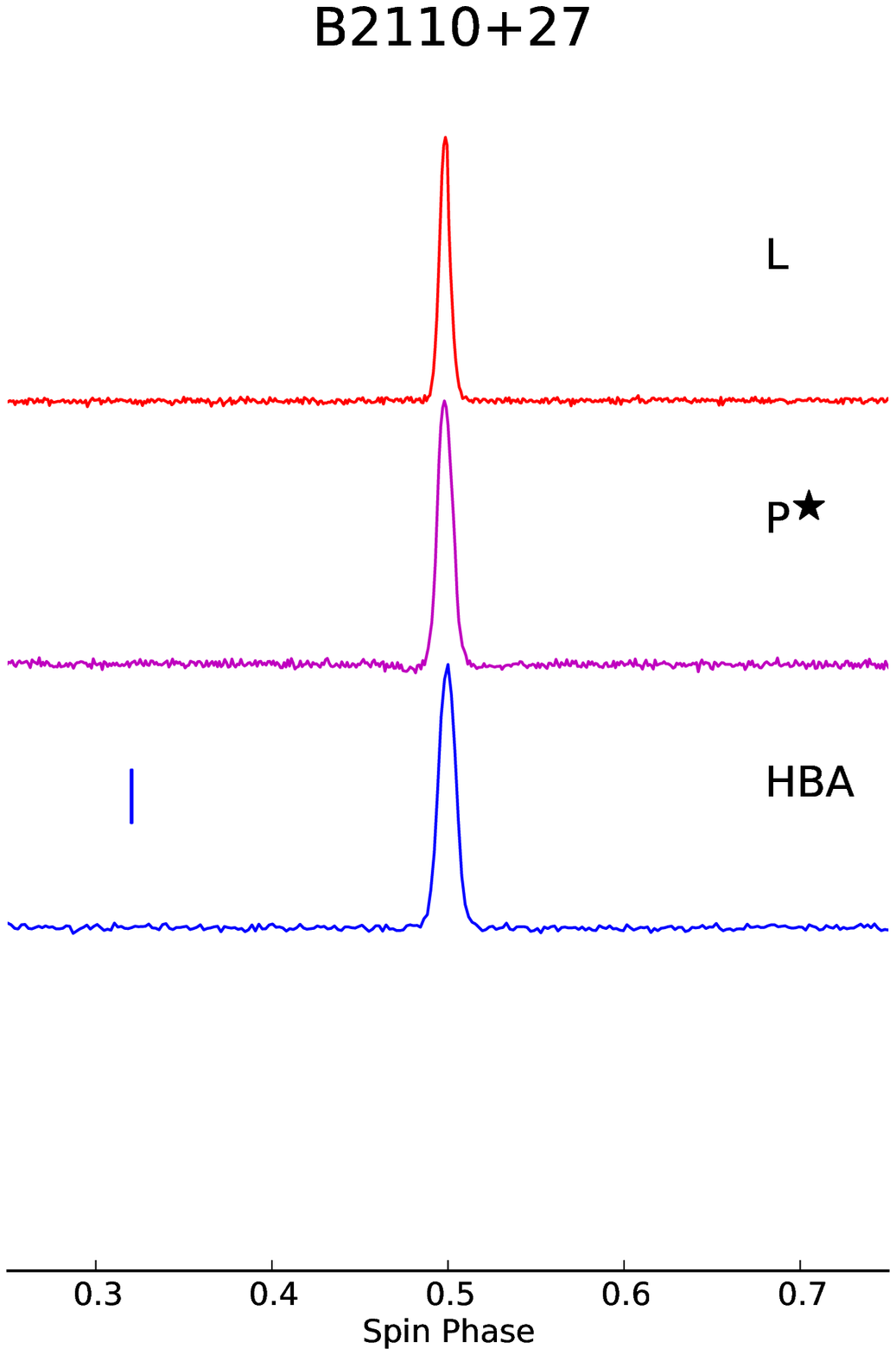} &  \includegraphics[width=32mm]{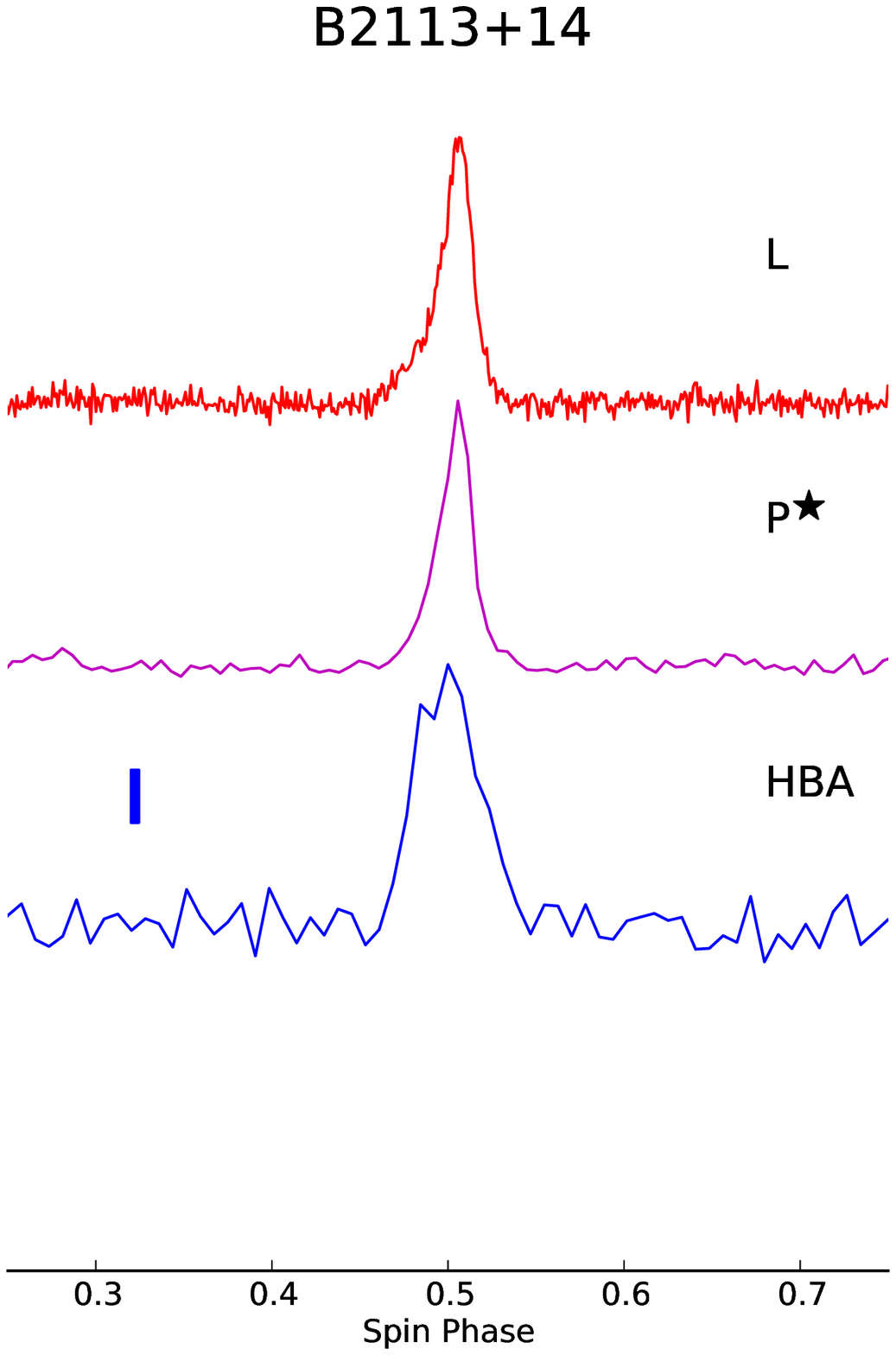} &   \includegraphics[width=32mm]{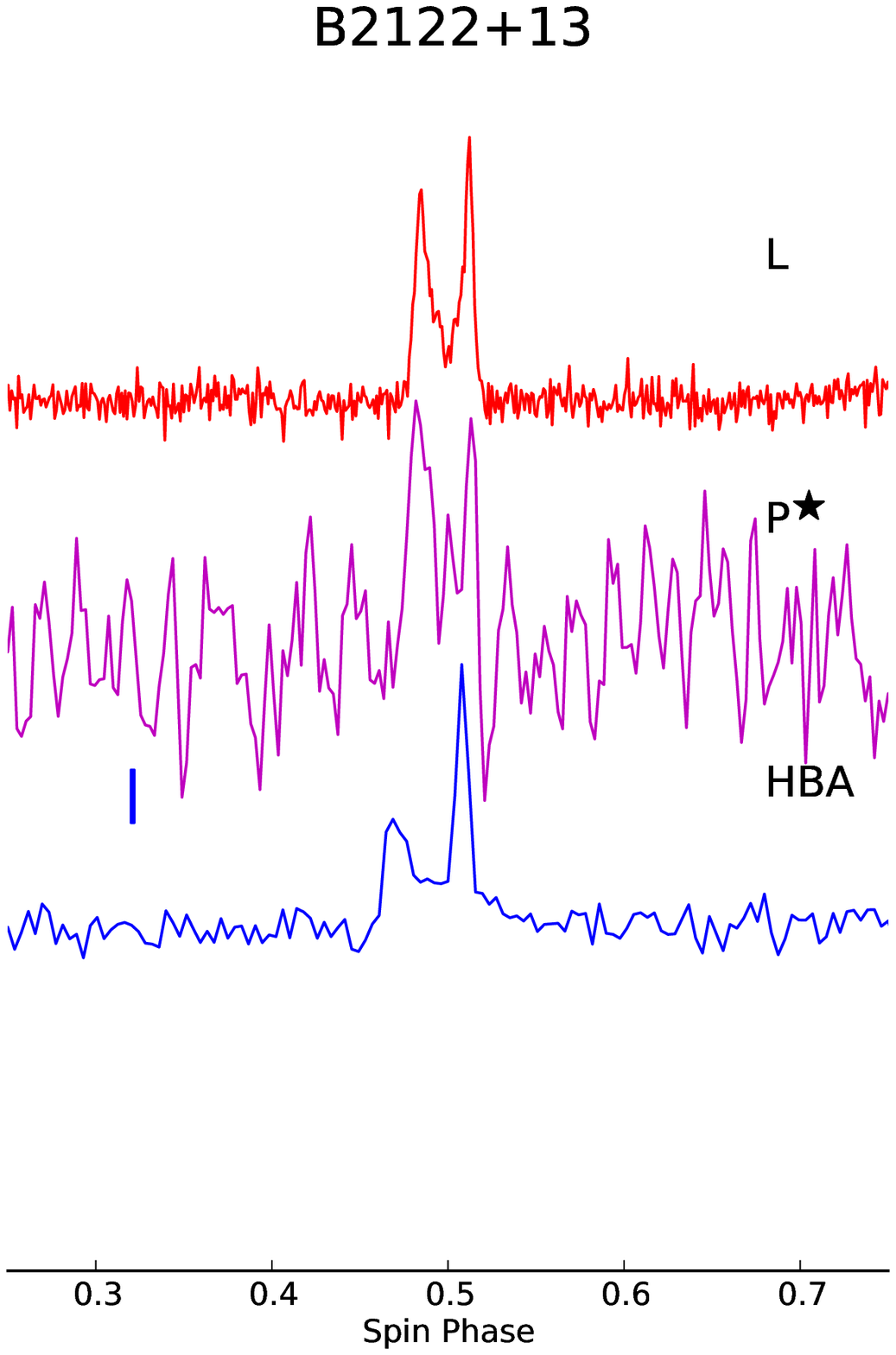} &   \includegraphics[width=32mm]{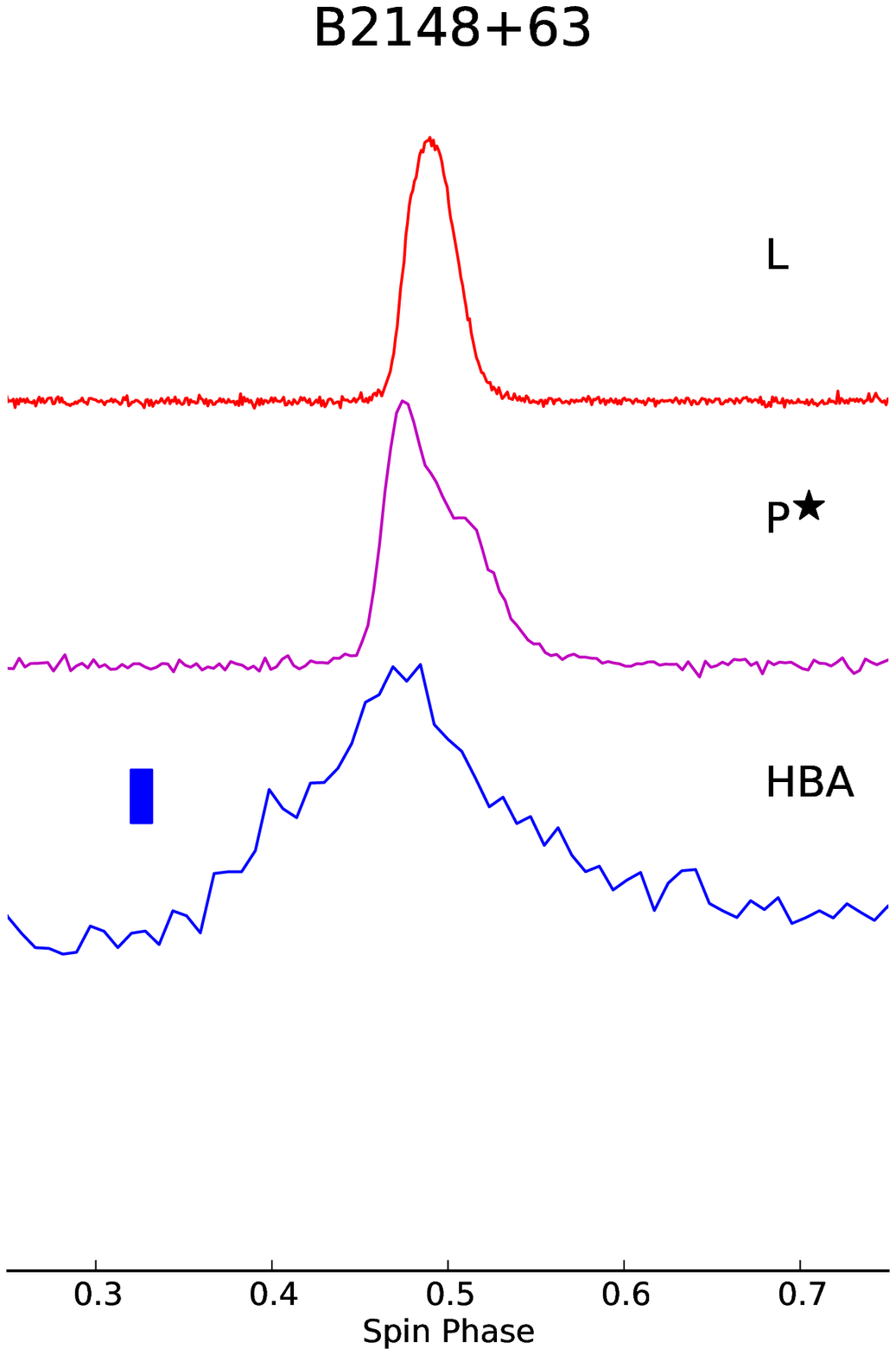} \\

 \includegraphics[width=32mm]{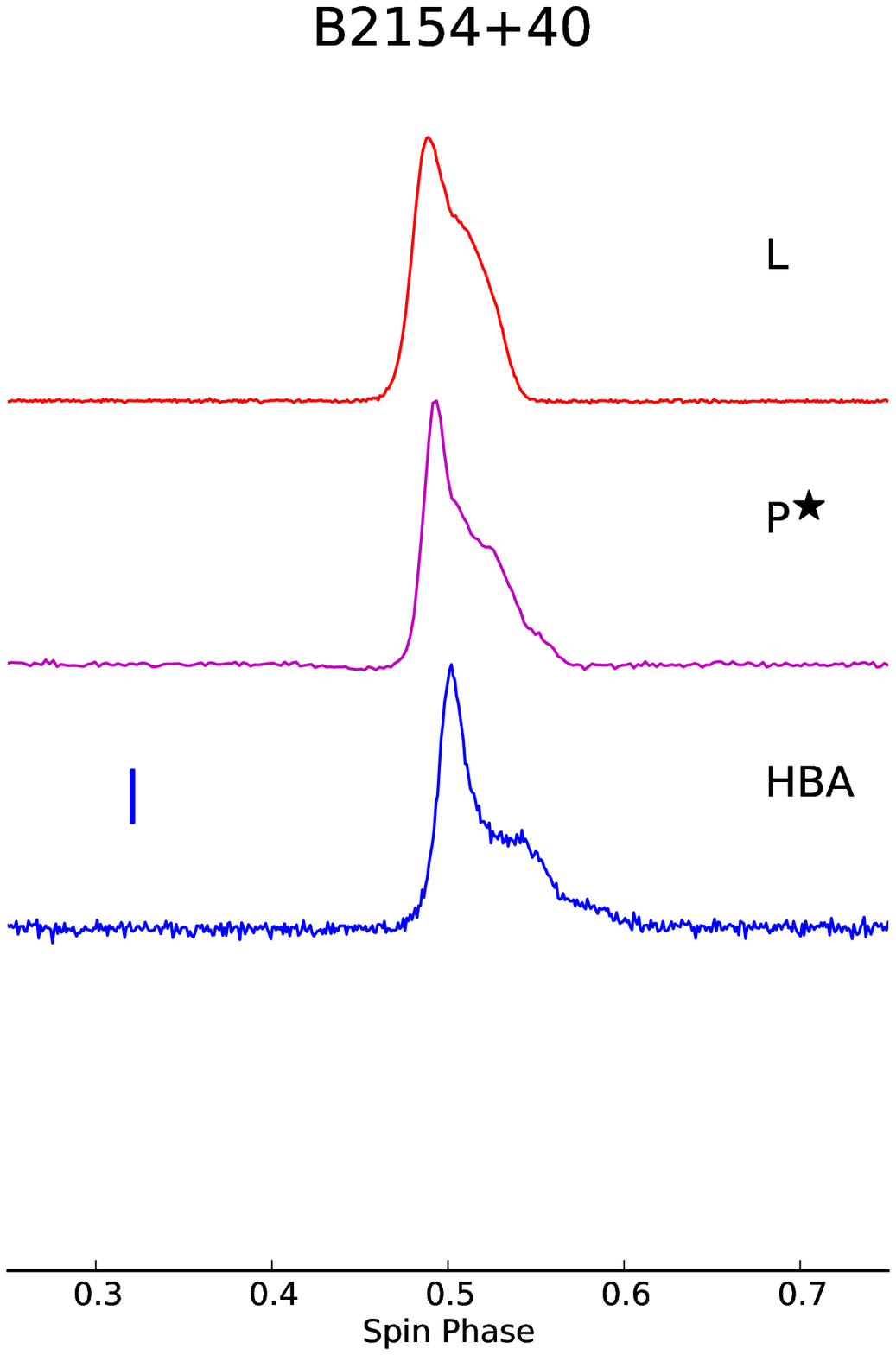} &  \includegraphics[width=32mm]{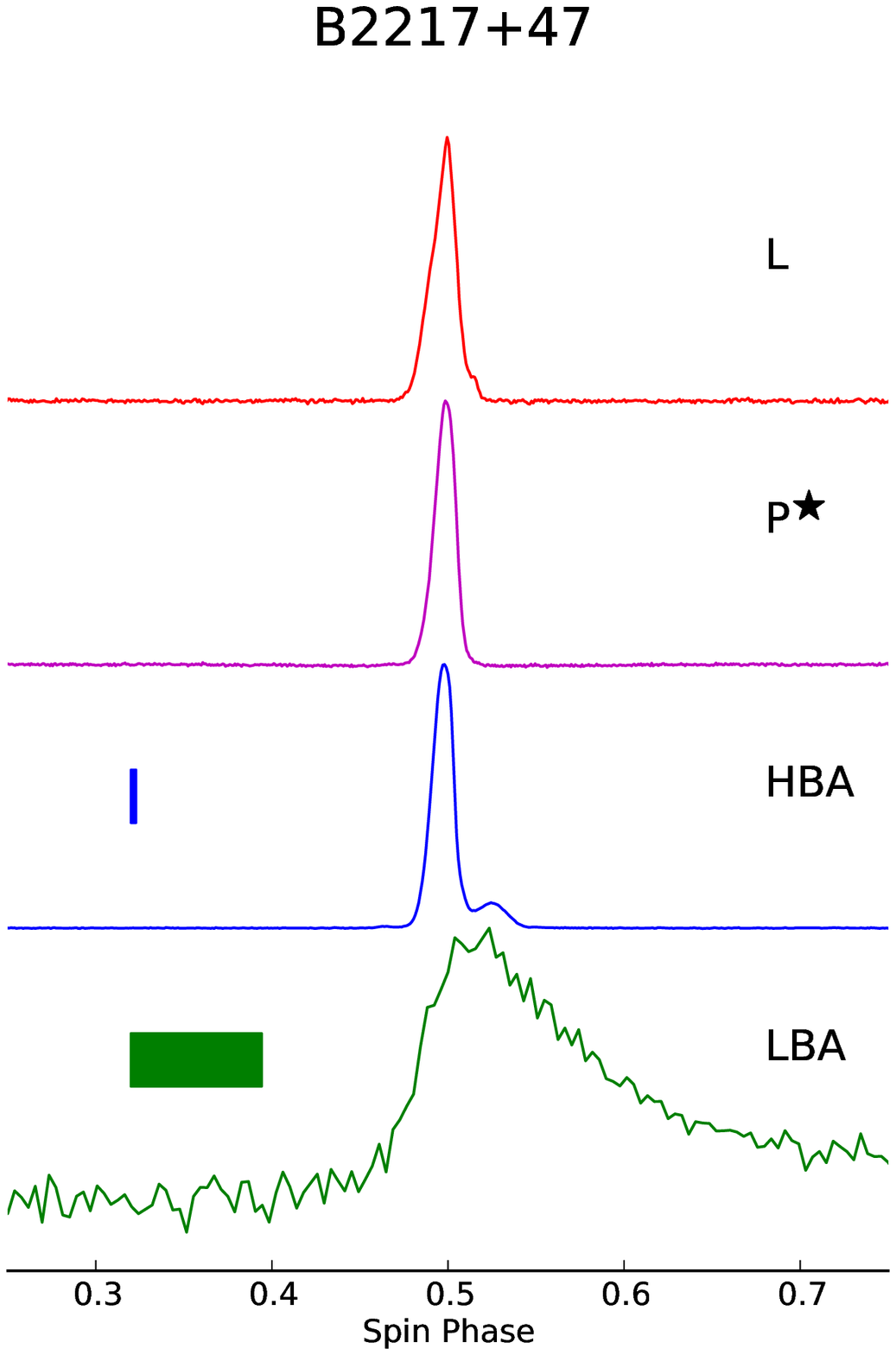} &   \includegraphics[width=32mm]{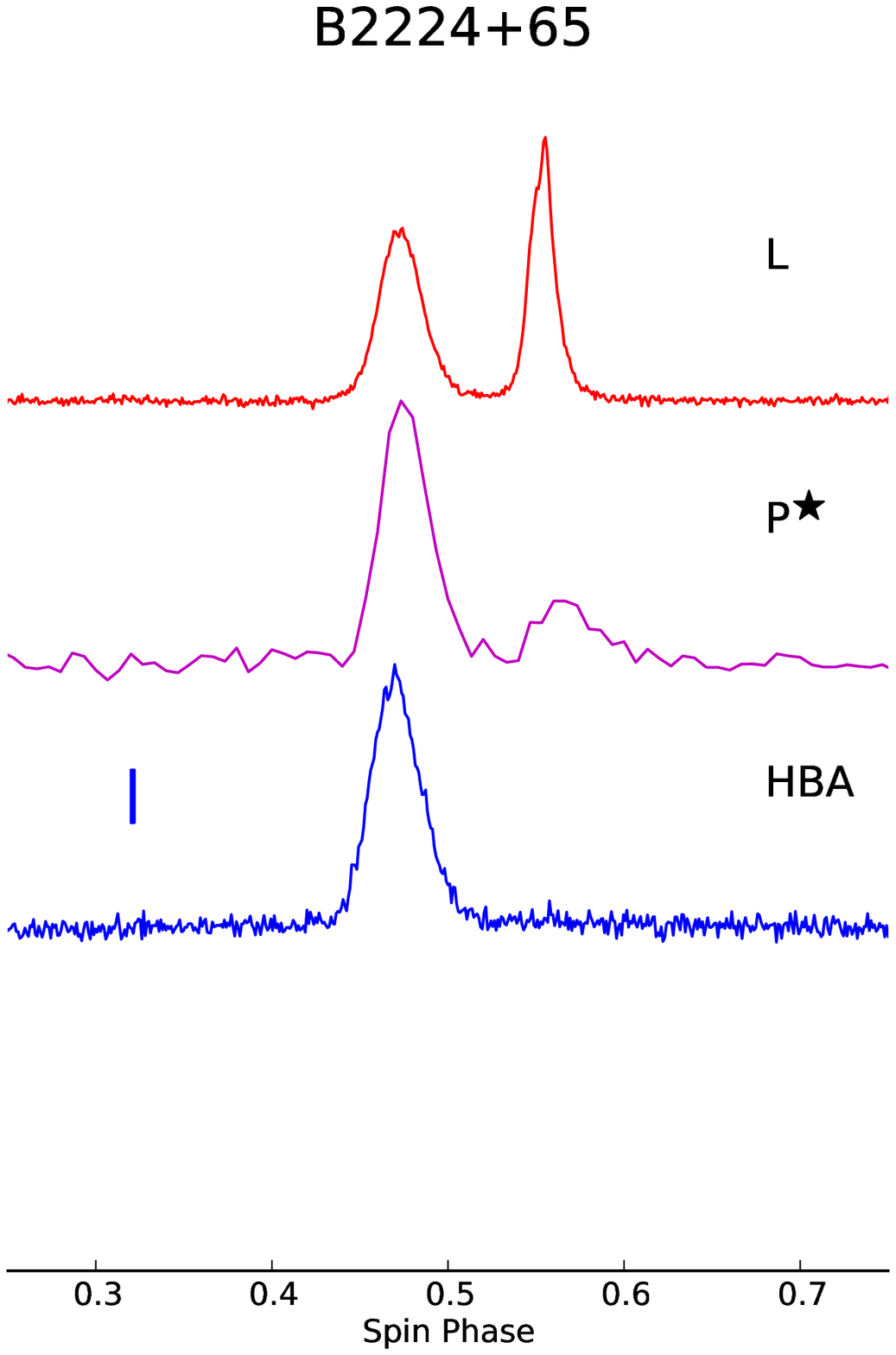} &   \includegraphics[width=32mm]{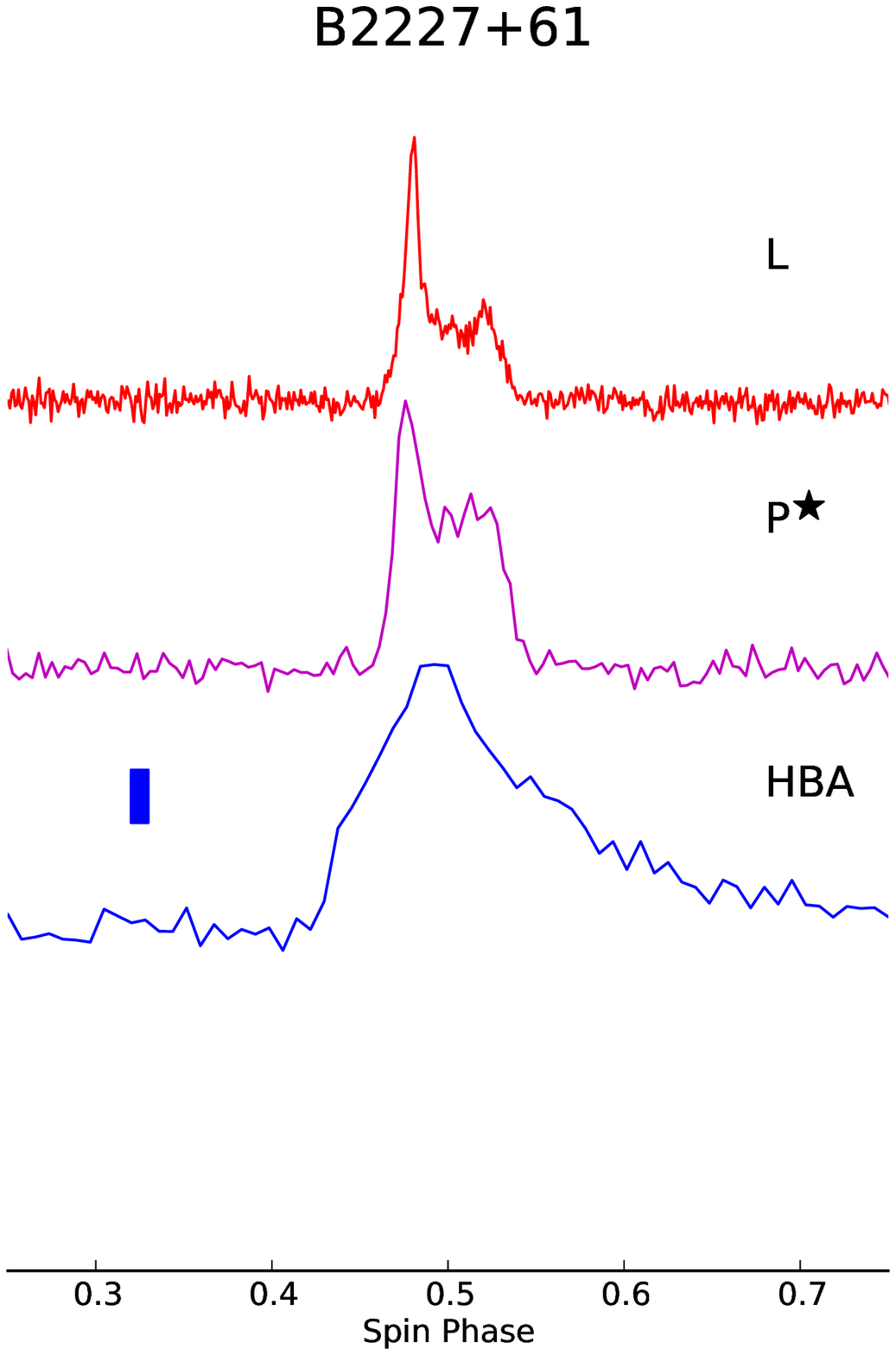} \\

\end{tabular}
\caption*{\label{fig:100_6}
Fig.\ref{fig:100}: continued
}
\end{figure*}

\begin{figure*}
\centering
\begin{tabular}{cccc}

 \includegraphics[width=32mm]{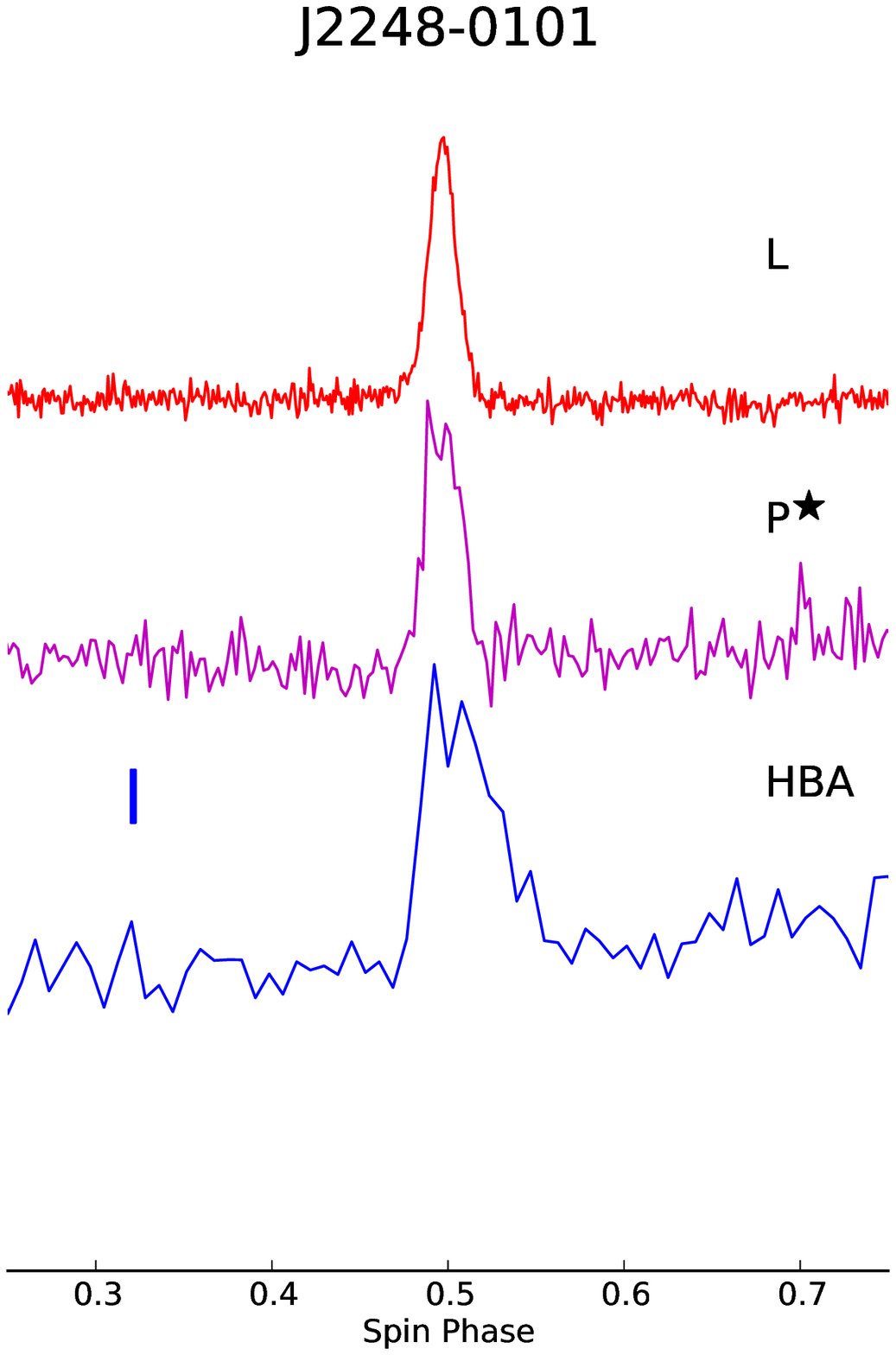} &  \includegraphics[width=32mm]{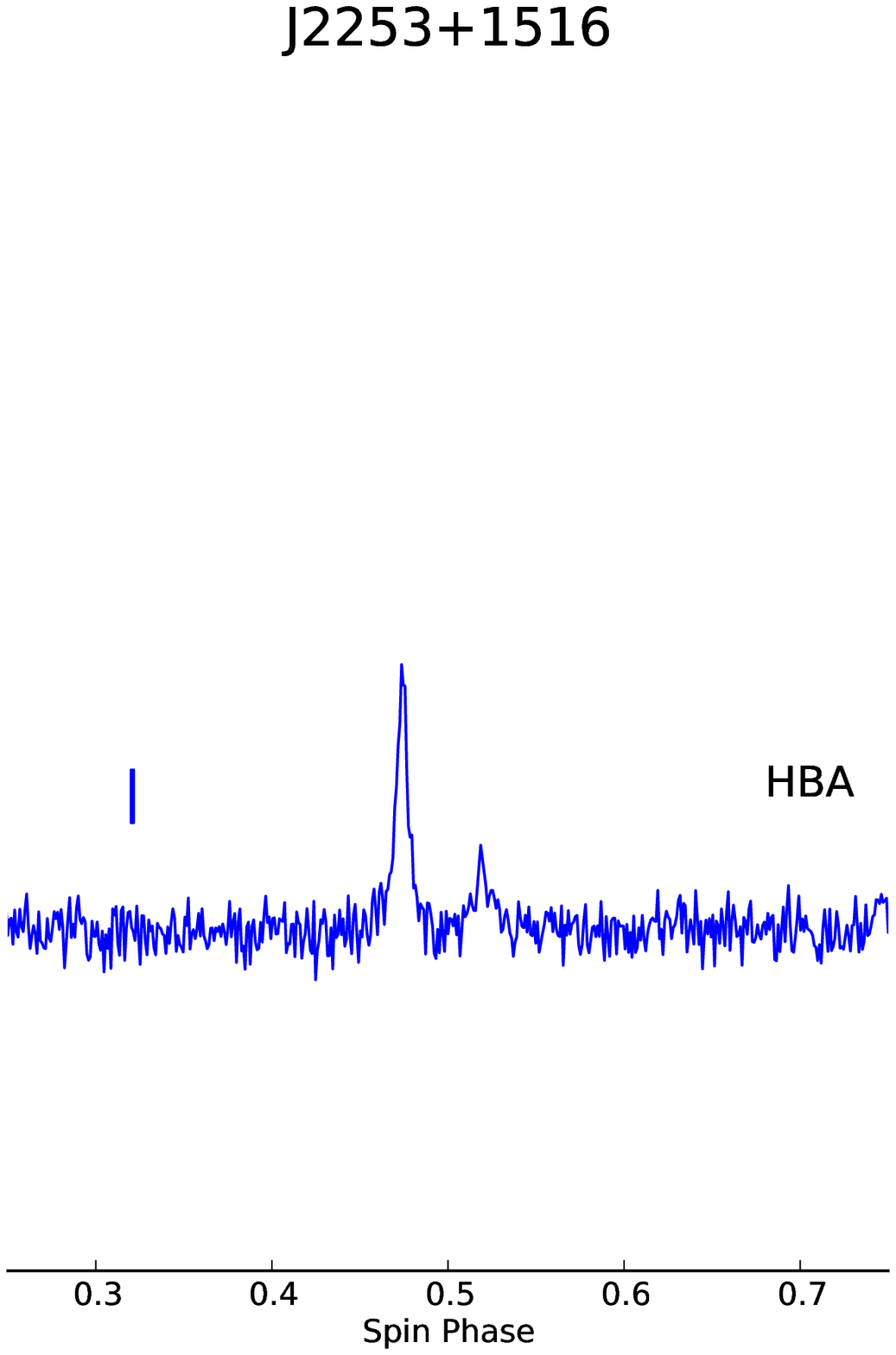} &   \includegraphics[width=32mm]{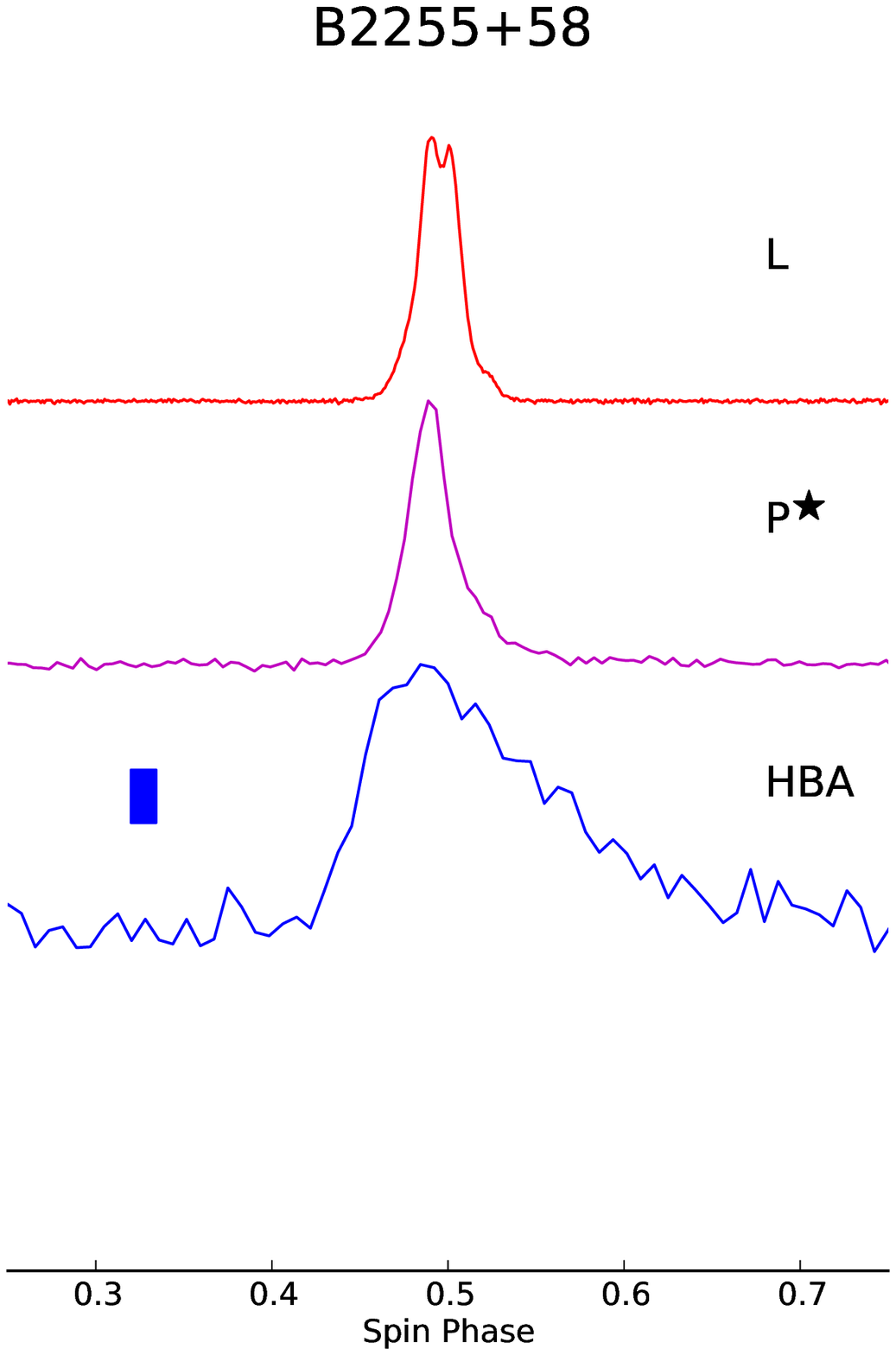} &   \includegraphics[width=32mm]{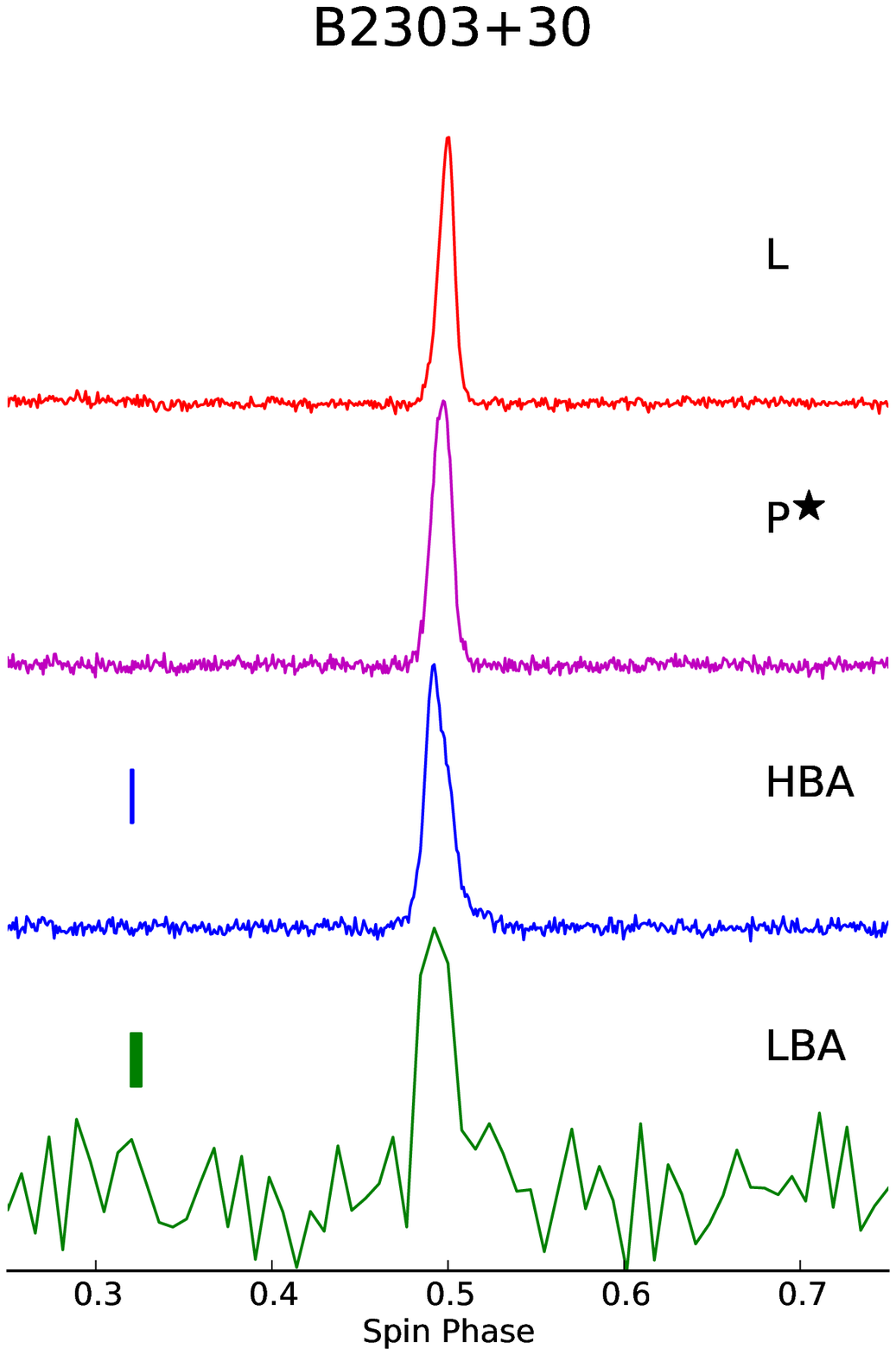} \\

  \includegraphics[width=32mm]{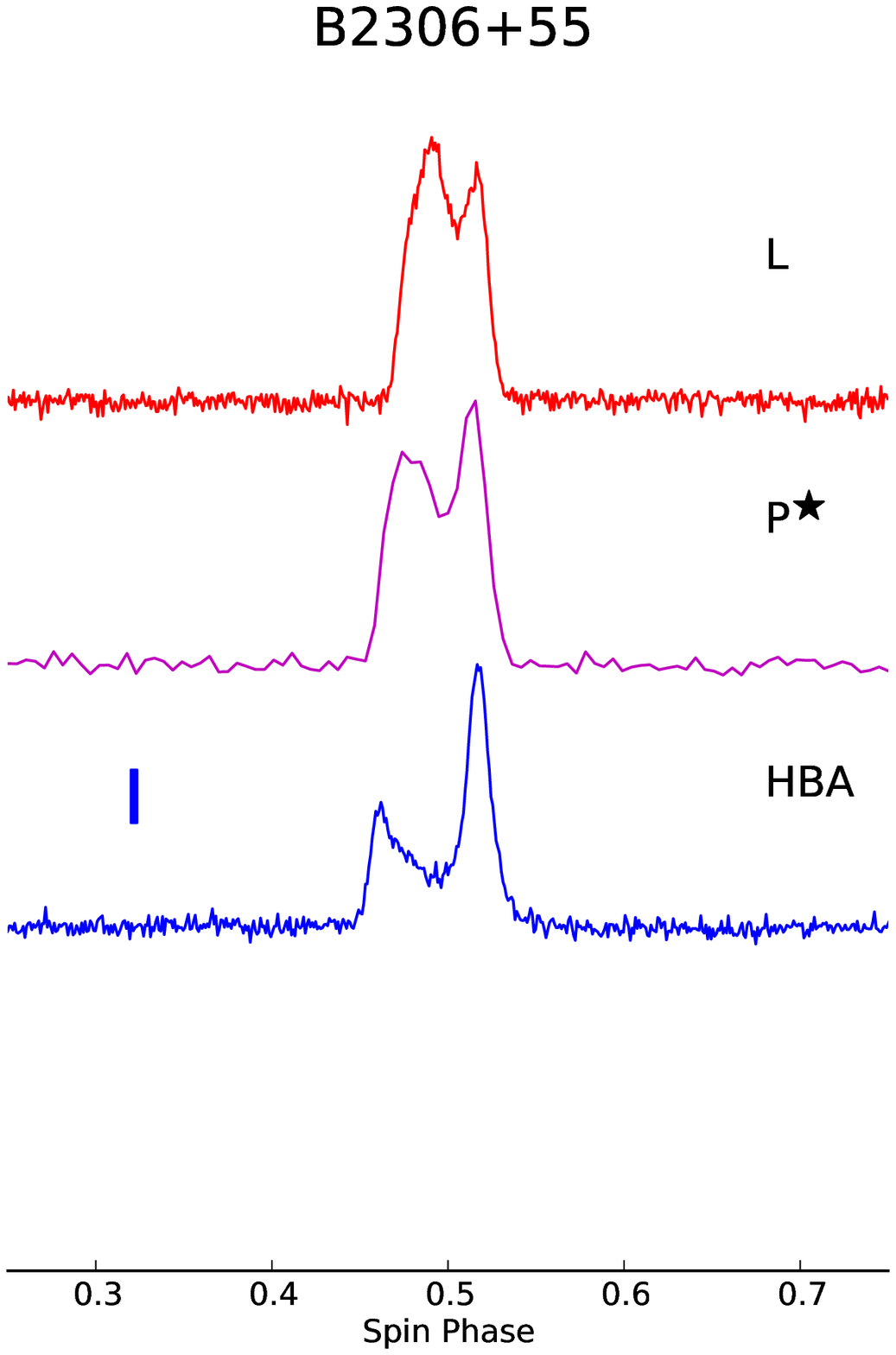} &   \includegraphics[width=32mm]{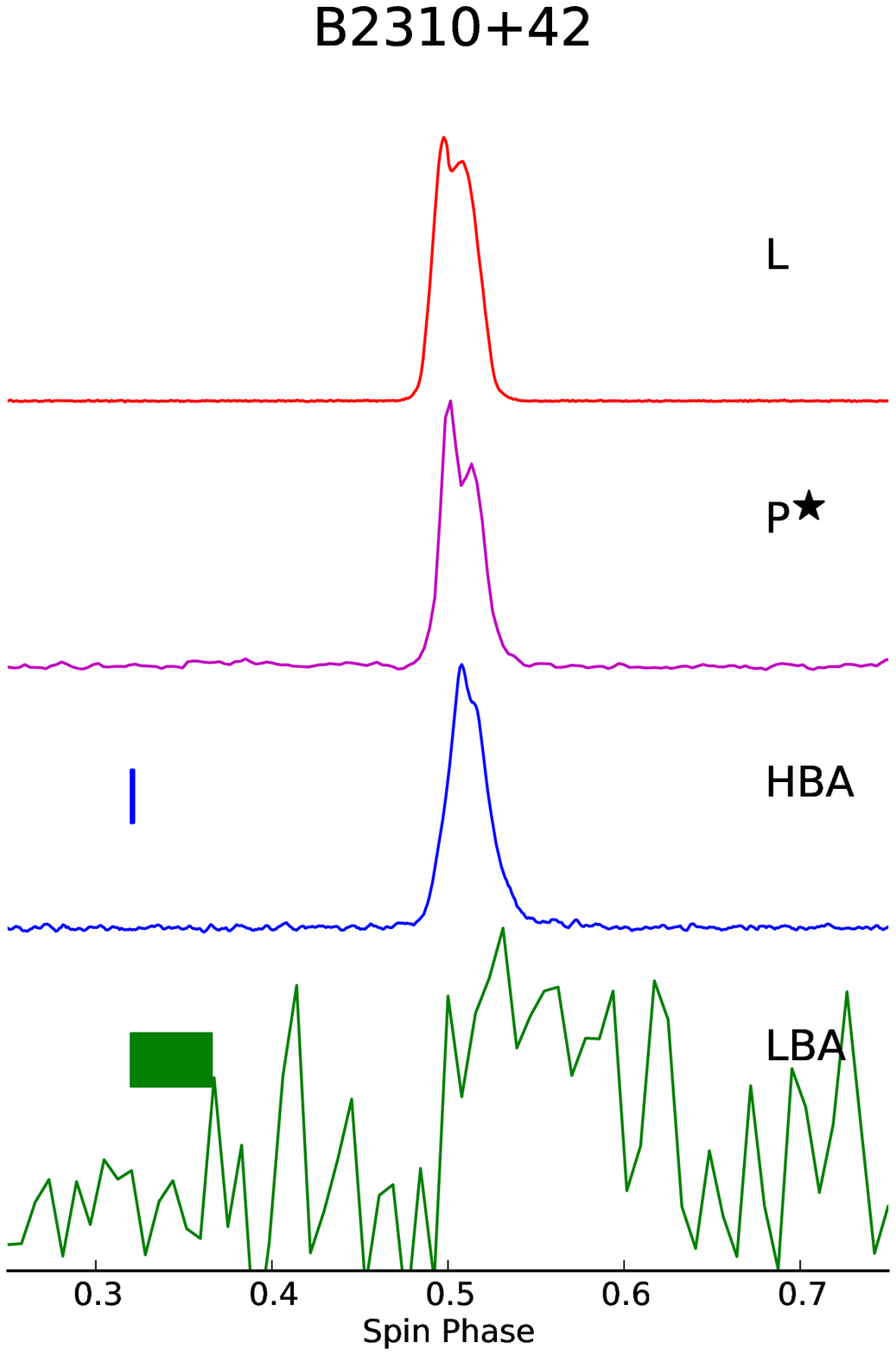} &   \includegraphics[width=32mm]{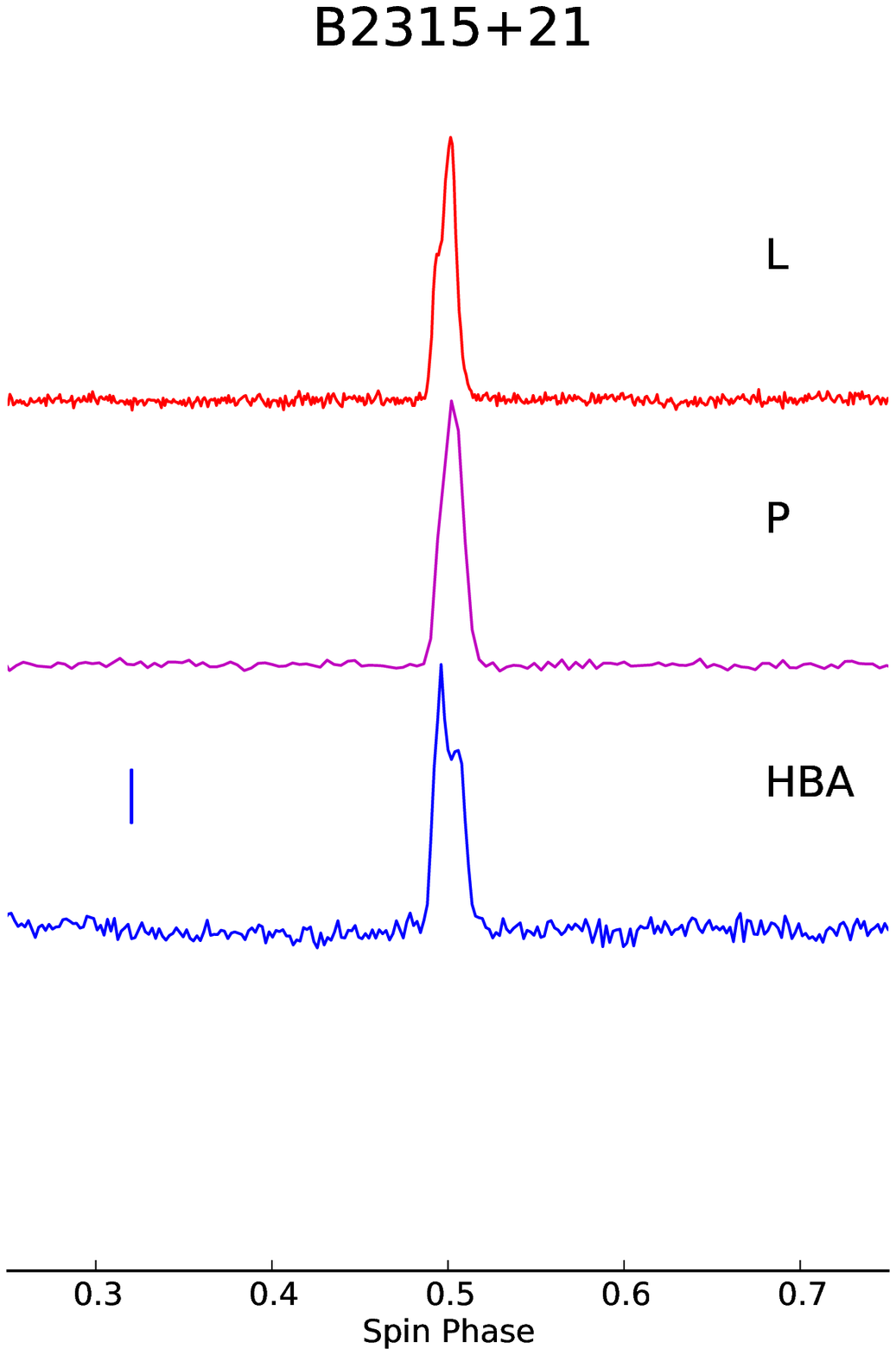} &   \includegraphics[width=32mm]{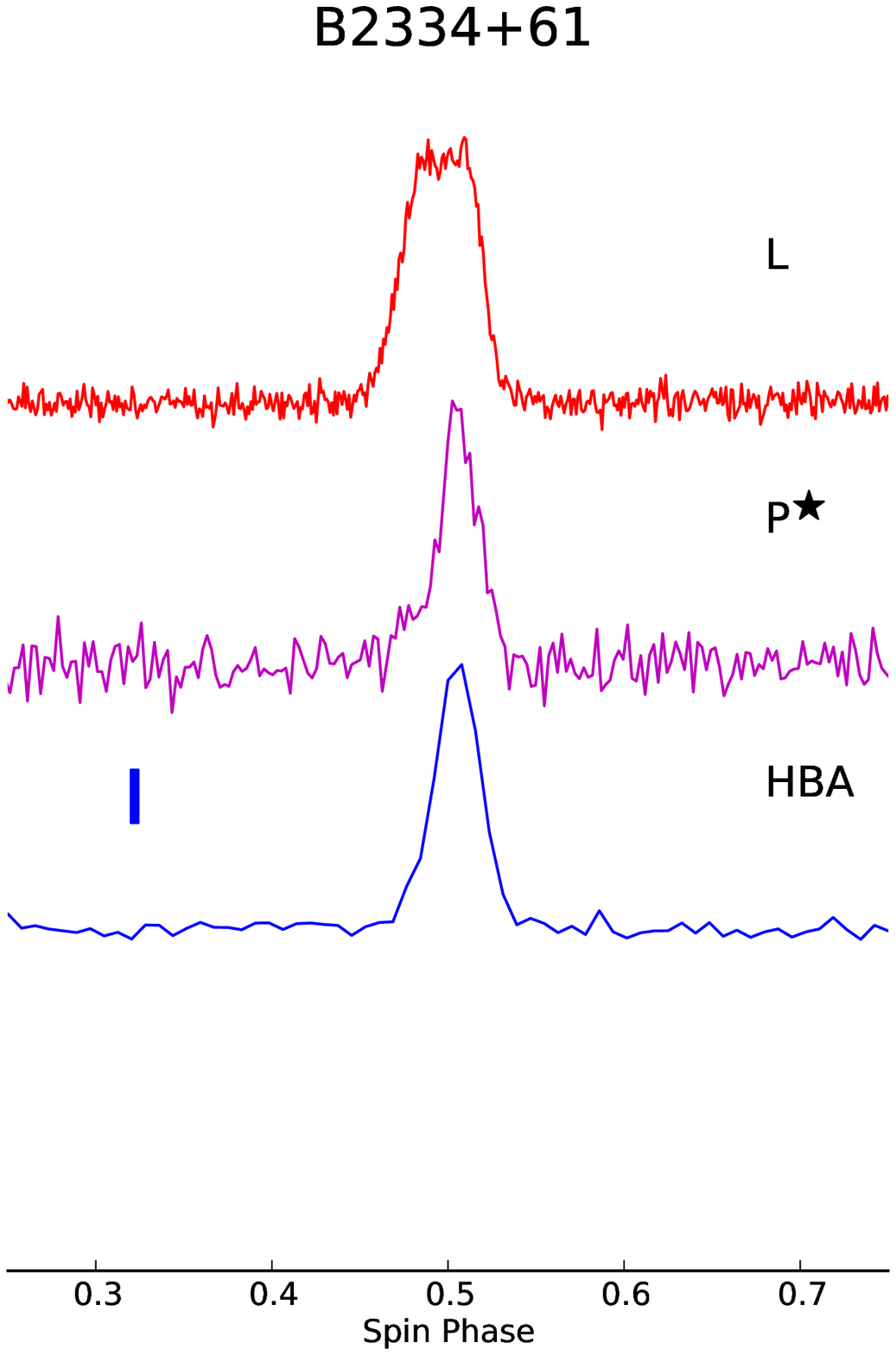} \\

\end{tabular}
\caption*{\label{fig:100_7}
Fig.\ref{fig:100}: continued
}
\end{figure*}

\begin{figure*}
\centering
\begin{tabular}{ccccc}
 
 \includegraphics[width=30mm]{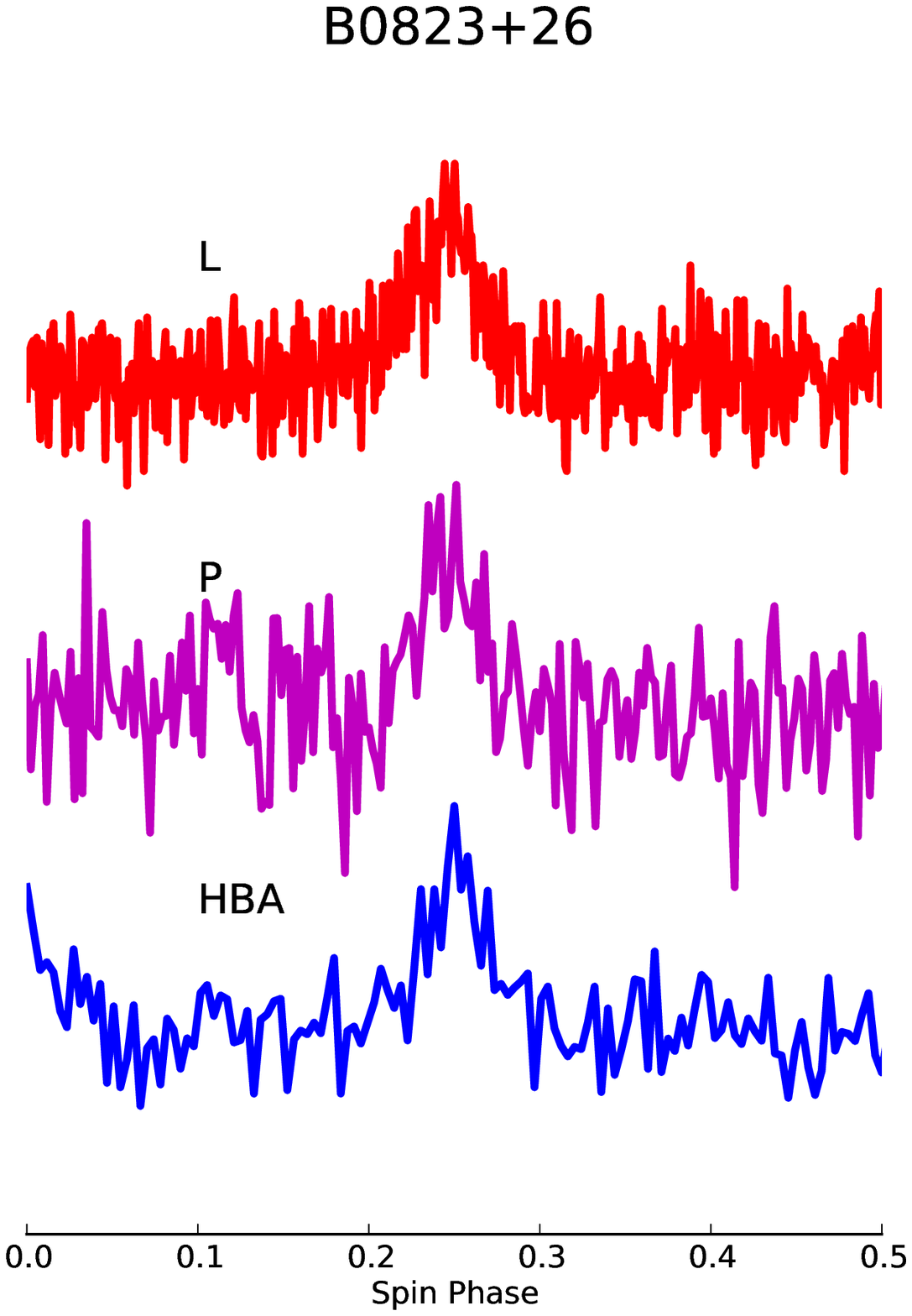} &   \includegraphics[width=30mm]{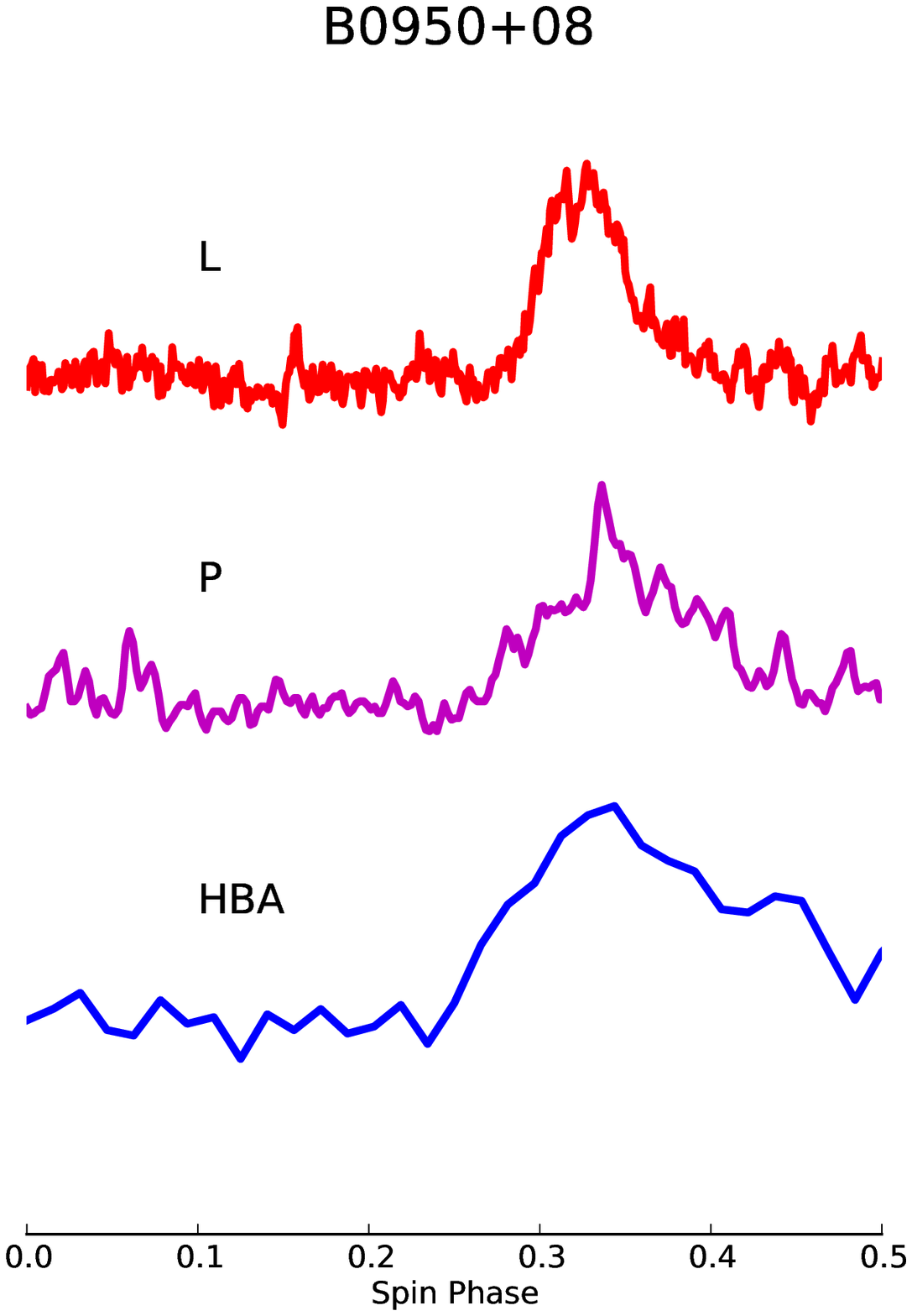} &   \includegraphics[width=30mm]{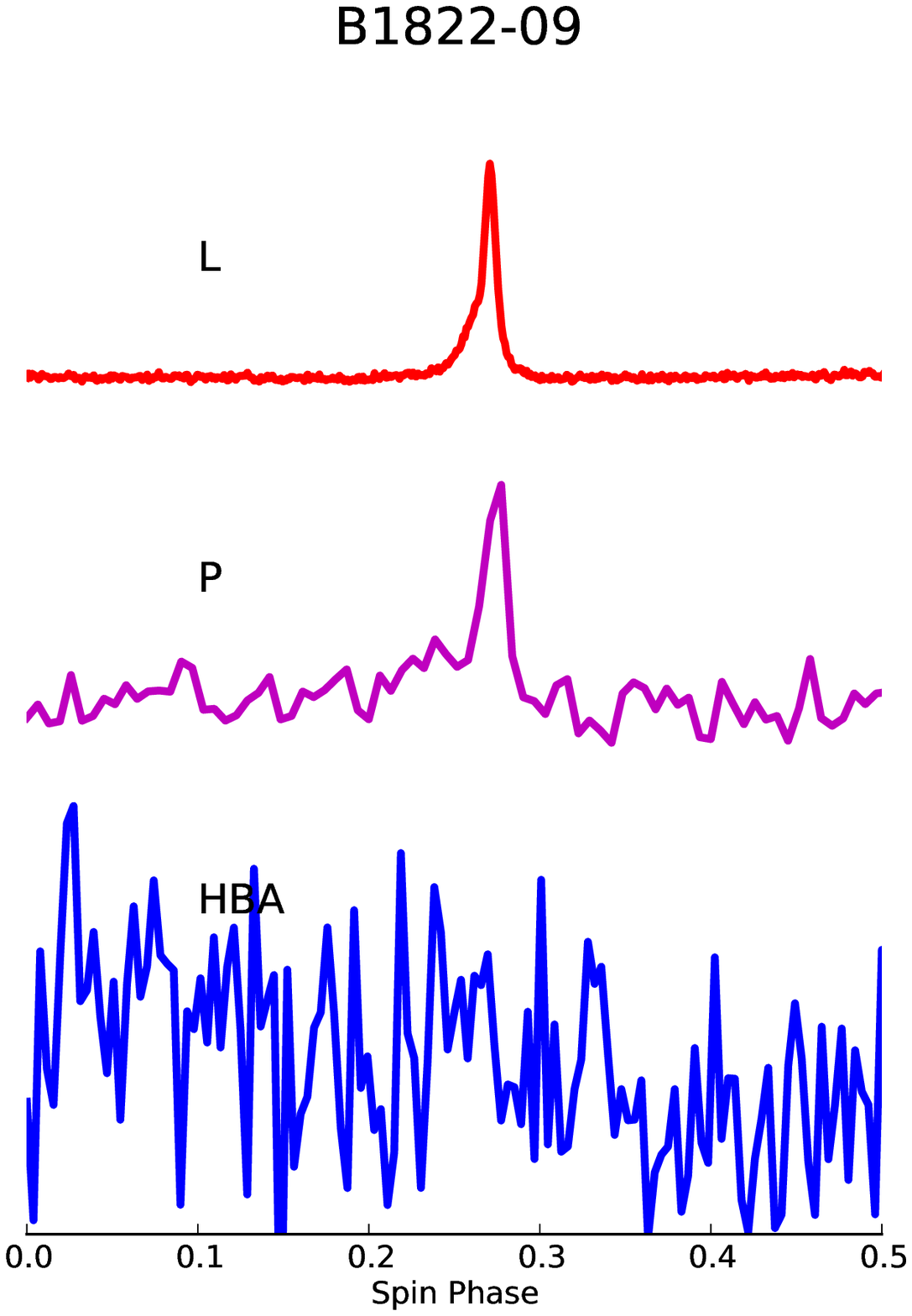} &   \includegraphics[width=30mm]{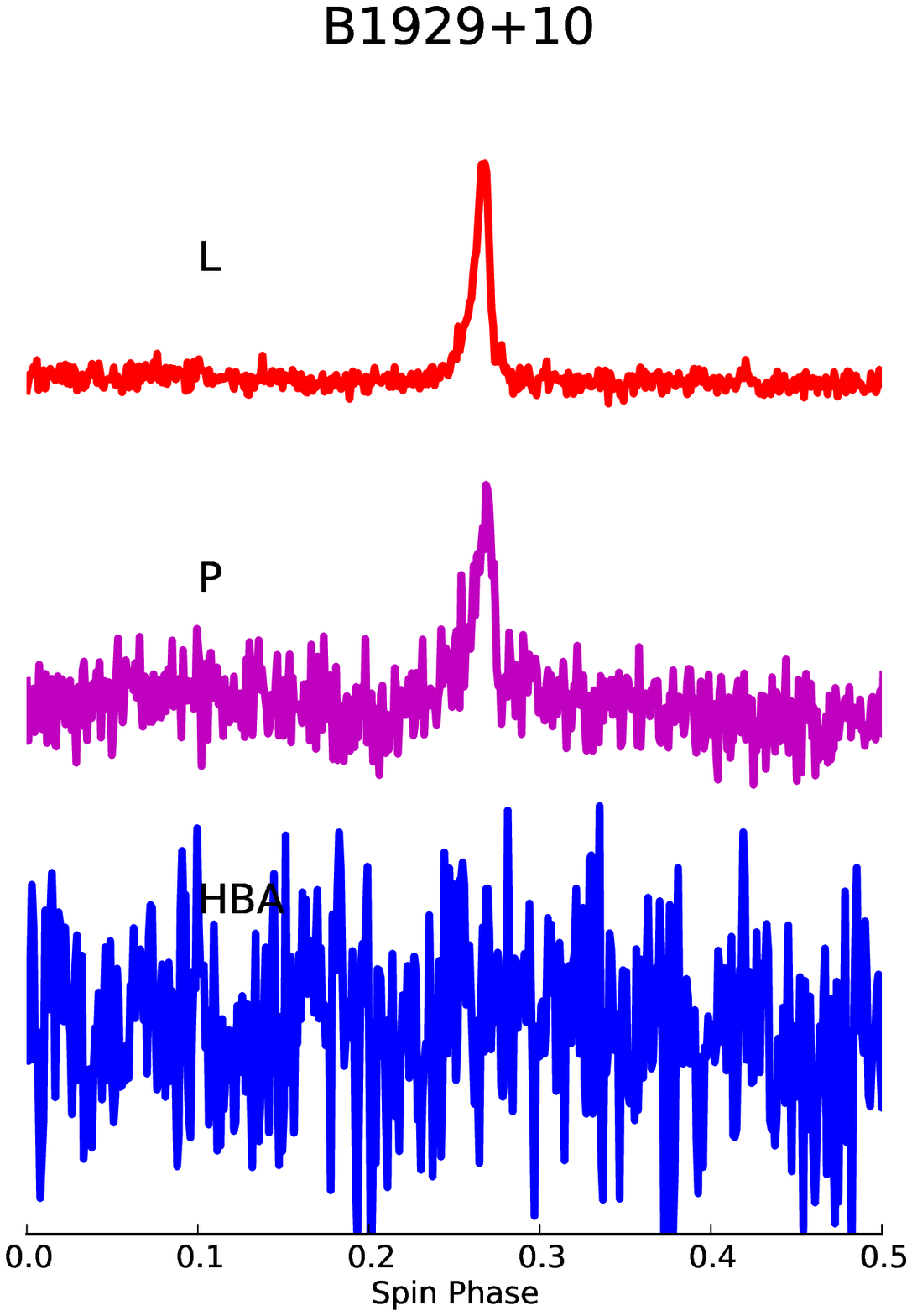} &   \includegraphics[width=30mm]{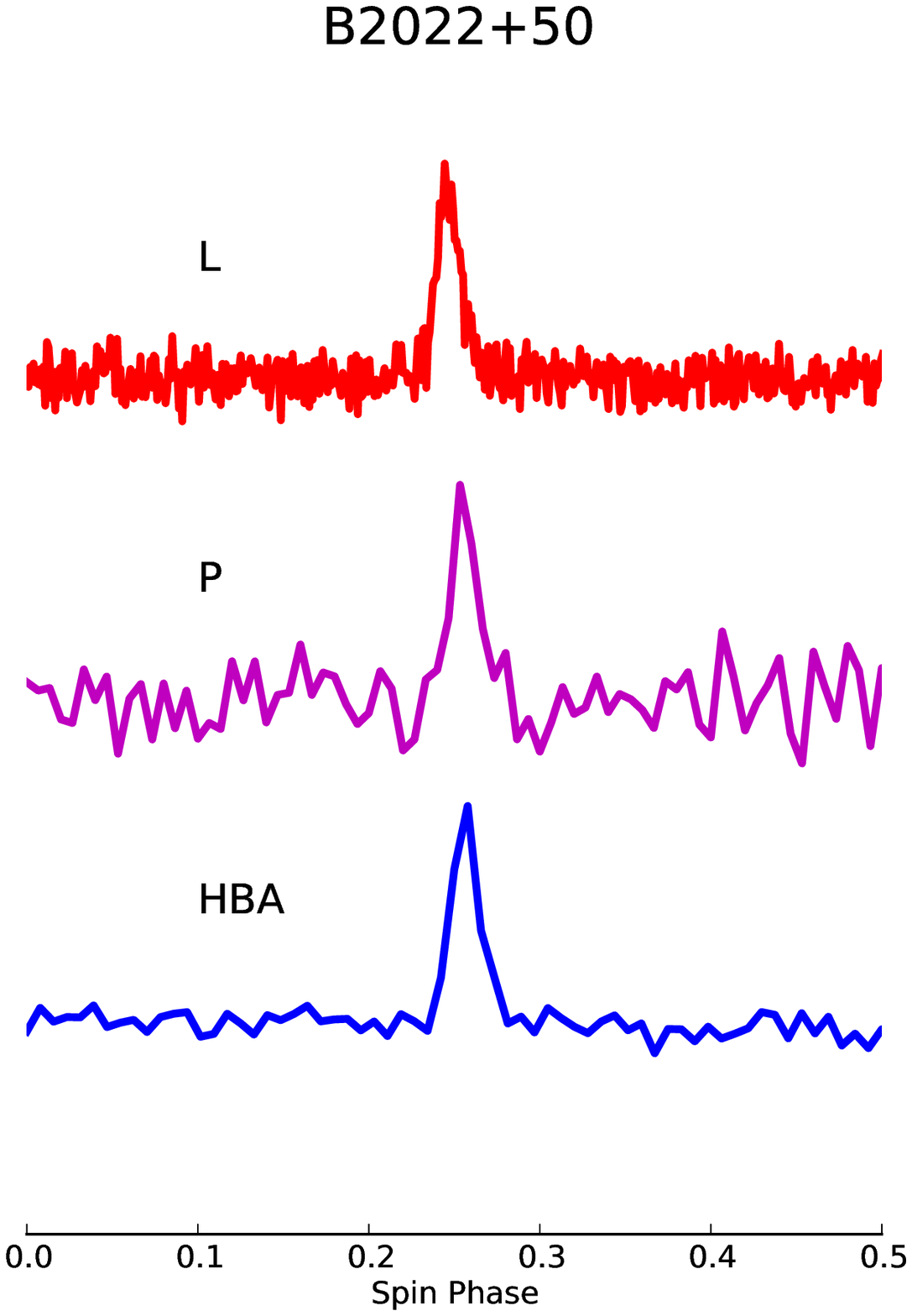} \\

\end{tabular}

\caption{\label{fig:ip}
Zoom-in on the $0.0-0.5$ phase longitude of the profile for the five pulsars in our sample that have interpulses.
The plots are scaled so that each profile has the same r.m.s. and are renormalised to the peak of the interpulse at each frequency.
In all cases LBA profiles were removed, as their signal-to-noise ratio and their channel smearing did not enable any estimate of the interpulse. 
The reference phase of the main pulse has been shifted by 0.75, relative to the main plots, to place the interpulse at the centre of the plot. 
}
\end{figure*}

\onllongtab{
\begin{landscape}
\begin{center}
\begin{longtable}{lccccccccccc}
\caption{List of the 100 pulsars and their rotational and derived parameters.}\\
\hline
\label{tab:100}
PSR Name & $P$ & $\dot P$ & Epoch & LBA Epoch & HBA Epoch & $DM_{\mathrm{eph}}$ & $DM_{\mathrm{HBA}}$ & $\log\tau$ & $\log B$  & $\log\dot{E}$  & Notes \\
 & {\footnotesize [s]}    &  {\footnotesize [s s$^{-1}$]} & {\footnotesize [MJD]}   & {\footnotesize [MJD]}   & {\footnotesize [MJD]}  &  {\footnotesize [cm$^{-3}$pc]}  &  {\footnotesize [cm$^{-3}$pc]}  & {\footnotesize [yr]} & {\footnotesize [G]} & {\footnotesize [s$^{-1}$erg]} & \\
\hline
\\
\endfirsthead
\caption{Continued.} \\
\hline
PSR Name & $P$ & $\dot P$ & Epoch & LBA Epoch & HBA Epoch  & $DM_{\mathrm{eph}}$\tablefootmark{a} & $DM_{\mathrm{HBA}}$ & $\log\tau$ & $\log B$  & $\log\dot{E}$  & Notes \\
 & {\footnotesize [s]}    &  {\footnotesize [s s$^{-1}$]} & {\footnotesize [MJD]}   & {\footnotesize [MJD]}   & {\footnotesize [MJD]}  &  {\footnotesize [cm$^{-3}$pc]}  &  {\footnotesize [cm$^{-3}$pc]}  & {\footnotesize [yr]} & {\footnotesize [G]} & {\footnotesize [s$^{-1}$erg]} & \\
\hline
\\
\endhead
\hline
\endfoot
 B0011+47 & 1.241 & $5.647\mbox{e-}16$ & 55316 & & 55858 & 30.67(7) $^{[1]}$ & 30.3(5) & $7.54$ & $11.93$ & $31.07$ & WES\tablefootmark{d} \\ 
 B0031$-$07 & 0.943 & $4.081\mbox{e-}16$ & 55857 & & 56630 & 11.28(8) $^{[1]}$ & 10.9(4) & $7.56$ & $11.80$ & $31.28$ & WES\tablefootmark{d} \\ 
 J0051+0423 & 0.355 & $7.047\mbox{e-}18$ & 49800 & & 56318 & 13.9(1) $^{[2]}$ & 13.93(1) & $8.90$ & $10.70$ & $30.79$ & LOTAAS\tablefootmark{b} \\ 
 B0059+65 & 1.679 & $5.970\mbox{e-}15$ & 56475 & & 55859 & 65.85(2) $^{[1]}$ & 65.8(8) & $6.65$ & $12.51$ & $31.70$ & EPN\tablefootmark{d} \\ 
 B0105+65 & 1.284 & $1.292\mbox{e-}14$ & 54578 & & 55859 & 30.6(5) $^{[1]}$ & 30.6(2) & $6.20$ & $12.62$ & $32.38$ & P- 2004\tablefootmark{c} \\ 
 B0114+58 & 0.101 & $5.846\mbox{e-}15$ & 55829 & & 55859 & 49.462(4) $^{[1]}$ & 49.42(1) & $5.44$ & $11.89$ & $35.34$ & EPN\tablefootmark{d} \\ 
 B0136+57 & 0.272 & $1.069\mbox{e-}14$ & 54589 & & 55859 & 73.776(6) $^{[1]}$ & 73.83(1) & $5.61$ & $12.24$ & $34.32$ & P-, L- 2004\tablefootmark{b} \\ 
 B0138+59 & 1.223 & $3.910\mbox{e-}16$ & 54266 & 55924 & 55859 & 34.74(1) $^{[1]}$ & 34.92(2) & $7.70$ & $11.84$ & $30.93$ & P- 2004\tablefootmark{c} \\ 
 B0301+19 & 1.388 & $1.295\mbox{e-}15$ & 55743 & & 55859 &  15.851(9) $^{[1]}$ & 15.65(4) & $7.23$ & $12.13$ & $31.28$ & WES\tablefootmark{d} \\ 
 B0320+39 & 3.032 & $6.358\mbox{e-}16$ & 54125 & 55924 & 55859 &  25.99(3) $^{[1]}$ & 26.19(1) & $7.88$ & $12.15$ & $29.95$ & P- 2002\tablefootmark{c} \\ 
 B0329+54 & 0.715 & $2.046\mbox{e-}15$ & 53850 & 55859 & 55859 &  26.7941(1) $^{[3]}$ & 26.7653($<1$) & $6.74$ & $12.09$ & $32.35$ & P- 2000\tablefootmark{c} \\ 
 B0331+45 & 0.269 & $7.331\mbox{e-}18$ & 55899 & & 55859 &  47.149(3) $^{[1]}$ & 47.15(1) & $8.76$ & $10.65$ & $31.17$ & EPN\tablefootmark{d} \\ 
 B0339+53 & 1.934 & $1.342\mbox{e-}14$ & 55804 & & 55859 &  67.28(6) $^{[1]}$ & 67.5(2) & $6.36$ & $12.71$ & $31.86$ & WES\tablefootmark{d} \\ 
 B0355+54 & 0.156 & $4.395\mbox{e-}15$ & 53813 & & 55859 &  57.1226(3) $^{[4]}$ & 57.14(1) & $5.75$ & $11.92$ & $34.66$ & P- 2003\tablefootmark{c} \\ 
 B0402+61 & 0.595 & $5.576\mbox{e-}15$ & 56097 & & 55859 &  65.303(7) $^{[1]}$ & 65.39(3) & $6.23$ & $12.27$ & $33.02$ &  \\ 
 B0410+69 & 0.391 & $7.656\mbox{e-}17$ & 54418 & & 55859 &  27.471(4) $^{[1]}$ & 27.44(1) & $7.91$ & $11.24$ & $31.70$ &  P- 2004\tablefootmark{c} \\ 
 B0447$-$12 & 0.438 & $1.028\mbox{e-}16$ & 55203 & & 56568 &  37.04(1) $^{[1]}$ & 37.01(1) & $7.83$ & $11.33$ & $31.68$ &  P- 2003\tablefootmark{c} \\ 
 B0450+55 & 0.341 & $2.377\mbox{e-}15$ & 54139 & 55924 & 55861 &  14.483(7) $^{[1]}$ & 14.59(7) & $6.36$ & $11.96$ & $33.38$ & P-, L- 2004\tablefootmark{c} \\ 
 B0450$-$18 & 0.549 & $5.755\mbox{e-}15$ & 55203 & & 56603 &  39.903(3) $^{[1]}$ & 39.91(2) & $6.18$ & $12.25$ & $33.14$ &  P- 2004\tablefootmark{c} \\ 
 B0523+11 & 0.354 & $7.362\mbox{e-}17$ & 55202 & & 56568 &  79.423(3) $^{[1]}$ & 79.43(4) & $7.88$ & $11.21$ & $31.81$ &  P- 2006\tablefootmark{c} \\ 
 B0525+21 & 3.746 & $4.004\mbox{e-}14$ & 54457 & 55924 & 56596 &  50.96(2) $^{[1]}$ & 50.87(2) & $6.17$ & $13.09$ & $31.48$ & P- 2003, L- 2004\tablefootmark{c}  \\ 
 B0540+23 & 0.246 & $1.542\mbox{e-}14$ & 54748 & & 55859 &  77.631(2) $^{[1]}$ & 77.7(9) & $5.40$ & $12.29$ & $34.61$ & P- 2006\tablefootmark{c} \\ 
 B0609+37 & 0.298 & $5.980\mbox{e-}17$ & 54590 & & 55859 &  27.136(4) $^{[1]}$ & 27.15(3) & $7.90$ & $11.13$ & $31.95$ & P- 2006\tablefootmark{c} \\ 
 B0611+22 & 0.335 & $5.911\mbox{e-}14$ & 55604 & & 56681 &  96.94(7) $^{[1]}$ & 96.92(1) & $4.95$ & $12.65$ & $34.79$ & WES\tablefootmark{d} \\ 
 B0626+24 & 0.477 & $1.985\mbox{e-}15$ & 55071 & & 55859 &  84.19(4) $^{[1]}$ & 84.18(3) & $6.58$ & $11.99$ & $32.86$ & P- 2004\tablefootmark{c} \\ 
 B0643+80 & 1.214 & $3.799\mbox{e-}15$ & 54855 & & 55859 &  33.33(2) $^{[1]}$ & 33.3(1) & $6.70$ & $12.34$ & $31.92$ &  P- 2004\tablefootmark{c} \\ 
 B0809+74 & 1.292 & $1.660\mbox{e-}16$ & 53508 & 56046 & 55916 &  5.733(1) $^{[3]}$ & 5.75(3) & $8.09$ & $11.67$ & $30.48$ & WES\tablefootmark{d} \\ 
 B0818$-$13 & 1.238 & $2.105\mbox{e-}15$ & 55507 & & 56603 &  40.949(3) $^{[1]}$ & 40.98(1) & $6.97$ & $12.21$ & $31.64$ & WES\tablefootmark{d} \\ 
 B0820+02 & 0.865 & $1.047\mbox{e-}16$ & 55508 & & 56301 &  23.717(6) $^{[1]}$ & 23.76(7) & $8.12$ & $11.48$ & $30.81$ & WES\tablefootmark{d} \\ 
 B0823+26 & 0.531 & $1.679\mbox{e-}15$ & 55419 & 55889 & 55966 &  19.464(4) $^{[1]}$ & 19.4711($<1$) & $6.70$ & $11.98$ & $32.65$ & WES\tablefootmark{d} \\ 
 B0834+06 & 1.274 & $6.799\mbox{e-}15$ & 55507 & 55883 & 55881 &  12.891(6) $^{[1]}$ & 12.8735($<1$) & $6.47$ & $12.47$ & $32.11$ & WES\tablefootmark{d} \\ 
 B0906$-$17 & 0.402 & $6.693\mbox{e-}16$ & 54887 & & 56603 &  15.885(3) $^{[1]}$ & 15.87(1) & $6.98$ & $11.72$ & $32.61$ & WES\tablefootmark{d} \\ 
 B0917+63 & 1.568 & $3.608\mbox{e-}15$ & 53883 & 55925 & 55860 &  13.21(2) $^{[1]}$ & 13.14(2) & $6.84$ & $12.38$ & $31.57$ & EPN\tablefootmark{d} \\ 
 B0919+06 & 0.431 & $1.372\mbox{e-}14$ & 55870 & & 56315 &  27.256(6) $^{[1]}$ & 27.3(1) & $5.70$ & $12.39$ & $33.83$ & WES\tablefootmark{d} \\ 
 B0943+10 & 1.098 & $3.546\mbox{e-}15$ & 54226 & 55925 & 55869 &  15.2(5) $^{[1]}$ & 15.32(1) & $6.69$ & $12.30$ & $32.02$ & EPN\tablefootmark{d} \\ 
 B0950+08 & 0.253 & $2.165\mbox{e-}16$ & 51541 & 55889 & 56603 &  2.961(3) $^{[1]}$ & 2.9696($<1$) & $7.27$ & $11.37$ & $32.72$ & EPN\tablefootmark{d} \\ 
 B1112+50 & 1.656 & $2.493\mbox{e-}15$ & 54121 & 55925 & 56233 &  9.202(8) $^{[1]}$ & 9.18(1) & $7.02$ & $12.31$ & $31.34$ & WES\tablefootmark{d} \\ 
 B1133+16 & 1.188 & $3.733\mbox{e-}15$ & 56235 & 55925 & 55916 &  4.8611(1)$^{[3]}$ & 4.8262($<1$) & $6.70$ & $12.33$ & $31.94$ & WES\tablefootmark{d} \\ 
 B1237+25 & 1.382 & $9.596\mbox{e-}16$ & 46498 & 55925 & 55860 &  9.254(6) $^{[1]}$ & 9.2538($<1$) & $7.36$ & $12.07$ & $31.16$ &  WES\tablefootmark{d}\\ 
 J1238+21 & 1.119 & $1.445\mbox{e-}15$ & 55507 & & 55860 &  17(2) $^{[1]}$ & 17.97(3) & $7.09$ & $12.11$ & $31.61$ & \\ 
 J1313+0931 & 0.849 & $8.000\mbox{e-}16$ & 50984 & & 55860 &  12.0(1) $^{[5]}$ & 12.05(2) & $7.23$ & $11.92$ & $31.71$ & \\ 
 B1322+83 & 0.670 & $5.675\mbox{e-}16$ & 53763 & & 55860 &  13.3(2) $^{[1]}$ & 13.31(3) & $7.27$ & $11.80$ & $31.87$ & WES\tablefootmark{d} \\ 
 B1508+55 & 0.740 & $5.007\mbox{e-}15$ & 55168 & 55906 & 55860 &  19.59(2) $^{[1]}$ & 19.6179($<1$) & $6.37$ & $12.29$ & $32.69$ & WES\tablefootmark{d} \\ 
 B1530+27 & 1.125 & $7.791\mbox{e-}16$ & 56508 & & 55859 &  14.76(2) $^{[1]}$ & 14.7(2) & $7.36$ & $11.98$ & $31.33$ & WES\tablefootmark{d} \\ 
 B1540$-$06 & 0.709 & $8.810\mbox{e-}16$ & 55962 & 55892 & 56605 &  18.403(4) $^{[1]}$ & 18.3788($<1$) & $7.11$ & $11.90$ & $31.99$ & WES\tablefootmark{d} \\ 
 B1541+09 & 0.748 & $4.328\mbox{e-}16$ & 55508 & 55892 & 55859 &  35.24(3) $^{[1]}$ & 34.97(3) & $7.44$ & $11.76$ & $31.61$ & WES\tablefootmark{d} \\ 
 B1604$-$00 & 0.422 & $3.063\mbox{e-}16$ & 54696 & 55892 & 56346 &  10.687(5) $^{[1]}$ & 10.69(1) & $7.34$ & $11.56$ & $32.21$ & WES\tablefootmark{d} \\ 
 B1612+07 & 1.207 & $2.360\mbox{e-}15$ & 55508 & & 56354 &  21.39(3) $^{[1]}$ & 21.41(1) & $6.91$ & $12.23$ & $31.72$ &  WES\tablefootmark{d} \\ 
 J1627+1419 & 0.491 & $3.930\mbox{e-}16$ & 48957 & & 55859 &  33.8(6) $^{[6]}$ & 32.18(9) & $7.30$ & $11.65$ & $32.12$ & \\ 
 B1633+24 & 0.491 & $1.192\mbox{e-}16$ & 55508 & 56517 & 55859 &  24.33(4) $^{[1]}$ & 24.27(1) & $7.81$ & $11.39$ & $31.60$ & EPN\tablefootmark{d} \\ 
 B1642$-$03 & 0.388 & $1.770\mbox{e-}15$ & 55643 & & 56605 &  35.738(3) $^{[1]}$ & 35.7556($<1$) & $6.54$ & $11.92$ & $33.08$ & WES\tablefootmark{d} \\ 
 J1645+1012 & 0.411 & $8.203\mbox{e-}17$ & 52814 & & 56276 &  36.4(6) $^{[6]}$ & 36.18(3) & $7.90$ & $11.27$ & $31.67$ & \\ 
 J1652+2651 & 0.916 & $6.542\mbox{e-}16$ & 55507 & & 55860 &  41.0(2) $^{[2]}$ & 40.79(2) & $7.35$ & $11.89$ & $31.53$ & WES\tablefootmark{d} \\ 
 B1737+13 & 0.803 & $1.441\mbox{e-}15$ & 55507 & & 55870 &  48.672(4) $^{[1]}$ & 48.67(1) & $6.95$ & $12.04$ & $32.04$ & WES\tablefootmark{d} \\ 
 B1749$-$28 & 0.563 & $8.000\mbox{e-}15$ & 54111 & & 56606 &  50.39(8) $^{[1]}$ & 50.38(1) & $6.05$ & $12.33$ & $33.25$ & WES\tablefootmark{d} \\ 
 B1818$-$04 & 0.598 & $6.347\mbox{e-}15$ & 56445 & & 56606 &  84.42(2) $^{[1]}$ & 84.34(3) & $6.17$ & $12.29$ & $33.07$ & WES\tablefootmark{d} \\ 
 B1821+05 & 0.753 & $2.267\mbox{e-}16$ & 55508 & & 56286 &  66.777(3) $^{[1]}$ & 66.79(2) & $7.72$ & $11.62$ & $31.32$ & LOTAAS\tablefootmark{b} \\ 
 B1822$-$09 & 0.769 & $5.249\mbox{e-}14$ & 56183 & 55892 & 56605 &  19.36(4) $^{[1]}$ & 19.36(2) & $5.37$ & $12.81$ & $33.66$ &  WES\tablefootmark{d}\\ 
 B1831$-$04 & 0.290 & $7.250\mbox{e-}17$ & 55508 & & 56605 &  79.302(8) $^{[1]}$ & 79.39(4) & $7.80$ & $11.17$ & $32.07$ & WES\tablefootmark{d} \\ 
 B1839+09 & 0.381 & $1.089\mbox{e-}15$ & 55541 & & 55870 &  49.145(8) $^{[1]}$ & 49.1501($<1$) & $6.74$ & $11.81$ & $32.89$ & WES\tablefootmark{d} \\ 
 B1839+56 & 1.653 & $1.495\mbox{e-}15$ & 55508 & 55892 & 55870 &  26.71(1) $^{[1]}$ & 26.7540($<1$) & $7.24$ & $12.20$ & $31.12$ & WES\tablefootmark{d} \\ 
 B1842+14 & 0.375 & $1.873\mbox{e-}15$ & 55817 & 55864 & 55870 &  41.512(4) $^{[1]}$ & 41.48(1) & $6.50$ & $11.93$ & $33.15$ & WES\tablefootmark{d} \\ 
 B1848+12 & 1.205 & $1.137\mbox{e-}14$ & 55835 & & 56276 &  70.61(2) $^{[1]}$ & 70.6(8) & $6.23$ & $12.57$ & $32.41$ &  LOTAAS\tablefootmark{b},WES\tablefootmark{d} \\ 
 B1848+13 & 0.346 & $1.493\mbox{e-}15$ & 55508 & & 56276 &  60.149(8) $^{[1]}$ & 60.1(1) & $6.56$ & $11.86$ & $33.15$ &  LOTAAS\tablefootmark{b},EPN\tablefootmark{d} \\ 
 B1857$-$26 & 0.612 & $2.050\mbox{e-}16$ & 55508 & & 55870 &  37.993(5) $^{[1]}$ & 37.92(3) & $7.68$ & $11.55$ & $31.55$ & WES\tablefootmark{d} \\ 
 B1905+39 & 1.236 & $5.408\mbox{e-}16$ & 55507 & & 55870 &  30.96(3) $^{[1]}$ & 30.98(5) & $7.56$ & $11.92$ & $31.05$ & WES\tablefootmark{d} \\ 
 B1907+00 & 1.017 & $5.513\mbox{e-}15$ & 55770 & & 56354 &  112.776(6) $^{[1]}$ & 113(1) & $6.47$ & $12.38$ & $32.32$ &  LOTAAS\tablefootmark{b},WES\tablefootmark{d} \\ 
 B1907+02 & 0.990 & $5.492\mbox{e-}15$ & 53773 & & 56339 &  171.722(9) $^{[1]}$ & 172(1) & $6.46$ & $12.37$ & $32.35$ &  LOTAAS\tablefootmark{b},WES\tablefootmark{d} \\ 
 B1907+10 & 0.284 & $2.644\mbox{e-}15$ & 55729 & & 55870 &  149.756(4) $^{[1]}$ & 149.9(1) & $6.23$ & $11.94$ & $33.66$ & WES\tablefootmark{d} \\ 
 B1911$-$04 & 0.826 & $4.069\mbox{e-}15$ & 55705 & & 56354 &  89.38(1) $^{[1]}$ & 89.38(3) & $6.51$ & $12.27$ & $32.46$ &  LOTAAS\tablefootmark{b},WES\tablefootmark{d} \\ 
 B1914+09 & 0.270 & $2.518\mbox{e-}15$ & 55931 & & 55870 &  60.981(6) $^{[1]}$ & 61.01(6) & $6.23$ & $11.92$ & $33.70$ & WES\tablefootmark{d} \\ 
 B1915+13 & 0.195 & $7.194\mbox{e-}15$ & 55745 & & 55870 &  94.54(4) $^{[1]}$ & 94.65(3) & $5.63$ & $12.08$ & $34.59$ & WES\tablefootmark{d} \\ 
 B1917+00 & 1.272 & $7.675\mbox{e-}15$ & 55695 & & 56354 &  90.27(2) $^{[1]}$ & 90(1) & $6.42$ & $12.50$ & $32.17$ & LOTAAS\tablefootmark{b},WES\tablefootmark{d} \\ 
 B1918+26 & 0.786 & $3.414\mbox{e-}17$ & 54142 & & 55870 &  27.68(9) $^{[1]}$ & 27.7(1) & $8.56$ & $11.22$ & $30.44$ & EPN\tablefootmark{d} \\ 
 B1919+21 & 1.337 & $1.349\mbox{e-}15$ & 55969 & 55860 & 55859 &  12.4632(1) $^{[3]}$ & 12.4453($<1$) & $7.20$ & $12.13$ & $31.35$ & WES\tablefootmark{d}  \\ 
 B1920+21 & 1.078 & $8.169\mbox{e-}15$ & 55507 & & 55859 &  217.08(2) $^{[1]}$ & 217.0(1) & $6.32$ & $12.48$ & $32.41$ & WES\tablefootmark{d} \\ 
 B1923+04 & 1.074 & $2.459\mbox{e-}15$ & 55445 & & 56339 &  102.24(2) $^{[1]}$ & 102.2(9) & $6.84$ & $12.22$ & $31.89$ & LOTAAS\tablefootmark{b},WES\tablefootmark{d} \\ 
 B1929+10 & 0.227 & $1.157\mbox{e-}15$ & 55830 & 55913 & 55859 &  3.178(4) $^{[1]}$ & 3.18(1) & $6.49$ & $11.71$ & $33.59$ & WES\tablefootmark{d} \\ 
 B1953+50 & 0.519 & $1.374\mbox{e-}15$ & 55508 & 55864 & 55858 &  31.974(3) $^{[1]}$ & 31.98(1) & $6.78$ & $11.93$ & $32.59$ & EPN\tablefootmark{d} \\ 
 B2021+51 & 0.529 & $3.066\mbox{e-}15$ & 55889 & & 55858 &  22.634(6) $^{[1]}$ & 22.55(1) & $6.44$ & $12.11$ & $32.91$ & WES\tablefootmark{d} \\ 
 B2022+50 & 0.373 & $2.515\mbox{e-}15$ & 55507 & & 55859 &  33.036(3) $^{[1]}$ & 33.0(2) & $6.37$ & $11.99$ & $33.28$ & WES\tablefootmark{d} \\ 
 B2043$-$04 & 1.547 & $1.471\mbox{e-}15$ & 55311 & & 56324 &  35.81(1) $^{[1]}$ & 36(1) & $7.22$ & $12.18$ & $31.20$ & WES\tablefootmark{d} \\ 
 B2044+15 & 1.138 & $1.823\mbox{e-}16$ & 55507 & & 56293 &  39.86(1) $^{[1]}$ & 39.8(1) & $8.00$ & $11.66$ & $30.69$ & LOTAAS\tablefootmark{b},P-2004\tablefootmark{c} \\ 
 B2106+44 & 0.415 & $8.606\mbox{e-}17$ & 55855 & & 56602 &  139.83(1) $^{[1]}$ & 140(1) & $7.88$ & $11.28$ & $31.68$ & WES\tablefootmark{d} \\ 
 B2110+27 & 1.203 & $2.623\mbox{e-}15$ & 55424 & & 56602 &  25.119(4) $^{[1]}$ & 25.1(1) & $6.86$ & $12.25$ & $31.77$ & WES\tablefootmark{d} \\ 
 B2113+14 & 0.440 & $2.895\mbox{e-}16$ & 55696 & & 55859 &  56.15(7) $^{[1]}$ & 56.2(1) & $7.38$ & $11.56$ & $32.13$ & WES\tablefootmark{d} \\ 
 B2122+13 & 0.694 & $7.677\mbox{e-}16$ & 55507 & & 56332 &  30.1(1) $^{[1]}$ & 30.3(2) & $7.16$ & $11.87$ & $31.96$ & LOTAAS\tablefootmark{b},EPN\tablefootmark{d} \\ 
 B2148+63 & 0.380 & $1.707\mbox{e-}16$ & 55769 & & 55858 &  129.704(1) $^{[7]}$ & 129.7(2) & $7.55$ & $11.41$ & $32.09$ & WES\tablefootmark{d} \\ 
 B2154+40 & 1.525 & $3.439\mbox{e-}15$ & 55334 & & 55858 &  70.86(1) $^{[1]}$ & 71.14(6) & $6.85$ & $12.37$ & $31.58$ & WES\tablefootmark{d} \\ 
 B2217+47 & 0.538 & $2.767\mbox{e-}15$ & 56228 & 55860 & 55858 &  43.49(1) $^{[1]}$ & 43.4811($<1$) & $6.49$ & $12.09$ & $32.84$ & WES\tablefootmark{d} \\ 
 B2224+65 & 0.683 & $9.647\mbox{e-}15$ & 55964 & & 55858 &  36.08(9) $^{[1]} $ & 36.33(5) & $6.05$ & $12.41$ & $33.08$ & WES\tablefootmark{d} \\ 
 B2227+61 & 0.443 & $2.255\mbox{e-}15$ & 55859 & & 55858 &  124.62(2) $^{[1]}$ & 124.7(1) & $6.49$ & $12.00$ & $33.01$ & WES\tablefootmark{d} \\ 
 J2248$-$0101 & 0.477 & $6.596\mbox{e-}16$ & 56405 & & 56304 &  29.06(3) $^{[1]}$ & 29.1(4) & $7.06$ & $11.75$ & $32.38$ & LOTAAS\tablefootmark{b},WES\tablefootmark{d} \\ 
 J2253+1516 & 0.792 & $6.653\mbox{e-}17$ & 49079 & & 56354 &  29.182(9) $^{[8]}$ & 29(1) & $8.28$ & $11.37$ & $30.72$ & LOTAAS\tablefootmark{b} \\ 
 B2255+58 & 0.368 & $5.749\mbox{e-}15$ & 55671 & & 55858 &  151.082(6) $^{[1]}$ & 151.2(1) & $6.01$ & $12.17$ & $33.66$ & WES\tablefootmark{d} \\ 
 B2303+30 & 1.576 & $2.893\mbox{e-}15$ & 54584 & 55892 & 55858 &  49.56(2) $^{[1]}$ & 49.58(5) & $6.94$ & $12.33$ & $31.47$ & WES\tablefootmark{d} \\ 
 B2306+55 & 0.475 & $1.995\mbox{e-}16$ & 55384 & & 55858 &  46.535(8) $^{[1]}$ & 46.54(2) & $7.58$ & $11.49$ & $31.87$ & WES\tablefootmark{d} \\ 
 B2310+42 & 0.349 & $1.123\mbox{e-}16$ & 54363 & 55860 & 56606 &  17.274(1) $^{[1]}$ & 17.27(1) & $7.69$ & $11.30$ & $32.02$ & EPN\tablefootmark{d} \\ 
 B2315+21 & 1.445 & $1.047\mbox{e-}15$ & 56407 & & 55858 &  20.874(7) $^{[1]}$ & 20.87(6) & $7.34$ & $12.10$ & $31.14$ & P- 2004\tablefootmark{c} \\ 
 B2334+61 & 0.495 & $1.933\mbox{e-}13$ & 55508 & & 55858 &  58.41(2) $^{[1]}$ & 58.43(3) & $4.61$ & $13.00$ & $34.80$ & WES\tablefootmark{d} \\
\end{longtable}
\tablefoot{
\tablefoottext{a}{For each DM measurement and its error we quote the reference paper:  1, 2, 3, 4, 5, 6, 7, and  8. }
\tablefoottext{b}{Observations from the LOTAAS survey \citep{coenen14}.}
\tablefoottext{c}{Year of the P- and/or L-band data from WSRT observations. L-band data were obtained from Lovell observations in all but the three cases where the year of the WSRT observations is indicated. All the L-band observations using the Lovell telescope were performed between October 2011 and November 2013.}
\tablefoottext{d}{Where P-band data were available from \cite{weltevrede05}, but no absolute reference was available to align the data with \lof\ data, reference is made to their paper as $WES$ and the corresponding profile in Fig. \ref{fig:100} was aligned by eye. For completeness, where the pulsar was not in $WES$, then $EPN$ indicates that the profile in P-band was obtained from the EPN pulsar database and was also only aligned by eye. In this latter case, the observations are not later than 2002. In the cases where nothing is specified, for the P-band the profile is either missing in all the databases used here or the Lovell data were used and the profile aligned with the standard procedure as described in the text.}
}
\tablebib{
(1)\citet{hobbs04}; (2) \citet{snt97}; (3) \citet{hassall12}; (4) \citet{js06};
(5) \citet{lzb00}; (6) \citet{lwf04}; (7) \citet{hl87}; (8) \citet{cn95}.
}
\end{center}
\end{landscape}
}

\onllongtab{
\begin{longtable}{lcccccc}
\caption{{\bf Width of the full profile and duty cycle.}
For each pulsar (Col. 1) at each observing band (Col. 2), we
list in Col. 3 the full width at half maximum, $w_{50}$, in Col. 4 the width of the full profile calculated as  $w_{10}$  (see text for details and Fig. \ref{fig:examplew}), an
d in Col. 5 is the duty cycle of the pulse,  $w_{10} /P$, indicated as percent
age of the total profile. In the last two columns we list the spectral indices $\delta$ of the evolution of $w_{50}$ and $w_{10}$ with observing frequency, modelled 
as  $w(\nu) \propto \nu^{\delta}$. In all cases the statistical error is quoted 
in parentheses on the last digit. Pulsars with an interpulse are marked with the
 notation (IP) next to their names. Where scattering affects the HB
A measurement, no spectrum was produced (see Fig. \ref{fig:w10a} for reference).
}\\
\hline
\label{tab:w10}
PSR Name & Band &  $w_{50}$ &  $w_{10}$ & $w_{10} / P$ & $\delta_{50}$ & $\delta
_{10}$  \\
 &   & {\footnotesize [deg]}  & {\footnotesize [deg]} & \% & & \\
\hline\hline
\endfirsthead
\caption{Continued.}\\
\hline
PSR Name & Band &  $w_{50}$ &  $w_{10}$ & $w_{10} / P$ & $\delta_{50}$ & $\delta
_{10}$  \\
 & & {\footnotesize [deg]}   & {\footnotesize [deg]} & \% & & \\
\hline\hline
\endhead
\hline
\endfoot
\\
 B0011+47 & & & & & $-$0.16(8) & 0.04(7) \\
 & HBA & 22(1) & 43(3) & 12.1(9) & \\  
 & P & 21(4) & 38(2) & 10.6(7) & \\  
 & L & 15.8(3) & 44.7(6) & 12.4(1) & \\  
 B0031$-$07 & & & & & $-$0.46(6) & $-$0.12(3) \\
 & HBA & 26(2) & 45(3) & 12.5(8) & \\  
 & P & 21.1(4) & 39.9(4) & 11.1(1) & \\  
 & L & 11.3(3) & 34.5(3) & 9.6(1) & \\  
 J0051+0423 & & & & & --(--) & --(--) \\
 & HBA & 15.5(3) & 36.6(5) & 10.2(1) & \\  
 B0059+65 & & & & & $-$0.08(9) & $-$0.15(7) \\
 & HBA & 2(1) & 26(1) & 7.4(5) & \\  
 & P & 18.3(8) & 24(2) & 6.8(6) & \\  
 & L & 16.2(3) & 19.4(3) & 5.38(9) & \\  
 B0105+65 & & & & & $-$0.07(9) & 0.01(9) \\
 & HBA & 8.4(6) & 16(1) & 4.5(4) & \\  
 & P & 4(1) & 15(1) & 4.3(3) & \\  
 & L & 7.0(3) & 16.2(4) & 4.5(1) & \\  
 B0114+58 & & & & & --(--) & --(--) \\
 & HBA & scattered & & & \\   
 & P & 10(2) & 31(2) & 8.8(6) & \\  
 & L & 10(1) & 16.5(7) & 4.6(2) & \\  
 B0136+57 & & & & & --(--) & --(--) \\
 & HBA & scattered & & & \\  
 & P & 4(1) & 9(1) & 2.7(3) & \\  
 & L & 4(1) & 9(1) & 2.7(3) & \\  
 B0138+59 & & & & & 0.3(6) & $-$0.04(3) \\
 & LBA & 5.6(7) & 50(4) & 14(1) & \\  
 & HBA & 5.6(3) & 32(1) & 9.1(3) & \\  
 & P & 8(1) & 31(1) & 8.6(3) & \\  
 & L & 12.3(3) & 32.7(5) & 9.1(1) & \\  
 B0301+19 & & & & & $-$0.21(5) & $-$0.18(4) \\
 & HBA & 17.6(6) & 22.5(6) & 6.3(1) & \\  
 & P & 15.3(6) & 15.5(6) & 4.3(1) & \\  
 & L & 10.9(3) & 14.8(3) & 4.11(9) & \\  
 B0320+39 & & & & & 0.1(1) & 0.05(4) \\
 & LBA & 8(2) & 15(2) & 4.2(6) & \\  
 & HBA & 4.6(3) & 9.8(3) & 2.73(9) & \\  
 & P & 4(1) & 8(1) & 2.4(3) & \\  
 & L & 6.7(8) & 11.6(3) & 3.23(9) & \\  
 B0329+54 & & & & & 0.1(1) & $-$0.09(2) \\
 & LBA & scattered & & & \\  
 & HBA & 8.8(3) & 32.7(3) & 9.08(9) & \\  
 & P & 2(1) & 29(1) & 8.2(3) & \\  
 & L & 1.4(3) & 26.8(3) & 7.45(9) & \\  
 B0331+45 & & & & & $-$0.9(3) & $-$0.4(1) \\
 & HBA & 5(2) & 19(2) & 5.5(7) & \\  
 & P & 7.6(9) & 17(1) & 4.7(3) & \\  
 & L & 1.8(3) & 9.9(4) & 2.7(1) & \\  
 B0339+53 & & & & & $-$0.2(1) & $-$0.3(1) \\
 & HBA & 7.7(6) & 16.2(7) & 4.5(2) & \\  
 & P & 7(1) & 11(1) & 3.1(3) & \\  
 & L & 5.3(5) & 8.4(7) & 2.3(2) & \\  
 B0355+54 & & & & & 0.1(4) & 0.09(3) \\
 & HBA & 13.0(4) & 30.2(8) & 8.4(2) & \\  
 & P & 5(1) & 18(1) & 5.1(3) & \\  
 & L & 16.2(3) & 35.5(3) & 9.87(9) & \\  
 B0402+61 & & & & & $-$0.8(1) & $-$0.23(5) \\
 & HBA & 16(1) & 28(1) & 7.8(3) & \\  
 & P & 3.5(9) & 15.5(6) & 4.3(1) & \\  
 & L & 2.8(3) & 14.8(3) & 4.11(9) & \\  
 B0410+69 & & & & & $-$0.4(4) & $-$0.4(1) \\
 & HBA & 5(2) & 19(2) & 5.5(7) & \\  
 & P & 2(1) & 12(1) & 3.5(3) & \\  
 & L & 2.1(3) & 8.4(8) & 2.3(2) & \\  
 B0447$-$12 & & & & & 0.8(1) & 0.1(1) \\
 & HBA & 2.8(2) & 25(2) & 7.0(7) & \\  
 & P & 12(1) & 23(1) & 6.7(3) & \\  
 & L & 20(1) & 27(1) & 7.7(5) & \\  
 B0450+55 & & & & & 0.1(1) & $-$0.07(5) \\
 & LBA & 5(2) & 39(4) & 10(1) & \\  
 & HBA & 25(4) & 39(2) & 10.9(7) & \\  
 & P & 8(1) & 28(2) & 7.8(6) & \\  
 & L & 17.9(3) & 31.0(3) & 8.6(9) & \\  
 B0450$-$18 & & & & & $-$0.1(1) & $-$0.1(1) \\
 & HBA & 21(1) & 32(2) & 9.0(7) & \\  
 & P & 16(1) & 22(1) & 6.3(3) & \\  
 & L & 16(1) & 22(1) & 6.3(3) & \\  
 B0523+11 & & & & & --(--) & --(--) \\
 & HBA & scattered & & & \\   
 & P & 13(1) & 21(1) & 5.9(3) & \\  
 & L & 13.4(3) & 17.2(3) & 4.79(9) & \\  
 B0525+21 & & & & & $-$0.17(6) & $-$0.12(6) \\
 & LBA & 25(1) & 28(2) & 8.0(6) & \\  
 & HBA & 21.3(1) & 25.3(1) & 7.03(4) & \\  
 & P & 16(1) & 21(1) & 5.9(3) & \\  
 & L & 14(1) & 19(1) & 5.5(3) & \\  
 B0540+23 & & & & & --(--) & --(--) \\
 & HBA & scattered & & & \\   
 & P & 7(1) & 23(1) & 6.7(3) & \\  
 & L & 7.4(3) & 24.6(3) & 6.84(9) & \\  
 B0609+37 & & & & & 0.1(2) & $-$0.23(8) \\
 & HBA & 2(2) & 30(2) & 8.6(7) & \\  
 & P & 4.5(3) & 24(1) & 6.7(3) & \\  
 & L & 4.9(3) & 17.9(4) & 5.0(1) & \\  
 B0611+22 & & & & & --(--) & --(--) \\
 & HBA & scattered & & & \\  
 & P & 5(1) & 14(1) & 3.9(4) & \\  
 & L & 6.7(3) & 14.1(3) & 3.91(9) & \\  
 B0626+24 & & & & & --(--) & --(--) \\
 & HBA & 7(1) & 23(1) & 6.6(4) & \\  
 & P & 5(1) & 14(1) & 3.9(3) & \\  
 & L & 9.9(3) & 17.2(3) & 4.79(9) & \\  
 B0643+80 & & & & & $-$0.3(1) & $-$0.2(1) \\
 & HBA & 4.9(6) & 9(1) & 2.7(3) & \\  
 & P & 1(1) & 11(1) & 3.1(3) & \\  
 & L & 2.5(3) & 6.7(3) & 1.86(9) & \\  
 B0809+74 & & & & & $-$0.32(1) & $-$0.17(1) \\
 & LBA & 33.4(3) & 52.7(9) & 14.6(2) & \\  
 & HBA & 12.3(3) & 29.5(3) & 8.2(1) & \\  
 & P & 12.1(2) & 25.6(2) & 7.1(6) & \\  
 & L & 14.4(3) & 26.4(3) & 7.33(9) & \\  
 B0818$-$13 & & & & & 0.0(1) & $-$0.06(6) \\
 & HBA & 4.9(6) & 10.5(6) & 2.9(1) & \\  
 & P & 6.2(1) & 9.8(1) & 2.71(3) & \\  
 & L & 6.0(3) & 9.1(3) & 2.54(9) & \\  
 B0820+02 & & & & & $-$0.17(9) & $-$0.12(6) \\
 & HBA & 12(1) & 19(1) & 5.5(3) & \\  
 & P & 9.2(3) & 14.3(3) & 3.98(9) & \\  
 & L & 7.7(3) & 13.0(3) & 3.62(9) & \\  
 B0823+26 (IP) & & & & & $-$0.39(8) & $-$0.47(6) \\
 & LBA & 10.5(3) & 34(2) & 9.7(7) & \\  
 & HBA & 4(1) & 12(1) & 3.5(3) & \\  
 & P & 3.3(7) & 9.2(7) & 2.6(2) & \\  
 & L & 2.8(3) & 6.0(3) & 1.66(9) & \\  
 B0834+06 & & & & & $-$0.05(4) & $-$0.16(2) \\
 & LBA & 7.7(3) & 15.5(3) & 4.3(9) & \\  
 & HBA & 6.0(3) & 13.7(3) & 3.81(9) & \\  
 & P & 6.4(1) & 8.8(1) & 2.44(3) & \\  
 & L & 6.7(3) & 8.7(3) & 2.41(9) & \\  
 B0906$-$17 & & & & & 0.0(3) & $-$0.06(8) \\
 & HBA & 2(2) & 25(2) & 7.0(7) & \\  
 & P & 6(1) & 19(1) & 5.5(3) & \\  
 & L & 6.7(3) & 20.1(4) & 5.6(1) & \\  
 B0917+63 & & & & & $-$0.18(7) & $-$0.21(5) \\
 & LBA & 14(2) & 20(2) & 5.7(7) & \\  
 & HBA & 9.8(6) & 14.1(6) & 3.9(1) & \\  
 & P & 7(2) & 11(1) & 3.3(4) & \\  
 & L & 6.7(3) & 8.8(3) & 2.44(9) & \\  
 B0919+06 & & & & & $-$0.0(1) & $-$0.31(3) \\
 & HBA & 4.6(3) & 22.9(3) & 6.35(9) & \\  
 & P & 8.2(6) & 19.2(6) & 5.3(1) & \\  
 & L & 4.2(3) & 10.9(3) & 3.03(9) & \\  
 B0943+10 & & & & & 0.3(1) & $-$0.18(3) \\
 & LBA & 6.3(6) & 26.0(8) & 7.2(2) & \\  
 & HBA & 7.0(6) & 20.4(6) & 5.7(1) & \\  
 & P & 13(1) & 23(3) & 6.6(8) & \\  
 & L & 7(2) & 13.4(4) & 3.7(1) & \\  
 B0950+08 (IP) & & & & & $-$0.31(2) & $-$0.22(2) \\
 & LBA & 33.0(3) & 65(1) & 18.1(4) & \\  
 & HBA & 16(5) & 50(5) & 14(1) & \\  
 & P & 12.3(7) & 28.5(7) & 7.9(2) & \\  
 & L & 11.6(3) & 29.2(4) & 8.1(1) & \\  
 B1112+50 & & & & & 0.1(1) & $-$0.08(6) \\
 & LBA & 4.2(9) & 12(1) & 3.5(4) & \\  
 & HBA & 4.2(6) & 9.8(6) & 2.7(1) & \\  
 & P & 2.9(6) & 7.1(6) & 2.0(1) & \\  
 & L & 5.3(3) & 8.4(3) & 2.35(9) & \\  
 B1133+16 & & & & & $-$0.16(3) & $-$0.17(2) \\
 & LBA & 14.1(3) & 18.3(3) & 5.08(9) & \\  
 & HBA & 10.9(3) & 14.1(3) & 3.91(9) & \\  
 & P & 9.8(1) & 12.7(1) & 3.52(3) & \\  
 & L & 1.4(3) & 8.8(3) & 2.44(9) & \\  
 B1237+25 & & & & & 0.2(1) & $-$0.15(2) \\
 & LBA & 1.4(6) & 22.5(6) & 6.3(1) & \\  
 & HBA & 1.8(3) & 17.6(3) & 4.88(9) & \\  
 & P & 12.4(7) & 14.6(7) & 4.1(1) & \\  
 & L & 10.9(3) & 13.0(3) & 3.62(9) & \\  
 J1238+21 & & & & & 0.4(4) & 0.0(1) \\
 & HBA & 1.4(6) & 9.8(6) & 2.7(1) & \\  
 & L & 3.2(4) & 10(1) & 2.8(3) & \\  
 J1313+0931 & & & & & --(--) & --(--) \\
 & HBA & 4(1) & 8(1) & 2.3(3) & \\  
 B1322+83 & & & & & $-$0.0(1) & $-$0.07(9) \\
 & HBA & 11(1) & 22(1) & 6.3(3) & \\  
 & P & 8(1) & 22(1) & 6.4(5) & \\  
 & L & 10.2(5) & 19(1) & 5.3(3) & \\  
 B1508+55 & & & & & 0.37(8) & 0.07(4) \\
 & LBA & scattered & & & \\  
 & HBA & 4.2(3) & 9.5(3) & 2.64(9) & \\  
 & P & 4.6(1) & 12.6(1) & 3.49(2) & \\  
 & L & 8.1(3) & 12.3(3) & 3.42(9) & \\  
 B1530+27 & & & & & $-$0.42(9) & $-$0.23(5) \\
 & HBA & 12.0(6) & 20.4(8) & 5.7(2) & \\  
 & P & 8(1) & 15.1(7) & 4.2(2) & \\  
 & L & 4.6(3) & 12.0(3) & 3.3(1) & \\  
 B1540$-$06 & & & & & $-$0.2(1) & $-$0.06(7) \\
 & LBA & 7(1) & 18(3) & 5.1(8) & \\  
 & HBA & 2(1) & 11(1) & 3.1(3) & \\  
 & P & 3.3(2) & 8.3(2) & 2.31(5) & \\  
 & L & 2.8(3) & 8.8(3) & 2.44(9) & \\  
 B1541+09 & & & & & 0.52(2) & 0.44(4) \\
 & LBA & scattered & & & \\  
 & HBA & 18.6(3) & 36.6(5) & 10.2(1) & \\  
 & P & 17(3) & 71(3) & 19(1) & \\  
 & L & 61(1) & 97(3) & 26(1) & \\  
 B1604$-$00 & & & & & $-$0.1(6) & $-$0.04(4) \\
 & LBA & 8(2) & 11(2) & 3.1(6) & \\  
 & HBA & 9.5(3) & 14.1(3) & 3.91(9) & \\  
 & P & 10.5(6) & 15.4(6) & 4.3(1) & \\  
 & L & 7.4(3) & 12.7(3) & 3.52(9) & \\  
 B1612+07 & & & & & $-$0.0(4) & $-$0.0(1) \\
 & HBA & 1.4(6) & 7.0(6) & 2.0(1) & \\  
 & P & 2.3(7) & 7.7(8) & 2.1(2) & \\  
 & L & 1.8(3) & 6.7(4) & 1.9(1) & \\  
 J1627+1419 & & & & & --(--) & --(--) \\
 & HBA & 23(1) & 32(5) & 8(1) & \\  
 B1633+24 & & & & & $-$0.1(1) & $-$0.5(7) \\
 & LBA & 14(2) & 18(2) & 5.1(7) & \\  
 & HBA & 8(2) & 28(2) & 7.8(7) & \\  
 & P & 9(1) & 33(2) & 9.4(7) & \\  
 & L & 9(1) & 20.1(4) & 5.6(1) & \\  
 B1642$-$03 & & & & & 0.1(1) & 0.17(5) \\
 & HBA & 2.8(3) & 7.7(3) & 2.15(9) & \\  
 & P & 3.0(1) & 6.3(1) & 1.74(5) & \\  
 & L & 3.5(3) & 10.2(3) & 2.83(9) & \\  
 J1645+1012 & & & & & $-$0.2(2) & $-$0.1(1) \\
 & HBA & 9(1) & 18(1) & 5.1(3) & \\  
 & L & 6(1) & 16(2) & 4.5(6) & \\  
 J1652+2651 & & & & & $-$0.17(8) & $-$0.16(7) \\
 & HBA & 16(1) & 21(1) & 5.9(3) & \\  
 & P & 14(1) & 16(1) & 4.7(2) & \\  
 & L & 11.6(3) & 14.4(4) & 4.0(1) & \\  
 B1737+13 & & & & & 0.2(2) & $-$0.02(7) \\
 & HBA & 5(1) & 19(1) & 5.5(3) & \\  
 & P & 13.8(5) & 22(1) & 6.3(3) & \\  
 & L & 13(1) & 19.7(7) & 5.5(2) & \\  
 B1749$-$28 & & & & & $-$0.3(1) & $-$0.25(5) \\
 & HBA & 6.0(3) & 12.0(3) & 3.32(9) & \\  
 & P & 4.2(1) & 8.0(1) & 2.22(3) & \\  
 & L & 3.5(3) & 7.4(3) & 2.05(9) & \\  
 B1818$-$04 & & & & & --(--) & --(--) \\
 & HBA & scattered & & & \\    
 & P & 5.9(7) & 16.3(7) & 4.5(1) & \\  
 & L & 6.0(3) & 11.3(3) & 3.13(9) & \\  
 B1821+05 & & & & & 0.95(9) & 0.49(4) \\
 & HBA & 3.5(3) & 8.8(3) & 2.44(9) & \\  
 & P & 3.2(3) & 19.3(8) & 5.4(2) & \\  
 & L & 22.5(3) & 28.9(5) & 8.0(1) & \\  
 B1822$-$09 (IP) & & & & & $-$0.2(1) & 0.4(1) \\
 & LBA & 8(1) & 33(2) & 9.4(6) & \\  
 & HBA & 4(1) & 9(1) & 2.7(3) & \\  
 & P & 4(1) & 9(1) & 2.6(5) & \\  
 & L & 4.2(3) & 21.5(3) & 5.96(9) & \\  
 B1831$-$04 & & & & & --(--) & --(--) \\
 & HBA & scattered & & & \\  
 & P & 50(7) & 122(15) & 34(4) & \\  
 & L & 108(2) & 132.0(9) & 36.7(2) & \\  
 B1839+09 & & & & & 0.3(2) & $-$0.0(1) \\
 & HBA & 8(2) & 19(2) & 5.5(7) & \\  
 & P & 3.9(7) & 13.2(8) & 3.7(2) & \\  
 & L & 8.8(3) & 14.8(9) & 4.1(2) & \\  
 B1839+56 & & & & & 0.4(1) & 0.21(6) \\
 & LBA & 2.1(6) & 5.6(6) & 1.6(1) & \\  
 & HBA & 1.8(3) & 3.5(3) & 0.98(9) & \\  
 & P & 4.3(8) & 11.4(6) & 3.2(1) & \\  
 & L & 6.3(3) & 9.9(3) & 2.74(9) & \\  
 B1842+14 & & & & & 0.19(6) & 0.12(4) \\
 & LBA & scattered & & & \\  
 & HBA & 33.8(3) & 61.9(3) & 17.19(9) & \\  
 & P & 6.7(5) & 12.0(5) & 3.3(1) & \\  
 & L & 7.7(3) & 12.4(3) & 3.44(9) & \\  
 B1848+12 & & & & & $-$0.2(2) & $-$0.3(1) \\
 & HBA & 4.2(6) & 11(1) & 3.3(3) & \\  
 & P & 2.4(4) & 5.4(4) & 1.5(1) & \\  
 & L & 2.8(3) & 6.0(5) & 1.7(1) & \\  
 B1848+13 & & & & & $-$0.2(2) & $-$0.2(1) \\
 & HBA & 8(2) & 19(2) & 5.5(7) & \\  
 & P & 5(1) & 14(2) & 4.1(6) & \\  
 & L & 4.9(3) & 12.0(4) & 3.3(1) & \\  
 B1857$-$26 & & & & & --(--) & --(--) \\
 & HBA & scattered & & & \\  
 & P & 19.3(9) & 39.6(9) & 11.0(2) & \\  
 & L & 30.3(3) & 37.0(4) & 10.3(1) & \\  
 B1905+39 & & & & & $-$0.11(4) & $-$0.14(6) \\
 & HBA & 19.7(6) & 26.0(9) & 7.2(2) & \\  
 & P & 16.5(7) & 20.1(7) & 5.6(1) & \\  
 & L & 15.1(3) & 19(1) & 5.5(3) & \\  
 B1907+00 & & & & & --(--) & --(--) \\
  & HBA & scattered & & & \\  
 & P & 1.7(4) & 3.9(4) & 1.1(1) & \\  
 & L & 1.8(3) & 8.8(7) & 2.4(1) & \\  
 B1907+02 & & & & & --(--) & --(--) \\
 & HBA & scattered & & & \\   
 & P & 2.7(8) & 9.9(8) & 2.7(2) & \\  
 & L & 2.1(3) & 12(1) & 3.4(2) & \\  
 B1907+10 & & & & & --(--) & --(--) \\
 & HBA & scattered & & & \\  
 & P & 5.7(5) & 15.6(5) & 4.3(1) & \\  
 & L & 6.0(3) & 16.9(3) & 4.69(9) & \\  
 B1911$-$04 & & & & & --(--) & --(--) \\
 & HBA & scattered & & & \\  
 & P & 2.95(9) & 5.71(9) & 1.59(2) & \\  
 & L & 3.2(3) & 7.0(3) & 1.95(9) & \\  
 B1914+09 & & & & & --(--) & --(--) \\
 & HBA & scattered & & & \\  
 & P & 4(1) & 13(1) & 3.6(5) & \\  
 & L & 12.0(3) & 16.9(9) & 4.7(2) & \\  
 B1915+13 & & & & & --(--) & --(--) \\
 & HBA & scattered & & & \\   
 & P & 4(1) & 14(1) & 4.1(3) & \\  
 & L & 7.4(3) & 15.5(3) & 4.3(9) & \\  
 B1917+00 & & & & & --(--) & --(--) \\
 & HBA & scattered & & & \\   
 & P & 1.4(6) & 9.1(6) & 2.5(1) & \\  
 & L & 1.4(3) & 10.2(3) & 2.83(9) & \\  
 B1918+26 & & & & & $-$0.1(2) & $-$0.2(1) \\
 & HBA & 1(1) & 9(1) & 2.7(3) & \\  
 & P & 7.5(7) & 11(2) & 3.1(5) & \\  
 & L & 6.0(6) & 7.4(3) & 2.1(1) & \\  
 B1919+21 & & & & & 0.11(5) & $-$0.02(3) \\
 & LBA & 5.3(3) & 11.3(3) & 3.13(9) & \\  
 & HBA & 7.0(3) & 10.2(3) & 2.83(9) & \\  
 & P & 7.0(6) & 9.8(6) & 2.7(1) & \\  
 & L & 8.4(3) & 10.6(3) & 2.93(9) & \\  
 B1920+21 & & & & & --(--) & --(--) \\
 & HBA & scattered & & & \\  
 & P & 2.7(6) & 13.0(6) & 3.6(1) & \\  
 & L & 5.3(3) & 17.2(5) & 4.8(1) & \\  
 B1923+04 & & & & & $-$0.3(1) & $-$0.15(9) \\
 & HBA & 9.1(9) & 13(1) & 3.7(3) & \\  
 & P & 4.9(7) & 10.7(7) & 3.0(2) & \\  
 & L & 4.2(3) & 9.1(3) & 2.54(9) & \\  
 B1929+10 (IP) & & & & & 0.13(8) & $-$0.18(3) \\
 & LBA & 11(5) & 50(7) & 14(2) & \\  
 & HBA & 6.3(4) & 28.1(9) & 7.8(2) & \\  
 & P & 9(1) & 21.8(3) & 6.06(9) & \\  
 & L & 8.8(3) & 17.9(3) & 4.99(9) & \\  
 B1953+50 & & & & & $-$0.2(1) & $-$0.13(5) \\
 & LBA & 16(4) & 14(3) & 3(1) & \\  
 & HBA & 3.2(3) & 10.9(3) & 3.03(9) & \\  
 & P & 2(1) & 6(1) & 1.9(3) & \\  
 & L & 3.2(3) & 8.1(3) & 2.25(9) & \\  
 B2021+51 & & & & & 0.3(1) & $-$0.22(5) \\
 & HBA & 5(1) & 25(1) & 7.0(3) & \\  
 & P & 4.5(5) & 21.2(5) & 5.9(1) & \\  
 & L & 8.1(3) & 15.8(3) & 4.4(9) & \\  
 B2022+50 (IP) & & & & & $-$0.4(4) & $-$0.5(1) \\
 & HBA & 5(2) & 22(3) & 6(1) & \\  
 & P & 4(2) & 14(2) & 4.0(5) & \\  
 & L & 2.5(3) & 8.1(3) & 2.25(9) & \\  
 B2043$-$04 & & & & & $-$0.1(1) & $-$0.0(1) \\
 & HBA & 4(1) & 9(1) & 2.5(4) & \\  
 & P & 4.2(3) & 7.6(3) & 2.1(1) & \\  
 & L & 3.9(3) & 7.7(3) & 2.15(9) & \\  
 B2044+15 & & & & & 0.1(3) & $-$0.07(7) \\
 & HBA & 2.1(6) & 9(1) & 2.7(3) & \\  
 & P & 2.8(9) & 16.9(6) & 4.7(1) & \\  
 & L & 2.8(3) & 14.1(3) & 3.91(9) & \\  
 B2106+44 & & & & & --(--) & --(--) \\
  & HBA & scattered & & & \\  
 & P & 18(3) & 50(3) & 14.0(9) & \\  
 & L & 21.1(3) & 31.7(3) & 8.8(9) & \\  
 B2110+27 & & & & & $-$0.3(2) & $-$0.2(1) \\
 & HBA & 3.5(6) & 6.3(6) & 1.8(1) & \\  
 & P & 3.3(3) & 6.3(3) & 1.74(9) & \\  
 & L & 1.8(3) & 4.6(3) & 1.27(9) & \\  
 B2113+14 & & & & & --(--) & --(--) \\
 & HBA & scattered & & & \\  
 & P & 6(1) & 14(1) & 3.9(5) & \\  
 & L & 7.0(3) & 19(1) & 5.3(3) & \\  
 B2122+13 & & & & & $-$0.2(2) & $-$0.2(1) \\
 & HBA & 1(5) & 22(2) & 6.2(6) & \\  
 & P & 15(2) & 16(2) & 4.7(6) & \\  
 & L & 11.6(3) & 14(1) & 4.1(3) & \\  
 B2148+63 & & & & & --(--) & --(--) \\
 & HBA & scattered & & & \\  
 & P & 18(1) & 31(1) & 8.8(3) & \\  
 & L & 11.3(3) & 19.0(3) & 5.28(9) & \\  
 B2154+40 & & & & & --(--) & --(--) \\
 & HBA & 7.4(3) & 33(1) & 9.4(3) & \\  
 & P & 8.5(7) & 26.4(7) & 7.3(2) & \\  
 & L & 14.8(3) & 23.9(3) & 6.65(9) & \\  
 B2217+47 & & & & & 0.1(1) & 0.1(5) \\
 & LBA & scattered & & & \\  
 & HBA & 4.6(3) & 9.1(3) & 2.54(9) & \\  
 & P & 4.7(2) & 8.8(2) & 2.44(7) & \\  
 & L & 5.3(3) & 11.3(3) & 3.13(9) & \\  
 B2224+65 & & & & & 0.48(3) & 0.22(4) \\
 & HBA & 11.6(3) & 22.9(9) & 6.3(2) & \\  
 & P & 7(2) & 48(2) & 13.3(6) & \\  
 & L & 34.5(3) & 44.0(4) & 12.2(1) & \\  
 B2227+61 & & & & & --(--) & --(--) \\
 & HBA & scattered & & & \\  
 & P & 20(1) & 26(1) & 7.4(3) & \\  
 & L & 2.8(3) & 23(1) & 6.6(3) & \\  
 J2248$-$0101 & & & & & --(--) & --(--) \\
 & HBA & scattered & & & \\  
 & P & 7.4(8) & 13(1) & 3.6(3) & \\  
 & L & 6.3(3) & 13(1) & 3.7(2) & \\  
 J2253+1516 & & & & & --(--) & --(--) \\
 & HBA & 2.1(7) & 22(2) & 6.3(6) & \\  
 B2255+58 & & & & & --(--) & --(--) \\
 & HBA & scattered & & & \\  
 & P & 6(1) & 24(1) & 6.7(4) & \\  
 & L & 9.1(3) & 20.4(3) & 5.67(9) & \\  
 B2303+30 & & & & & $-$0.2(1) & $-$0.2(6) \\
 & LBA & 5(2) & 11(2) & 3.3(6) & \\  
 & HBA & 5.6(3) & 12.3(5) & 3.4(1) & \\  
 & P & 4.5(2) & 8.1(2) & 2.26(7) & \\  
 & L & 3.2(3) & 7.4(3) & 2.05(9) & \\  
 B2306+55 & & & & & 0.43(7) & $-$0.14(2) \\
 & HBA & 4.9(3) & 30.6(4) & 8.5(1) & \\  
 & P & 20(1) & 24(1) & 6.8(4) & \\  
 & L & 16.9(3) & 22.2(4) & 6.2(1) & \\  
 B2310+42 & & & & & 0.07(5) & $-$0.07(3) \\
 & HBA & scattered & & & \\  
 & HBA & 8.4(3) & 18.3(3) & 5.08(9) & \\  
 & P & 8(1) & 13(1) & 3.9(2) & \\  
 & L & 9.9(3) & 15.5(3) & 4.3(9) & \\  
 B2315+21 & & & & & $-$0.1(1) & $-$0.0(1) \\
 & HBA & 5.6(6) & 7.7(6) & 2.1(1) & \\  
 & P & 2(1) & 7(1) & 2.0(3) & \\  
 & L & 4.2(3) & 7.0(3) & 1.96(9) & \\  
 B2334+61 & & & & & 0.5(1) & 0.0(1) \\
 & HBA & 8(2) & 25(2) & 7.0(7) & \\  
 & P & 8.1(8) & 22(2) & 6.2(7) & \\  
 & L & 17.2(4) & 25(1) & 7.1(3) & \\  
\\
\hline
\end{longtable}
}

\onllongtab{
\centering
\begin{longtable}{lcccccc}
\caption{{\bf Peak ratios}. For pulsars with double and multiple peaks (Col. 
1), for each frequency band (Col. 2) the amplitudes of the two most prominent 
peaks normalised to the amplitude of the main peak of the profile are indicate
d (Cols. 3 and 4), ordered by spin phase $\phi$ (Cols. 5 and 6), and their r
atio $\rm{P}2/\rm{P}1$ (Col. 7).
The horizontal line in the table marks the change between previously studied cas
es and new (as in Fig. \ref{fig:PRPL}).}\\
\hline
\label{tab:peaks}
PSR Name & Band & $\rm{P}1$ &$\rm{P}2$ & $\phi_1$ &$\phi_2$  & $\rm{P}2/\rm{P}1$
  \\
\\
\hline\hline
\endfirsthead
\caption{Continued.} \\
\hline
PSR Name & Band & $\rm{P}1$ &$\rm{P}2$ & $\phi_1$ &$\phi_2$  & $\rm{P}2/\rm{P}1$
 \\
\\
\hline
\hline
\\
\endhead
\hline
\endfoot
\\
B0059+65 & HBA & 0.28 & 0.96 & 0.47 & 0.52 & 3.41 \\ 
& P & 0.96 & 0.64  & 0.48& 0.52 & 0.66 \\ 
& L & 0.99 & 0.75  & 0.48& 0.52 & 0.77 \\ 
B0301+19 & HBA & 1.00 & 0.67 & 0.48 & 0.52 & 0.68 \\ 
& P & 0.89 & 1.00  & 0.48& 0.52 & 1.12 \\ 
& L & 0.66 & 0.98  & 0.49& 0.51 & 1.49 \\ 
B0320+39 & LBA & 0.57 & 0.68 & 0.48 & 0.50 & 1.19 \\ 
& HBA & 0.54 & 0.79  & 0.49& 0.50 & 1.46 \\ 
& L & 0.95 & 0.32  & 0.50& 0.51 & 0.33 \\ 
B0329+54 & HBA & 1.00 & 0.12 & 0.50 & 0.53 & 0.12 \\ 
& P & 1.00 & 0.16  & 0.50& 0.53 & 0.17 \\ 
& L & 0.99 & 0.30  & 0.50& 0.52 & 0.30 \\ 
B0525+21 & LBA & 0.87 & 0.70 & 0.52 & 0.58 & 0.80 \\ 
& HBA & 0.99 & 0.75  & 0.52& 0.58 & 0.76 \\ 
& P & 0.98 & 0.92  & 0.53& 0.57 & 0.93 \\ 
& L & 0.73 & 1.01  & 0.53& 0.57 & 1.39 \\ 
B0834+06 & LBA & 0.97 & 0.70 & 0.50 & 0.51 & 0.72 \\ 
& HBA & 1.01 & 0.62  & 0.50& 0.51 & 0.62 \\ 
& P & 1.00 & 0.60  & 0.50& 0.51 & 0.60 \\ 
& L & 0.93 & 0.65  & 0.49& 0.51 & 0.70 \\ 
B0917+63 & LBA & 0.62 & 0.85 & 0.48 & 0.52 & 1.37 \\ 
& HBA & 0.83 & 0.95  & 0.49& 0.51 & 1.15 \\ 
& P & 0.58 & 1.00  & 0.49& 0.51 & 1.72 \\ 
& L & 1.01 & 0.93  & 0.49& 0.50 & 0.92 \\ 
B1133+16 & LBA & 0.71 & 0.97 & 0.49 & 0.52 & 1.35 \\ 
& HBA & 0.98 & 0.81  & 0.49& 0.52 & 0.82 \\ 
& P & 0.98 & 0.71  & 0.49& 0.51 & 0.72 \\ 
& L & 0.99 & 0.30  & 0.49& 0.51 & 0.31 \\ 
B1237+25 & LBA & 0.99 & 0.34 & 0.49 & 0.53 & 0.34 \\ 
& HBA & 1.00 & 0.44  & 0.48& 0.52 & 0.45 \\ 
& P & 1.00 & 0.61  & 0.49& 0.52 & 0.61 \\ 
& L & 0.91 & 0.98  & 0.48& 0.51 & 1.07 \\ 
B1604$-$00 & LBA & 1.10 & 0.76 & 0.49 & 0.51 & 0.69 \\ 
& HBA & 0.52 & 0.82  & 0.49& 0.51 & 1.57 \\ 
& P & 0.73 & 0.61  & 0.49& 0.51 & 0.83 \\ 
& L & 0.56 & 0.79  & 0.49& 0.51 & 1.42 \\ 
B2044+15 & HBA & 0.45 & 0.97 & 0.47 & 0.51 & 2.18 \\ 
& P & 0.37 & 1.00  & 0.48& 0.51 & 2.70 \\ 
& L & 0.34 & 0.99  & 0.49& 0.51 & 2.95 \\ 
B0402+61 & HBA & 0.66 & 0.71 & 0.49 & 0.52 & 1.08 \\ 
& P & 0.86 & 0.44  & 0.49& 0.51 & 0.51 \\ 
& L & 0.48 & 0.71  & 0.51& 0.52 & 1.50 \\ 
\hline
B0447$-$12 & HBA & 0.95 & 0.43 & 0.48 & 0.50 & 0.45 \\ 
& P & 0.81 & 0.75  & 0.49& 0.52 & 0.92 \\ 
& L & 0.70 & 0.87  & 0.48& 0.51 & 1.24 \\ 
B0450+55 & LBA & 1.02 & 0.22 & 0.47 & 0.53 & 0.22 \\ 
& HBA & 0.89 & 0.67  & 0.49& 0.55 & 0.75 \\ 
& P & 0.99 & 0.24  & 0.49& 0.53 & 0.24 \\ 
& L & 0.48 & 0.60  & 0.46& 0.50 & 1.27 \\ 
B0523+11 & HBA & 0.94 & 0.84 & 0.48 & 0.52 & 0.89 \\ 
& P & 1.00 & 0.67  & 0.48& 0.52 & 0.67 \\ 
& L & 0.61 & 0.98  & 0.48& 0.51 & 1.60 \\ 
B0609+37 & HBA & 0.36 & 1.01 & 0.48 & 0.50 & 2.77 \\ 
& P & 0.26 & 0.76  & 0.49& 0.51 & 2.91 \\ 
& L & 0.32 & 0.63  & 0.50& 0.51 & 2.00 \\ 
B0626+24 & HBA & 0.89 & 0.27 & 0.48 & 0.50 & 0.31 \\ 
& P & 0.95 & 0.41  & 0.49& 0.50 & 0.43 \\ 
& L & 0.48 & 0.75  & 0.49& 0.50 & 1.57 \\ 
B0643+80 & HBA & 0.83 & 0.95 & 0.49 & 0.50 & 1.14 \\ 
& P & 0.95 & 0.48  & 0.49& 0.50 & 0.50 \\ 
B0906$-$17 & HBA & 0.95 & 0.19 & 0.49 & 0.51 & 0.20 \\ 
& P & 0.95 & 0.21  & 0.49& 0.51 & 0.22 \\ 
& L & 0.32 & 0.95  & 0.48& 0.49 & 3.00 \\ 
B0943+10 & LBA & 0.95 & 0.24 & 0.50 & 0.54 & 0.25 \\ 
& HBA & 0.95 & 0.41  & 0.50& 0.52 & 0.43 \\ 
& P & 0.91 & 0.76  & 0.49& 0.51 & 0.83 \\ 
B1530+27 & HBA & 0.83 & 0.74 & 0.49 & 0.52 & 0.90 \\ 
& P & 1.02 & 0.48  & 0.50& 0.51 & 0.47 \\ 
J1652+2651 & HBA & 0.89 & 1.00 & 0.48 & 0.52 & 1.13 \\ 
& P & 0.82 & 1.01  & 0.48& 0.51 & 1.23 \\ 
& L & 0.65 & 0.84  & 0.49& 0.51 & 1.30 \\ 
B1737+13 & HBA & 0.21 & 1.03 & 0.47 & 0.51 & 4.88 \\ 
& P & 0.63 & 1.01  & 0.48& 0.51 & 1.61 \\ 
& L & 0.93 & 0.57  & 0.48& 0.51 & 0.62 \\ 
B1905+39 & HBA & 0.71 & 0.38 & 0.48 & 0.49 & 0.53 \\ 
& P & 0.95 & 0.60  & 0.48& 0.51 & 0.63 \\ 
& L & 0.95 & 1.00  & 0.49& 0.51 & 1.05 \\ 
B1918+26 & HBA & 1.00 & 0.26 & 0.50 & 0.51 & 0.26 \\ 
& P & 1.07 & 1.03  & 0.49& 0.51 & 0.96 \\ 
& L & 0.92 & 0.72  & 0.50& 0.51 & 0.78 \\ 
B1919+21 & LBA & 0.57 & 0.59 & 0.49 & 0.50 & 1.03 \\ 
& HBA & 0.95 & 0.71  & 0.49& 0.51 & 0.75 \\ 
& P & 0.95 & 0.59  & 0.49& 0.50 & 0.62 \\ 
& L & 0.74 & 0.95  & 0.50& 0.51 & 1.29 \\ 
B2122+13 & HBA & 0.40 & 0.99 & 0.47 & 0.51 & 2.47 \\ 
& P & 0.89 & 0.96  & 0.48& 0.51 & 1.09 \\ 
& L & 0.74 & 0.91  & 0.48& 0.51 & 1.23 \\ 
B2154+40 & HBA & 0.78 & 0.35 & 0.50 & 0.53 & 0.45 \\ 
& P & 0.89 & 0.46  & 0.49& 0.52 & 0.52 \\ 
& L & 0.62 & 0.69  & 0.49& 0.50 & 1.12 \\ 
B2306+55 & HBA & 0.46 & 1.00 & 0.46 & 0.52 & 2.18 \\ 
& P & 0.79 & 0.98  & 0.47& 0.52 & 1.24 \\ 
& L & 0.97 & 0.87  & 0.49& 0.52 & 0.89 \\ 
B2315+21 & HBA & 0.85 & 0.67 & 0.50 & 0.51 & 0.79 \\ 
& P & 0.32 & 1.11  & 0.49& 0.50 & 3.50 \\ 
& L & 0.38 & 0.95  & 0.49& 0.50 & 2.50 \\ 

\hline
\end{longtable}
}
\end{document}